\documentclass{aa}

\newcommand{\kms}{\,km\,s$^{-1}$~}

\defcitealias{wlpn2007}{W2007}
\usepackage{graphicx}
\usepackage{txfonts}
\usepackage{mathtools}
\usepackage{natbib}
\usepackage{hyperref}
\bibpunct{(}{)}{;}{a}{}{,}
\begin{document} 

\title{Molecules, shocks, and disk in the axi-symmetric wind of the MS-type
  AGB star RS Cancri}

\author{
       J.M. Winters \inst{1} %{https://orcid.org/0000-0001-6114-9173}
      \and 
       D.T. Hoai \inst{2}
      \and
       K.T. Wong \inst{1}
      \and
       W. Kim \inst{3,4} 
      \and 
       P.T. Nhung \inst{2}
       \and 
       P. Tuan-Anh \inst{2} 
       \and
       P. Lesaffre \inst{5}
      \and
       P. Darriulat \inst{2}
      \and
       T. Le Bertre 
      \inst{6}
      }

\institute{Institut de Radioastronomie Millim{\'e}trique (IRAM), 
           300 rue de la Piscine, Domaine Universitaire, 
           F-38406 St. Martin d’H{\`e}res, France email: winters@iram.fr
          \and
           Department of Astrophysics, Vietnam National Space Center 
           (VNSC), Vietnam Academy of Science and Technology (VAST),
           18 Hoang Quoc Viet, Cau Giay, Ha Noi, Viet Nam
          \and
           Instituto de Radioastronom{\'i}a Milim{\'e}trica (IRAM), 
           Av. Divina Pastora 7, N{\'u}cleo Central, E-18012, Granada, Spain
          \and
           I. Physikalisches Institut, Universit\"at zu K\"oln, 
           Z\"ulpicher Str. 77, 50937 K\"oln, Germany
          \and
           Laboratoire de Physique de l'{\'E}cole Normale Sup{\'e}rieure, 
           24 rue Lhomond, F-75231 Paris, France
          \and
           LERMA, UMR 8112, CNRS and Observatoire de Paris, 
           PSL Research University, 61 av. de l’Observatoire, 
           F-75014 Paris, France
          }

   \date{Received ; accepted }

% \abstract{}{}{}{}{} 
% 5 {} token are mandatory

  \abstract
  % context heading (optional)
  % {} leave it empty if necessary  
  {The latest evolutionary phases of low- and intermediate mass stars
    are characterized by complex physical processes like turbulence,
    convection, stellar pulsations, magnetic fields, condensation of
    solid particles, and the formation of massive outflows that inject
    freshly produced heavy elements and dust particles into the
    interstellar medium.} 
  % aims heading (mandatory)
  {By investigating individual objects in detail we wish to analyze and
    disentangle the effects of the interrelated physical processes on
    the structure of the wind forming region around these objects.}
  % methods heading (mandatory)
  {We use the Northern Extended Millimeter Array (NOEMA) to obtain spatially
    and spectrally resolved observations of the semi-regular
    Asymptotic Giant Branch star RS Cancri to shed light on the
    morpho-kinematic structure of its inner, wind forming
    environment by applying detailed 3-D reconstruction modeling and LTE
    radiative transfer calculations.}
  % results heading (mandatory)
  {We detect 32 lines of 13 molecules and isotopologs (CO, SiO, SO,
   SO$_2$,  H$_2$O, HCN, PN), including several transitions from
   vibrationally excited states.  HCN, H$^{13}$CN, millimeter
   vibrationally excited H$_2$O, SO, $^{34}$SO, SO$_2$, and PN are
   detected for the first time in RS Cnc.  Evidence for rotation is seen
   in HCN, SO, SO$_2$, and SiO(v=1). From CO and SiO channel maps, we
   find an inner, equatorial density enhancement, and a bipolar outflow
   structure with a mass  loss rate of $1 \times 10^{-7}M_\odot{\rm
   yr}^{-1}$ for the equatorial region  and of $2 \times
   10^{-7}M_\odot{\rm yr}^{-1}$ for the polar outflows.  The
   $^{12}$CO/$^{13}$CO ratio is measured to be $\sim20$ on average,
   $24\pm2$ in the polar outflows and $19\pm3$ in the equatorial region.
   We do not find direct evidence of a companion that might explain this
   kind of kinematic structure, and explore the possibility that a
   magnetic field might be the cause of it. The innermost molecular gas
   is influenced by stellar pulsation and possibly by convective cells
   that leave their imprint on broad wings of certain molecular lines,
   such as SiO and SO.}
  % conclusions heading (optional), leave it empty if necessary 
  {RS Cnc is one of the few nearby, low
   mass-loss-rate, oxygen-rich AGB stars with a wind displaying both an
   equatorial disk and bipolar outflows. Its orientation with respect to
   the line of sight is particularly favorable for a reliable study of
   its morpho-kinematics. The mechanism causing early spherical symmetry
   breaking remains however uncertain, calling for additional high
   spatial and spectral resolution observations of the emission of
   different molecules in different transitions, along with a deeper 
   investigation of the coupling among the different physical processes at
   play.}


   \maketitle
%
%-------------------------------------------------------------------

\section{Introduction}

Mass-loss in red giants is due to a combination of stellar pulsations
and radiation pressure on dust forming in dense shocked regions in the
outer stellar atmosphere (e.g., \citet{ho18}).  Even if the basic
principles are understood, a fully consistent picture, including the
role of convection, time-dependent chemistry and a consistent
description of dust formation, still needs to be developed.  In
particular, the contribution of transparent grains to the acceleration
of matter close to the stellar photosphere \citep{ntietal2012}  still
needs to be assessed.

The mechanisms shaping circumstellar environments around Asymptotic
Giant Branch (AGB) stars are vividly debated. Among them, magnetic
fields \citep{mbwg2000,dhwbetal2017}, binarity
\citep{tj93,mm99,dmrgh2020}, stellar rotation \citep{dh96}, or
common-envelope evolution \citep{ovmetal2015,gp2018} have been
considered. 

A major difficulty is to explain the observed velocity field in
axi-symmetrical sources, with larger velocities at high latitudes than
at low latitudes \citep{hmwng14,nhwetal2015}. Also recent
observations of rotating structures and streams bring additional
conundrums \citep{thnetal2019,hntetal2019}. 

We have concentrated our efforts on two sources, relatively close
(d$\sim$130\,pc), that show composite profiles in CO rotational lines
\citep{wljne2003}: EP Aqr \citep{wlpn2007}, henceforth called W2007),
and RS Cnc \citep{lwlgm2010}.  Data obtained at IRAM show that these
two sources have an axi-symmetrical structure with a low-velocity
($\sim$ 2\kms) wind close to the equatorial plane, and faster ($\sim$
8\kms) outflows around the polar axes \citep{hmwng14,nhwetal2015}.
For EP Aqr, \citetalias{wlpn2007} find a radial dependence of the
density showing intermediate maxima.  Additional data have been
obtained with ALMA \citep{nht2019,hrddk2018}.  They reveal a spiral
structure explaining the earlier \citetalias{wlpn2007} results. 

RS Cnc is one of the best examples for the interaction between the
stellar wind from an AGB star and the surrounding interstellar medium
\citep{hmwng14}.  Its high declination makes RS Cnc an ideal target
for the Northern Extended Millimeter Array (NOEMA).  Previous studies,
based on IRAM data, have shown that it is a twin of EP Aqr, but
observed at a different angle, with a polar axis inclined at about
30$^\circ$ with respect to the line of sight
\citep{lwlgm2010,hmwng14,nhwetal2015}.  This is favorable for studying
simultaneously polar and equatorial structures, whereas the different
viewing angle between EP Aqr and RS Cnc can be exploited to
discriminate between different models in explaining the observed
composite CO line profiles \citep{lhnw2016}.  In contrast to EP Aqr,
Technetium is detected in the atmosphere of RS Cnc \citep{lh99},
proving that it is evolving along the Thermal Pulsing Asymptotic Giant
Branch (TP-AGB) in the Hertzsprung Russell Diagram.  From a chemical
point of view, RS Cnc is in a slightly more advanced evolutionary
stage on the AGB, as indicated by its spectral classification as MS
star (see below) and by a higher photospheric ratio of
$^{12}$C/$^{13}$C ($\sim$35, \citet{sl86}, see Sect.~\ref{cokinesec}
for an improved evaluation  based on CO rotational lines from the
circumstellar environment).

RS Cnc is a semi-regular variable star with periods of $\approx$
122\,d and $\approx$ 248\,d \citep{ad2005}, located at a distance of
$\approx 150$\,pc \citep{GaiaeDR3,gaiabaylerjones2021}.  It is listed
as S-star CSS~589 in \citet{ste84}, based on its spectral
classification M6S given in \citet{kee54}. With its weak ZrO bands,
its chemical type is intermediate between M and S \citep{kee54}.  The
stellar temperature is estimated to T$_*\approx 3200$\,K and its
luminosity is L$_* \approx 4950 $L$_\odot$ \citep{ds98a}. From CO
rotational line observations two circumstellar wind components were
identified: an equatorial structure expanding at about 2\kms and a
bipolar outflow reaching a terminal velocity of v$_{\rm exp} \approx
8$\kms \citep{lwlgm2010,hmwng14}, carrying mass-loss rates of $4
\times 10^{-8} M_\odot$yr$^{-1}$ and $8 \times
10^{-8}M_\odot$yr$^{-1}$, respectively (see Sect.~\ref{cokinesec} for
an improved value of the mass-loss rate derived here). From previous
observations at mm and radio wavelengths, lines of $^{12}$CO,
$^{13}$CO, SiO, and HI have been detected
\citep{nbchhssvw92,dtjetal2015,dbdetal2016,gl2003,mr2007}.

NOEMA was recently equipped with the wide band correlator PolyFiX,
covering a total bandwidth of 15.6\,GHz and therefore offering the
potential to observe several lines from different species
simultaneously. In this paper we present new data obtained with NOEMA
in D and A-configuration, complemented by short spacing observations
obtained at the IRAM 30m telescope. Observational details are
summarized in Sect.~\ref{obssec} and our results are presented in
Sect.~\ref{resusec}. Sect.~\ref{discusec} contains a discussion of the
morphological structures and compares them to similar structures found
in EP Aqr. Our conclusions are summarized in Sect.~\ref{conclusec}.

\section{Observations} \label{obssec}

New observations of RS Cnc have been obtained in CO(2-1) with
NOEMA/WideX in the (extended) 9 antenna A-configuration in December
2016 \citep{nht2018}    and with NOEMA/PolyFiX in the (compact) 9
antenna D-configuration during  the science verification phase of
PolyFiX in December 2017 and in the 10 antenna A configuration in
February 2020.    The WideX correlator covered an instantaneous
bandwidth of 3.8\,GHz  in  two orthogonal polarizations with a channel
spacing of 2\,MHz. Additionally, up to 8 high-spectral resolution
units could be placed on spectral lines, providing channel spacings
down to 39\,kHz. WideX was decommissioned in September 2017 and
replaced in December 2017 by the new correlator PolyFiX. This new
correlator simultaneously  covers 7.8\,GHz in two sidebands and for
both polarizations, and provides a channel spacing of 2\,MHz
throughout the 15.6\,GHz total bandwidth.  In addition, up to 128 high
spectral resolution ``chunks'', providing a fixed channel spacing of
62.5\,kHz over their 64\,MHz bandwidth each, can be  placed in the
15.6\,GHz wide frequency range covered by PolyFiX for both
polarizations.

RS Cnc was observed with two individual frequency setups covering a
total frequency range of $\approx$ 32\,GHz in the 1.3\,mm atmospheric
window (see Fig.~\ref{polyfixcoveragefig}).  We used the two quasars
J0923+282 and 0923+392 as phase and amplitude calibrators that were
observed every $\approx$20 min.  Pointing and focus of the telescopes
was checked about every hour, and corrected when necessary.  The
bandpass was calibrated on the strong quasars 3C84 and 3C273,  and the
absolute flux scale was fixed on MWC349 and LkHa101, respectively. The
accuracy of the absolute flux calibration at 1.3\,mm is estimated to
be better than 20\%.

In order to add the short spacing information filtered out by the
interferometer, in May and July 2020 we observed at the IRAM 30m
telescope maps of 1$\arcmin$ by 1$\arcmin$ using the On-The-Fly (OTF)
mode. This turned out to be necessary for the $^{12}$CO(2-1) and
$^{13}$CO(2-1) lines but was not needed for the SiO lines, whose
emitting region was found to be smaller than ${\sim}3\arcsec$.  In
case of the $^{12}$CO(2-1) and $^{13}$CO(2-1) lines, the
interferometer filters out large scale structures that account for
about a factor of three and four, respectively, of the total line
flux, information that is recovered by adding the short spacing data
from the OTF map. A comparison of the respective line profiles is
shown in Fig.~\ref{resolvedfluxfig}.

The data were calibrated and imaged within the
GILDAS\footnote{https://www.iram.fr/IRAMFR/GILDAS} suite of software
packages using CLIC for the NOEMA data calibration and the uv table 
creation, CLASS for calibrating the OTF maps, and the MAPPING package
for merging and subsequent uv fitting, imaging, and self-calibration of the
combined data sets. Continuum data were extracted for each sideband
of the two frequency setups individually by filtering out spectral
lines, and then averaging over 400\,MHz bins to properly rescale the
uv coordinates to the mean frequency of  each bin. Phase
self-calibration was performed on the corresponding continuum
data. The gain table containing the self-calibration solutions was
then applied to the spectral line uv tables using the SELFCAL
procedures provided in MAPPING.

The resulting data sets were imaged applying either natural weighting,
or, on the high signal-to-noise (S/N) cubes, by applying  robust
weighting with a threshold of 0.1 to increase the spatial resolution
by typically a factor 2. The resulting dirty maps were then CLEANed
using the {\tt Hogbom} algorithm \citep{hog74}. 

The beam characteristics and sensitivities of the individual
combined data sets from A- and D-configuration (and including the
pseudo-visibilities from the  OTF maps, where appropriate) are listed
in table~\ref{obsparalinestab} for all detected lines.

\begin{sidewaystable*}
  \caption{Properties of the combined data sets for all detected
lines. Line frequencies and upper level energies are from the CDMS
(M{\"u}ller et al. 2005), unless otherwise stated. The quoted flux
uncertainties include the rms of the fits and the absolute flux
calibration accuracy of 20\%, the uncertainties quoted for the source
sizes refer to the rms errors of the Gaussian fits (see text).}
  \begin{center}
\begin{tabular}{llrrrrlcrl}
\hline
\hline
Line                        & Frequency & E$_u/k$ & Peak flux         & FWHP          & beam size          & PA  & 1$\sigma$ noise         & vel.res & Comments\tablefootmark{a}            \\
                            & GHz       &      K  & Jy                & arcsec        & arcsec$^2$         & deg & mJy/beam                &   \kms  &                                      \\
\hline		                                                     
$^{12}$CO(2-1)               & 230.538000 &  16.6 & 53.971$\pm$10.841 & 6.16$\pm$0.01 & 0.48 $\times$ 0.30 & 36  &    2.88                 &    0.5  & A+D+30m, rw                    \\ % 12CO21-ADconf-shift-merged-base-robust
$^{13}$CO(2-1)               & 220.398684 &  15.9 &  4.693$\pm$ 0.948 & 7.20$\pm$0.01 & 0.50 $\times$ 0.31 & 35  &    2.79                 &    0.5  & A+D+30m, rw                          \\ % 13CO21-05kms-ADconf-shift-merged-base-robust
\hline			                                                     
SiO(v=0,5-4)                & 217.104919 &  31.3 & 17.464$\pm$ 3.523 & 1.71$\pm$0.01 & 0.51 $\times$ 0.32 & 36  &    3.38                 &    0.5  & A+D, rw, sc                          \\ % SiO54-merged-filt-shift-base-robust-selfcal, uvt is 217.104938 = 0.025km/s
SiO(v=1,5-4)                & 215.596018 &1800.2 &  0.105$\pm$ 0.025 & 0.19$\pm$0.02 & 0.58 $\times$ 0.43 & 38  &    1.71                 &    1.0  & A, nw, sc, Feb 2020: no maser        \\ % w19ax001-siov154-1kms-shift-base-selfcal, uvt is 215.596047= 0.04km/s
SiO(v=1,5-4)                & 215.596018 &1800.2 &  0.105$\pm$ 0.025 & 0.19$\pm$0.02 & 2.10 $\times$ 1.80 & 0   &    2.71                 &    0.5  & D, nw, sc, Dec 2017: maser           \\ % tintsv95-siov1-newred-shift-base-selfcal, uvt is 215.596047= 0.04km/s
SiO(v=2,5-4)                & 214.088575 &3552.1 &  0.013$\pm$ 0.005 & 0.35$\pm$0.09 & 1.00 $\times$ 0.74 & 35  &    1.03                 &    3.0  & A+D, nw, double peak profile (?)     \\ % SiOv2-merged-3kms-hr-shift-base  
SiO(v=0,6-5)                & 260.518009 &  43.8 & 23.906$\pm$ 4.817 & 1.62$\pm$0.01 & 0.43 $\times$ 0.26 & 32  &    3.39                 &    0.5  & A+D, rw, sc                          \\ % SiO65-merged-shift-base-robust-selfcal, uvt at 260.518031 = 0.025km/s
SiO(v=1,6-5)                & 258.707324 &1812.7 &  0.168$\pm$ 0.038 & 0.11$\pm$0.01 & 0.60 $\times$ 0.42 & 26  &    1.96                 &    1.0  & A+D, nw, sc                          \\ % SiOv165-1kms-merged-shift-base-selfcal, uvt at 258.707344 = 0.023km/s
$^{29}$SiO(v=0,5-4)          & 214.385752 &  30.9 &  5.372$\pm$ 1.083 & 1.19$\pm$0.01 & 0.52 $\times$ 0.32 & 37  &    1.12                 &    3.0  & A+D, rw, sc                          \\ % 29SiO-merged-shift-base-robust-selfcal
Si$^{17}$O(v=0,6-5)          & 250.744695 &  42.1 &  0.340$\pm$ 0.076 & 0.88$\pm$0.04 & 1.90 $\times$ 1.50 & 36  &    4.15\tablefootmark{b}&    3.0  & D, nw, sc    tentative identification\\ % Si17O65-Dconf-3kms-shift-base-selfcal
$^{29}$Si$^{17}$O(v=0,6-5)    & 247.481525 &  41.6 &  0.020$\pm$ 0.008 & 0.73$\pm$0.44 & 1.90 $\times$ 1.50 & 26  &    2.10                 &    3.0  & D, nw, sc, tentative detection       \\ % d17ie002-u247481-newred-3kms-shift-base-selfcal
\hline			                                                     
SO(5(5)-4(4))               & 215.220653 &  44.1 &  0.455$\pm$ 0.093 & 0.79$\pm$0.01 & 0.51 $\times$ 0.32 & 36  &    1.16                 &    3.0  & A+D, rw, sc                          \\ 
SO(6(5)-5(4))               & 219.949442 &  35.0 &  0.634$\pm$ 0.130 & 0.80$\pm$0.01 & 0.50 $\times$ 0.31 & 36  &    1.17                 &    3.0  & A+D, rw, sc                          \\ 
SO(6(6)-5(5))               & 258.255826 &  56.5 &  0.870$\pm$ 0.178 & 0.74$\pm$0.01 & 0.43 $\times$ 0.27 & 32  &    1.59                 &    3.0  & A+D, rw, sc                          \\ % at band edge, increased noise! uvt at 258.255813 = 0.015km/s 
SO(7(6)-6(5))               & 261.843721 &  47.6 &  1.168$\pm$ 0.238 & 0.78$\pm$0.01 & 0.43 $\times$ 0.26 & 32  &    1.38                 &    3.0  & A+D, rw, sc                          \\ % SO3sig76-merged-3kms-shift-filt-base-robust-selfcal, uvt at 261.843684 = 0.0423 
$^{34}$SO(6(5)-5(4))         & 215.839920 &  34.4 &  0.030$\pm$ 0.009 & 0.92$\pm$0.11 & 0.91 $\times$ 0.80 & 69  &    0.86                 &    3.0  & A+D, nw                              \\ %34SO65-merged-3kms-shift-base
$^{34}$SO(5(6)-4(5))         & 246.663470 &  49.9 &  0.026$\pm$ 0.009 & 0.93$\pm$0.14 & 0.69 $\times$ 0.48 & 26  &    0.99                 &    3.0  & A+D, nw                              \\ %34SO54-merged-3kms-shift-base
\hline			                                                     
SO$_2$ (16(3,13)-16(2,14))  & 214.689394 & 147.8 & 0.021$\pm$ 0.006 & 0.50$\pm$0.06 & 0.90 $\times$ 0.68 & 37  &    1.06                 &    3.0  & A+D, nw, sc                          \\
SO$_2$ (22(2,20)-22(1,21))  & 216.643304 & 248.4 & 0.023$\pm$ 0.007 & 0.38$\pm$0.05 & 0.89 $\times$ 0.67 & 36  &    1.11                 &    3.0  & A+D, nw, sc                          \\
SO$_2$ (28(3,25)-28(2,26))  & 234.187057 & 403.0 & 0.022$\pm$ 0.006 & 0.19$\pm$0.05 & 0.71 $\times$ 0.56 & 46  &    1.28                 &    3.0  & A+D, nw, sc                          \\
SO$_2$ (14(0,14)-13(1,13))  & 244.254218 &  93.9 & 0.043$\pm$ 0.011 & 0.43$\pm$0.03 & 0.69 $\times$ 0.49 & 27  &    1.00                 &    3.0  & A+D, nw, sc                          \\
SO$_2$ (10(3, 7)-10(2, 8))  & 245.563422 &  72.7 & 0.025$\pm$ 0.007 & 0.36$\pm$0.04 & 0.69 $\times$ 0.49 & 28  &    1.00                 &    3.0  & A+D, nw, sc                          \\
SO$_2$ (15(2,14)-15(1,15))  & 248.057402 & 119.3 & 0.015$\pm$ 0.005 & 0.26$\pm$0.06 & 0.69 $\times$ 0.48 & 27  &    1.07                 &    3.0  & A+D, nw, sc                          \\
SO$_2$ (32(4,28)-32(3,29))  & 258.388716 & 531.1 & 0.020$\pm$ 0.006 & 0.21$\pm$0.04 & 0.63 $\times$ 0.45 & 25  &    1.10                 &    3.0  & A+D, nw, sc                          \\
SO$_2$ ( 9(3, 7)- 9(2, 8))  & 258.942199 &  63.5 & 0.026$\pm$ 0.008 & 0.49$\pm$0.05 & 0.64 $\times$ 0.45 & 26  &    1.08                 &    3.0  & A+D, nw, sc                          \\
SO$_2$ (30(4,26)-30(3,27))  & 259.599448 & 471.5 & 0.022$\pm$ 0.005 & 0.12$\pm$0.03 & 0.63 $\times$ 0.45 & 25  &    1.03                 &    3.0  & A+D, nw, sc                          \\
SO$_2$ (30(3,27)-30(2,28))  & 263.543953 & 459.0 & 0.019$\pm$ 0.006 & 0.16$\pm$0.04 & 0.61 $\times$ 0.42 & 26  &    1.25                 &    3.0  & A+D, nw, sc                          \\
SO$_2$ (34(4,30)-34(3,31))  & 265.481972 & 594.7 & 0.020$\pm$ 0.006 & 0.19$\pm$0.04 & 0.61 $\times$ 0.42 & 26  &    1.27                 &    3.0  & A+D, nw, sc                          \\
\hline			                                                     
H$_2$O(v$_2$=1,5(5,0)-6(4,3))  & 232.686700\tablefootmark{c} &3462.0 &  0.029$\pm$ 0.007 & unresolved    & 0.71 $\times$ 0.57 & 47  &    1.17                 &    3.0  & A+D, nw, sc, JPL \\ %h2o-232-merged-3kms-shift-base-selfcal
H$_2$O(v$_2$=1,7(7,0)-8(6,3))  & 263.451357\tablefootmark{d} &4474.7 &  0.021$\pm$ 0.005 & unresolved    & 0.61 $\times$ 0.42 & 26  &    1.17                 &    3.0  & A+D, nw, sc, JPL \\ %h2o-263-merged-3kms-shift-base-selfcal
\hline			                                                     
HCN(3-2)                    & 265.886434 &  25.5 &  1.116$\pm$ 0.234 & 0.76$\pm$0.01 & 0.42 $\times$ 0.26 & 32  &    4.80                 &    0.5  & A+D, rw, sc                          \\ % HCN-merged-shift-base-robust-selfcal
H$^{13}$CN(3-2)              & 259.011798 &  24.9 & 0.041$\pm$ 0.011 & 0.71$\pm$0.05 & 0.64 $\times$ 0.45 & 26  &    0.95                 &    3.0  & A+D, nw, sc                          \\ % H13CN-merged-3kms-shift-base-selfcal, new detection
\hline			                                                      
PN(N=5-4, J=6-5)            & 234.935694 &  33.8 & 0.028$\pm$ 0.009 & 0.80$\pm$0.10 & 0.70 $\times$ 0.56 & 47  &    1.00                 &    3.0  & A+D, nw, sc                          \\ %PN-merged-3kms-shift-base-selfcal
\hline
\label{obsparalinestab}
\end{tabular}
\tablefoot{
\tablefoottext{a}{A: NOEMA A-configuration, D: NOEMA D-configuration, 
                  30m: short spacing data, rw: robust weighting, 
                  nw: natural weighting, sc: self-calibrated,
                  JPL: Spectral line catalog by NASA/JPL \citep{JPL}}
\tablefoottext{b}{Increased noise at band edge}
\tablefoottext{c}{\citet{belovetal87}}
\tablefoottext{d}{\citet{pearsonetal1991}}
}
\end{center}
\end{sidewaystable*}

\section{Results} \label{resusec}

\begin{figure*}
    \centering
    \vspace{-3.0cm}
    \includegraphics[width=200mm]{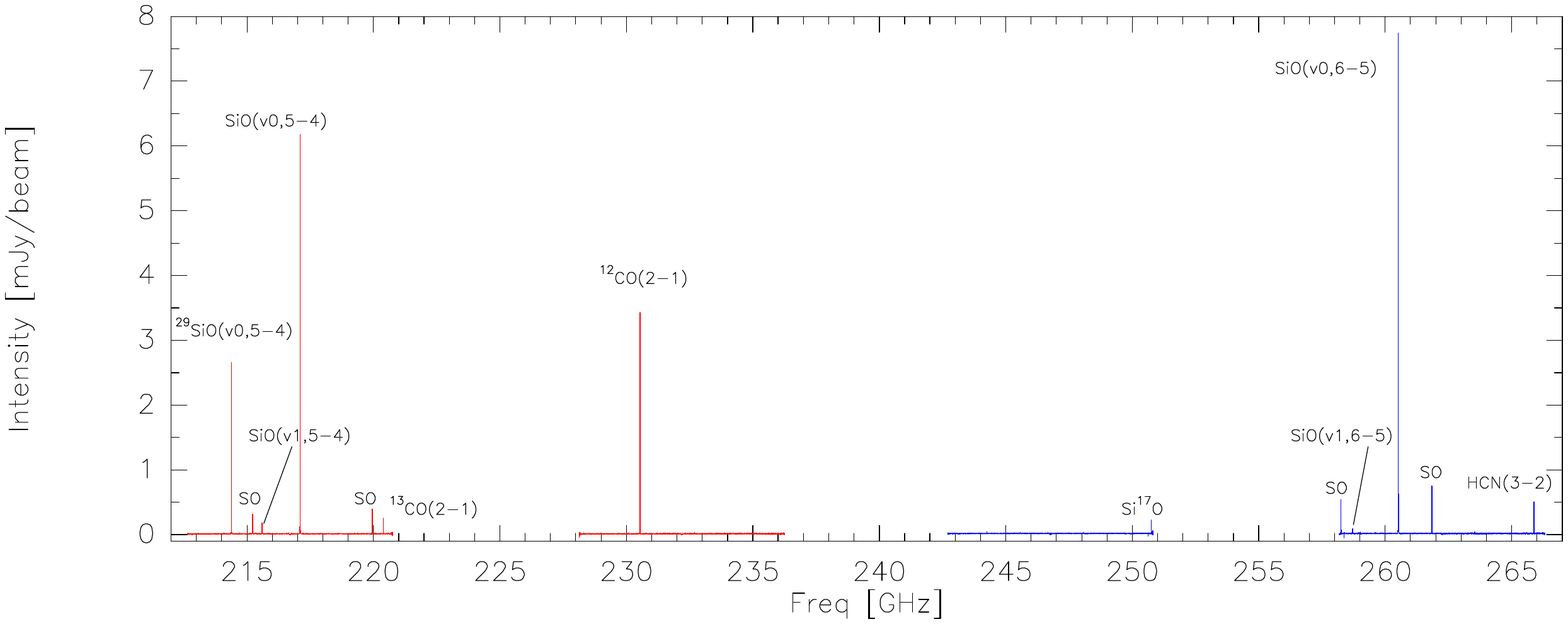}
    \vspace{-2.00cm} 
    \includegraphics[width=90mm]{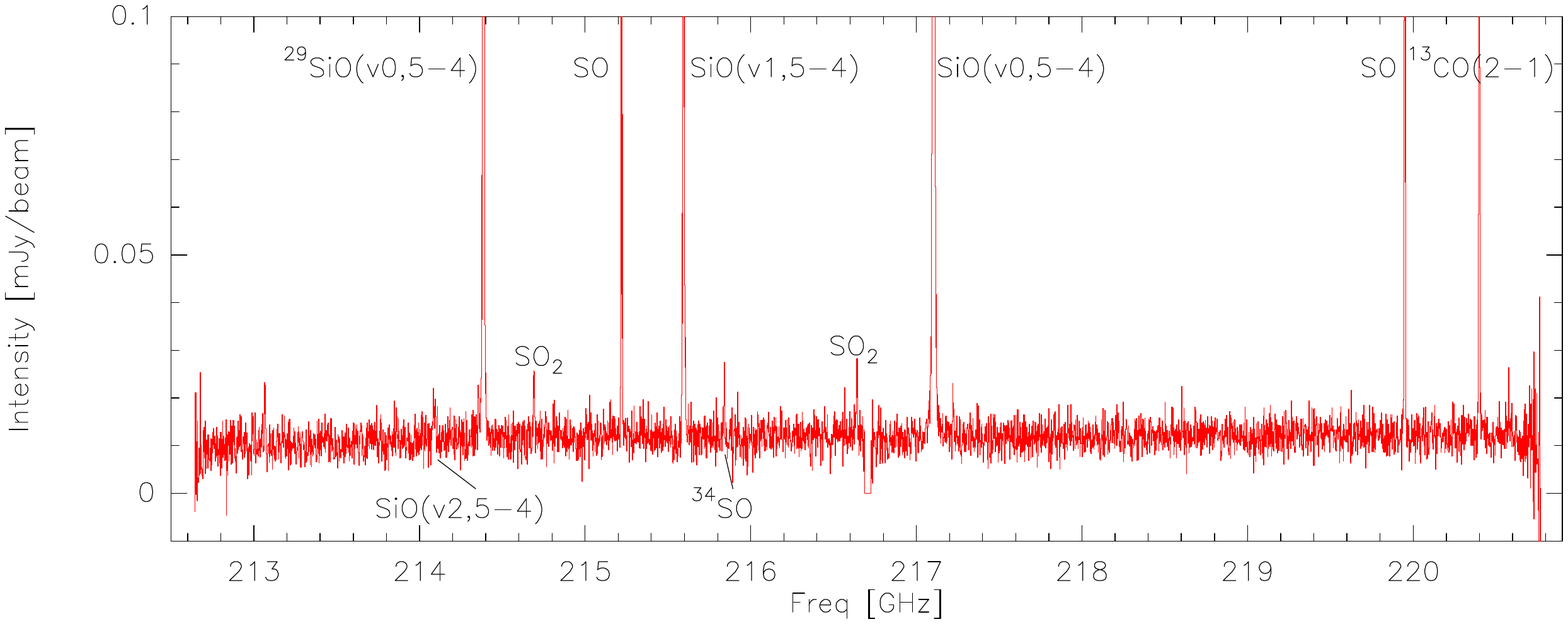}
    \includegraphics[width=90mm]{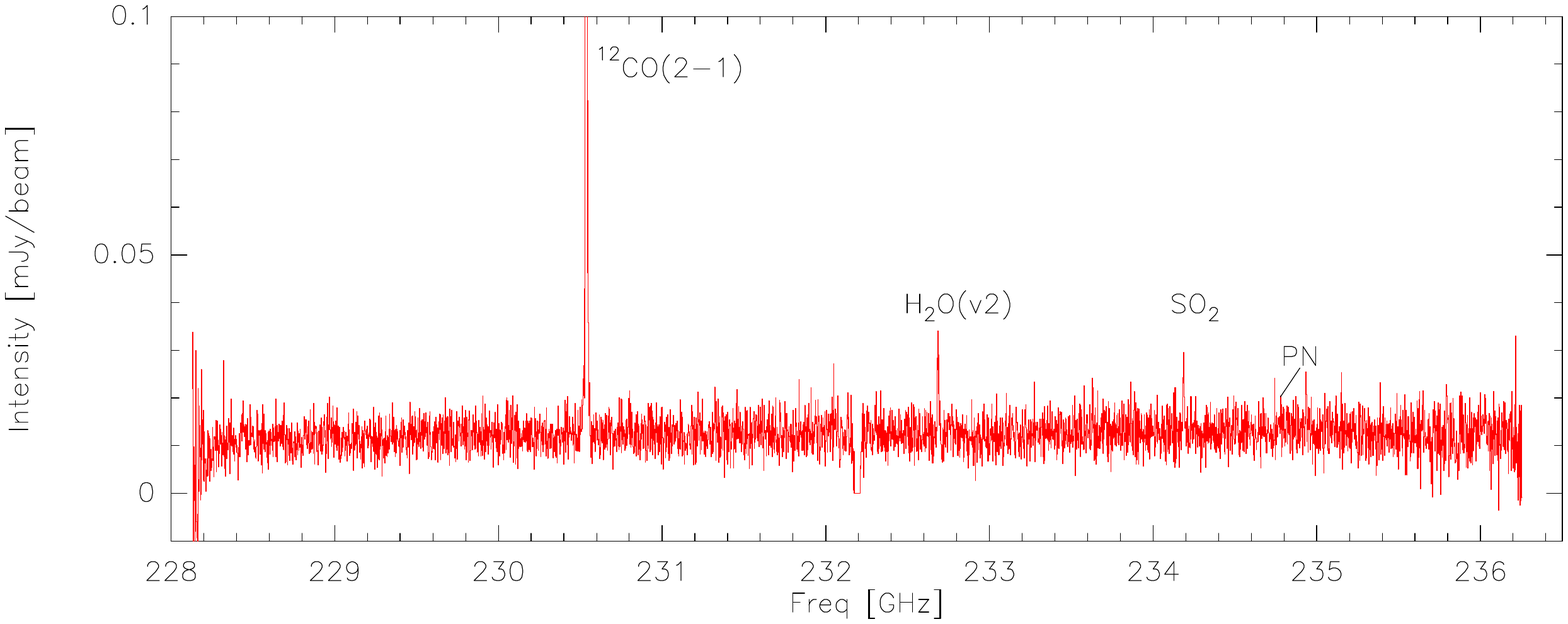}
    \includegraphics[width=90mm]{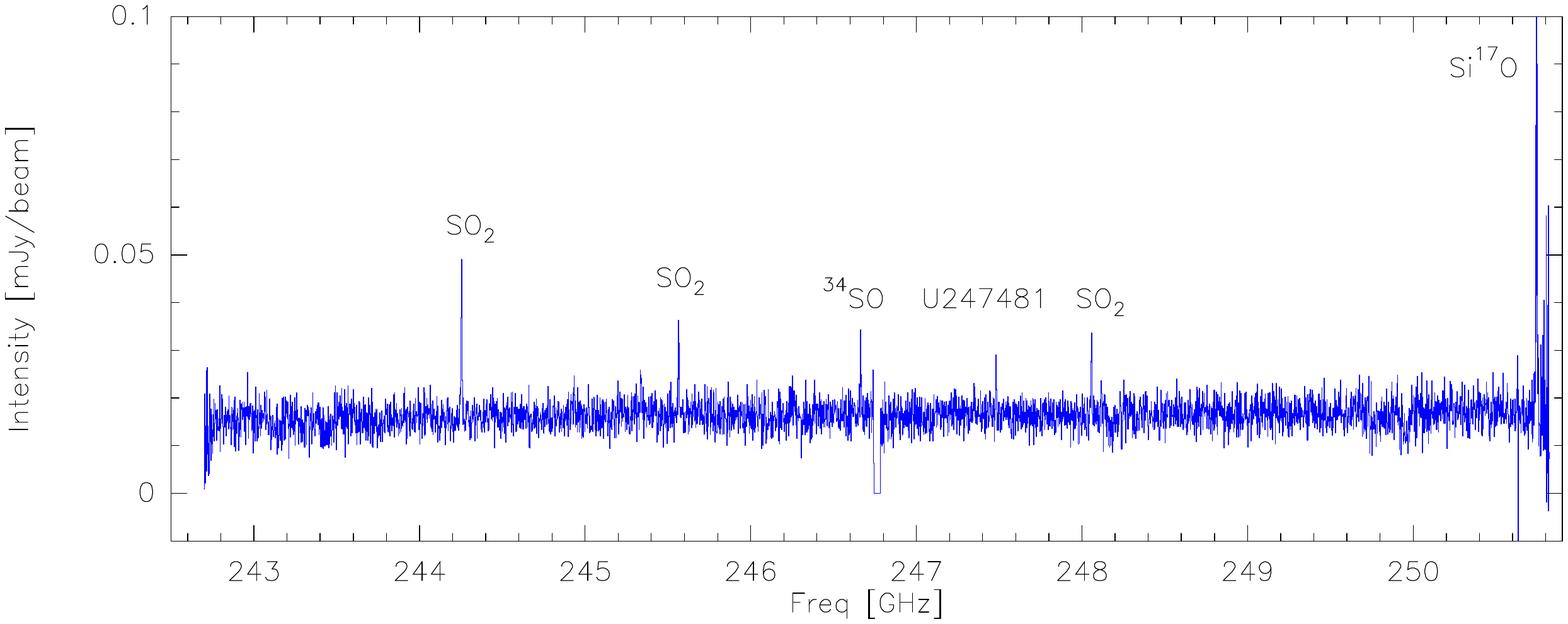}
    \includegraphics[width=90mm]{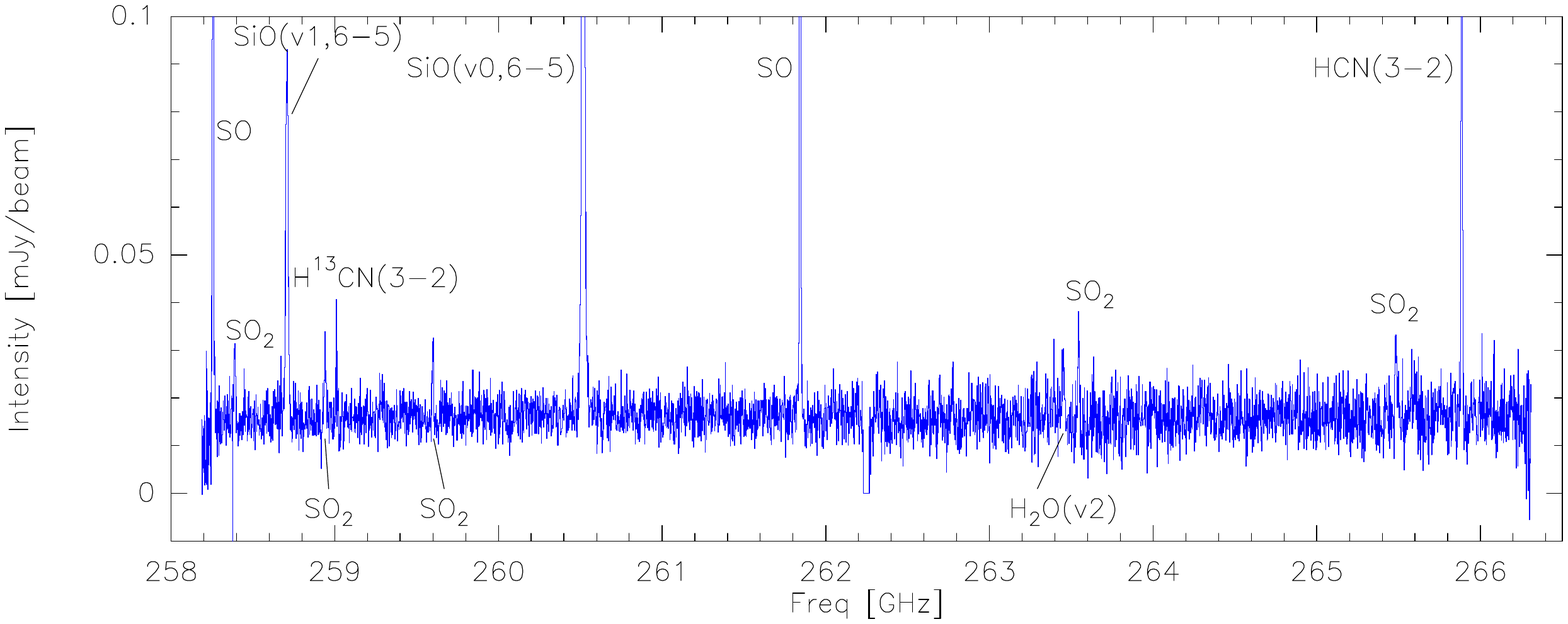}
    \caption{Overview of the frequency ranges observed with PolyFiX using two 
             spectral setups (setup1: red and setup2: blue, respectively).
             Lower diagrams: Zoom on the individual spectra covering 7.8\,GHz 
             each.
             Upper row: setup1, 
             lower row: setup2. The central 20\,MHz at the border between 
             inner and outer baseband are blanked out, i.e., set to zero, as 
             this region is contaminated by the LO2 separation of the 8\,GHz 
             wide IF in the IF processor ("LO2 zone").}
    \label{polyfixcoveragefig}
% /data/RSCnc/maps/rscnc-hires.greg 
\end{figure*}

The PolyFiX data, covering with two setups the frequency ranges
213-221\,GHz (setup1, LSB), 228-236\,GHz (setup1, USB), 243-251\,GHz
(setup2, LSB), and 258-266\,GHz (setup2, USB) (see
Fig.~\ref{polyfixcoveragefig}) showed different lines of CO and SiO,
and, for the first time, many lines of species like SO, SO$_2$, HCN,
PN and some of their isotopologs. Furthermore, the data confirmed
the H$_2$O line at 232.687\,GHz already detected serendipitously with
WideX in 2016, with a second H$_2$O line at 263.451\,GHz seen for the
first time in RS Cnc.

All lines covered by the same setup (1 or 2, see
Fig.~\ref{polyfixcoveragefig}) share the same phase-, amplitude-, and
flux calibration. All 32 detected lines are listed in
table~\ref{obsparalinestab}.

\subsection{Continuum}\label{contsec}

\begin{figure}[t]
    \centering
    \includegraphics[width=110mm]{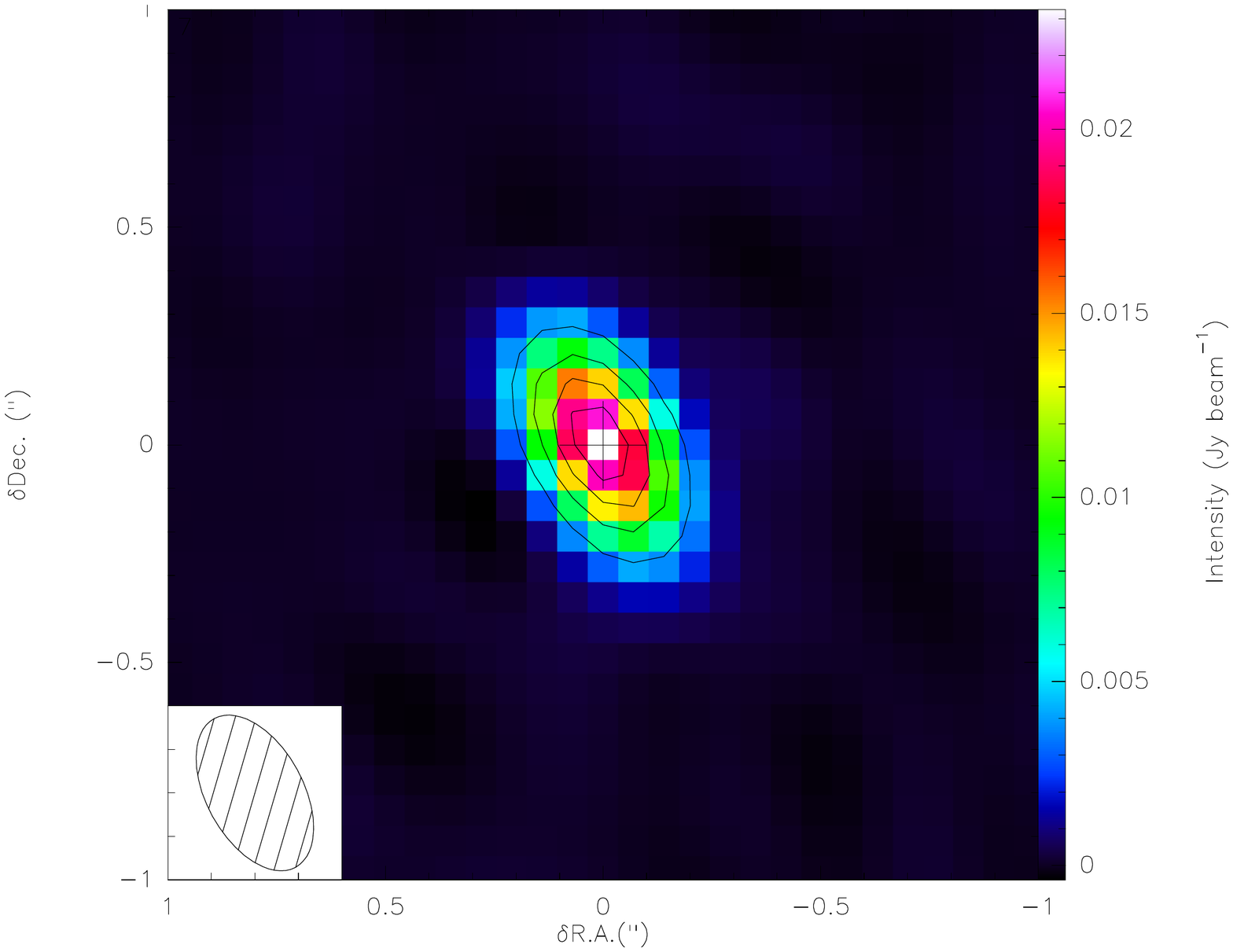}
    \caption{Continuum map around 247\,GHz from A-configuration. 
             Contours are plotted in 100$\sigma$ steps, 
             where 1$\sigma$ is 47.6\,$\mu$Jy\,beam$^{-1}$.
             The synthesized beam is indicated in the lower left corner.}
    \label{contmapfig}
% /data/RSCnc/maps/Newreduction/Selfcal/w19ax002-LSB-shift-filt-cont-selfcal
\end{figure}

Fig.~\ref{contmapfig} shows the continuum map from A-configuration
only, using robust weighting to increase the spatial resolution to
$0.39\arcsec\times0.22\arcsec$ at PA 28$^{\circ}$. After
self-calibration, S/N = 492 is
obtained.  The continuum source is unresolved, a point source fit
results in a flux at $\approx$ 247\,GHz of 23.65$\pm$4.7\,mJy (where
the quoted error accounts for the accuracy of the  absolute flux
calibration of 20\%)  and a source position at RA = 09:10:38.780 and
DEC = 30:57:46.62 in February 2020. All line data cubes discussed in
the remainder of this paper are re-centered on this continuum
position.\\
The source position is offset from the J2000 coordinates by
-0.26$\arcsec$ in RA and by -0.68$\arcsec$  in DEC, consistent with
the proper motion of RS Cnc (-10.72\,mas\,yr$^{-1}$ in RA and
-33.82\,mas\,yr$^{-1}$ in DEC, \citet{GaiaeDR3,gaiabaylerjones2021}).\\

From the PolyFiX data, spanning a total frequency range of about
53\,GHz, we determine a spectral index of 1.99$\pm$0.09 for RS Cnc in
the 1\,mm range, fully consistent with a black body spectrum of the
continuum (see also \citet{lwlgm2010}).

\subsection{Detected molecules and lines}\label{linessec}

Within the total frequency coverage of about 32\,GHz we detect 32
lines of 13 molecules and isotopologs, including several transitions
from vibrationally excited states. All these lines are listed in
table~\ref{obsparalinestab} and are presented in the following
sections. Peak flux and FWHP of the line-emitting regions, as listed
in table~\ref{obsparalinestab}, are determined by circular Gaussian
fits in the uv-plane to the central channel (if the source is
(partially) spatially resolved) or by  point source fits to the
central channel (if the source is unresolved). All line profiles shown
in the following subsections in figures \ref{codetecfig} and
\ref{siodetecfig} through \ref{siowingfig}  are integrated over square
apertures whose sizes are given in each figure caption. Two-component
profiles are seen in CO and $^{13}$CO, only, but not in any other of
the lines detected here.

We looked for, but did not detect the vibrationally excited
$^{12}$CO(v=1,2-1) line, nor do we detect C$^{18}$O(2-1), resulting in
3$\sigma$ upper limits for the line peaks of 6\,mJy/beam and
3\,mJy/beam, respectively (the $^{12}$CO(v=1,2-1) line was not covered
in our A-configuration data). 

\subsubsection{CO}\label{cosec}

\begin{figure}[h]
    \centering
    \vspace{-3.0cm}
    \includegraphics[width=110mm]{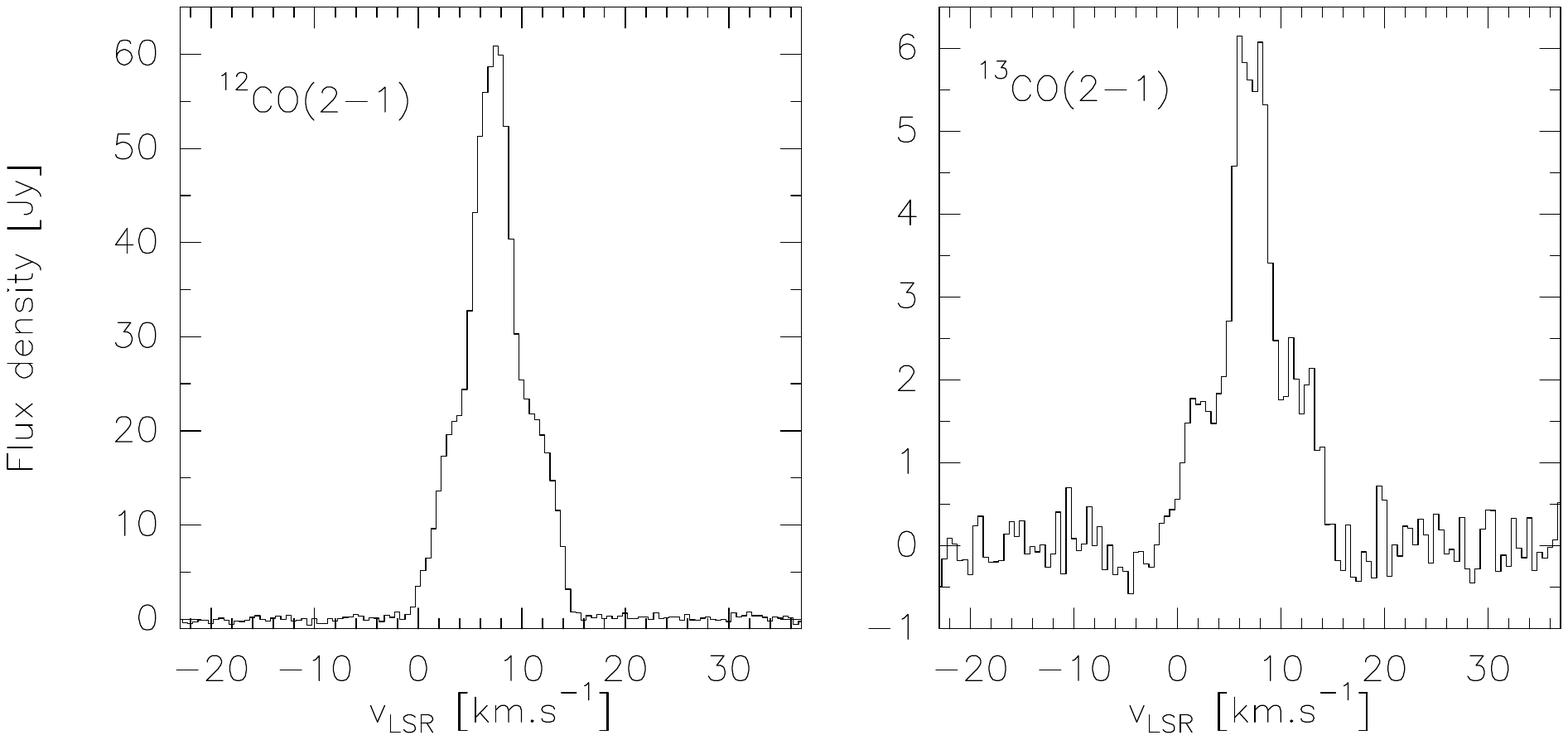}
    \caption{Left:  $^{12}$CO(2-1).
             Right: $^{13}$CO(2-1).
             A and D-configuration are merged, OTF data are added,
             spectral resolution is 0.5\kms. 
             The CO emission is integrated over the central 
             $22\arcsec \times 22\arcsec$, 
             i.e., over the full field of view of the NOEMA antennas at 
             230\,GHz.} 
    \label{codetecfig}
% read-and-plot-COlines.greg
\end{figure}

\begin{figure}[h]
    \centering
    \includegraphics[width=110mm]{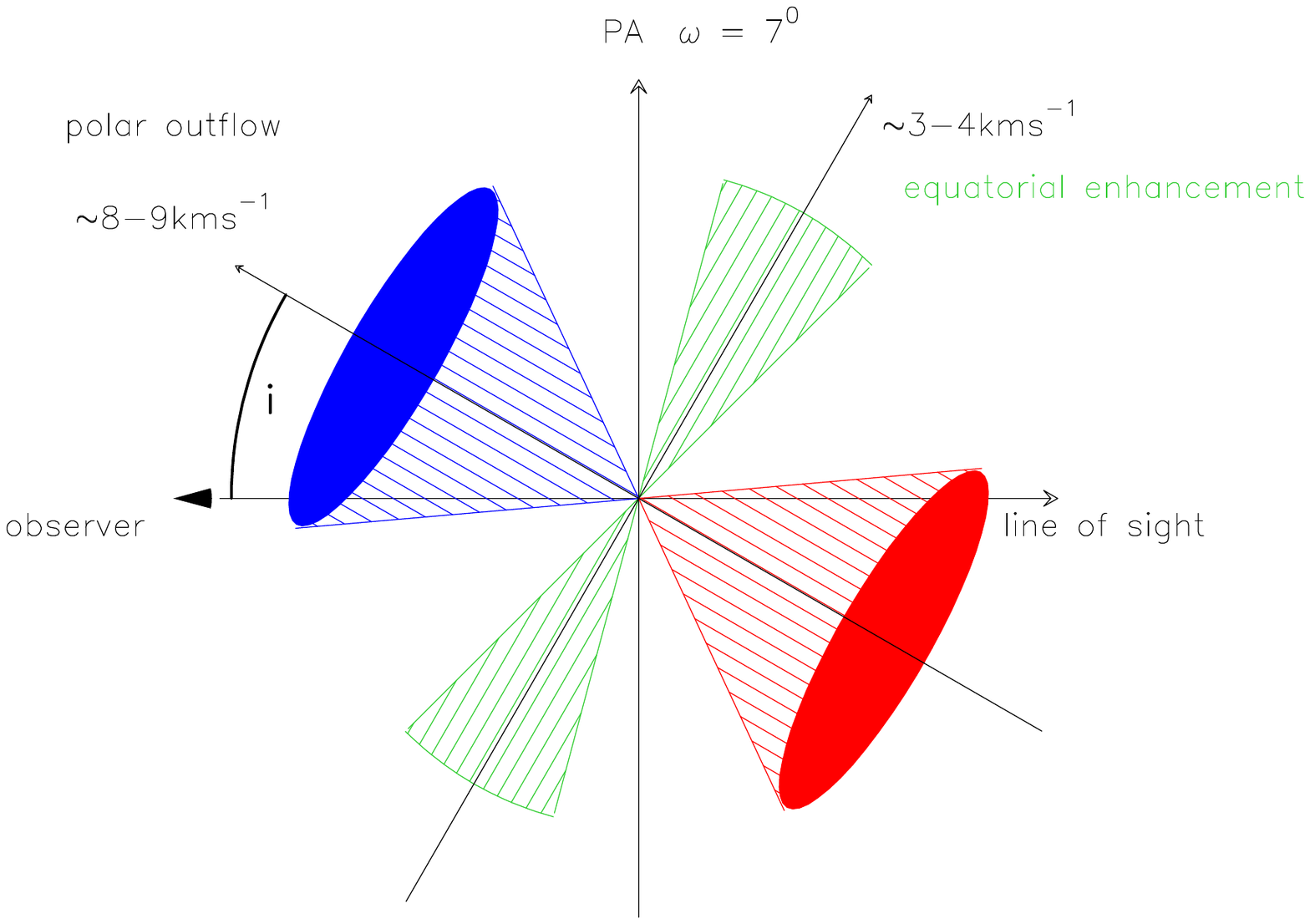}
    \caption{Sketch of the geometrical structure of the wind components
        as inferred from the current data (see Sect.~\ref{cokinesec}).
        The sketch is not to scale: there is a smooth transition between the
        equatorial enhancement and the polar outflows.}
    \label{geomsketch}
% sketch.greg
\end{figure}

The profiles of $^{12}$CO(2-1), $^{13}$CO(2-1) (see
Fig.~\ref{codetecfig}), but also of the $^{12}$CO(1-0) line (see
\citet{lwlgm2010}) show a very distinct shape composed of a broad
component that extends out to v$_{\rm lsr,*}\pm$8\kms and a narrow
component indicating velocities of $\pm$2\kms with respect to v$_{\rm
lsr,*}=7$\kms. Velocity-integrated intensity maps of CO are shown
in Fig.~\ref{COSiOoutflowfig}, indicating a clear kinematic structure
in north-south direction.  In Fig.~\ref{geomsketch} we present a
schematic representation  of the geometrical structure of RS Cnc as
implied by the data, see  Sect.~\ref{cokinesec}. The CO emitting
region is spatially extended, consisting of a dense equatorial
structure that corresponds to the low velocity expansion and an
inclined, bipolar structure corresponding to an outflow at a
projected velocity of 8\kms. These structures were discussed already
in \citet{hmwng14}, based on Plateau de Bure data obtained on
$^{12}$CO(2-1) and $^{12}$CO(1-0) that had a spatial resolution of
about $1\arcsec$. Their model was later refined by \citet{nht2018},
based on $^{12}$CO(2-1) data obtained with the WideX correlator in
NOEMA's A-configuration, providing a spatial  resolution of
$0.44\arcsec\times0.28\arcsec$. They find a position angle of the
projected bipolar outflow axis of $\omega = 7^{\circ}$ (measured
counter-clockwise from north) and an inclination angle of the outflow
axis with respect to the line-of-sight of $i = 30^{\circ}$. The CO
distribution is further investigated in Sect.~\ref{cokinesec} below.

Such a structure had already been found in the S-type star $\pi^1$ Gru
\citep{sah92}, which was later confirmed by higher spatial resolution
observations using ALMA \citep{2017A&A...605A..28D}. This object has
a G0V companion \citep{fea53} and possibly a second, much closer
companion \citep{2020A&A...644A..61H}. In \citet{hmwng14} we reported
for RS Cnc the possible presence of a companion seen in the
$^{12}$CO(1-0) channel maps at velocities around 6.6\kms and located
about $1\arcsec$ west-northwest of the continuum source.  The new data
allow for a more detailed study of this feature, which is presented in
Sect.~\ref{cokinesec}.

\subsubsection{SiO}\label{siosec}

\begin{figure}[h]
    \centering
    \includegraphics[width=110mm]{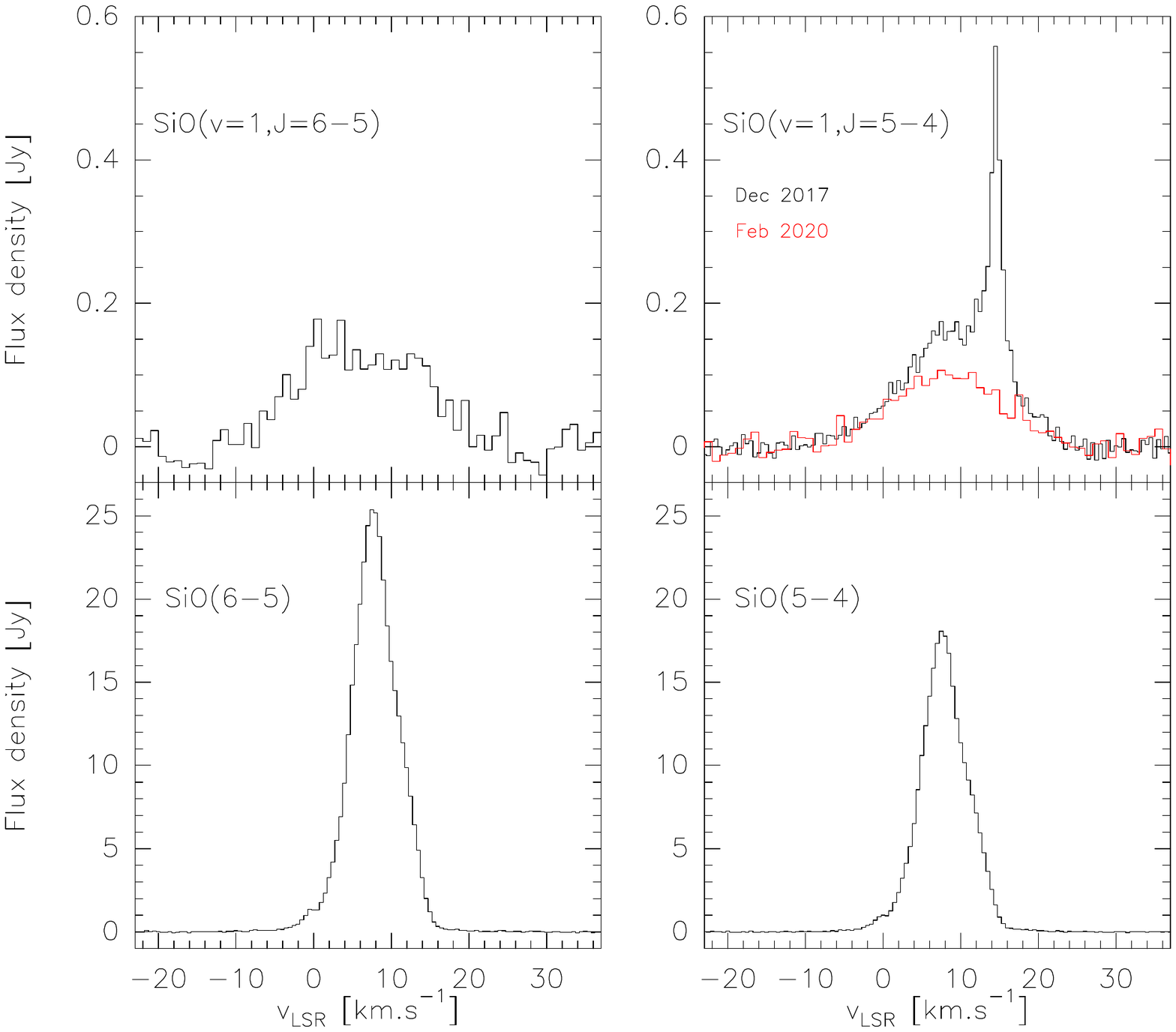}
    \caption{Left: SiO(6-5): upper: v=1, lower: v=0. A and D-configuration 
             merged. 
             Right: SiO(5-4): upper: v=1, D-configuration 
             (black) and A-configuration (red), lower: v=0,  A and 
             D-configuration merged.
             Spectral resolution is 1\kms for (v=1) and 0.5\kms for the (v=0) 
             lines, respectively. The emission
             is integrated  over the central $5\arcsec \times 5\arcsec$
             aperture.}
    \label{siodetecfig}
% read-and-plot-SiOlines.greg
\end{figure}

\begin{figure}[h]
    \centering
    \vspace{-3.0cm}
    \includegraphics[width=110mm]{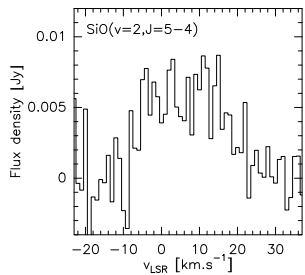}
    \caption{Profile around the SiO(v=2, J=5-4) line frequency. A and 
             D-configuration merged.
             Spectral resolution is 1\kms and the emission is integrated 
             over the central $1\arcsec \times 1\arcsec$ aperture.}
    \label{siov2linefig}
% read-and-plot-siov2line.greg
\end{figure}

\begin{figure}[h]
    \centering
    \includegraphics[width=110mm]{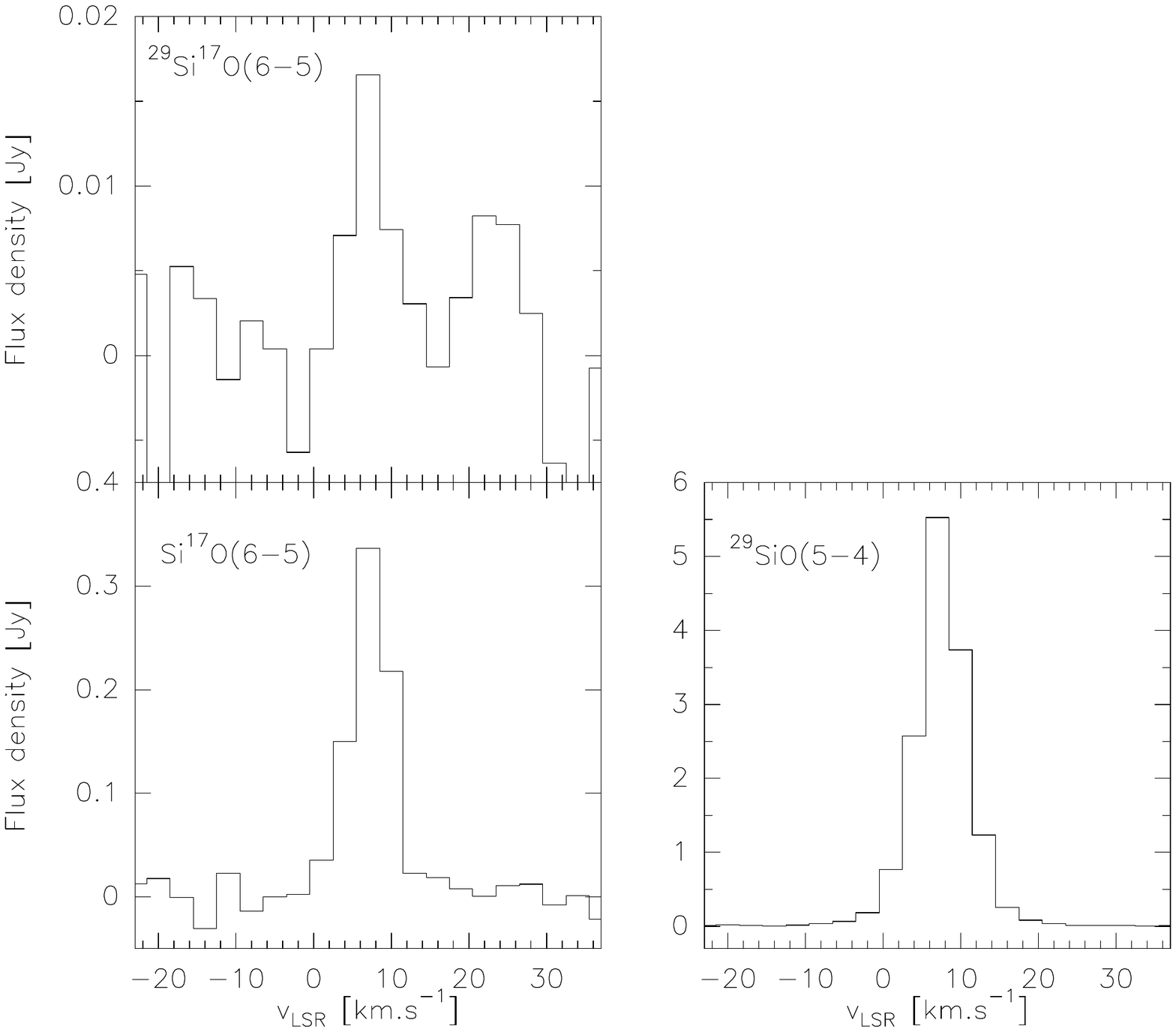}
    \caption{Upper left: Profile of the 247.482\,GHz line, possibly
             $^{29}$Si$^{17}$O(6-5); D-configuration, only.  
             Lower left: Si$^{17}$O(6-5): D-configuration, only (line was 
                    not covered in A configuration).  
             Right: $^{29}$SiO(5-4); A and D-configuration merged.  
             Spectral resolution in all cases is 3\kms and the emission is
             integrated over the central $5\arcsec \times 5\arcsec$ aperture.}
    \label{sioisodetecfig}
% read-and-plot-SiOiso4.greg
\end{figure}

We detect a suite of $^{28}$Si$^{16}$O (henceforth SiO) transitions,
including the vibrational ground state lines of SiO(5-4) and SiO(6-5),
the first and second vibrationally excited state of SiO(5-4), and the
first vibrationally excited state of SiO(6-5). All SiO profiles are
shown in Figs.~\ref{siodetecfig} and \ref{siov2linefig}. 
The spatial region emitting the vibrational ground state lines
extends out to about $2\arcsec$ from the continuum peak (see
table~\ref{obsparalinestab}, Fig.~\ref{COSiOoutflowfig}, and
Sect.~\ref{siokinesec}).
Interestingly, we detect a strong maser
component on the SiO(v=1,J=5-4) line at v$_{\rm lsr}\approx$14\kms in
the data obtained in December 2017, which has completely disappeared
when we re-observed RS Cnc in February 2020 (see
Fig.~\ref{siodetecfig}). Such behavior is well known for pulsating AGB
stars, and lends support to the idea that the SiO masers are excited by
infrared pumping as opposed to collisional pumping (see, e.g.,
\citet{pardoetal2004}).

The SiO(v=2,J=5-4) line is detected above the 3$\sigma$ level of
3\,mJy/beam over a broad range of Doppler velocities from at least -5
to 18\kms (Fig.~\ref{siov2linefig}).  Given its high excitation energy
($\sim$3500\,K), we expect this line to trace exclusively the
innermost region around RS Cnc, as was the case in $o$ Cet, where
SiO(v=2) absorption/emission was spatially resolved by ALMA
\citep{wongetal2016}.  Its broad line width suggests that it may trace
the same high-velocity wings seen in other detected SiO lines
(Sect.~\ref{siowingsec}).  Our detection is however too weak to allow
for a detailed study of the morpho-kinematics of the emission.

At the upper edge of the LSB of setup 2 at 250.744\,GHz, we
serendipitously detect a strong line that we identify as ground-state
Si$^{17}$O(6-5) at 250.7446954 GHz \citep{siocdms2013} from the
Cologne Database of Molecular Spectroscopy
(CDMS\footnote{https://cdms.astro.uni-koeln.de}, \citet{cdms2005});
the profile is shown in Fig.~\ref{sioisodetecfig}. This line and other
transitions of Si$^{17}$O have already been detected in a number of
well-studied objects, such as the S-type star W Aql \citep{do2020},
the M-type star R Dor \citep{do2018}, and the evolved, high mass-loss
rate oxygen-rich star IK Tau \citep{vsc2017}. No other Si$^{17}$O
transitions are covered in our setups. There is however a highly
excited H$_2$O line at 250.7517934\,GHz (v$_2$=2,
J(K$_a$,K$_c$)=9(2,8)-8(3,5); E$_u/k$ = 6141 K) listed in the JPL
catalog\footnote{https://spec.jpl.nasa.gov/ftp/pub/catalog/catform.html}
and predicted by \citet{yuetal2012} from the Bending-Rotation approach
analysis. If the detected line was H$_2$O emission, it would be
redshifted from the systemic velocity by about 9\kms. As indicated by
the modeling of \citet{gmretal2016}, the 250.752\,GHz line may exhibit
strong maser action in regions of hot gas (T$_{\rm kin} = 1500$\,K)
with cool dust  (T$_{\rm d} \leq 1000$\,K). While we cannot
unequivocally exclude some contamination from a potential new,
redshifted H$_2$O maser, we consider Si$^{17}$O a more likely
identification of the 250.744 GHz emission. From the respective
integrated line intensities of Si$^{16}$O(6-5) and Si$^{17}$O(6-5),
which are $\sim$ 163\,Jy\kms and $\sim$ 3\,Jy\kms, and taking the
difference of the Einstein coefficients of the transitions into
account, we estimate the isotopolog ratio $^{16}$O/$^{17}$O $\sim
50$, assuming equal excitation conditions for both transitions and
optically thin emission of both lines. This value is much lower than
the solar isotopic ratio of $\sim2700$ \citep{lpg2009} due to
dredge-up events \citep{kl2014,hls2016} and is broadly consistent with
those obtained in the M-type star R Dor and the S-type star W Aql
(61-74; \citet{do2018,do2020}). The initial mass of RS Cnc is about
$1.5M_{\odot}$ \citep{lwlgm2010}\footnote{As quoted in
\citet{lwlgm2010}, the value of $1.5 M_\odot$ was estimated by Busso
\& Palmerini (their private comms.)  using the FRANEC stellar
evolution code \citep{2011ApJS..197...17C} and the molecular
abundances determined by Smith \& Lambert. \citet{1990ApJS...72..387S}
reported oxygen isotopic ratios of $^{16}$O/$^{17}$O=710 and
$^{16}$O/$^{18}$O=440 in RS Cnc (their Table 9).  The
$^{17}$O/$^{18}$O ratio of 0.62 corresponds to an initial mass of
$1.4-1.5 M_\odot$ in the comparative study of \citet{denutteetal2017},
who investigated the $^{17}$O/$^{18}$O isotopic ratio as a sensitive
function of initial mass of low-mass stars based on the models of
\citet{2004MNRAS.352..984S}, \citet{kl2014}, and the FRANEC model.},
which is in the same range as R Dor ($1.4M_{\odot}$; \citet{do2018})
and W Aql ($1.6 M_{\odot}$; \citet{denutteetal2017}) that gives a
$^{16}$O/$^{17}$O ratio of $<1000$ \citep{hls2016}. We note however,
that the oxygen isotopic ratio ($^{16}$O/$^{17}$O) derived from the
line intensity ratio is likely underestimated if the Si$^{16}$O line
is not optically thin, as has been shown in \citet{do2018}, who
obtained a value of $\sim400$ in R Dor with radiative transfer
modeling.  Indeed, as we will demonstrate in Sect.~\ref{siokinesec},
the Si$^{16}$O emission in RS Cnc is optically thick, especially
within a projected radius of $\sim1\arcsec$. A photospheric
$^{16}$O/$^{17}$O ratio of 710 in RS Cnc (=HR 3639) has previously
been estimated by \citet{1990ApJS...72..387S} from the spectra of
near-infrared overtone band transitions of C$^{16}$O and C$^{17}$O,
which is probably a more realistic ratio.  We do not cover
C$^{17}$O(2-1) in our setups and, therefore, cannot give an
independent estimate of the $^{16}$O/$^{17}$O ratio. Since
Si$^{18}$O(6-5) and C$^{18}$O(2-1) are either not covered or not
detected, there is not enough information from our data to obtain a
meaningful constraint on the initial stellar mass from oxygen isotopic
ratios (e.g. from the $^{17}$O/$^{18}$O ratio;
\citet{denutteetal2017}).

We detect a line at 247.482\,GHz at low signal-to-noise ratio that
might be identified as $^{29}$Si$^{17}$O(v=0,J=6-5) at
247.4815250\,GHz, based on the same line list of \citet{siocdms2013}
in the CDMS (see Fig.~\ref{sioisodetecfig}). In contrast to
Si$^{17}$O(6-5) however, $^{29}$Si$^{17}$O(6-5) has never been
detected, only higher-J lines of $^{29}$Si$^{17}$O have been
tentatively detected in R Dor (J=7-6 and J=8-7, \cite{do2018}). In
fact, the 247.482\,GHz line is seen with an integrated line intensity
of $\sim 0.08$\,Jy\kms in our D-configuration data only, observed in
December 2017, but it does not show up in the A-configuration data,
taken in February 2020.  This may largely be due to the much reduced
brightness sensitivity in the A configuration, which is a factor $\sim
15$ smaller, due to the smaller synthesized beam area, rather than
being due to variable maser action in this line. Based on the
D-configuration data, the source position of the 247.482\,GHz emission
appears slightly offset toward the northwest direction from the
Si$^{17}$O(6-5) emission. Further data on $^{29}$Si$^{17}$O, possibly
covering the J=6-5, J=7-6 and J=8-7 transitions, would be needed to
draw any firm conclusion.

\subsubsection{HCN}\label{hcnsec}

\begin{figure}[h]
    \centering
    \vspace{-3.0cm}
    \includegraphics[width=100mm]{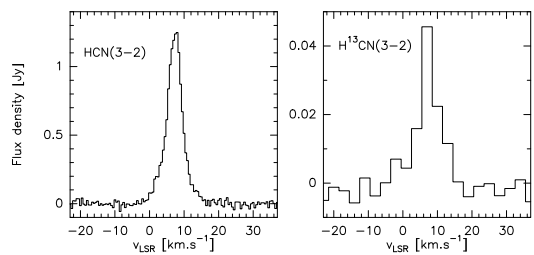}
    \caption{Left: HCN(3-2); A and D-configuration merged, spectral resolution 
             is 0.5\kms.
             Right: H$^{13}$CN(3-2); A and D-configuration merged, spectral 
             resolution is 3\kms.
             The emission of both lines is integrated over the
             central $2\arcsec \times 2\arcsec$.} 
    \label{hcndetecfig}
    % read-and-plot-HCNlines.greg
\end{figure}

We clearly detect the HCN(3-2) and H$^{13}$CN(3-2) lines, the profiles
are displayed in  Fig.~\ref{hcndetecfig}, and velocity-integrated
intensity maps of both species are shown in Fig.~\ref{HCNstructfig}.
Both lines are slightly spatially resolved, a circular Gaussian fit to
HCN(3-2) gives a peak flux of 1.12\,Jy and a FWHP size of
0.76$\arcsec$ on the merged data. To our knowledge, this is the first
detection of HCN and H$^{13}$CN  in RS Cnc (see
Sect.~\ref{nontechemistrysec}).  From the first moment map (shown in
Fig.~\ref{rotationfig}, left), a clear velocity pattern is evident
that indicates possible rotation in the HCN emitting region (see
Sect.~\ref{rotdiskmasssec}). Also, the velocity-integrated intensity
maps presented in Fig.~\ref{HCNstructfig} show a clear kinematic
structure in east-west direction.

The formation of the HCN molecule in oxygen-rich environments is
further discussed in Sect.~\ref{nontechemistrysec}. A modeling using
the 1-D LTE radiative transfer code XCLASS \citep{moeller2017}, see
Appendix~\ref{xclassmodels}, gives a column density for HCN in RS Cnc
of N$_{\rm HCN}=1.6\times10^{15}$cm$^{-2}$, corresponding to an
abundance of $X($HCN/H$_2) = 6.6\times10^{-7}$. This value is well
within the range found for other M- and S-type stars as modeled by
\citet{sroetal2013}, who find $X($HCN/H$_2) =$ a few $10^{-7}$ (for
more details see Sect.~\ref{nontechemistrysec} and
Appendix~\ref{xclassmodels}).

\subsubsection{H$_2$O}\label{watersec}

\begin{figure}[h]
    \centering
    \vspace{-3.0cm}
    \includegraphics[width=110mm]{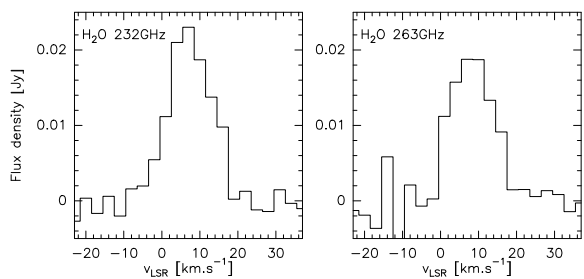}
    \caption{Left: H$_2$O line at 232.687\,GHz.  
             Right: H$_2$O line at 263.451\,GHz. Data are merged from A and 
             D-configuration, spectral resolution is 3\kms and the emission 
             of both lines is integrated over the central
             $1\arcsec \times 1\arcsec$ aperture.}
    \label{h2odetecfig}
\end{figure}

The WideX spectrum, obtained in A-configuration in December 2016,
serendipitously revealed a line at 232.687\,GHz that we ascribe to the
J(K$_a$,K$_c$)=5(5,0)-6(4,3) transition of o-H$_2$O in the v$_2=1$
vibrational state. The H$_2$O source is weak and seems still
unresolved within the synthesized beam of $0.5\arcsec \times
0.34\arcsec$ obtained in the A-configuration in February 2020,
consistent with its high upper-state energy of 3462\,K. The line
profile is shown in Fig.~\ref{h2odetecfig}, left. With the follow-up
observations employing PolyFiX in D and A-configuration we covered,
and detected, also the 263.451\,GHz o-H$_2$O v$_2=1$,
J(K$_a$,K$_c$)=7(7,0)-8(6,3) line (Fig.~\ref{h2odetecfig}, right;
E$_u/k$ = 4475\,K). Both lines are resampled to a resolution of 3\kms,
data are merged from A and D configuration, and the emission is
integrated over an aperture of $1\arcsec \times 1\arcsec$.  Intensity
maps of both lines are shown in Fig.~\ref{H2Ointensfig}, testifying to
the compactness of the H$_2$O emitting region.

These are the first detections of millimeter vibrationally excited
H$_2$O emission in RS Cnc. We note that the 22\,GHz H$_2$O maser in
the ground state has been tentatively detected by
\citet{1995A&A...296..727S} in one of the two epochs they covered, but
the 22\,GHz line is not detected in other observations
\citep{1973ApJ...180..831D,lew97,1995PPMtO..14..185H,2014ApJS..211...15Y}.
RS Cnc also shows clear photospheric H$_2$O absorption at $2.7\,\mu$m
\citep{ms76a,nk93}, and at $1.3\,{\mu}$m
\citep[$7500$\,cm$^{-1}$;][]{jhwdl98}, although the H$_2$O band near
900\,nm is not detected \citep{1966ApJ...143..291S}.

Both the 232 and 263\,GHz water lines have the upper levels belonging
to the so-called "transposed backbone" in the v$_2=1$ vibrationally
excited state of H$_2$O, that is K$_a$ = J and K$_c$ = 0 or 1 (see Fig. 1
of \citet{1993LNP...412..399A}).  The 232\,GHz line was first detected
in evolved stars together with the 96\,GHz line from another
transposed backbone upper level by \citet{mm89} toward the Red
Supergiant VY CMa and the AGB star W Hya.  The latter is an M-type
star with a similar mass-loss rate as RS Cnc. The authors find that
the 232\,GHz line emission in both stars may be of (quasi-)thermal
nature while the 96\,GHz line clearly showed maser action.  The
(unpublished) detection of the 263\,GHz line was mentioned in
\citet{1993LNP...412..399A}, who also described a mechanism that may
lead to a systematic overpopulation of the transposed backbone upper
levels in the v$_2=1$ state of H$_2$O in the inner region of
circumstellar envelopes. If the vibrational decay routes (to the
ground state) of the transposed backbone upper levels become more
optically thick than the lower levels in the v$_2=1$ state, then
differential radiative trapping may cause population inversion of
these lines. Additional vibrationally excited H$_2$O emission lines
from transposed backbone upper levels were predicted and later
detected in VY CMa by \citet{2006A&A...454L.107M} and
\citet{2013ApJS..209...38K}.  We observed the 232\,GHz line in RS Cnc
at three epochs (December 2016, December 2017, and February 2020) and
the 263\,GHz line at the latter two epochs, and the emission appears
to be stable in time for both lines. The profiles appear to be very
similar, both are broad, even broader than the (ground-state) lines of
other species reported here, and there is no sign for any narrow
component in either of the two profiles at any of the epochs. As the
lines should arise from a region very close to the star, compatible
with their broad widths, see Sect.~\ref{siowingsec}, one might expect
to see time variations due to the varying density and radiation field
caused by the stellar pulsation, in particular if the emission would
be caused by maser action, as seen on  the SiO(v=1,5-4) line observed
in December 2017 (see Fig.~\ref{siodetecfig}). Also, the modeling of
\citet{gmretal2016} shows only very little inversion of the involved
level populations for the 263\,GHz H$_2$O transition. We therefore
think that both lines could be thermally excited. A definite
assessment of the nature of the vibrationally excited H$_2$O emission
would however require some detailed modeling of the emission, together
with  high sensitivity monitoring of the line profiles with high
spectral resolution, possibly including other H$_2$O lines from
transposed backbone upper levels and/or known maser lines for
comparison, which is beyond the scope of the present paper.

\subsubsection{SO}
\label{sosec}

\begin{figure}[h]
    \centering
    \includegraphics[width=110mm]{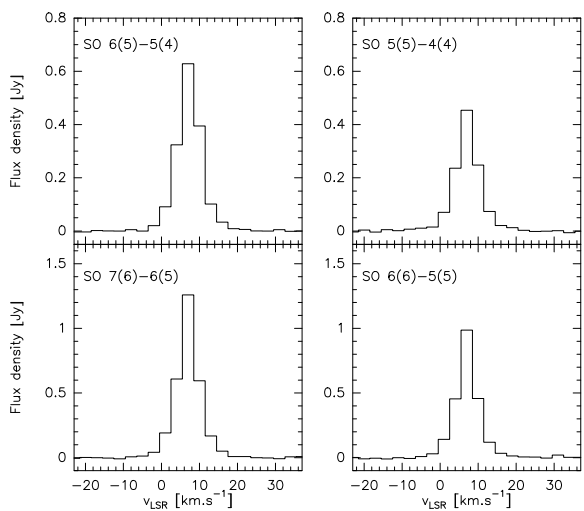}
    \caption{Profiles of the four detected SO lines, 
             A and D-configuration merged.
             Spectral resolution is 3\kms and the emission is integrated 
             over the central $2\arcsec \times 2\arcsec$ aperture.}
    \label{solinesfig}
% read-and-plot-SO4lines.greg
\end{figure}

\begin{figure}[h]
    \centering
    \vspace{-3.0cm}
    \includegraphics[width=110mm]{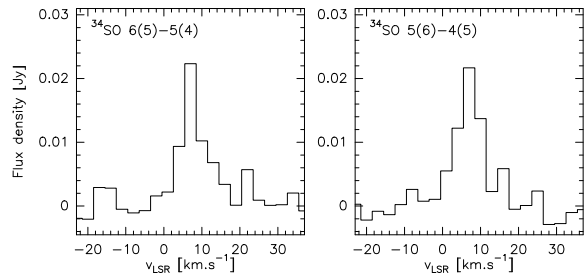}
    \caption{Profile of the two $^{34}$SO lines detected here, A and 
             D-configuration merged.
             Spectral resolution is 3\kms and the emission is integrated 
             over the central $2\arcsec \times 2\arcsec$ aperture.}
    \label{soisolinesfig}
% read-and-plot-SOisolines.greg
\end{figure}

Four lines of SO are detected (see Fig.~\ref{solinesfig}) and two
lines of the isotopolog $^{34}$SO (Fig.~\ref{soisolinesfig}).  These
represent the first detections of SO and $^{34}$SO in RS Cnc.  SO has
been observed in several M-type stars, including R Dor and W Hya,
\citep{ddbetal2016}), but remains undetected in S-type stars (e.g., W
Aql, \citet{2008A&A...480..431D,do2020}). All SO lines detected here
are slightly spatially resolved with a FWHP around $0.8\arcsec$ and
therefore seem to be emitted from the same region as
HCN. Velocity-integrated intensity maps of SO are shown in
Fig.~\ref{SOstructfig}. In fact, the SO lines show the same velocity
pattern (indicating rotation) as does HCN, although the velocity
resolution of the SO lines is only 3\kms, see  Fig.~\ref{SOstructfig}
and the first moment map in the right panel of Fig.~\ref{rotationfig}.

Using the integrated line strengths of SO(6(5)-5(4)) and
$^{34}$SO(6(5)-5(4)) found here ($\sim$ 4.69\,Jy\kms and $\sim$
0.20\,Jy\kms, respectively), and taking the difference of the Einstein
coefficients of the transitions into account, we estimate the
isotopolog ratio $^{32}$SO/$^{34}$SO $\sim 23$, assuming equal
excitation conditions for both transitions and optically thin emission
of both lines. This value is in good agreement with the values of
21.6$\pm8.5$ or 18.5$\pm5.8$ as derived from the radiative transfer
models for M-type stars by \citet{ddbetal2016,2020MNRAS.494.1323D},
respectively.  We note that for the S-type star W Aql an
Si$^{32}$S/Si$^{34}$S isotopolog ratio of 10.6$\pm2.6$ was derived
by \citet{do2020}. As $^{32}$S is mainly produced by oxygen-burning in
massive stars and, to a lesser extent, in type Ia supernovae, and as
$^{34}$S is formed by subsequent neutron capture
\citep[e.g.,][]{1984ApJ...286..644N,1992A&ARv...4....1W,1995ApJS...98..617T,2008MNRAS.390.1710H},
the $^{32}$S/$^{34}$S isotopic ratio remains virtually unaltered
during AGB evolution \citep[see, e.g. tables in the
FRUITY\footnote{http://fruity.oa-teramo.inaf.it/}
database,][]{2011ApJS..197...17C} and, therefore, should reflect the
chemical initial conditions of the natal cloud from which the star has
formed. The spread in the isotopic ratio seen among the different AGB
stars mentioned above would then rather be indicative of the Galactic
environment in which the star  has formed \citep[see,
e.g.,][]{1996A&A...305..960C,2020A&A...642A.222H} instead of
reflecting any evolutionary effect. For the low mass-loss rate M-type
stars R Dor and W Hya, \citet{ddbetal2016} reproduce their observed
line profiles best with centrally peaked SO (and SO$_2$)
distributions, consistent with the maps presented in
Fig.~\ref{SOstructfig}.

\subsubsection{SO$_2$}\label{so2sec}

In SO$_2$, 11 lines are detected, their parameters are summarized in
table~\ref{so2poptab}, and all profiles are shown in
Fig.~\ref{so2profiles}. These are the first detections of SO$_2$ in RS
Cnc. A previous survey with the IRAM 30m telescope by \cite{olmg93}
did not detect SO$_2$ in RS Cnc with an rms noise of 0.052\,K (or
$\sim 0.25$\,Jy at 160.8\,GHz).  As an example we show the SO$_2$
(14(0,14)-13(1,13)) line at 244.3\,GHz, only in
Fig.~\ref{so2244fig}. A first moment map of the SO$_2$
(14(0,14)-13(1,13)) line is  shown in Fig.~\ref{rotationfig} on the
middle left diagram. Although the source remains barely resolved
(source size $\sim 0.43\arcsec$) by the beam
($0.69\arcsec\times0.49\arcsec$), there is a signature of a rotating
structure in SO$_2$, as was also seen in EP Aqr
\citep{hrddk2018,thnetal2019}.  Integrated intensity maps of three
SO$_2$ lines  (SO$_2$(9(3, 7)-9(2, 8)), which has the lowest upper
level energy of the SO$_2$ lines detected here ($E_u = 64$\,K);
SO$_2$(14(0,14)-13(1,13)), the strongest line, and
SO$_2$(34(4,30)-34(3,31)), which has the highest upper level energy of
the detected lines, $E_u = 595$\,K) are shown in
Fig.~\ref{SO2structfig}. All lines show kinematic structure in the E-W
direction, about orthogonal to the   outflow structure seen in CO and
SiO,  cf. Fig.~\ref{COSiOoutflowfig}.

We derive the rotational temperature and column density of the SO$_2$
emitting region with a population diagram analysis
(Sect.~\ref{thermalsec}) and by an XCLASS modeling
(Appendix~\ref{xclassmodels}).  Both methods give a similar rotational
temperature of $\sim 320-350$\,K and a column density of $\sim
3.5\times10^{15}$\,cm$^{-2}$.

\subsubsection{PN}\label{pnsec}

\begin{figure}[h]
  \begin{minipage}[b]{0.45\linewidth}
  \centering
  \vspace{-3.0cm}
  \includegraphics[width=110mm]{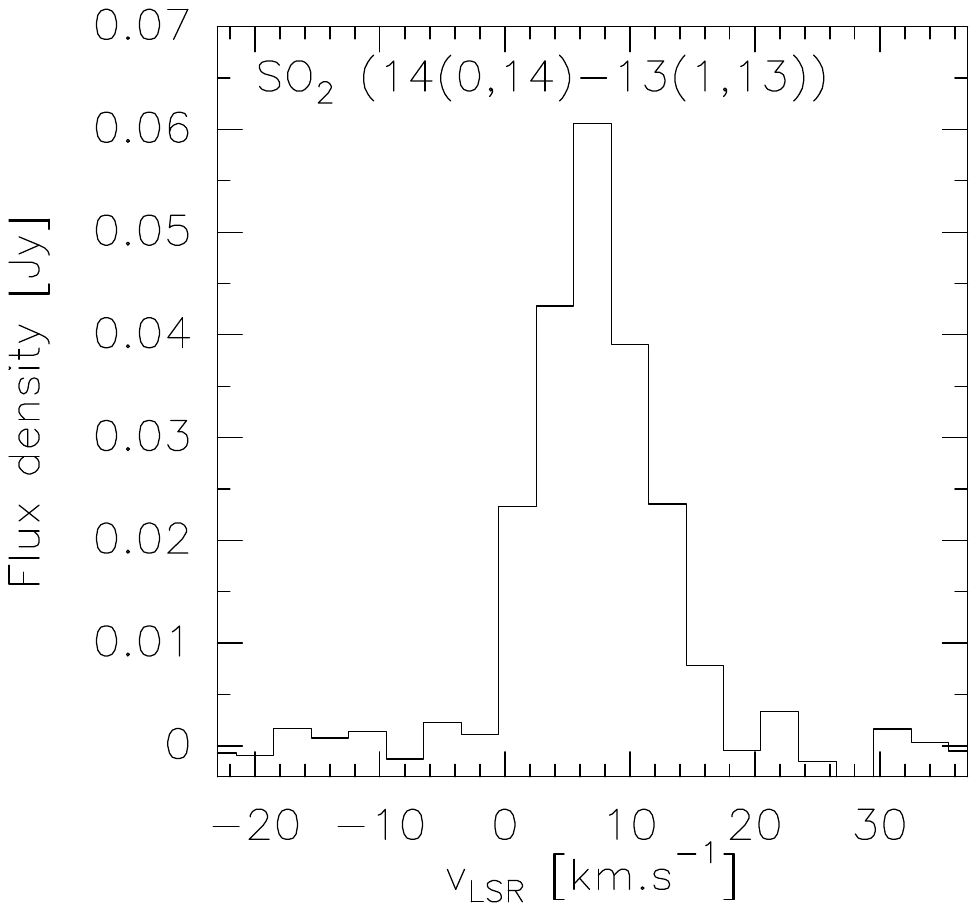}
  \caption{Profile of SO$_2$ (14(0,14)-13(1,13)). A and D-configuration merged,
           spectral resolution is 3\kms and the emission is integrated 
           over the central $2\arcsec \times 2\arcsec$ aperture.}
  \label{so2244fig}
  % read-and-plot-so2indivline.greg
  \end{minipage}
  \hspace{0.25cm}
  \begin{minipage}[b]{0.45\linewidth}
  \centering
  \vspace{-3.0cm}
  \includegraphics[width=110mm]{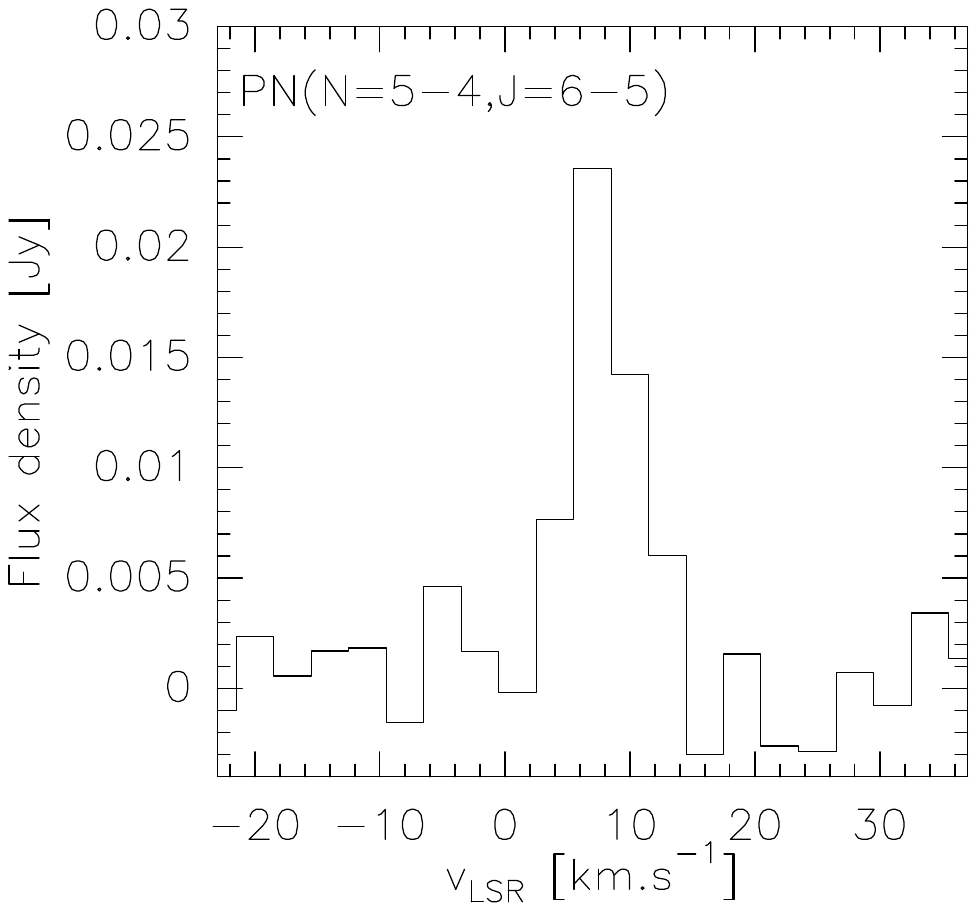}
  \caption{Profile of PN(N=5-4, J=6-5). A and D-configuration merged, 
           spectral resolution is 3\kms and the emission is integrated 
           over the central $2\arcsec \times 2\arcsec$ aperture.}
  \label{pnlinefig}
  % read-and-plot-pnline.greg
  \end{minipage}
\end{figure}

We detect a line at 234.936\,GHz that we ascribe to the PN molecule,
which would be the first detection of PN in RS Cnc. PN has been
detected in several M-type stars (e.g.,
\citet{debecketal2013,zsb2018}), and in the C-rich envelopes of
IRC~+10216 and CRL~2688
\citep{guelinetal2000,cernicharoetal2000,milametal2008}. The presence
of PN in an MS-type star therefore does not seem to come as a
surprise. RS Cnc however appears to be the lowest mass-loss rate
source, in which this molecule has been reported so far.  The PN line
profile is shown in Fig.~\ref{pnlinefig}. The line is spatially
resolved at $0.8\arcsec$, which places it in about the same region as
HCN and SO. The first moment map of this line also shows signatures of
rotation but due to the weakness of the line, the evidence is low.  An
integrated intensity map of PN is presented in Fig.~\ref{PNstructfig},
showing that the line emitting region is slightly spatially resolved.
The $3\sigma$ feature seen about $1.5\arcsec$ south of the phase
center should not be considered as a detection but rather as a noise
peak, as long as this structure is not confirmed by higher sensitivity
observations.

\subsection{High-velocity wings in SiO, and in other molecules}
\label{siowingsec}

\begin{figure}[ht]
    \centering
    \includegraphics[width=90mm]{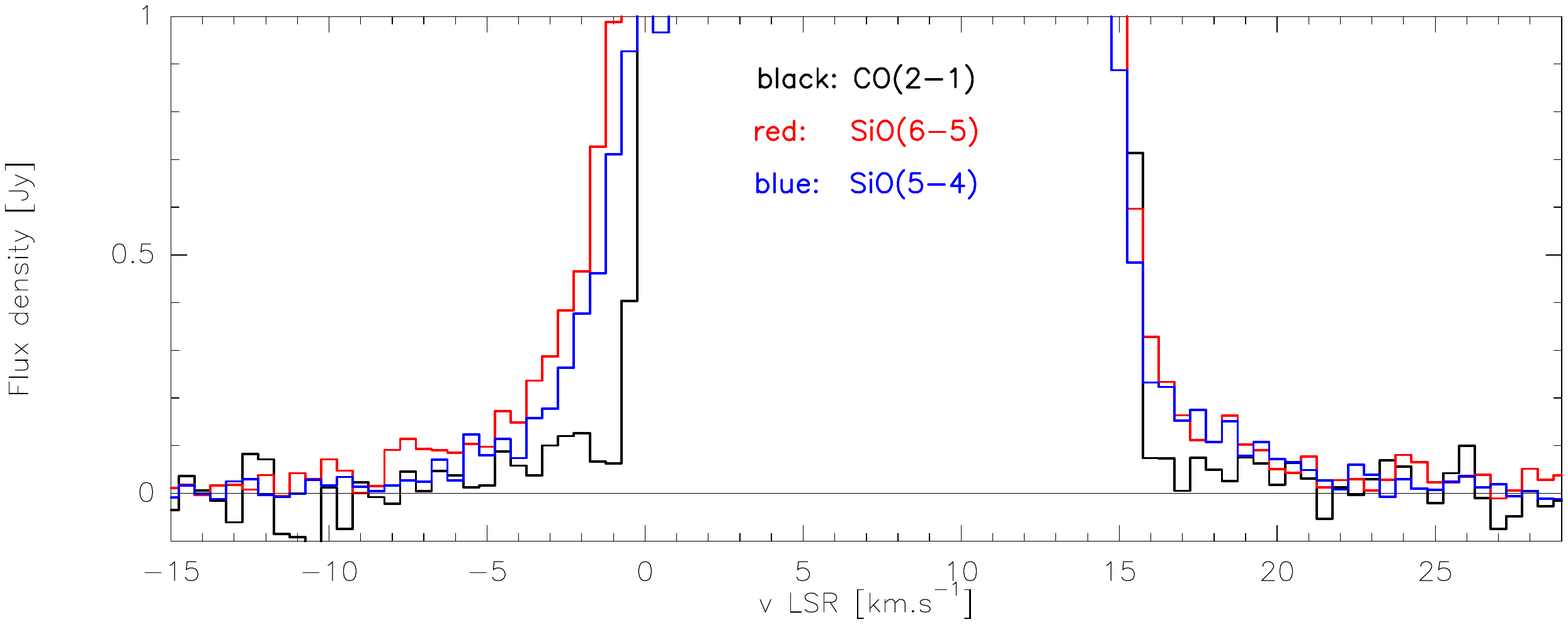}
    \caption{Line wings in SiO(5-4) and SiO(6-5) compared to CO(2-1). The 
             emission is integrated over the central 
             $5\arcsec \times 5\arcsec$ aperture.}
% winters:winters@winters:/data/RSCnc/maps/February-2020/high-velo-wings.greg
    \label{siowingfig}
\end{figure}

\begin{figure}[ht]
    \centering
    \includegraphics[width=40mm]{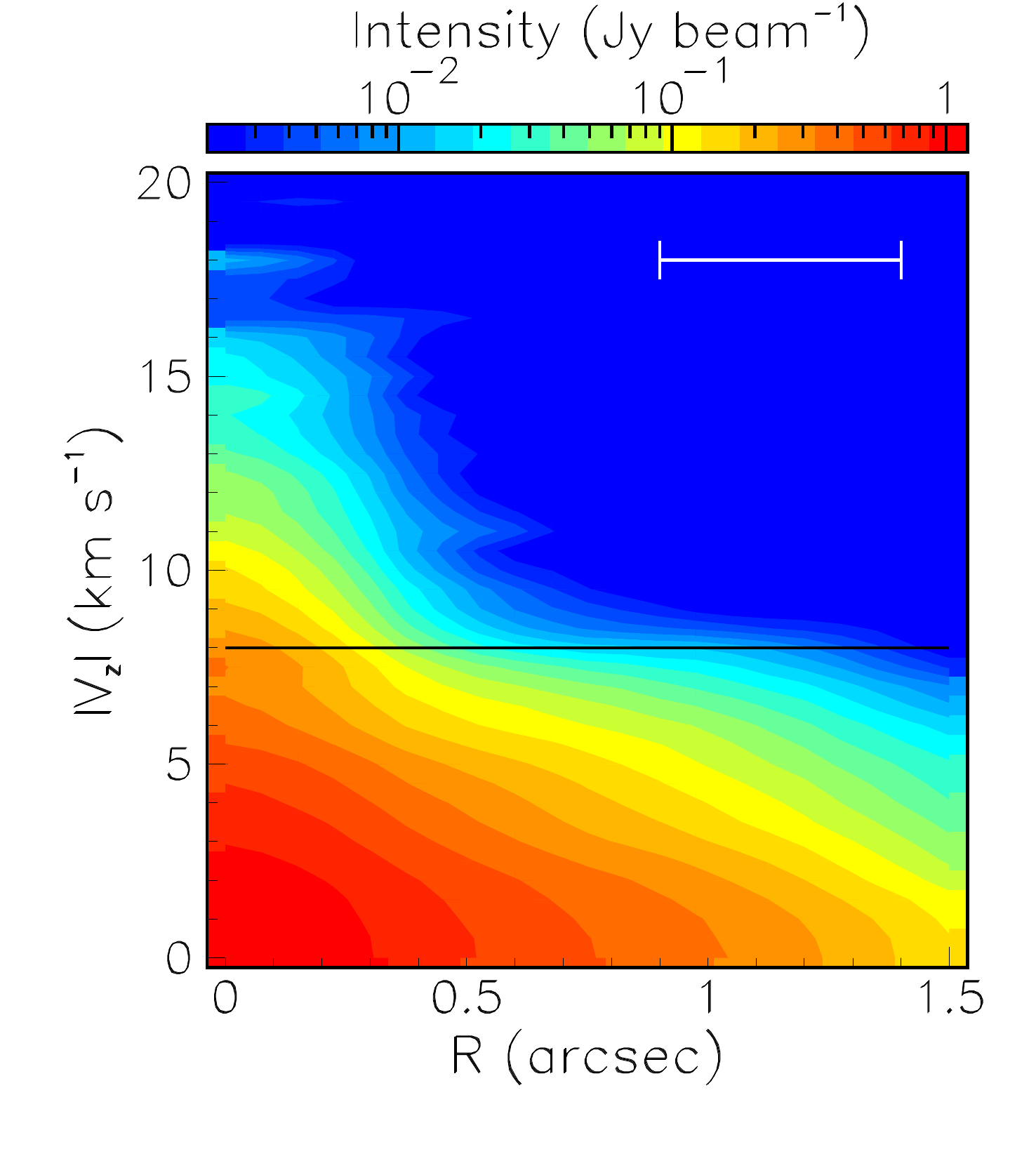}
    \includegraphics[width=40mm]{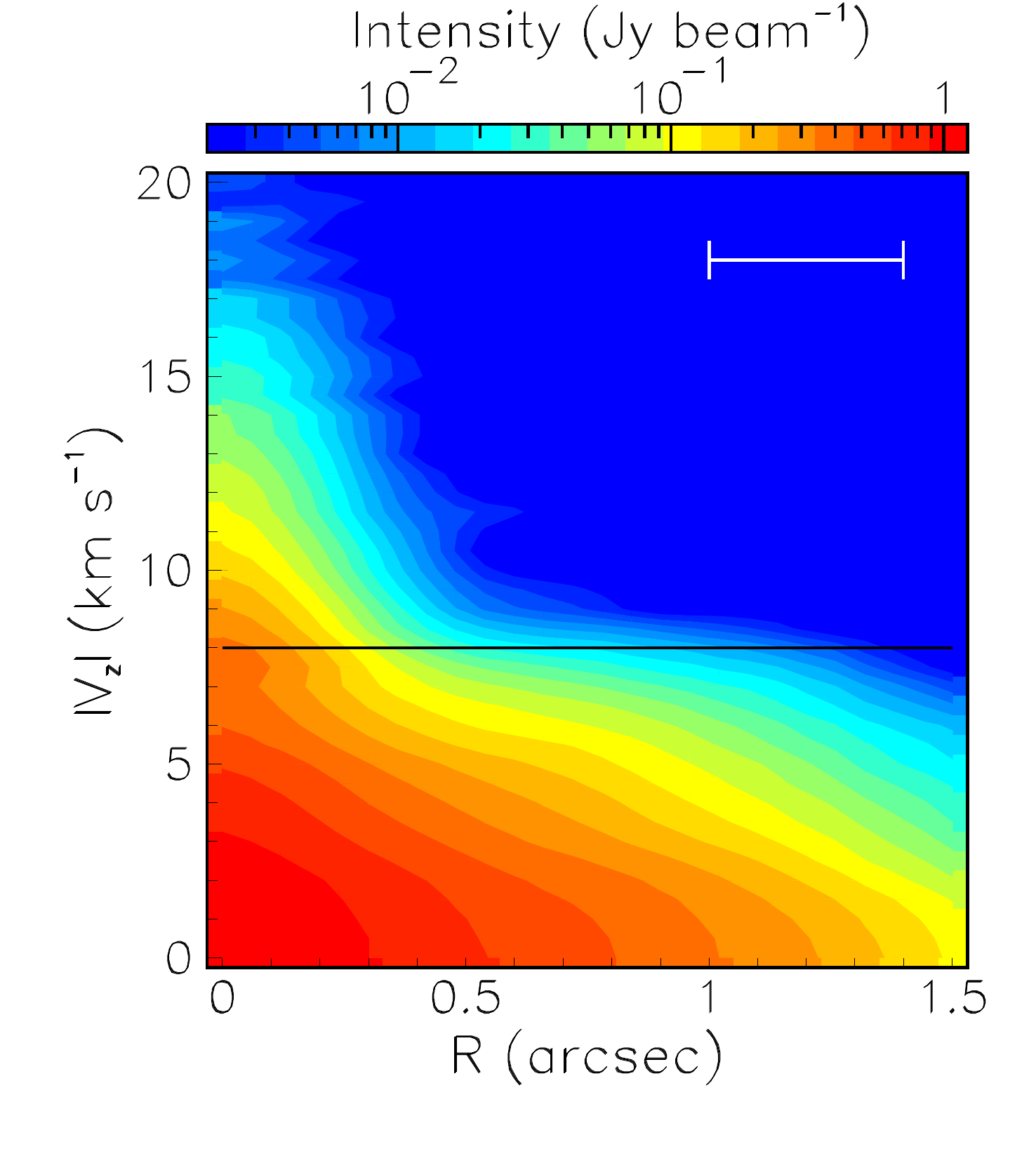}
    \caption{High velocities close to the line of sight as seen in
             SiO. PV maps are shown in the ${\rm V_{z}}$ vs. $R$ plane 
             for SiO(5-4) (left) 
             and SiO(6-5) (right). The horizontal black line indicates 
             the wind terminal velocity as traced in CO and the white scale 
             bar indicates the spatial resolution.  
             $R = \sqrt{(\delta {\rm DEC})^2+(\delta {\rm RA})^2}$, 
             ${\rm |V_{z}|=|v_{\rm lsr}-v_{\rm lsr,*}|}$}
    \label{siostreamfig}
\end{figure}

\begin{figure}[ht]
    \centering
    \includegraphics[width=90mm]{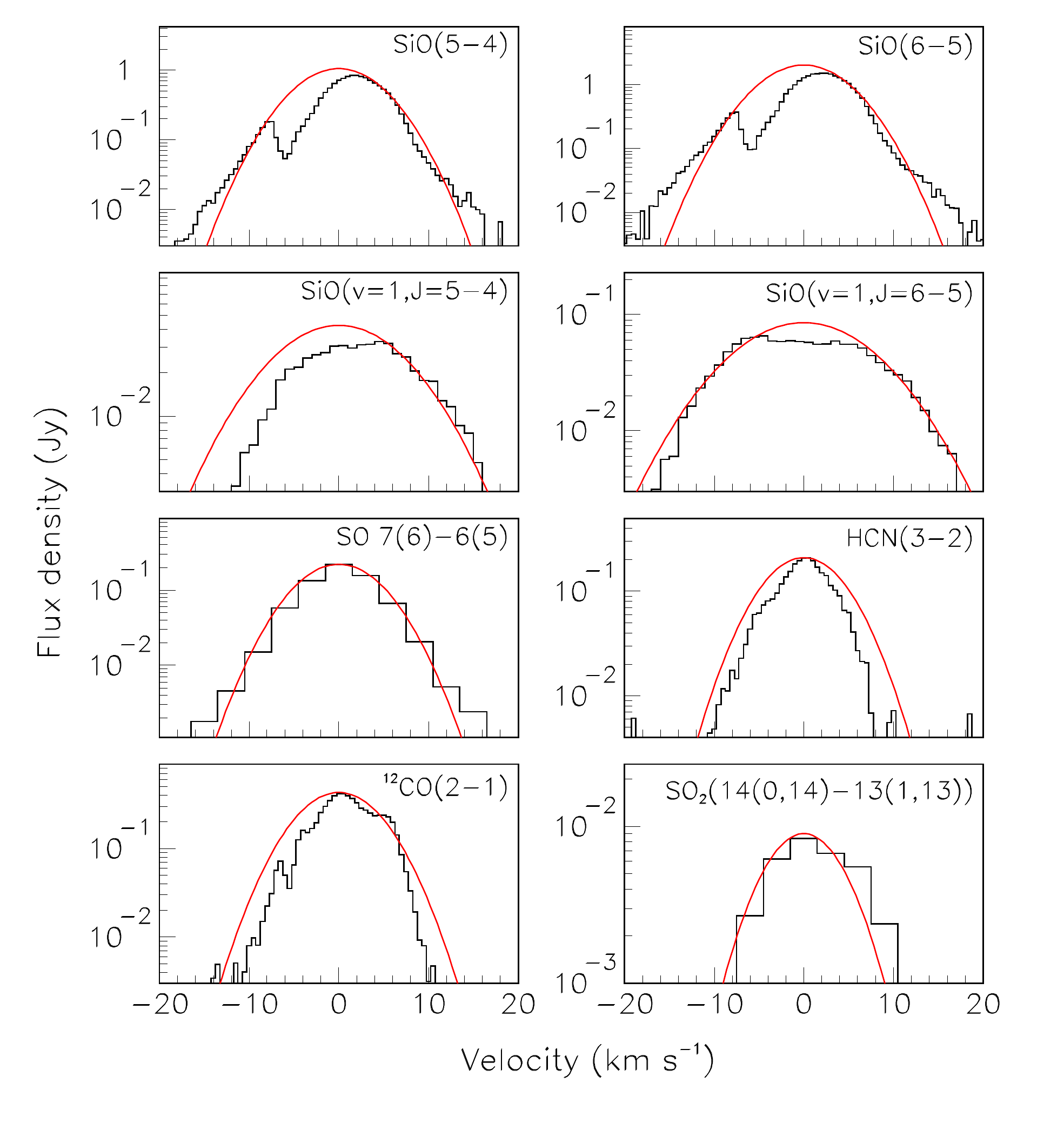}
    \caption{Line profiles of different molecules on a logarithmic
      intensity scale. Gaussian profiles are shown for comparison,
      FWHM$=10$\kms for the ground state lines of all molecules and
      FWHM$=14$\kms for the (v=1) lines of SiO. All observed profiles
      are integrated over $R<0.2\arcsec$.}
    \label{wingfig}
\end{figure}

In SiO, five lines in three different vibrational states (v=0,1,2) are
detected (see  Figs.~\ref{siodetecfig} and \ref{siov2linefig}). The
vibrational ground state lines clearly indicate the presence of
material at velocities much higher than the wind terminal velocity of
$\sim$ 8\kms as traced by CO lines  at this stellar latitude (see
Sect.~\ref{cokinesec}). This is illustrated in  Fig.~\ref{siowingfig}
and in Fig.~\ref{siostreamfig}, where we have defined v$_{\rm
z}=$v$_{\rm lsr}-$v$_{\rm lsr,*}$, the Doppler velocity relative to
the star. The high-velocity region is centered on the line of sight
and is confined to the inner $\approx 0.3\arcsec$, see
Fig.~\ref{siostreamfig}.  A similar feature has been previously seen
in high spatial resolution observations of other oxygen-rich, low
mass-loss rate AGB stars, for example, W Hya \citep{2017NatAs...1..848V}, EP
Aqr \citep{thnetal2019}, $o$ Cet \citep{htnetal2020}, R Dor
\citep{drdhn2018,nhtmnras2019,nhungetal2021}, and in 15 out of 17
sources observed in the ALMA Large Program ATOMIUM
\citep{dmrgh2020,atomium1}, calling for a common mechanism causing
high-velocity wings in this type of objects. In the case of EP Aqr,
where the bipolar outflow axis almost coincides with the line of sight
(with an inclination angle of $i \approx 10^{\circ}$), the high
velocity wings were interpreted in terms of narrow polar jets. For R
Dor and $o$ Cet, that do not show obvious signs of axial symmetry of
their winds, such an interpretation could not be retained and it was
argued instead that the high velocity wings were caused by (a mixture
of) turbulence, thermal broadening, and some effect of shocks, acting
at distances below some 10 to 15\,AU from the central star. The
presence of broad wings in the SiO lines emitted from RS Cnc, whose
symmetry axis is inclined by $\approx 30^{\circ}$ with respect to the
line of sight (see Sect.~\ref{cokinesec}), lends support to the latter
type of interpretation and casts serious doubts on the polar jet
interpretation proposed earlier for EP Aqr, which shows a
morpho-kinematics similar to that of RS Cnc \citep{nhwetal2015}.
Indeed, if the broad line widths are  present independent of the
orientation of a possible symmetry axis with the line of sight, they
must be caused by a mechanism of non-directional (accounting for the
resolving beam) nature.  A possible candidate, whose action is limited
to the close vicinity of the star, is pulsation-driven shocks that
dissipate their energy relatively close to the star and imply positive
and negative velocities in the shocked region that can be much higher
than the terminal outflow velocity of the wind. Such structures could
be explained by the B-type models discussed in \citet{wljhs2000} as
presented in \citet{wlnoj2002}, see their figure~3.  Recent 3-D
model calculations that self-consistently describe convection and
fundamental-mode radial pulsations in the stellar mantle would provide
the physical mechanism that leads to the development of such shocks
close to the star surface \cite[e.g.,][]{flh2017} and could therefore
replace the simplified inner boundary condition (the so-called
``piston approximation'') that was used in the earlier 1-D models
mentioned above.

In the data presented here, high Doppler velocity wings are seen in
nearly all lines detected with sufficient sensitivity to probe the
profile over at least v$_{\rm lsr,*}\pm10$\kms. This is illustrated in
Fig.~\ref{wingfig}, where v$_z$ profiles are integrated over a circle
of  radius $0.2\arcsec$ centered on the star. Gaussian profiles
centered at the origin are shown as visual references (not fits),
showing how absorption produces asymmetric profiles.  A major
difference is seen between vibrational ground state lines, which have
a Gaussian FWHM of $\sim 10$\kms, and vibrationally excited state
lines, which have a Gaussian FWHM of $\sim14$\kms. Such difference is
not  surprising, assuming that the high velocity wings are formed in
the inner layer of the circumstellar envelope (CSE), which is
preferentially probed by the (v=1) lines. In this context we note that
\citet{rcg2021} recently reported the detection of a narrow
SiO(v=1,1-0) maser line in RS Cnc at a velocity of +14\kms with
respect to the star's lsr velocity.  The effect of shocks on line
profiles has first been observed in the near-infrared range on CO
ro-vibrational lines, probing the stellar photosphere and the
innermost circumstellar region within $\sim 10$\,R$_*$ (e.g., $\chi$
Cyg, an S-type star, \cite{hhr82}). Very high angular resolution
observations obtained over the past decade using VLT, VLTI, and ALMA
show that the effect of shocks from pulsations and convection cell
ejections is confined within some 10 AU from the star  \citep[see, for
example,][and references
therein]{2018A&A...620A..75K,ho18,2019ApJ...883...89O}.  Rotation,
when observed, is instead found to extend beyond this distance,
typically up to $\sim$20\,AU
\citep[e.g.,][]{2018A&A...613L...4V,hdtetal2018,nhungetal2021}.  The
angular resolution of the present data is insufficient to detect such
difference directly; however, the effect of rotation and shocks on
lines of sight contained within a beam centered on the star depends on
the region probed by each specific line: lines that probe the inner
layers exclusively, as the (v=1) lines, are mostly affected by shocks,
and somewhat by rotation; CO lines, for which the probed region
extends very far out, see little of rotation and even less of shocks
because the emission from the inner envelope provides too small a
fraction of the total emission. Between these two extremes the
relative importance of the contributions of shocks and rotation
depends on the radial extent of the region probed by the line. Such an
interpretation is consistent with the data displayed in
Fig.~\ref{wingfig}.

\subsection{Rotation}\label{rotdiskmasssec}

\begin{figure*}[ht]
    \centering
    \includegraphics[width=45mm]{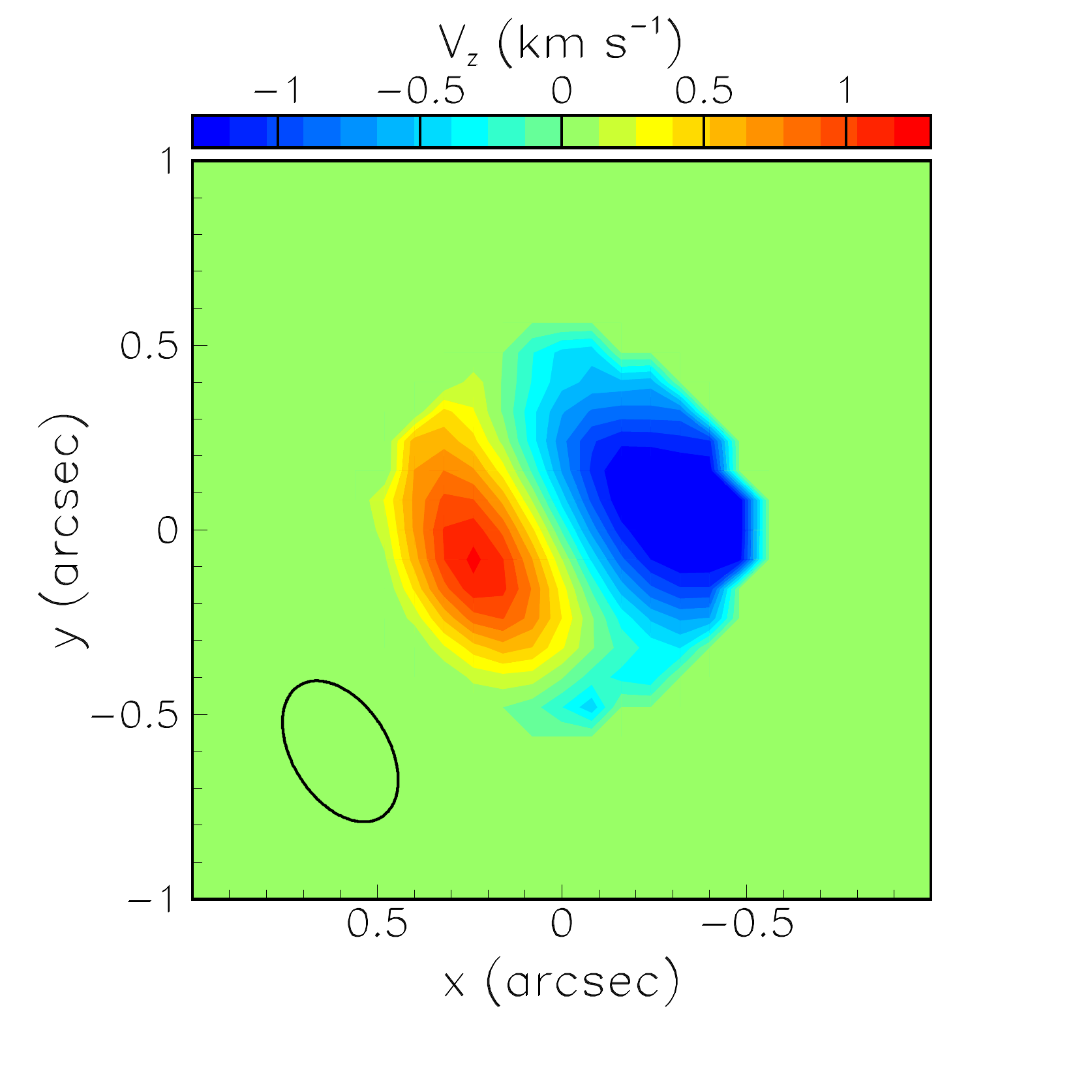}
    \includegraphics[width=45mm]{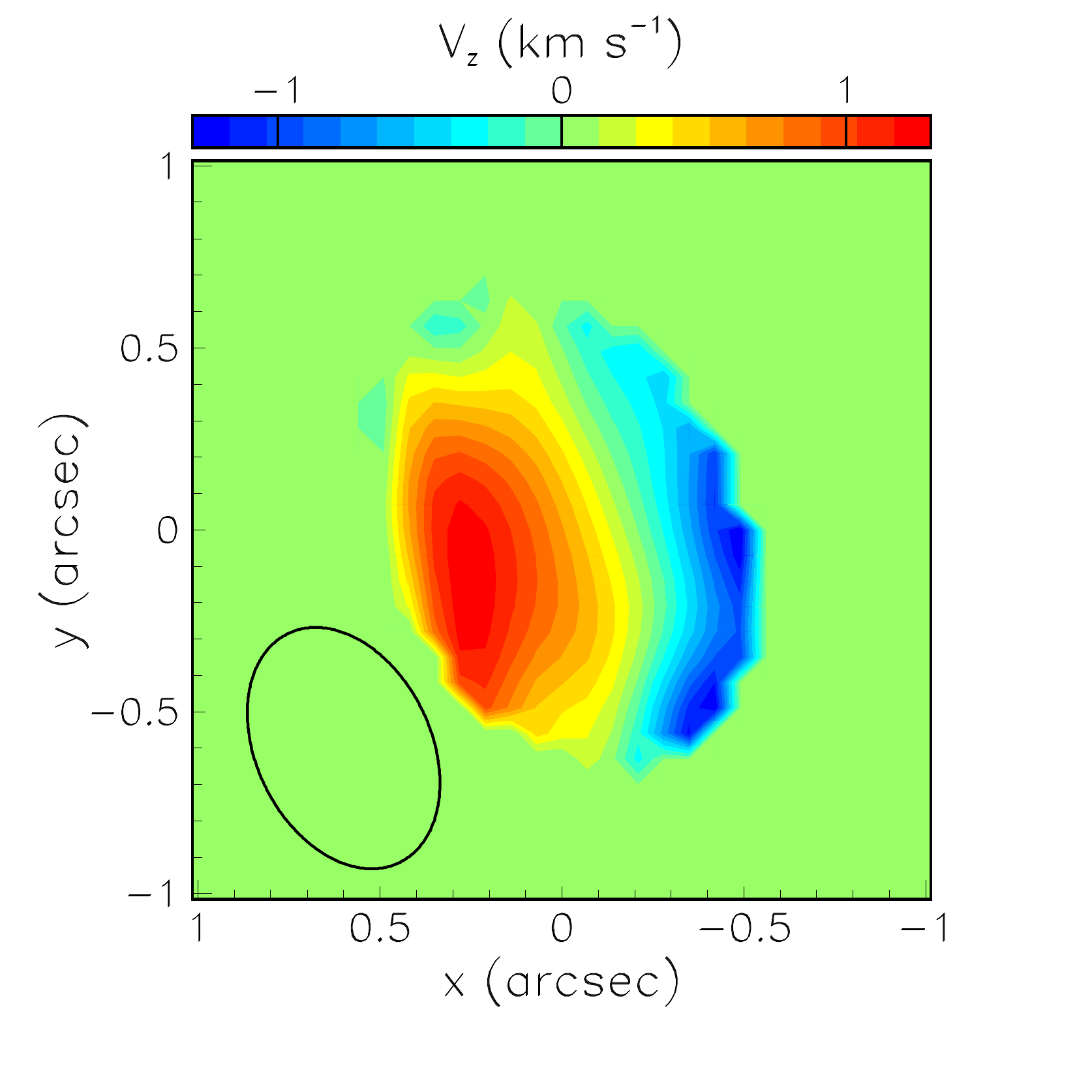}
    \includegraphics[width=45mm]{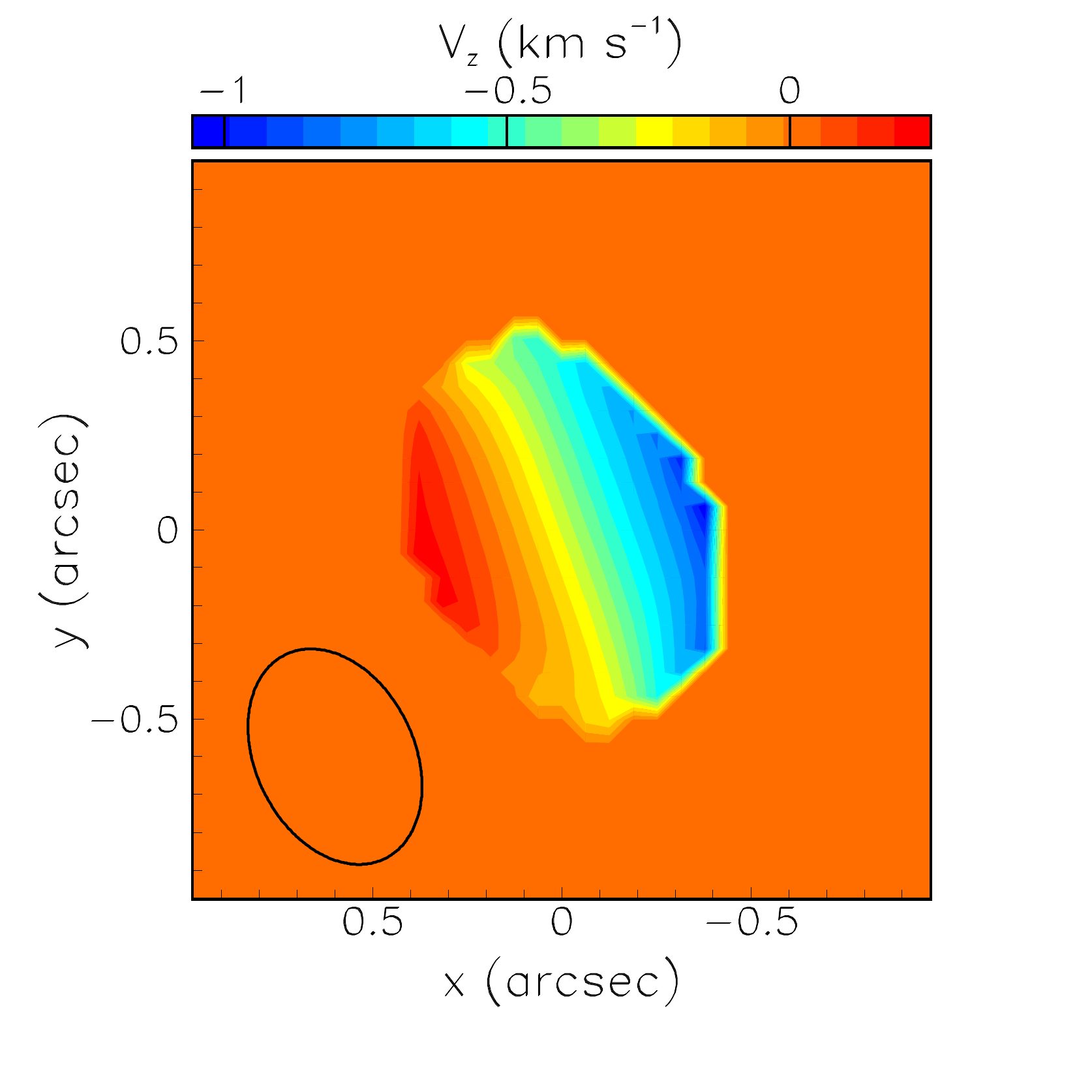}
    \includegraphics[width=45mm]{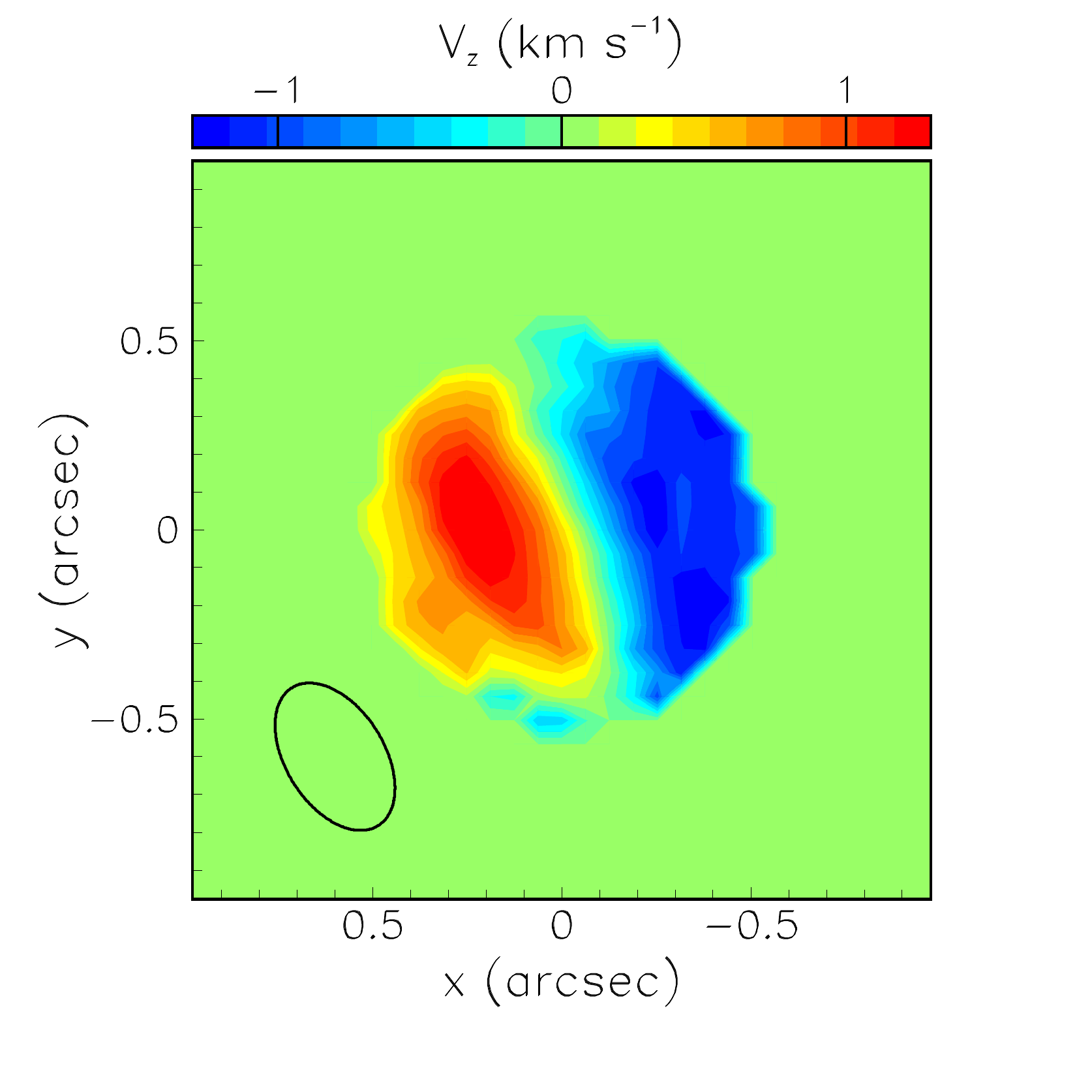}
    \caption{First moment maps of different lines, indicating a possibly
             rotating structure (see Sect.~\ref{rotdiskmasssec}). 
             Left:         HCN(3-2), 
             middle left:  SO$_2$ (14(0,14)-13(1,13)),
             middle right: SiO(v=1,6-5), 
             right:        SO(7(6)-6(5)).
             The black ellipses indicate the synthesized beam.}
    \label{rotationfig}
% from Hoai's report, mail of 09/06/2020 
\end{figure*}

In Fig.~\ref{rotationfig} we present first moment maps of HCN(3-2)
(left), SO$_2$ (14(0,14)-13(1,13)) (middle left), SiO(v=1,6-5) (middle
right), and SO(7(6)-6(5)) (right).

At projected distances from the star not exceeding $0.5\arcsec$, all
four tracers display approximate anti-symmetry with respect to a line
at PA $\sim 10^\circ$.  This is suggestive of the presence of rotation
in the inner CSE layer around an axis that projects on this line in
the plane of the sky.  Such a morpho-kinematic structure has also been
observed in other stars, notably in R Dor
\citep{2018A&A...613L...4V,hdtetal2018,nhungetal2021}.  The angular
resolution of the present data does not allow for a detailed
exploration of this region and prevents commenting on its possible
cause. Yet, the anti-symmetry axis of the velocity pattern projected
on the plane of the sky at a PA that approximately coincides with the
projected symmetry axis of the polar outflows (see
Sect.~\ref{cokinesec}) is remarkable and suggests that rotation is
taking place about this same polar axis in the inner CSE layer.

The line-of-sight velocities of these structures are small, on the
order of the velocities derived from CO for the equatorial region, and
which we interpret here as possible signs of rotation (rather than
indicating another bipolar outflow oriented perpendicular to the
larger-scale outflow traced in CO and SiO (v=0) lines).  Note that out
of these four lines, the HCN(3-2) line is detected with the highest
signal-to-noise ratio (S/N = 233 in the line peak,
cf. table~\ref{obsparalinestab}).

The  mean Doppler velocity $<{\rm v_z}>$, averaged over the inner $0.5
\arcsec$, of the HCN line can be fit in position angle $\omega$,
measured counter-clockwise from north, by

\begin{equation}
  \left< {\rm v_z} \right>_{\rm HCN} = -0.19\,{\rm km\,s}^{-1} +
                                        1.0\,{\rm km\,s}^{-1}\, 
                                                     \sin(\omega-19^{\circ})\, ,
\label{hcnroteq}
\end{equation}  
whereas the SiO(v=1,6-5) velocity is well fit by
\begin{equation}
  \left< {\rm v_z} \right>_{\rm SiO(v=1,6-5)} = -0.37\,{\rm km\,s}^{-1} + 
                                               0.46\,{\rm km\,s}^{-1}\, 
                                                     \sin(\omega-26^{\circ}) .
\label{siov165roteq}
\end{equation}  

The small offsets, $\sim -0.3$\kms on average, are within the
uncertainty attached to the measurement of the star's LSR
velocity. The coefficients of the sine terms measure the projected
rotation velocity, namely the rotation velocity divided by the sine of
the angle made by the rotation axis with the line of sight. Assuming
that the rotation axis is the axi-symmetry axis of the CSE, this angle
is $i \sim 30^\circ$ (see Sect.~\ref{cokinesec}),  meaning rotation
velocities of $\sim 2$ and $\sim 1$\kms for HCN and SiO
respectively. Better angular resolution observations are needed to
confirm the presence of rotation within a projected distance of
$0.5\arcsec$ from the star and we prefer to summarize the results
presented in this subsection in the form of an upper limit to the mean
rotation velocity of a few \kms.

\subsection{Global outflow structure traced by CO and SiO}
\label{kinestrucsec}

\begin{figure*}[h]
    \centering
    \includegraphics[width=60mm]{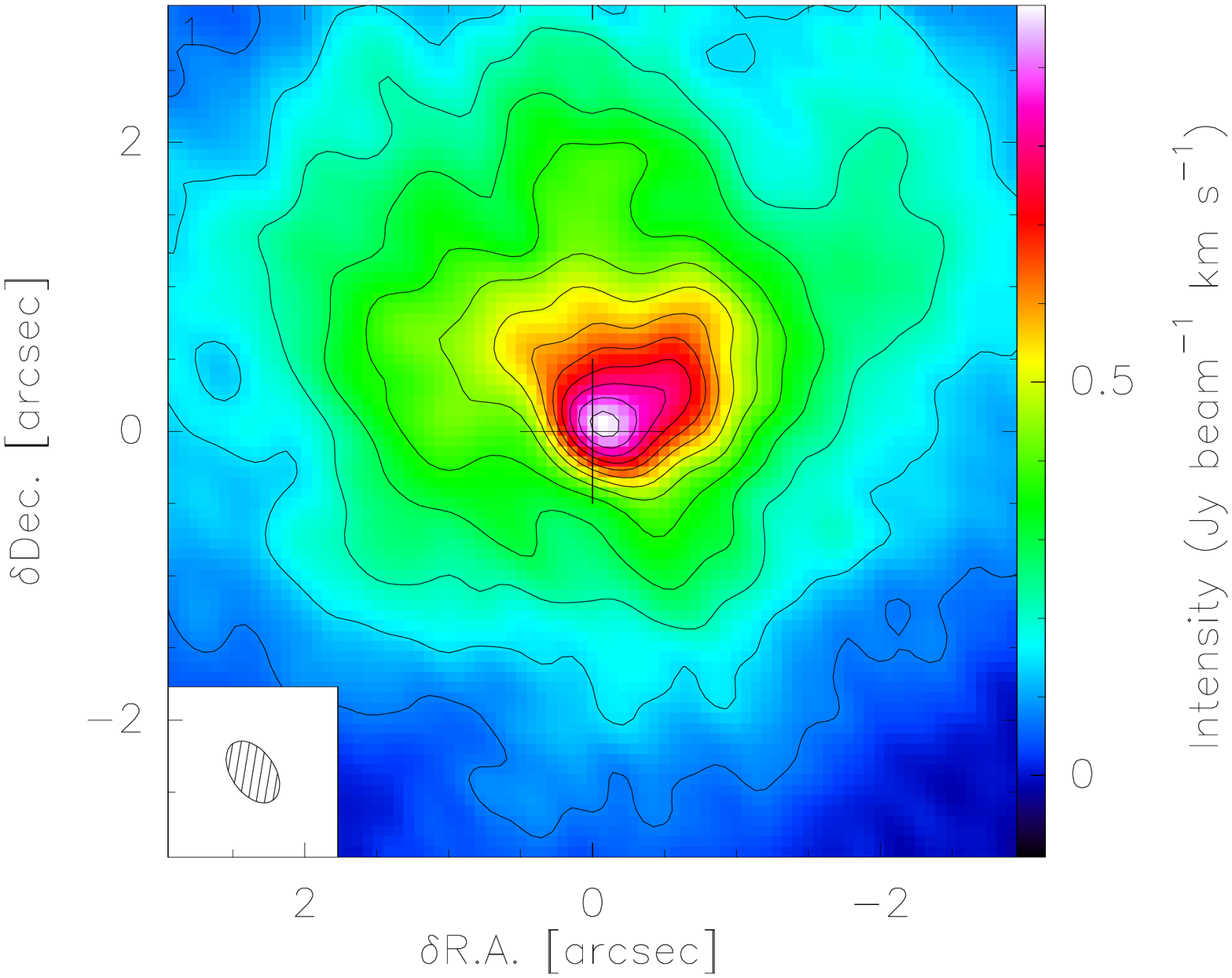}
    \includegraphics[width=60mm]{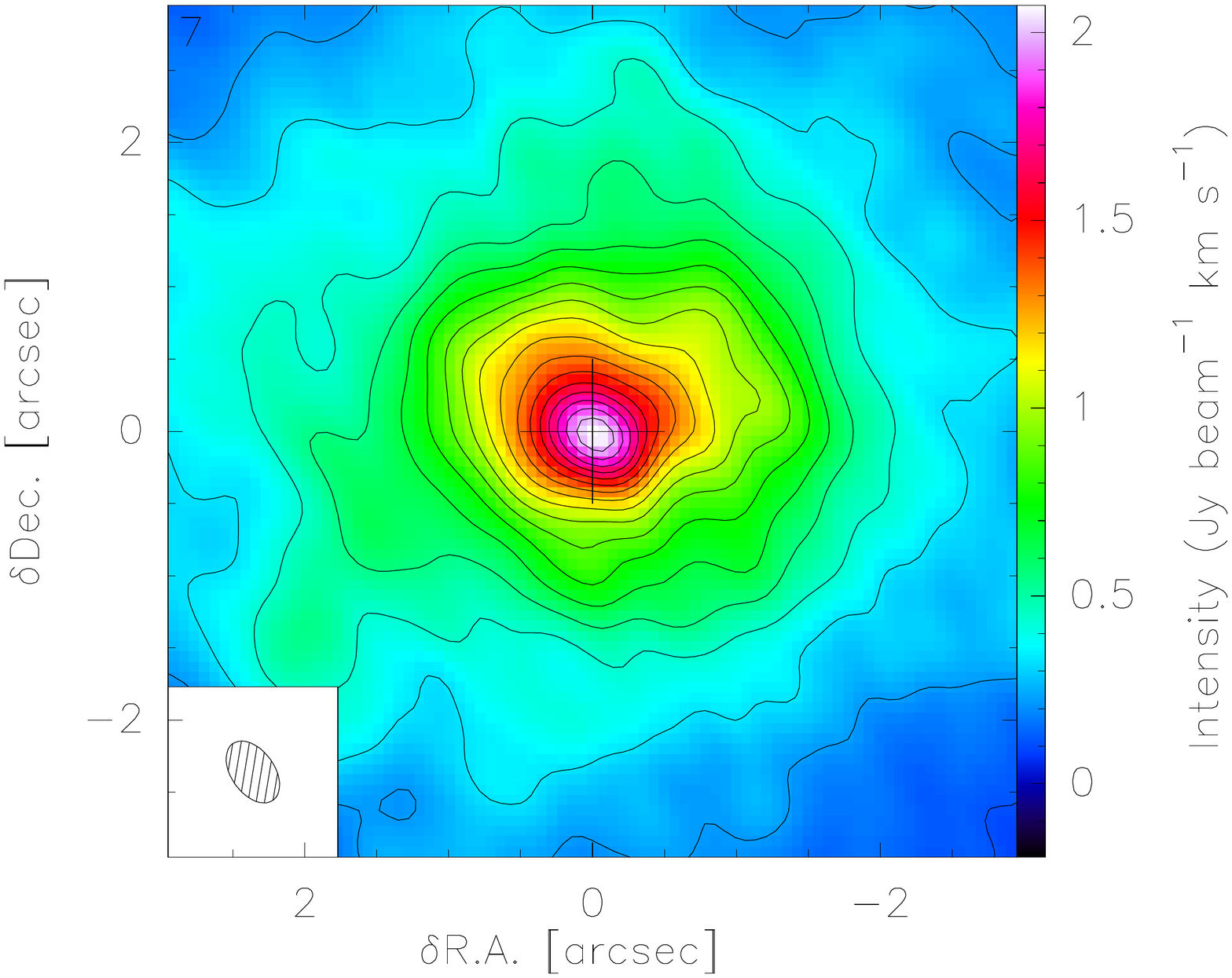}
    \includegraphics[width=60mm]{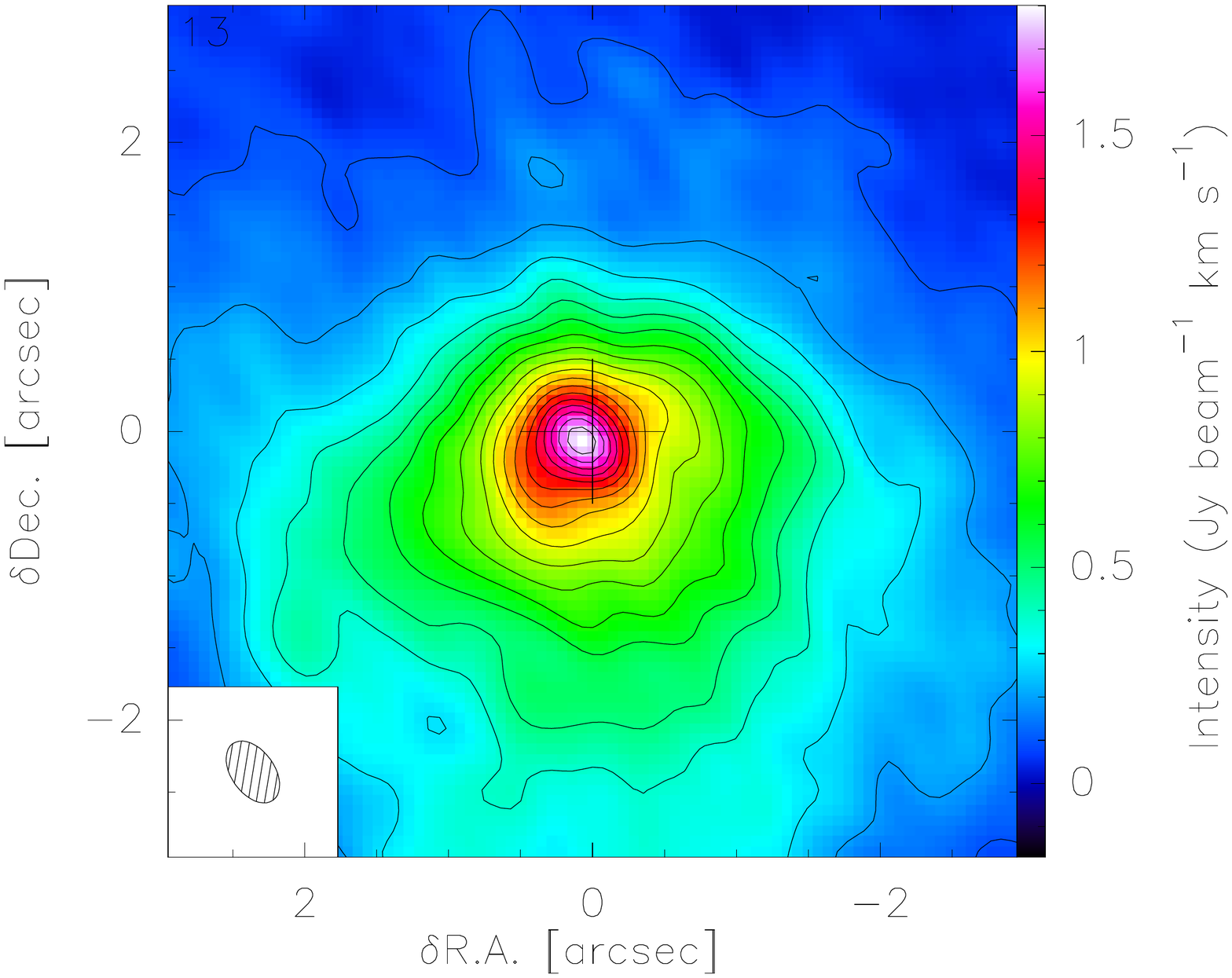}
    \includegraphics[width=60mm]{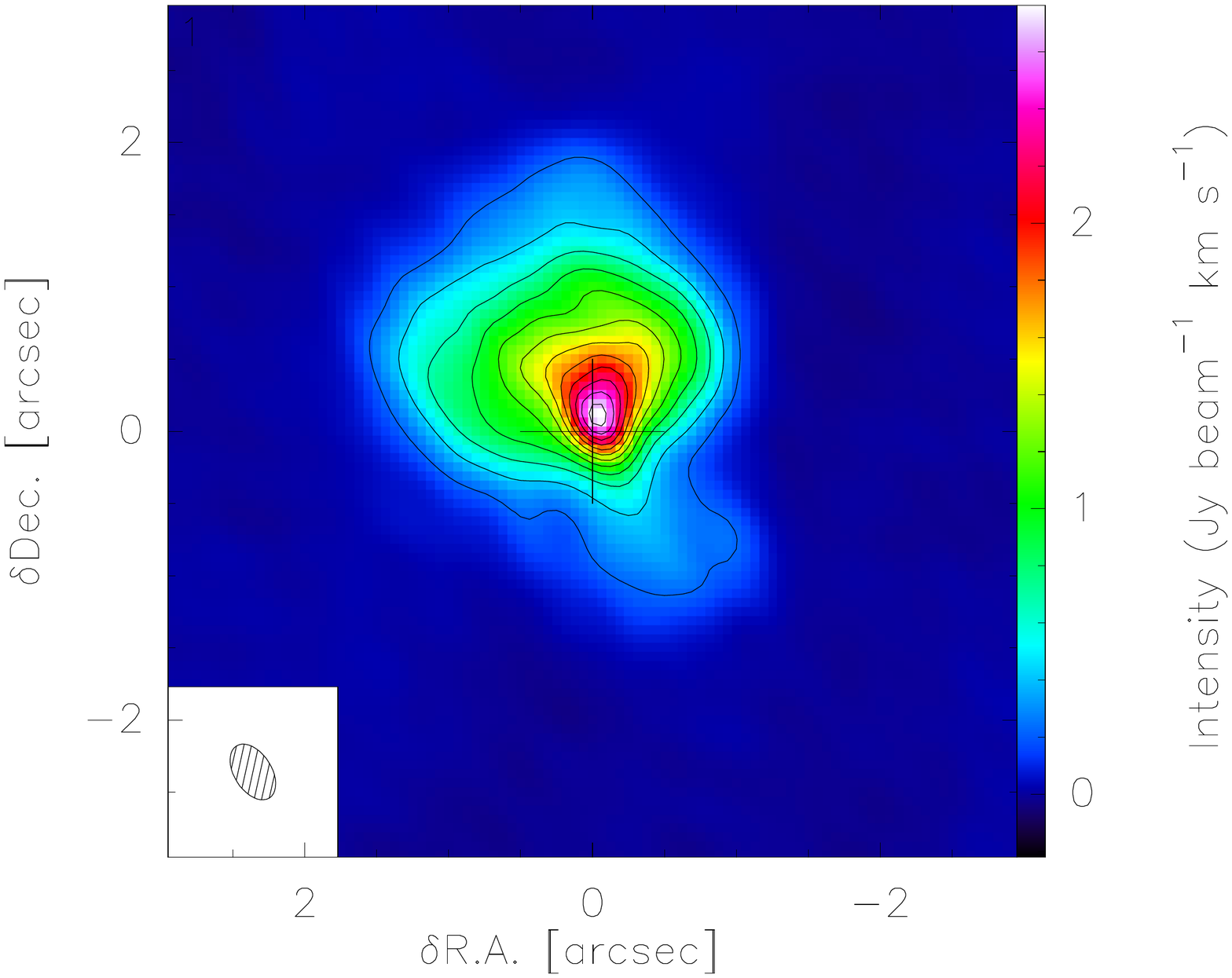}
    \includegraphics[width=60mm]{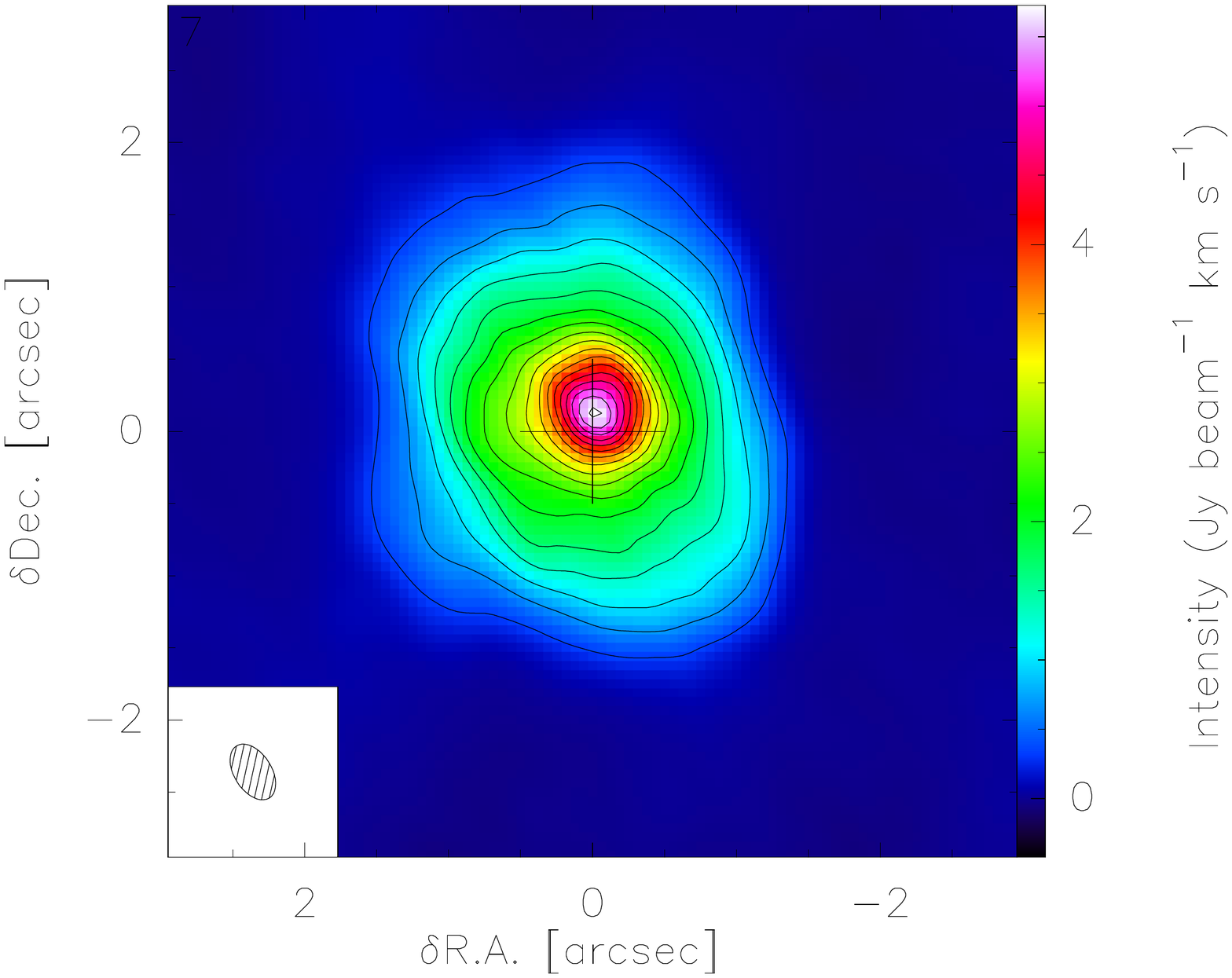}
    \includegraphics[width=60mm]{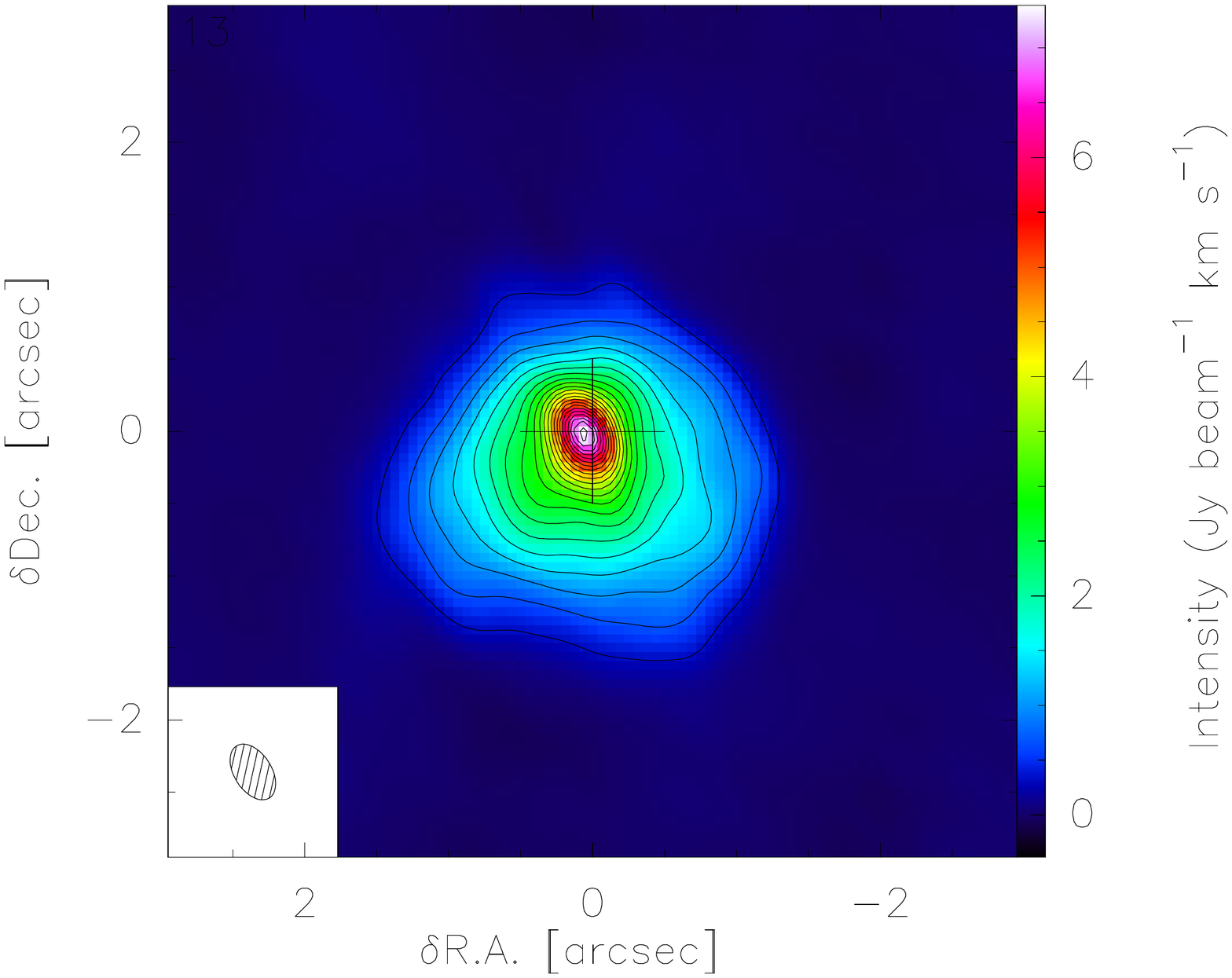}
    % make-blue-center-red_CO_SiO.map
    \caption{Velocity-integrated intensity maps of the $^{12}$CO(2-1) line 
      (upper row) and the vibrational ground state line of 
      SiO(6-5) (lower row),
      covering three velocity intervals.
      Left: blue line wing [v$_{\rm lsr,*}-10$,v$_{\rm lsr,*}-2$] \kms,
      Middle: line center [v$_{\rm lsr,*}-2$,v$_{\rm lsr,*}+2$] \kms,
      Right: red line wing [v$_{\rm lsr,*}+2$,v$_{\rm lsr,*}+10$] \kms.
      North is up and east is to the left. Note the different color scales.
      Contours are plotted every $5 \sigma$ for CO and every $20
      \sigma$ for SiO, where (from left to right) $1 \sigma = 14.6, 22.0,
      19.2$ mJy/beam$\cdot$\kms for CO(2-1) and $1 \sigma = 11.1, 16.7,
      16.4$ mJy/beam$\cdot$\kms for SiO(6-5). The black ellipse in the lower
      left corner indicates the synthesized beam.}
    \label{COSiOoutflowfig}
\end{figure*}

The detailed structure of the morpho-kinematics of the CSE has been
studied earlier using observations of the $^{12}$CO(1-0) and
$^{12}$CO(2-1) molecular line emission.  The analyses of
\citet{hmwng14} and \citet{nhwetal2015} have confirmed the
interpretation of the two-component nature of the Doppler velocity
spectrum originally given by \citet{lwlgm2010}.  The CSE is
axi-symmetric about an axis making an angle of $i \sim 30^\circ$ with
the line of sight and projecting  on the plane of the sky at a
position angle $\omega \sim 7^\circ$ east of north  (see also the
sketch in Fig.~\ref{geomsketch}).  The expansion velocity reaches
$\sim 8$ to $9$\kms along the axis,  we refer to this part of the CSE
as bipolar outflow, and $\sim 3$ to $4$\kms in the plane perpendicular
to the axis, we refer to this part of the CSE as equatorial
enhancement. The transition from the equator to the poles of the CSE
is smooth.  Section~\ref{cokinesec} below, using observations of the
$^{12}$CO(2-1) and $^{13}$CO(2-1) molecular lines, will confirm and
significantly refine this picture. It shows in the right panels of
Fig.~\ref{COomegdeproj}  projections of the CSE on the plane
containing the axis and perpendicular  to the plane of the sky, which
give a good qualitative idea of the  global structure.
 
Velocity-integrated channel maps of the CO(2-1) and SiO(6-5)
observations  analyzed in the present article are displayed in
Fig.~\ref{COSiOoutflowfig}.  They show clearly the bipolar outflows,
inclined toward the observer in the north and receding in the south.
We note that the red wings are brighter than the blue wings as a
result of  absorption (see Sects.~\ref{cokinesec} and
\ref{siokinesec}) The SiO emitting region is seen to be  significantly
more compact than the CO emitting region; this is in conformity with
observations on many other oxygen-rich AGB stars and is generally
interpreted as the result of SiO molecules condensing on dust grains
and being ultimately dissociated by the interstellar radiation at some
200\,AU from the star, much before CO molecules are (see for example
\citet{sowlk2004}). 

\subsection{Temperature and SO$_2$ abundance}
\label{thermalsec} 

\begin{figure}[ht]
    \centering
    \vspace{-4.5cm}
    \includegraphics[width=90mm]{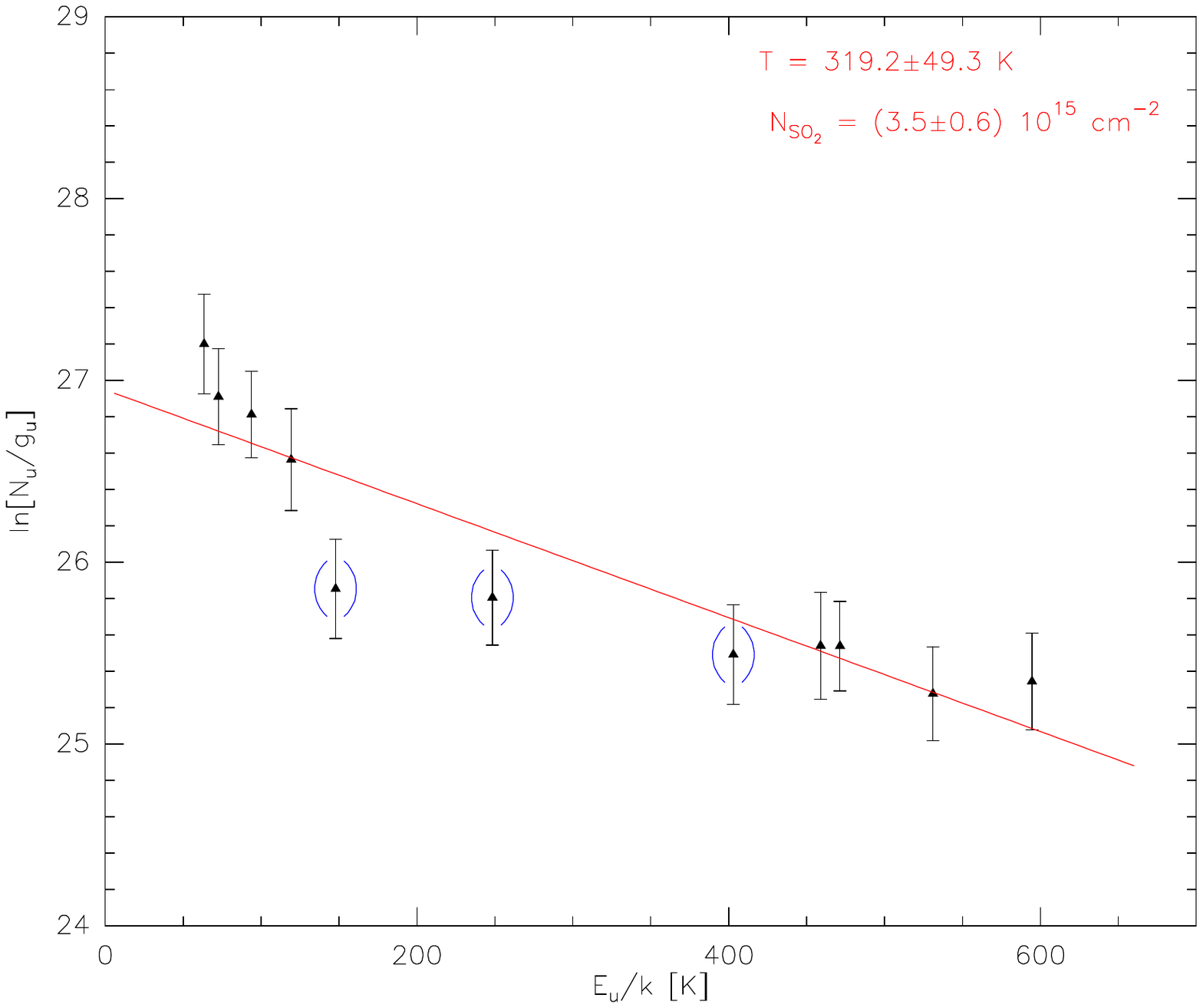}
    \caption{Population diagram for SO$_2$. The three data points in
        brackets correspond to the three lowest frequency SO$_2$ lines,
        observed with setup 1, which, due to a different uv coverage, resulted
        in a comparably larger beam than the setup 2 observations
        (see table~\ref{so2poptab} and Eq.~(\ref{zetaeq})).}
    \label{so2popfig}
\end{figure}

\begin{table*}
  \caption{Parameters of the detected SO$_2$ lines used for the
    population  diagram analysis. A and D-configuration data are merged. Quoted
    errors include the rms errors of the Gaussian fits in the uv plane
    and the absolute flux calibration accuracy of 20\%. The SO$_2$
    line parameters are retrieved from the CDMS, based on the calculations
    by \citet{lov1985} and \citet{mb2005}.
  }
  \begin{center}
  \begin{tabular}{rrrrrc}
\hline
\hline
Frequency & W$_I=\int S(v) {\rm dv}$ & g$_u$ & log$_{10}$(A$_{ul}$) & E$_u/k$ & $\theta_a \times \theta_b$ \\
 GHz      &  Jy km s$^{-1}$          &       &  $s^{-1}$       &    K    &  arcsec$^2$     \\    
 % This is for A+D configuration merged and selfcalibrated
 % reduc01:RS_Cnc/PolyFiX/maps/SO2/so2-merged-data-output-50chan-24_26-selfcal-fluxerr20.txt
\hline
214.6894  &  0.141$\pm$0.0385  & 33  & -4.0043   & 147.843 & 0.90$\times$0.68 \\
216.6433  &  0.166$\pm$0.0434  & 45  & -4.0329   & 248.442 & 0.89$\times$0.67 \\
234.1871  &  0.160$\pm$0.0439  & 57  & -3.8401   & 403.033 & 0.71$\times$0.56 \\
244.2542  &  0.293$\pm$0.0698  & 29  & -3.7855   &  93.901 & 0.69$\times$0.49 \\
245.5634  &  0.170$\pm$0.0451  & 21  & -3.9240   &  72.713 & 0.69$\times$0.49 \\
248.0574  &  0.119$\pm$0.0333  & 31  & -4.0939   & 119.328 & 0.69$\times$0.48 \\
258.3887  &  0.153$\pm$0.0396  & 65  & -3.6773   & 531.100 & 0.63$\times$0.45 \\
258.9422  &  0.192$\pm$0.0524  & 19  & -3.8800   &  63.472 & 0.64$\times$0.45 \\
259.5994  &  0.182$\pm$0.0448  & 61  & -3.6835   & 471.496 & 0.63$\times$0.45 \\
263.5440  &  0.152$\pm$0.0448  & 61  & -3.7227   & 459.038 & 0.61$\times$0.42 \\
265.4820  &  0.168$\pm$0.0448  & 69  & -3.6426   & 594.661 & 0.61$\times$0.42 \\
\hline
\label{so2poptab}
% reduc01:RS_Cnc/PolyFiX/maps/SO2/rscnc-so2-indiv.map
\end{tabular}
\end{center}
\end{table*}

In this subsection, we use the 11 detected SO$_2$ lines to derive an
approximate temperature and column density of the SO$_2$ emitting
region by means of a population diagram.  Following \citet{gl99}), in
the optically thin case,  the column density of the upper level
population $N_u$ of a transition u->l can be expressed as
\begin{equation}
N_u = \frac{8 \pi \rm{k} \nu ^2}{{\rm h} c^3 A_{ul}}  \int T_b {\rm dv} \, .
\end{equation}
$N_u$ is the column density of the upper level population of the
transition, $k$ and $h$ are the Boltzmann and Planck constant,
respectively, $\nu$ is the line frequency, $c$ the speed of light,
$A_{ul}$ is the Einstein coefficient for spontaneous emission of the
transition, and $\int T_b$\,dv is the velocity-integrated main-beam
brightness temperature.  The latter is converted to the surface
brightness distribution of the source $S_\nu$ per beam,  measured by
the interferometer, by means of 
\begin{equation}
T_b = \frac{\lambda^2}{2k\Omega_b} S_\nu
\end{equation}
with $\lambda = \frac{c}{\nu}$ the observing wavelength, and $\Omega_b
=\frac{\pi \theta_a \theta_b}{4 \ln 2}$ with $\theta_a$ and $\theta_b$
the major and minor axis of the synthesized beam.

We determine $\int S(v) {\rm dv} =\vcentcolon W_I$ from a circular
Gaussian fit to the velocity integrated emission in the uv-plane,
where the integration is taken from $({\rm v_{lsr,*}-4.5})$\,\kms to
$({\rm v_{\rm lsr,*}+4.5})$\,\kms, that is over the three central
channels of the  SO$_2$ lines. 

For the population diagram we then get
\begin{equation}
\ln \left(\frac{N_u}{g_u} \right)          = 
\ln \left( \frac{\zeta_u W_I}{g_u} \right) = 
\ln \left( \frac{N_{\rm SO_2}}{Q_{\rm SO_2,rot}} \right) - \frac{E_u}{kT}
\label{popdiageq}
\end{equation}
where we define $\zeta_u$ as
\begin{equation}
\zeta_u = \frac{4.784 \times 10^{-7}}{{\rm h} c A_{ul}}
\frac{1}{\theta_a \theta_b} \, .
\label{zetaeq}
\end{equation}

In Eq.~(\ref{popdiageq}), $W_I$ is expressed in Jy$\times$\kms,
$\theta_a$ and $\theta_b$ are given in arcsec, $E_u$ is the upper
level energy, $g_u$ the statistical weight of the upper level,  and
$T$ the excitation temperature. All relevant parameters of the SO$_2$
transitions used here are listed in table~\ref{so2poptab}.

In fitting a straight line to the population diagram (shown in
Fig.~\ref{so2popfig}) to determine a rotational temperature according
to Eq.~(\ref{popdiageq}), we assume that the SO$_2$ level populations
are dominated by collisions, i.e., that local thermodynamic equilibrium
(LTE) holds for the rotational excitation of SO$_2$, and that the
lines are optically thin. The assumption of LTE populations may be
questionable for SO$_2$, see \citet{ddbetal2016}, and essentially
could result in an underestimation of the kinetic gas temperature in
the SO$_2$ emitting region (cf. the discussion in \citet{gl99}, their
section~5).  The effect on the derived column density is more
difficult to assess without a detailed non-LTE modeling. We note
however that our result is consistent with the SO$_2$ abundance
derived for similar objects by means of a comprehensive non-LTE
description (see below). The assumption of the lines being optically
thin,  on the other hand, is justified by the results of our XCLASS
modeling, see below and Appendix~\ref{xclassmodels}.

With the temperature of $T\approx320\,$K, resulting from a linear fit
to the population diagram shown in Fig.~\ref{so2popfig}, we get the
partition function $Q_{\rm SO_2,rot}$ (interpolated from values given
in the CDMS), from which an effective SO$_2$ column density of $N_{\rm
  SO_2} = 3.5\times10^{15}$\,cm$^{-2}$ is determined  by means of
Eq.~(\ref{popdiageq}).  The SO$_2$ source is compact (see
table~\ref{obsparalinestab} and Fig.~\ref{SO2structfig}, FWHP
$\approx0.5\arcsec$, corresponding to 75\,AU, or $\approx 70\,R_*$),
consistent with the estimated temperature in this inner (possibly
rotating) region.  Assuming a mass-loss rate of $1 \times 10^{-7}
M_\odot$yr$^{-1}$ in the equatorial region
(cf. Sect.~\ref{cokinesec}), an outflow velocity of at maximum 8\kms
as indicated by the line widths (but excluding the high-velocity wings
discussed in Sect.~\ref{siowingsec}), an inner radius of $\approx
10^{14}$\,cm, and an outer radius of that region of $\approx
10^{15}$\,cm,  corresponding to $0.5\arcsec$, we estimate upper limits
of the SO$_2$ abundance of $X$(SO$_2/$<H>) = $7.3\times10^{-7}$, or,
if hydrogen would be completely bound in H$_2$, $X$(SO$_2$/H$_2$) =
$1.5\times10^{-6}$.  For comparison, \citet{ddbetal2016} find an
SO$_2$ abundance of $\sim 5\times10^{-6}$ for the low mass-loss rate
($\sim 1-2 \times 10^{-7} M_\odot$yr$^{-1}$) oxygen-rich stars R Dor
and W Hya, about a factor $\sim 3$ higher than the value found here
for the MS-type star RS Cnc. We note that 
\citet{2008A&A...480..431D} and \citet{do2020} do not detect SO$_2$ in
the S-type star W Aql. Our result therefore appears to be consistent
with an intermediate chemical state of RS Cnc.
 
In Appendix~\ref{xclassmodels}, an XCLASS modeling is presented that
results in the same SO$_2$ column density of
$3.5\times10^{15}$\,cm$^{-2}$ as derived from the population diagram,
and in a slightly higher rotational temperature of 350\,K, still
within the uncertainties {of $\sim\pm$50\,K} derived here (see
Fig.~\ref{so2popfig}). The XCLASS modeling of the 11 SO$_2$ lines
results in an average optical depth of 0.1, confirming that the lines
are optically thin.

\section{Discussion}\label{discusec}

\subsection{$^{12}$CO(2-1) and $^{13}$CO(2-1): morpho-kinematics of 
  the circumstellar envelope}\label{cokinesec}

Rotational CO lines in the vibrational ground state are known to probe
the circumstellar envelope of AGB stars up to distances on the
1000\,AU scale, where the CO molecules are dissociated by the
interstellar UV radiation \citep{mgh88}.  In this section, we use
observations of the $^{12}$CO(2-1) and $^{13}$CO(2-1) lines to probe
the morpho-kinematics of the wind at distances from the star in excess
of $\sim 50$\,AU, in a region where the wind is expected to evolve
smoothly toward the constant expansion regime.
  
Both $^{12}$CO(2-1) and $^{13}$CO(2-1) data have a Doppler velocity
(${\rm v_z}$) spectrum that shows clearly a two-component structure,
where the blue-shifted wing of the $^{12}$CO line is partly absorbed,
very similar to the corresponding situation in EP Aqr
\citep{thnetal2019}.  CO rotational lines from higher J levels (5-4
and 9-8) observed with  Herschel \citep{dtjetal2015} show triangular
profiles consistent with the double-component wind structure that we
are proposing.

\begin{figure*}
  \centering
  \includegraphics[height=5cm,trim=1cm .5cm .7cm .3cm,clip]{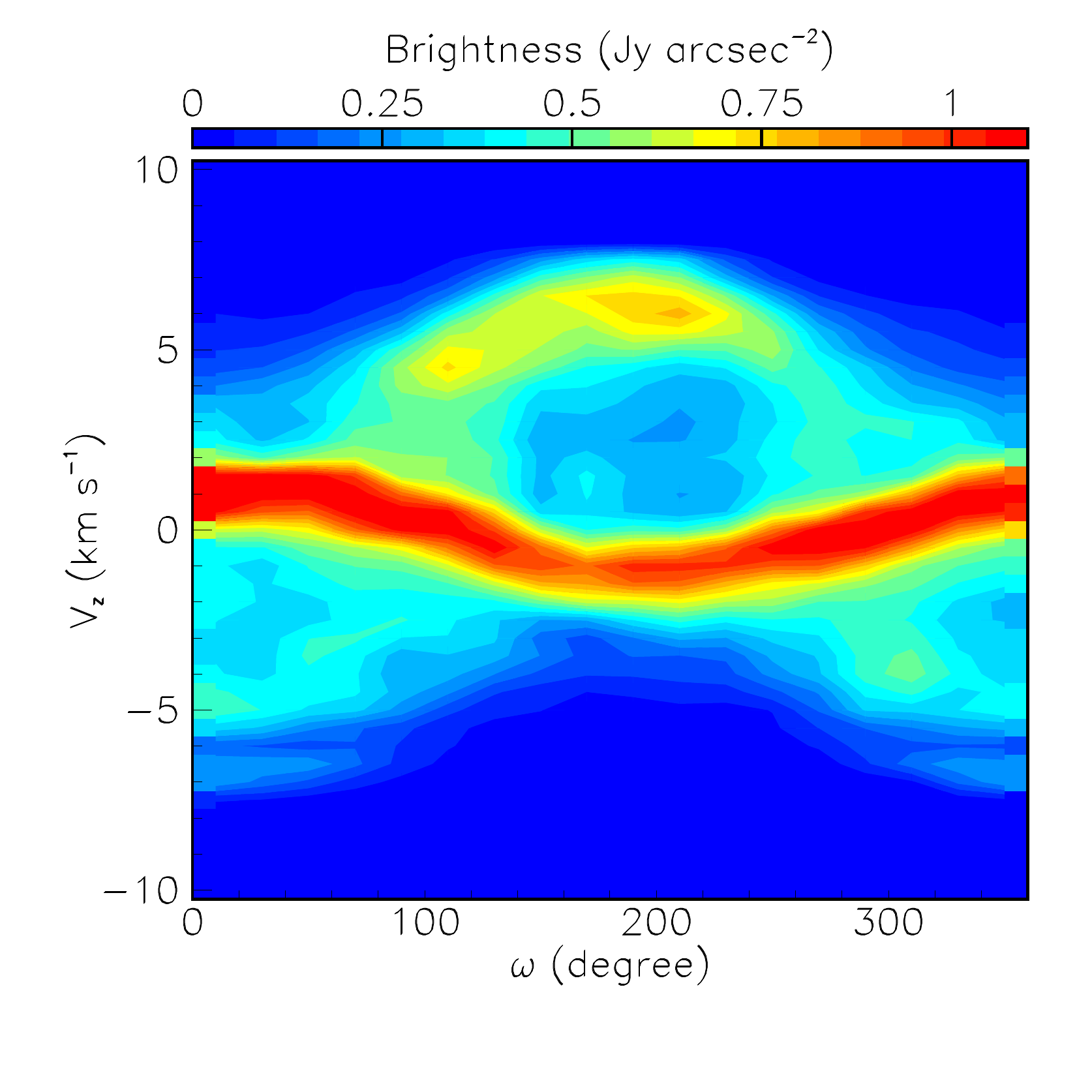}
  \includegraphics[height=5cm,trim=1cm .5cm .7cm .3cm,clip]{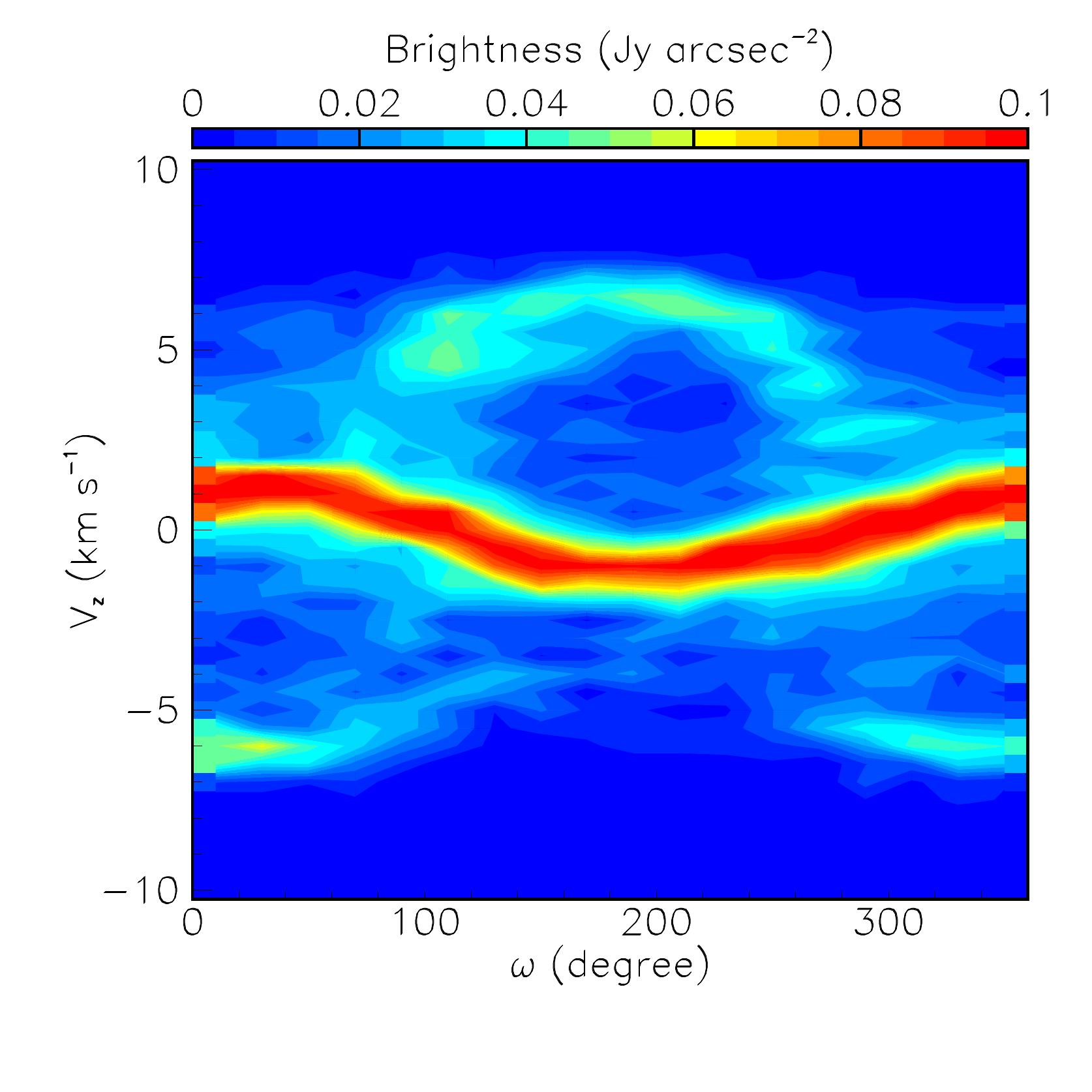}
  \includegraphics[height=5cm,trim=1cm .5cm 2.cm .3cm,clip]{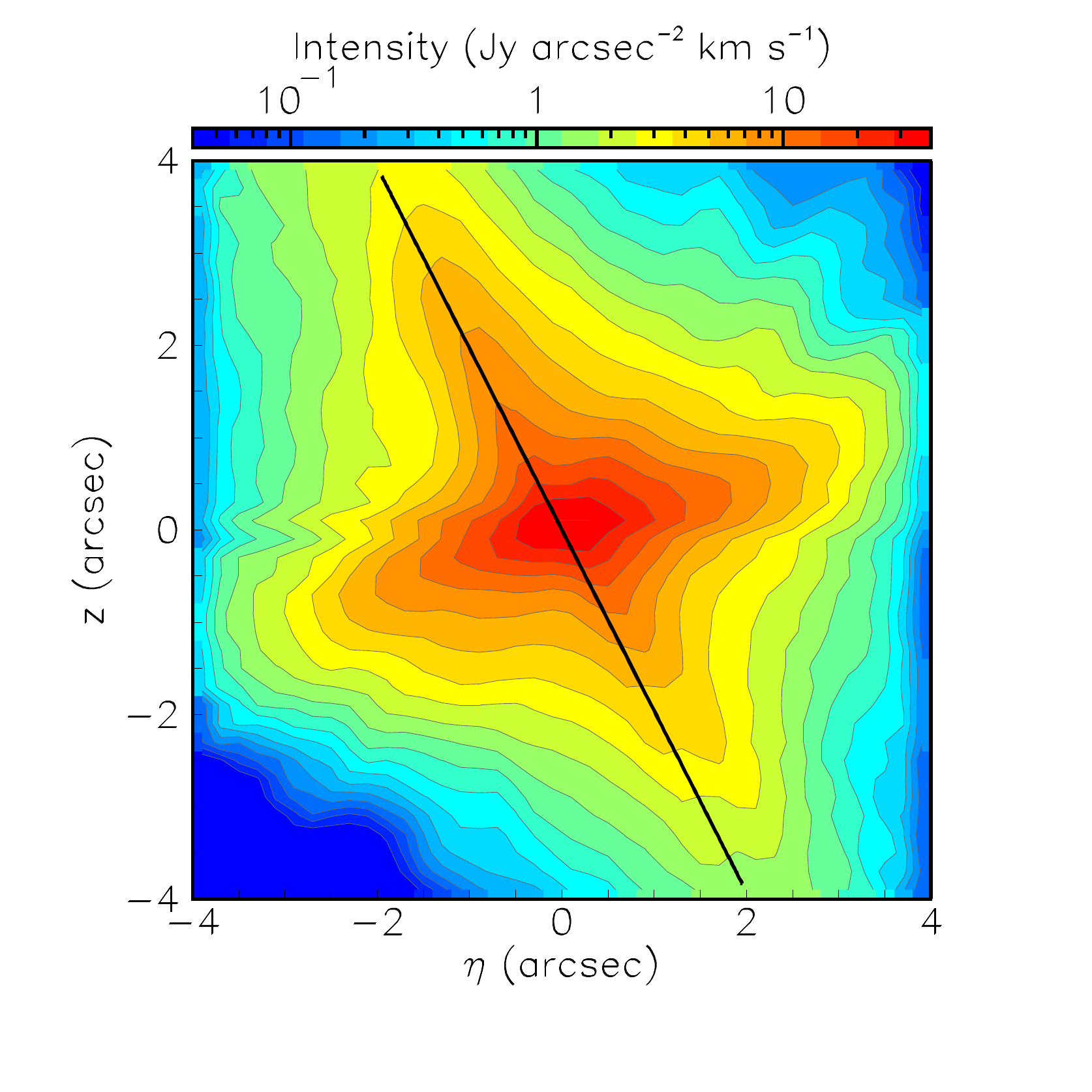}
  \includegraphics[height=5cm,trim=1cm .5cm 2.cm .3cm,clip]{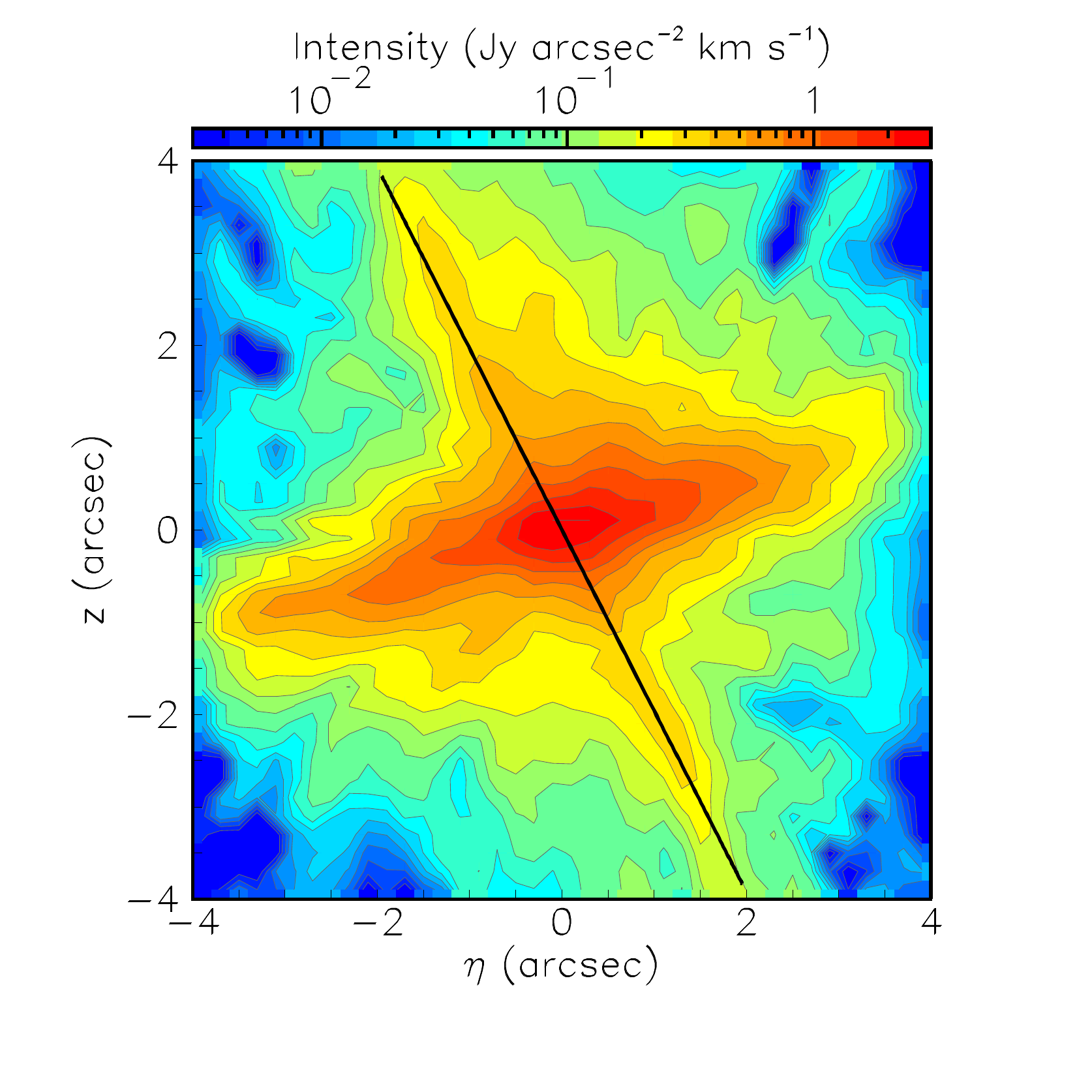}
  \caption{Left panels: flux density averaged over
    $0.4\arcsec<R<2.5\arcsec$ in the ${\rm v_z}$  vs $\omega$ plane for
    $^{12}$CO(2-1) (left) and $^{13}$CO(2-1) (center-left).
    Right panels: de-projected flux-density projected on
    the plane perpendicular to the sky and containing the polar outflow
    axis, shown as black lines, for $^{12}$CO(2-1) (center-right)
    and $^{13}$CO(2-1) (right).}
    \label{COomegdeproj}
\end{figure*}

In the spatial distribution of the CO emission, a clear separation is
observed between the equatorial region and the polar outflows. This is
illustrated in Fig.~\ref{COomegdeproj}, which displays in the leftmost
panels the maps of the flux density in the Doppler velocity (${\rm
v_z}$) vs position angle ($\omega$) plane. The flux density is
integrated over an interval of projected distance ($R$) from the star
between 0.4\arcsec and 2.5\arcsec. The equatorial region is seen as an
intense oscillation at low values of $|{\rm v_z}|$ while the  polar
outflows are seen as emission at larger $|{\rm v_z}|$ values, in phase
opposition. From these maps we estimate, in agreement with earlier
findings, that the outflow axis projects on the plane of the sky at a
position angle of $\omega = 7^{\circ}\pm5^{\circ}$, at which the
equatorial and bipolar outflow oscillations are maximal and minimal,
respectively. The right panels show de-projections of the data cubes
projected on the plane perpendicular to the plane of the sky that
contains the polar outflow axis. They are drawn assuming a dependence
of the wind velocity on stellar latitude $\alpha$ of the form ${\rm
v_{term}} = 4 + 5\sin^4\alpha$ \kms, increasing from 4 to 9\kms from
the equator to the poles. The velocity field is assumed to be radial
and to have reached its terminal value in the $R$ interval considered
here.

\begin{figure*}
  \centering
  \includegraphics[height=5cm,trim=1cm .5cm .7cm .3cm,clip]{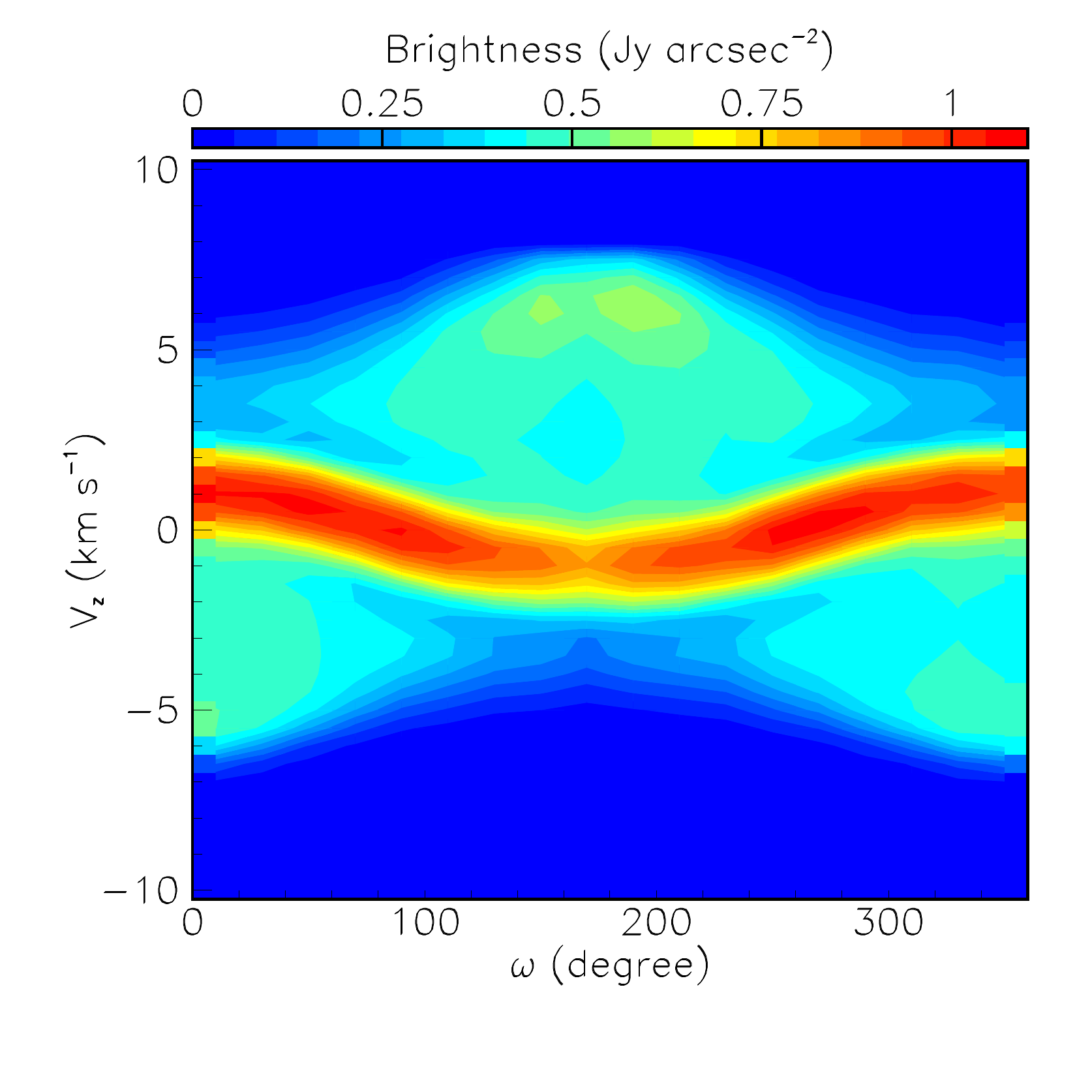}
  \includegraphics[height=5cm,trim=1cm .5cm .7cm .3cm,clip]{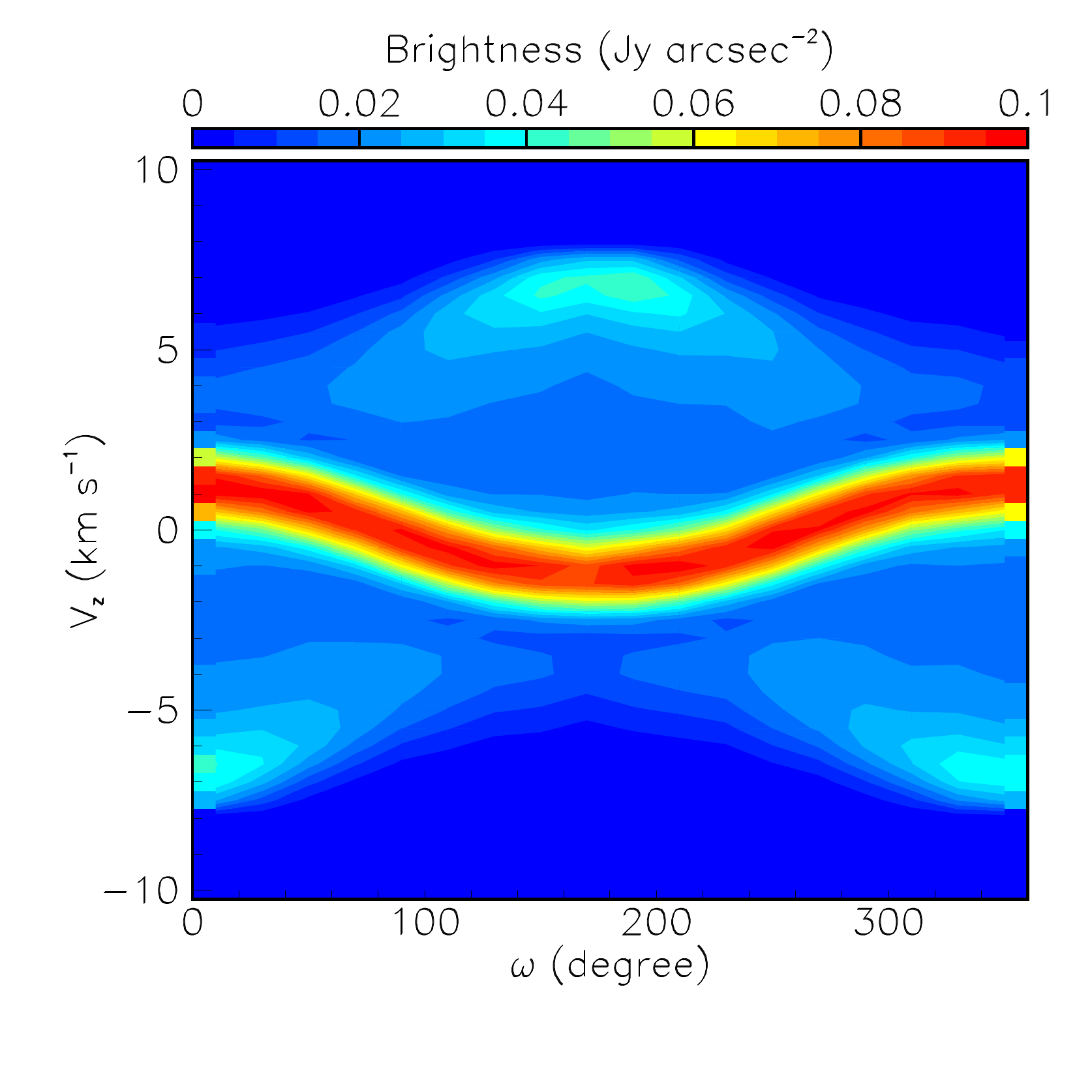}
  \includegraphics[height=5cm,trim=1cm .5cm 2.cm .3cm,clip]{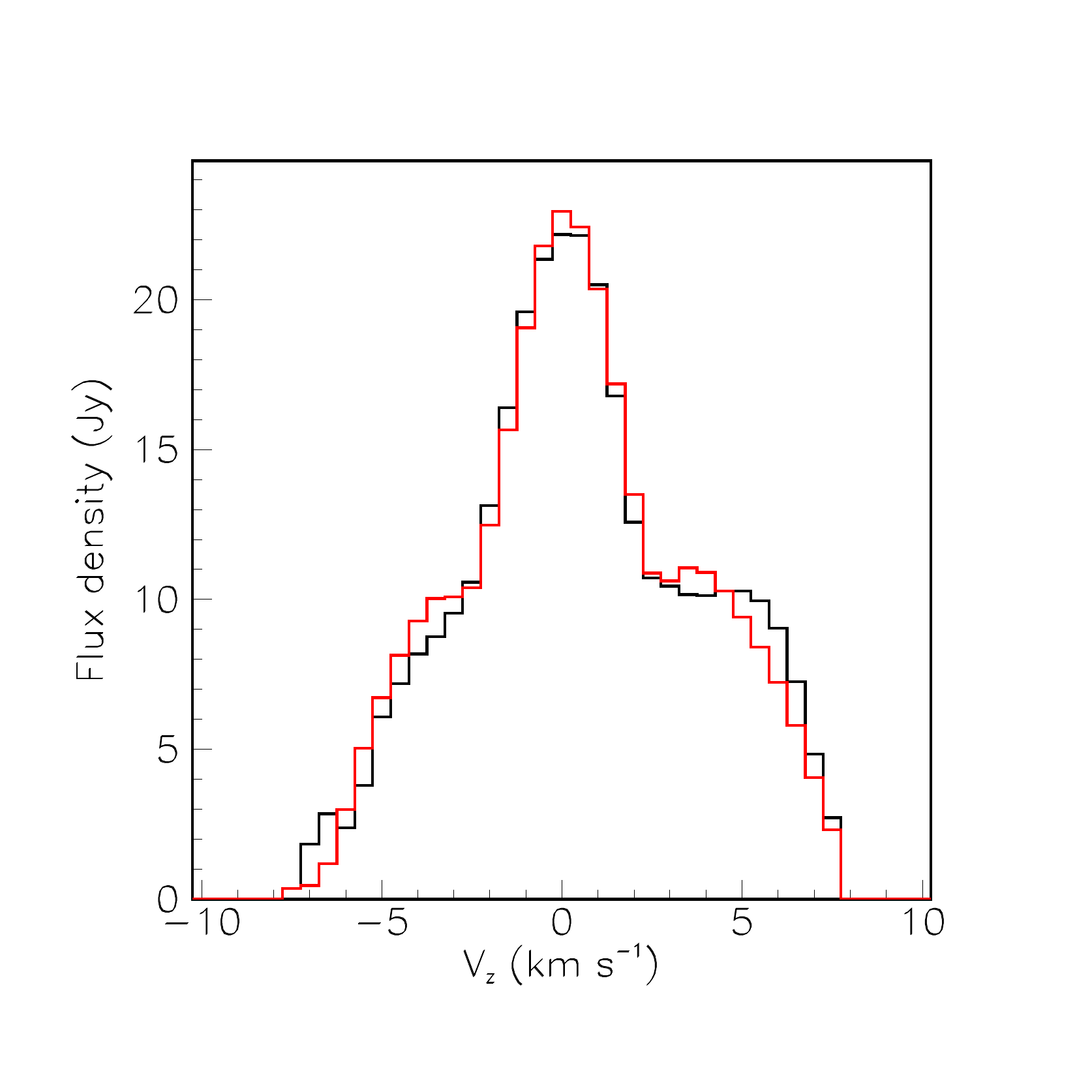}
  \includegraphics[height=5cm,trim=1cm .5cm 2.cm .3cm,clip]{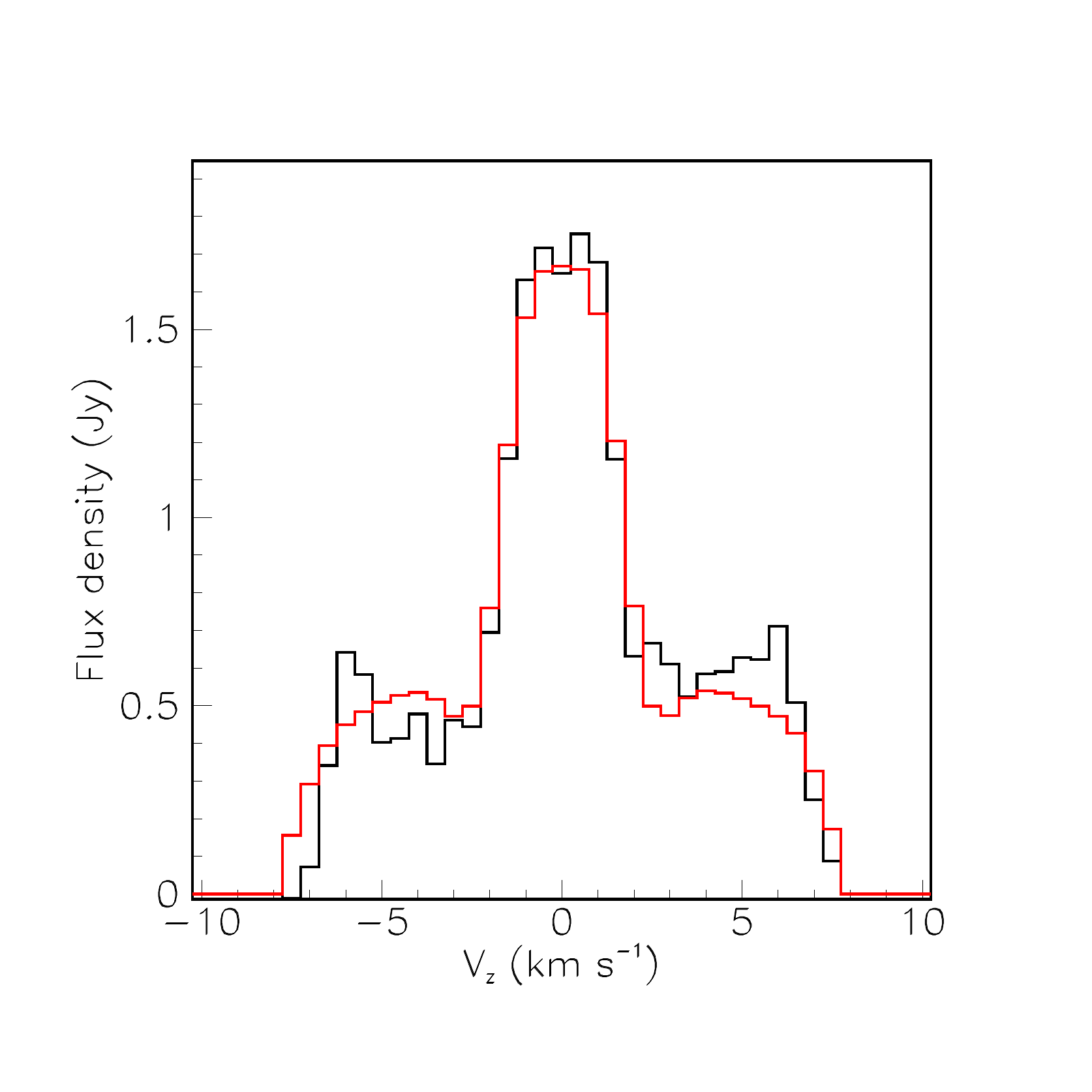}
  \caption{Best fit results for CO. Left panels: same as in
      Figure~\ref{COomegdeproj}, but for the model best fit. Right
      panels: modeled ${\rm v_z}$ spectra (red) are compared with
      observations (black) for $^{12}$CO(2-1) (center-right) and
      $^{13}$CO(2-1) (right).}
  \label{COmodelfig}
\end{figure*}

In order to interpret these observations, in particular possible
differences of the $^{12}$CO/$^{13}$CO ratio between equatorial region
and polar outflows, we need to take into account the effect of
absorption. We choose to consider only regions outside 0.4\arcsec
$\sim$60\,AU from the star, which allows for major simplifications: we
can ignore the very complex kinematics that governs the inner CSE
layers, including shocks and possible rotation.  As we cannot expect
the data to strongly constrain the temperature distribution, we take
it of the form $T=T_0\exp(-r/r_0)$ and fit the observed data cubes by
means of a $\chi^2$ minimization scheme for different values of the
($T_0,r_0$) pair consistent with earlier estimates
\citep{2015RAA....15..713N}.  We find that the morpho-kinematics is
best described by modeling separately polar outflows and equatorial
region, with a separation at stellar latitude $\alpha_0 \sim
30^{\circ}$. We parametrize the radial dependence of the wind velocity
as ${\rm v=v_{term}} r/(r+r_{1/2})$: it increases from 0 at $r=0$ to
${\rm v_{term}}$ at $r=\infty$, reaching $\frac{1}{2} {\rm v_{term}}$
at $r=r_{1/2}$. In order to account for different opening angles of
the polar outflows and different flaring angles of the equatorial
enhancement, we allow,  within each region, for latitudinal
dependences of the velocity described as Gaussian functions in $\sin
\alpha$ in the equatorial region and in $\cos \alpha$ in the polar
outflows with adjustable Gaussian widths $\sigma$.

We do the same for the number densities, which vary radially  as
$\rho_0 ({\rm v_{term}/v})r^{-2}$ with different $\rho_0$ values for
the equatorial region and the polar outflows, and of course for
$^{12}$CO and $^{13}$CO. The radiative transfer equation is integrated
in a sphere of $10\arcsec$ radius and the calculated flux densities
are then smeared with the synthesized beam of the observations.
Reasonably good fits of the data cubes are obtained given the
crudeness of the model (see Fig.~\ref{COmodelfig}).

\begin{table*}
  \caption{Parameters of the best fit model for CO. The quoted uncertainties
    correspond to a 10\% increase of the best fit $\chi^2$, leaving
    all other parameters fixed at their best fit value: they should
    not be understood as uncertainties but as indicators of the
    sensitivity of the quality of the fit to each parameter
    separately.}
  \begin{center}
  \begin{tabular}{cccccccccc}
    \hline
    \hline
  Region  & ${\rm v_{term}}$ & $r_{1/2}$       & $T_0$ & $r_0$  & opening angle & $\sigma(\rho)$ & $\sigma({\rm v})$ & $\rho_0$($^{12}$CO) & $\rho_0$($^{13}$CO) \\
          &   \kms          &  arcsec       &  K    & arcsec & (FWHM) deg   &                &                   & cm$^{-3}$           & cm$^{-3}$    \\
  \hline
  equator &$3.5\pm0.7$      & $0.6\pm0.2$   & 100   & 2.0    &  30           & $0.21\pm0.03$  & $>0.4$            & $86\pm10$   & $4.2\pm0.4$ \\
  poles   &$8.6\pm0.5$      & $0.29\pm0.10$ & 130   & 2.0    &  70           & $0.50\pm0.02$  & $0.70\pm0.04$     & $70\pm6$    & $2.9\pm0.4$ \\
  \hline
  \label{fitparatab}
  \end{tabular}
  \end{center}
\end{table*}

We find that the best fits to the morpho-kinematics of the two lines
are relatively insensitive to the exact form assumed for the
temperature structure. For $r_0=2$\arcsec, we find that values of
$T_0$ between 70\,K and 170\,K in the equatorial region and between
100\,K and 200\,K in the polar outflows are acceptable. Changing the
value of $r_0$ modifies the values of $T_0$ accordingly but does not
improve the quality of the fit. Parameters of the best fit model
include an angle of the polar outflow axis with the line of sight $i
\approx 30^{\circ}$ as expected, a separation in latitude between
polar and equatorial regions, $\alpha_0 = 29\pm4\deg$, terminal
velocities of 3 to 4\kms in the equatorial region, and 8 to 9\kms in
the polar outflows, outflow opening angle and equatorial flaring angle
of $\sim 70^{\circ}$ and $\sim 30^{\circ}$, respectively (FWHM
$=2.35\times\arccos\left(\sigma(\rho)\right)$ and
$2.35\times\arcsin\left(\sigma(\rho)\right)$, respectively), these are
listed in table~\ref{fitparatab}.  The wind is still being accelerated
(the escape velocity of a $1.5M_{\odot}$ star at a distance of 150\,AU
($\approx 1\arcsec$ for RS Cnc) is 4.2\kms), having reached half
terminal velocity at the inner edge of the observed radial range,
earlier in the poles than in the equatorial region. The number
densities of CO molecules correspond to mass-loss rates of $1.0 \times
10^{-7}M_{\odot}$yr$^{-1}$ in the equatorial region and $2.0 \times
10^{-7}M_{\odot}$yr$^{-1}$ in the outflows\footnote{This is about a
factor 2 larger than the values quoted in \citet{hmwng14}. Their data
were affected by a pointing offset of the old 30m OTF maps that lead
to an underestimation of the CO line flux of about a factor 2.} for a
CO/H$_2$ ratio of $2 \times 10^{-4}$.  The $^{12}$CO/$^{13}$CO ratio
is measured to be $\sim 20$ on average, but larger in the polar
outflows (24$\pm$2) than in the equatorial region (19$\pm$3). This is
a barely significant difference, but a similar asymmetry seems to be
present in EP Aqr \citep{thnetal2019}: such a result is unexpected and
needs to be confirmed by higher sensitivity observations before being
accepted. Indeed, if it were, it might suggest that the polar outflows
are fed in part from material freshly produced in the 3-$\alpha$
process and mixed into the atmosphere by the third dredge-up following
a He shell flash.  This process would not only increase the $^{12}$C
abundance in the atmosphere, but also the $^{12}$C/$^{13}$C isotope
ratio as indicated, e.g., by \citet{1990ApJS...72..387S}, see in
particular their figure~9.

The presence of an equatorial density enhancement with a rather small
flaring angle, $\sim 30^\circ$ FWHM, suggests that it may be in  fact
a disk, which may be expected to be rotating and to have an inner
rim. The size of the beam is however too large to study this
reliably. From a close inspection of the $< {\rm v_z} >$ distribution
near the star,  using the method described in
Sect.~\ref{rotdiskmasssec}  for HCN and SiO, we infer a rotation
velocity at $r \sim 0.5\arcsec$ of ${\rm v_{rot}}=|{\rm v_z}|~/\sin i
\sim 2.5$\kms.  This is however a very crude estimate given the size
of the beam and the lack of precise knowledge of the morpho-kinematics
in the innermost radial range.

\begin{figure*}
  \centering
  \includegraphics[height=6.5cm,trim=0.3cm 0.5cm 0cm 1cm,clip]{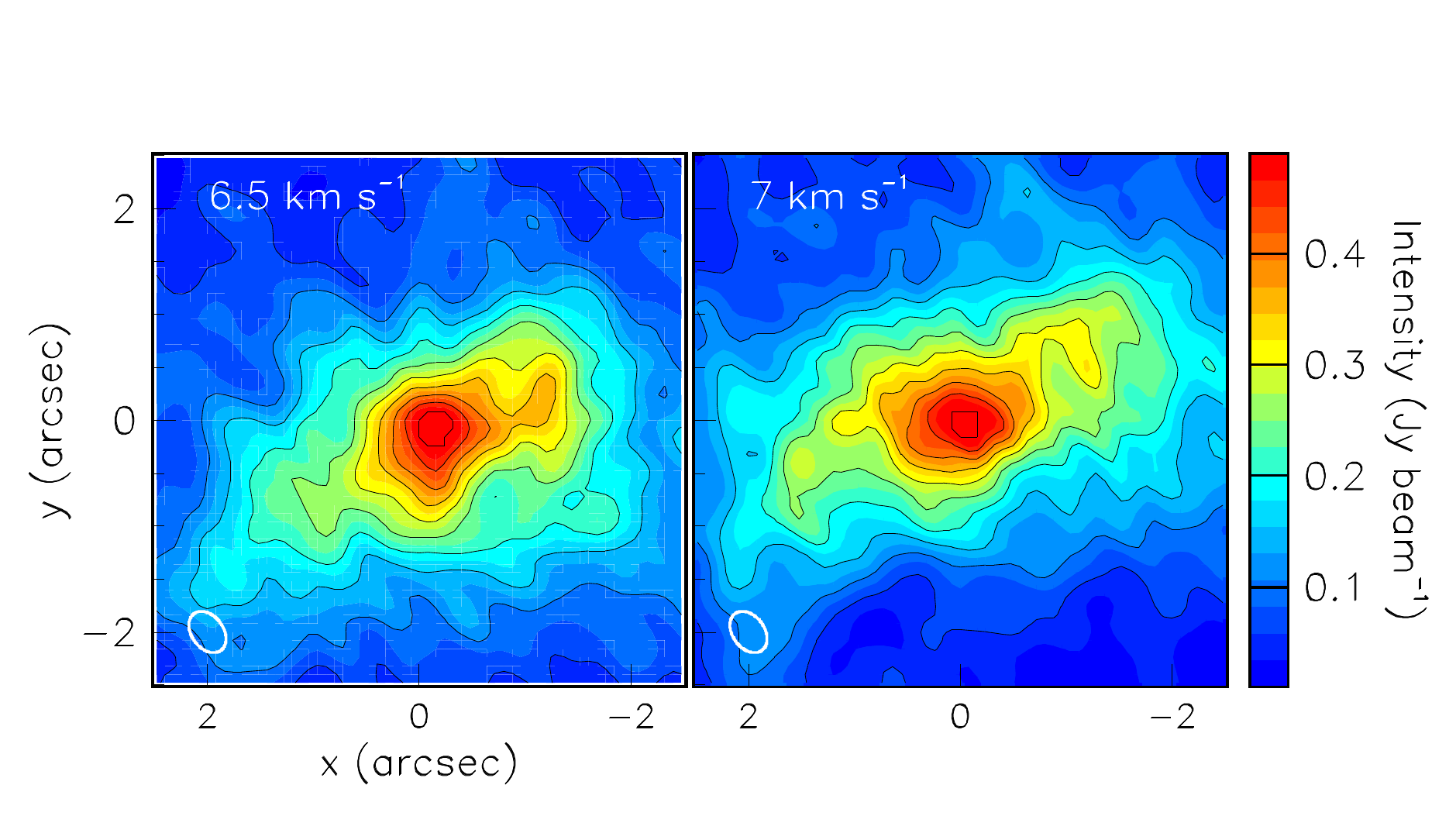}
      \includegraphics[width=60mm]{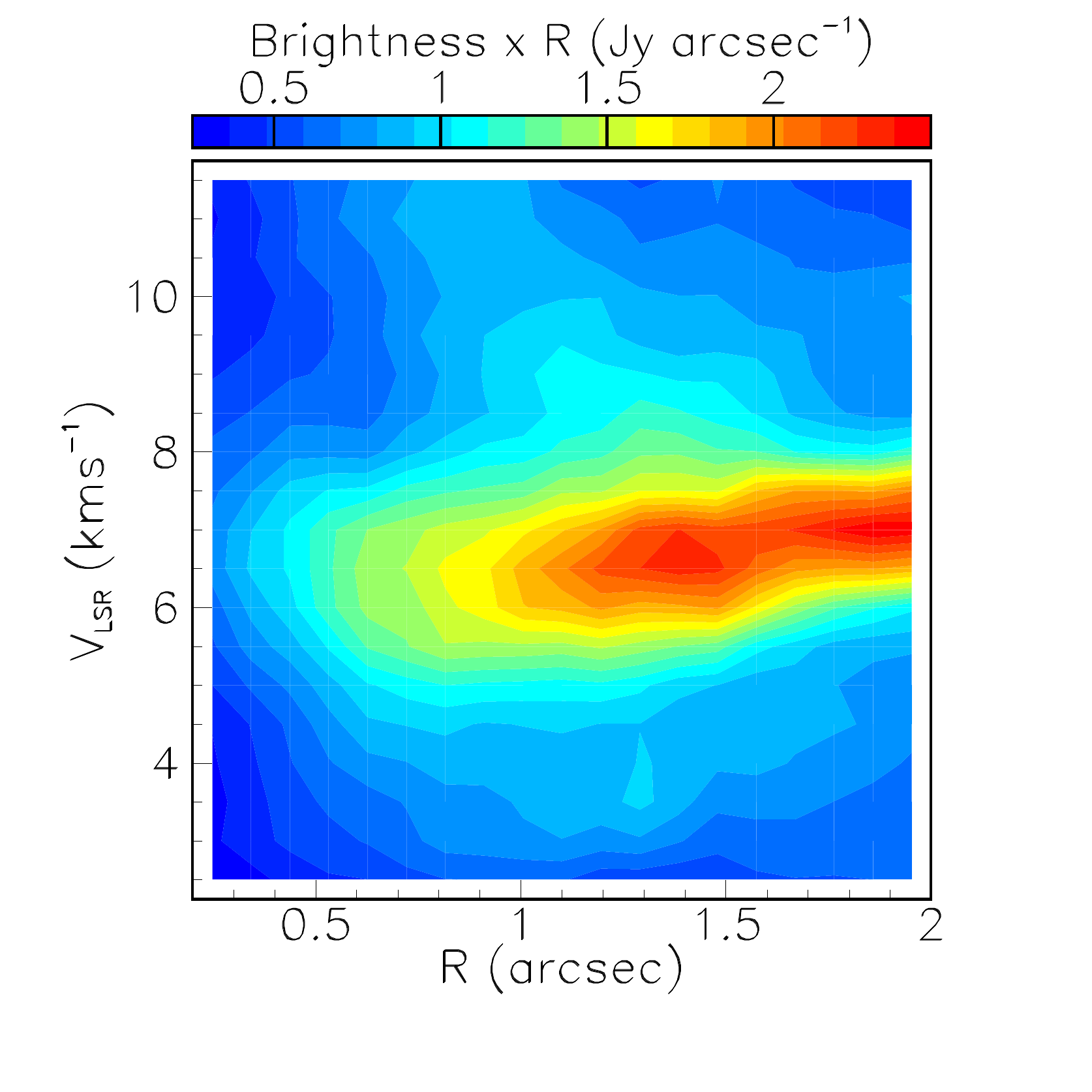}
      \includegraphics[width=60mm]{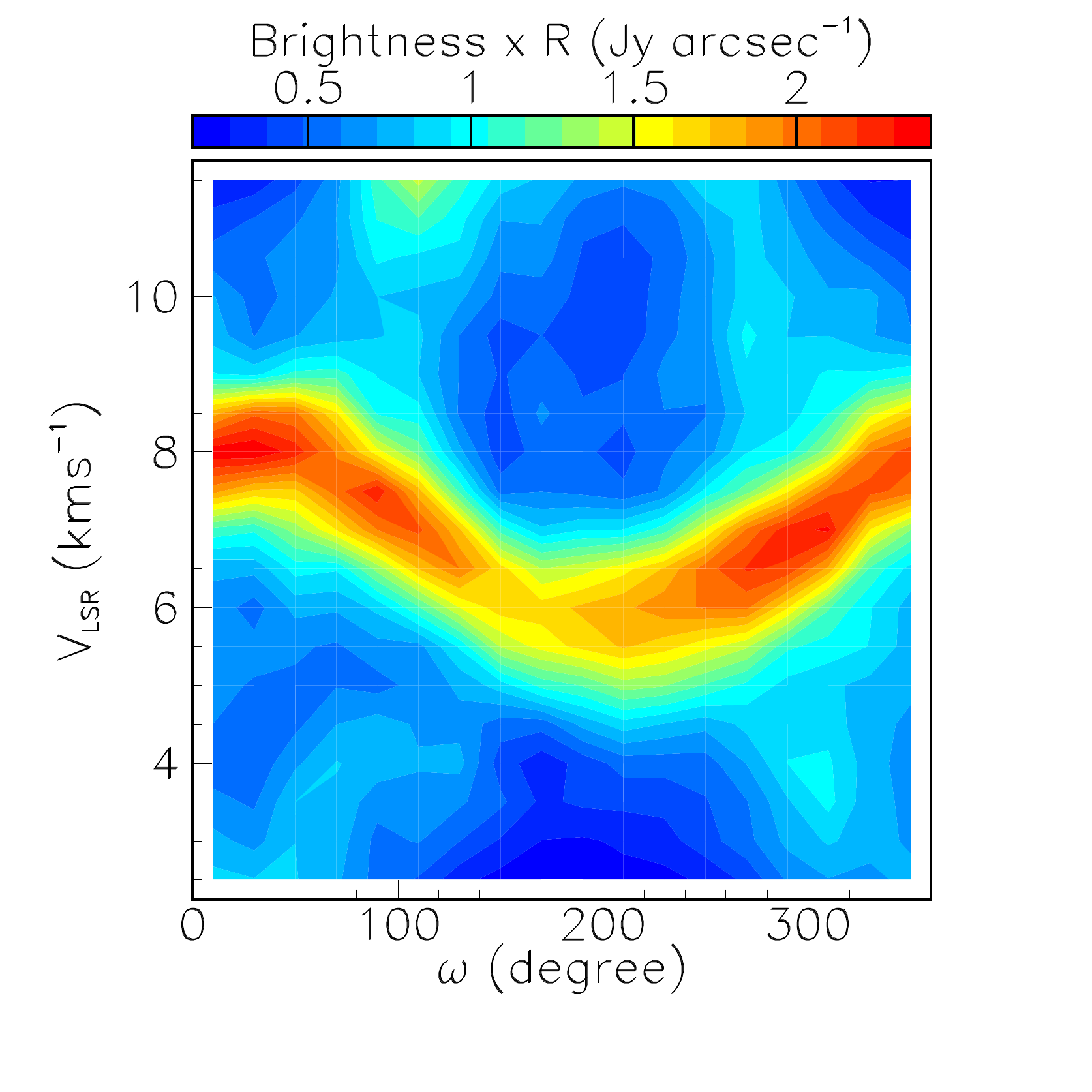}
      \includegraphics[width=60mm]{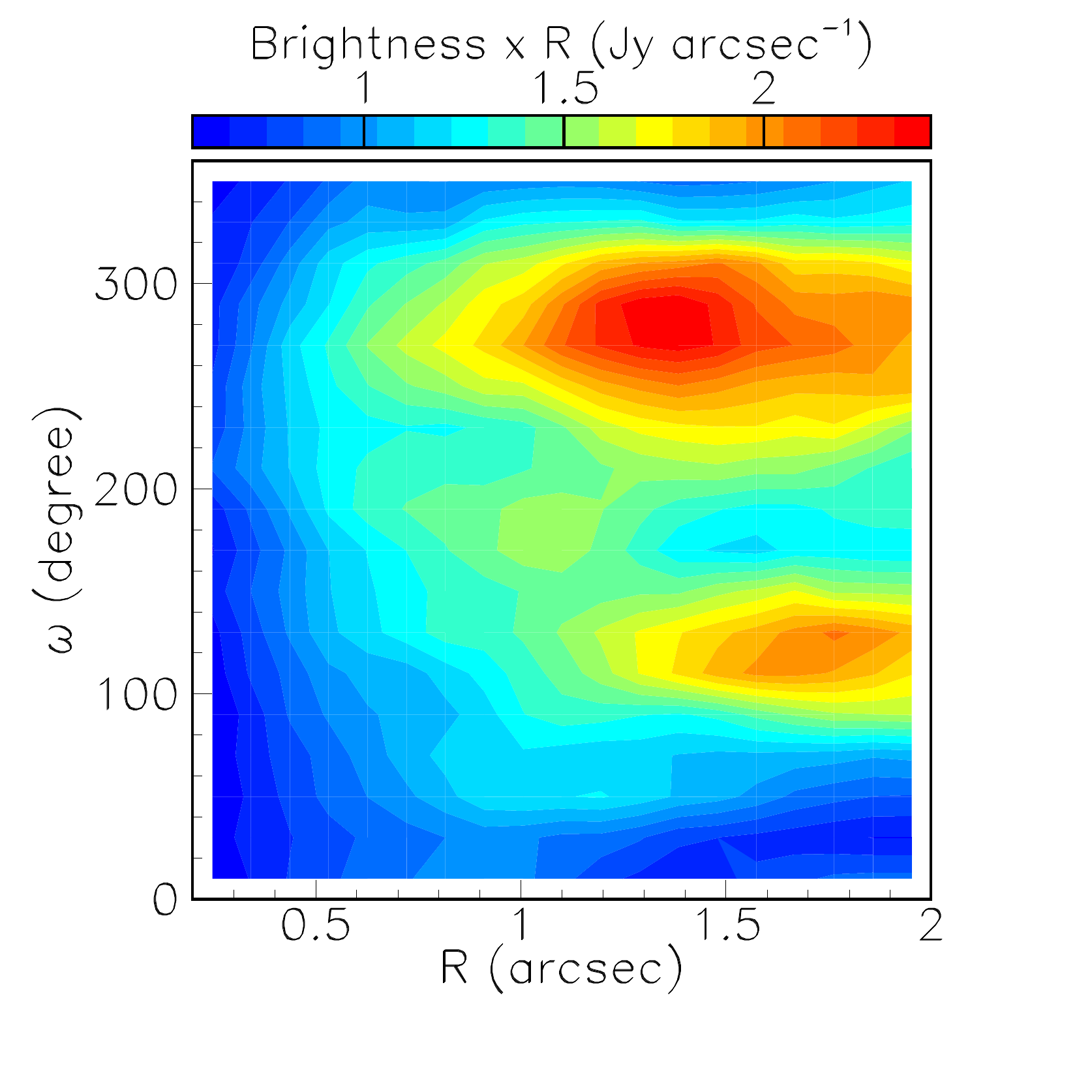}
 \caption{Upper panels: Channel maps of the $^{12}$CO(2-1) line
                        emission between 6.25 and 7.25 \kms. 
          Lower panels:
                       projection of the flux density multiplied by $R$ on the 
                       v$_{\rm lsr}$ vs $R$ plane for 
                       250$^{\circ}<\omega<$330$^{\circ}$ (left), on the
                       v$_{\rm lsr}$ vs $\omega$ plane for 0.5$<R<$2\,arcsec 
                       (middle), and on the $\omega$ vs $R$ plane for
                       5.75$<$v$_{\rm lsr}<$7.25\kms (right).}
 \label{companionfig}
\end{figure*}

The blob of enhanced $^{12}$CO(2-1) line emission that had been
identified in \citet{hmwng14} as possibly suggesting the presence of a
companion is seen on the channel maps in the v$_{\rm lsr} =$ 6 to
7\kms range as an elongation in the west-north-west/east-south-east
direction (Fig.~\ref{companionfig}).  Projections of the data cube on
different planes (v$_z$ vs $\omega$, v$_z$ vs $R$ and $\omega$ vs $R$)
in its neighborhood show that it can be described as a pair of
elongations at position angles of $\sim120^\circ$ and $\sim270^\circ$,
the latter being significantly more  intense than the former and
covering a broad range of $R$ between $1$ and $2\arcsec$.  While these
features provide no justification for a possible identification of a
companion, they cannot be used either as arguments against the
presence of an unobserved companion.

\subsection{SiO(5-4) and SiO(6-5): evidence for strong absorption}
\label{siokinesec}

In the present section we compare the SiO(5-4) and SiO(6-5) line
emission with the $^{12}$CO results described in the previous section.
      
\begin{figure*}
  \centering
  \includegraphics[height=6.2cm,trim=0.5cm .5cm .7cm .3cm,clip]{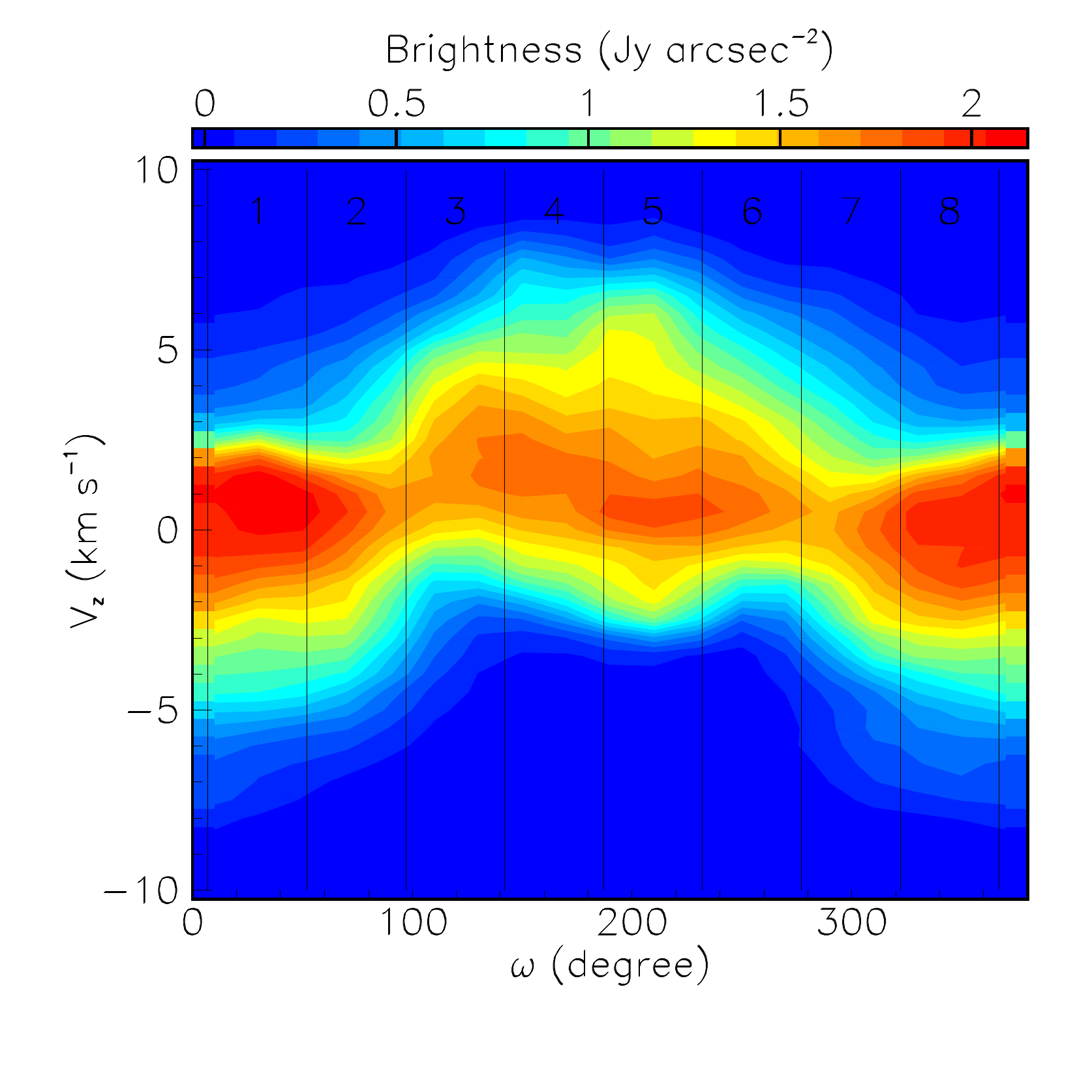}
  \includegraphics[height=6.2cm,trim=0.5cm .5cm .7cm .3cm,clip]{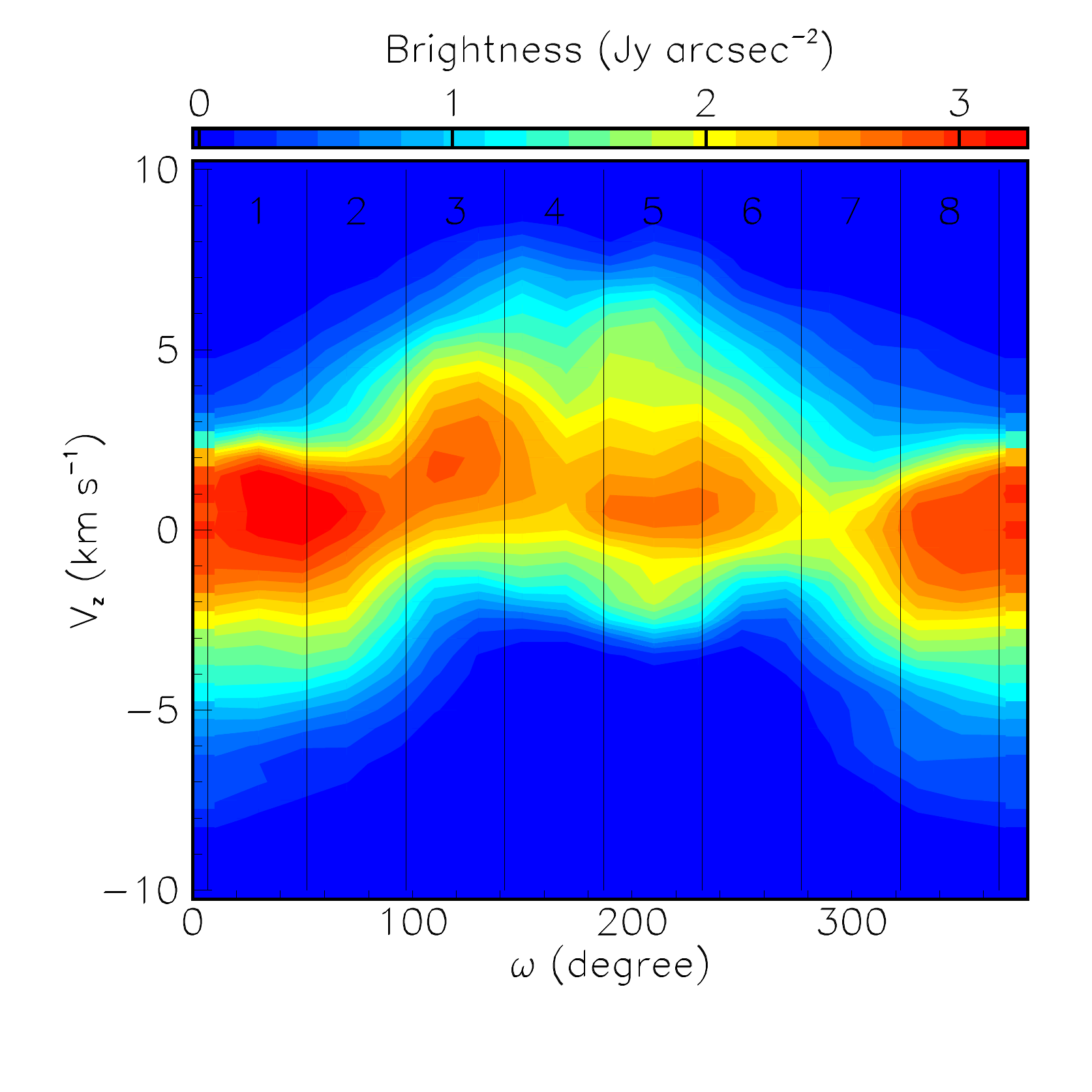}
  \includegraphics[height=6.2cm,trim=1cm .5cm 2.cm .3cm,clip]{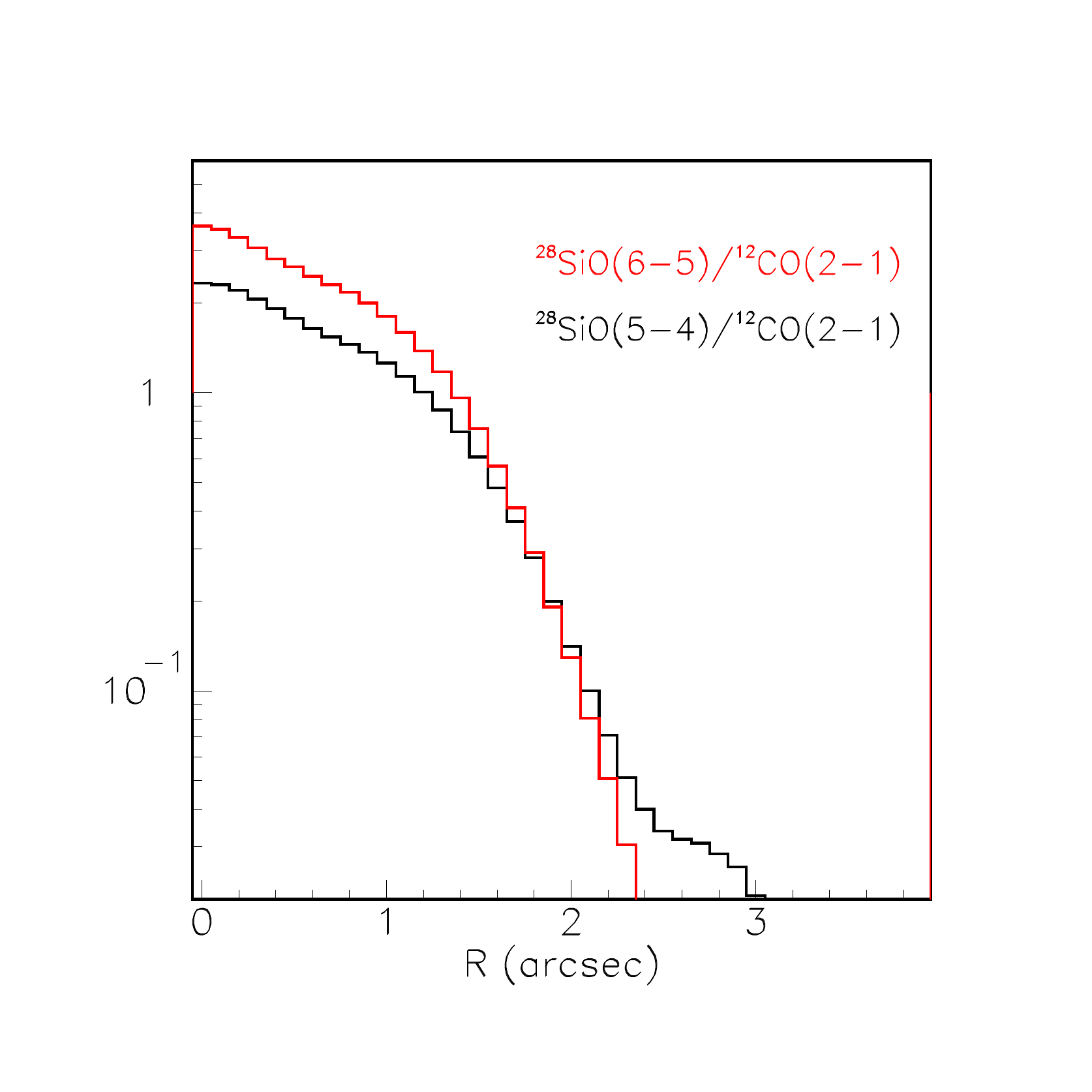}  
  \caption{${\rm v_z}$ vs $\omega$ maps for
    $0.5\arcsec<R<1.5\arcsec$. Left: SiO(5-4), Middle: SiO(6-5),
    octant intervals in $\omega$ as discussed in the text are indicated.
    Right: intensity ratio SiO(5-4)/$^{12}$CO(2-1) (black) and
    SiO(6-5)/$^{12}$CO(2-1) (red) as a function of $R$ for $|{\rm v_z}| <
    8$\kms.}
  \label{SiOwavefig}
\end{figure*}

In contrast to CO, the SiO emission does not resolve the equatorial
region from the polar outflows, as illustrated in the leftmost panels
of Fig.~\ref{SiOwavefig}.  Part of the reason is the much smaller
radial range being probed, as illustrated in the right panel of
Fig.~\ref{SiOwavefig}. As mentioned earlier in
Sect.~\ref{kinestrucsec}, the radial extent of the SiO emission is
often significantly smaller than that of the CO emission, usually
interpreted as evidence for the progressive condensation of SiO
molecules on dust grains \citep[e.g.,][]{sowlk2004}. This would cause
the progressive decline of the SiO/CO ratio observed in the right
panel of Fig.~\ref{SiOwavefig}  up to $R\sim1.5\arcsec$, followed by a
more abrupt cut-off around $R\sim2\arcsec$, caused by the dissociation
of the SiO molecules by the interstellar UV radiation.
         
\begin{figure*}
  \centering
  \includegraphics[height=6.5cm,trim=1cm .5cm 2.cm .5cm,clip]{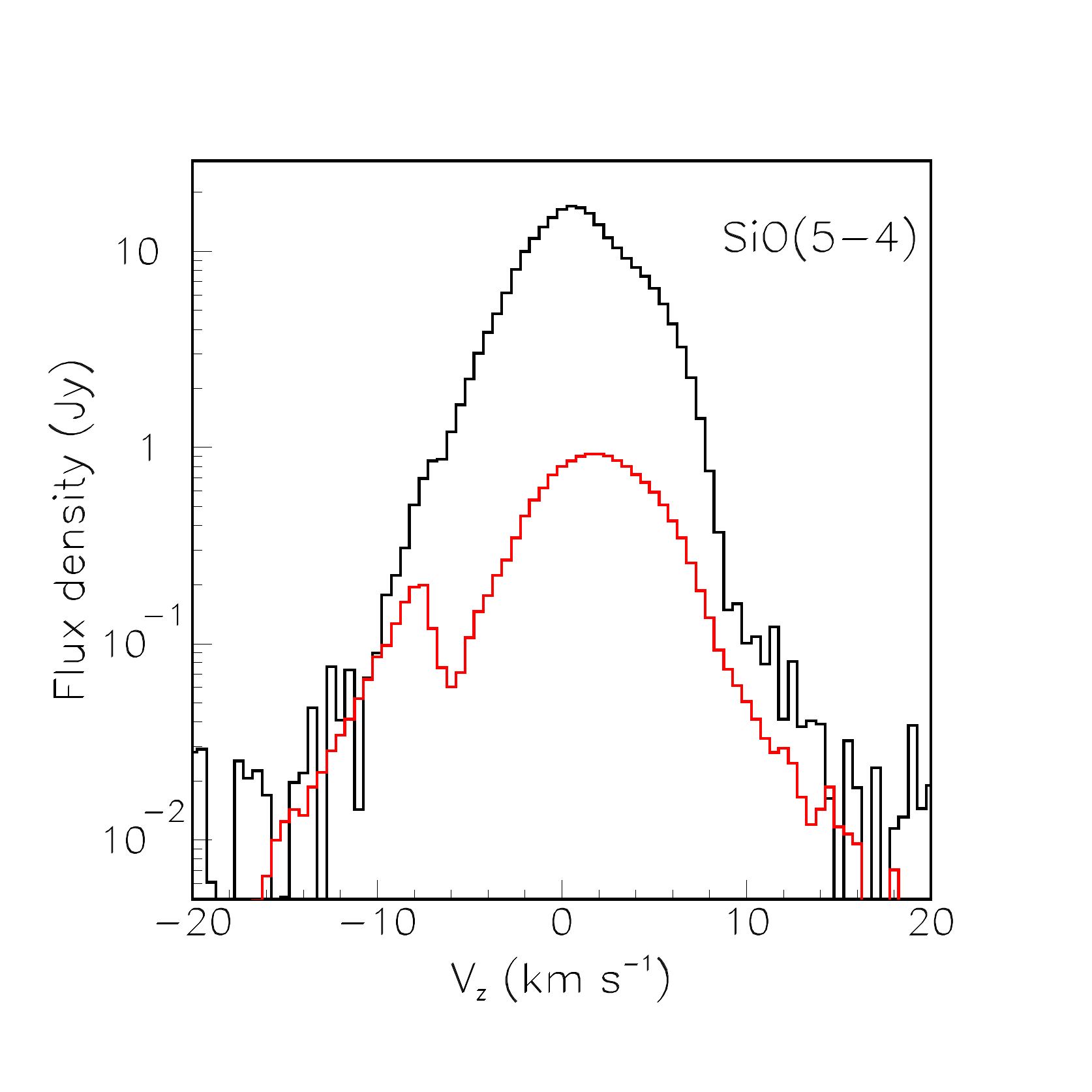}
  \includegraphics[height=6.5cm,trim=1cm .5cm 2.cm .5cm,clip]{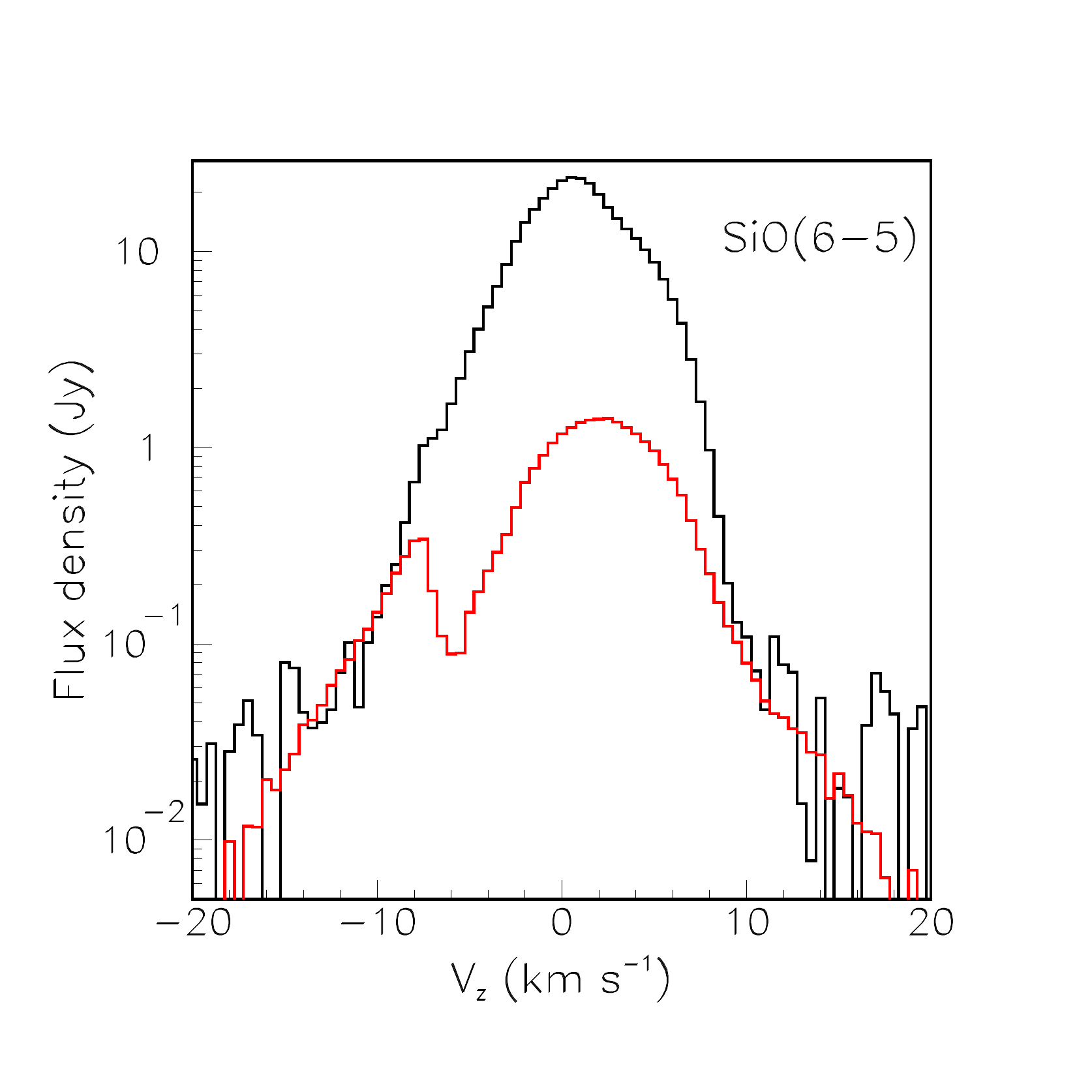}
  \includegraphics[height=6.5cm,trim=1cm .5cm 2.cm .5cm,clip]{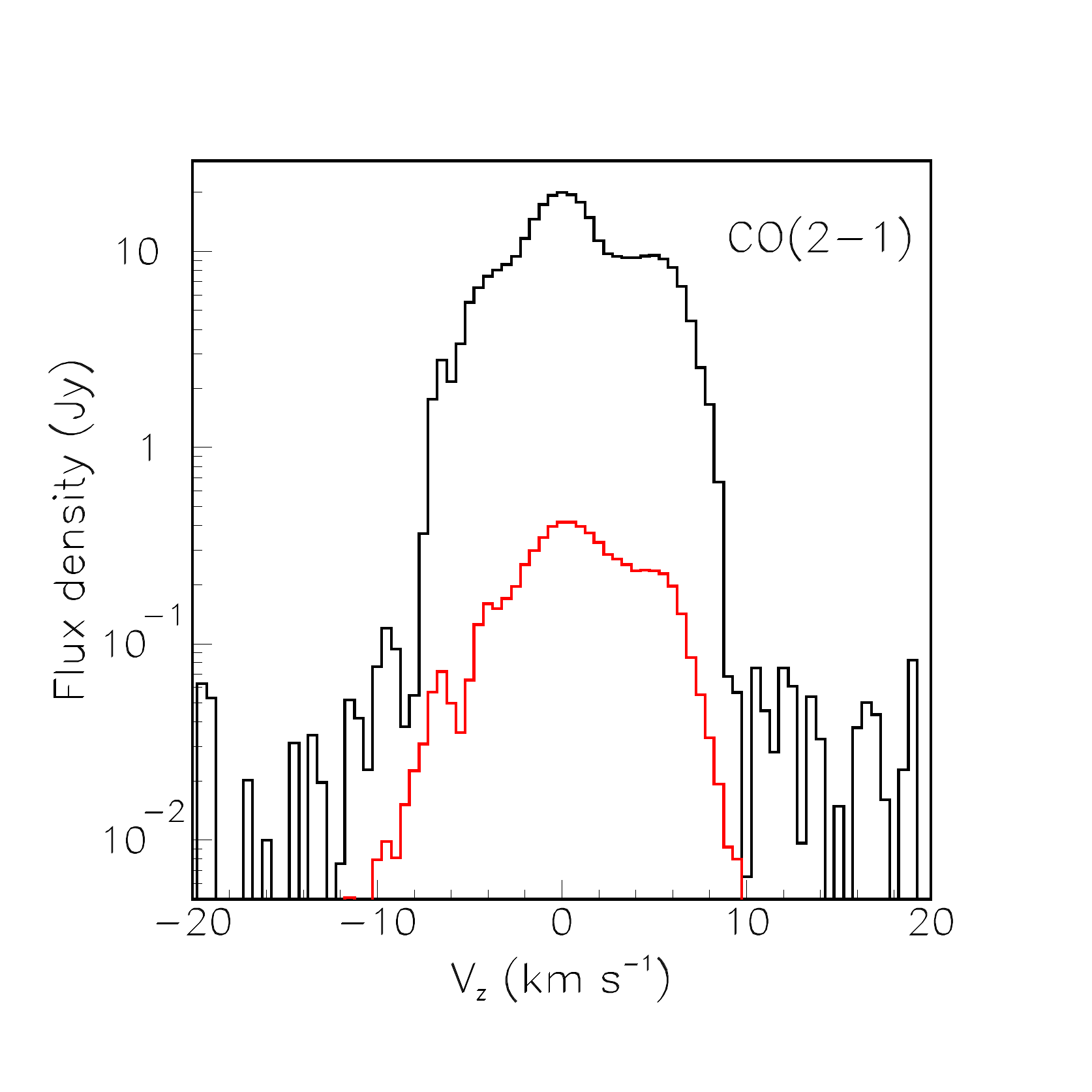}
  \caption{Doppler velocity spectra integrated over $0.2<R<2.5$\,arcsec (black) and
    within $R<0.2$\,arcsec (red) for SiO(5-4), SiO(6-5) and $^{12}$CO(2-1) from
    left to right}
  \label{SiOCOspecfig}
\end{figure*}

Another striking difference is the presence of high velocity wings in
the SiO data close to the star, that are essentially absent in the CO
data (see Figs.~\ref{SiOCOspecfig} and \ref{wingfig}). This may be
understood as the SiO data probing much more efficiently the close
neighborhood of the star, where pulsation shocks and possibly
convection cells are known to play an important role, as demonstrated
by a host of observations at shorter wavelengths, from far infrared to
near UV, and in agreement with theoretical modeling \citep[see,
e.g.,][and references
therein]{hhr82,wkgs2000,rwwbs2003,nlhh2005,flh2017,2017ApJ...841...33M,ho18}. 
Also worth noting on Fig.~\ref{SiOCOspecfig} is the significant absorption
in front of the stellar disk seen in the SiO data at $\sim -6$\kms,
reminiscent of similar observations in stars such as R Dor
\citep{nhungetal2021}, W Hya \citep{takigawaetal2017} and $o$ Ceti
\citep{wongetal2016}.  Of these three stars, two are found to be
surrounded by an optically thick SiO layer that extends well beyond
the stellar disk, while in the third, $o$ Ceti, SiO emission decreases
abruptly beyond some $10$ to $50$\,AU from the center of the star. It
is therefore important, in order to understand the mechanism of
condensation of SiO molecules on dust grains, to study the SiO
emission region in oxygen-rich AGB stars. We devote accordingly the
remainder of the present section to this task. In a first part, we
state some qualitative comments that help with unraveling the general
picture; in a second part, we account for radiative transfer, using
the model of the morpho-kinematics developed for CO emission in the
preceding section.  The morpho-kinematics of the SiO(5-4) and SiO(6-5)
lines are so similar that we cannot expect their comparison to be very
sensitive to temperature.  At temperature $T$, in the optically thin
LTE approximation, the ratio $R_T$ of the SiO(5-4)/SiO(6-5) line
emission depends only on the ratio of the Einstein coefficients,
$5.2\times10^{-4}/9.1\times10^{-4}=0.57$ and the ratio of the level
populations, $(2J+1)(\exp(-E_u/T)$, where the upper level energies
$E_u$ are $31.3$\,K and $43.8$\,K, respectively (line parameters from
the CDMS, based on \citet{siocdms2013}).\\
Namely:
\begin{equation}
R_T = 0.48\exp(-31.3/T)/\exp(-43.8/T)\, ,
\end{equation}
from which we obtain:
\begin{equation}
T[K] = 12.5/(\ln(R_T) + 0.74\,.
\end{equation}

\begin{figure*}
  \centering
 \includegraphics[height=5.cm,trim=1.cm .5cm 2.cm .5cm,clip]{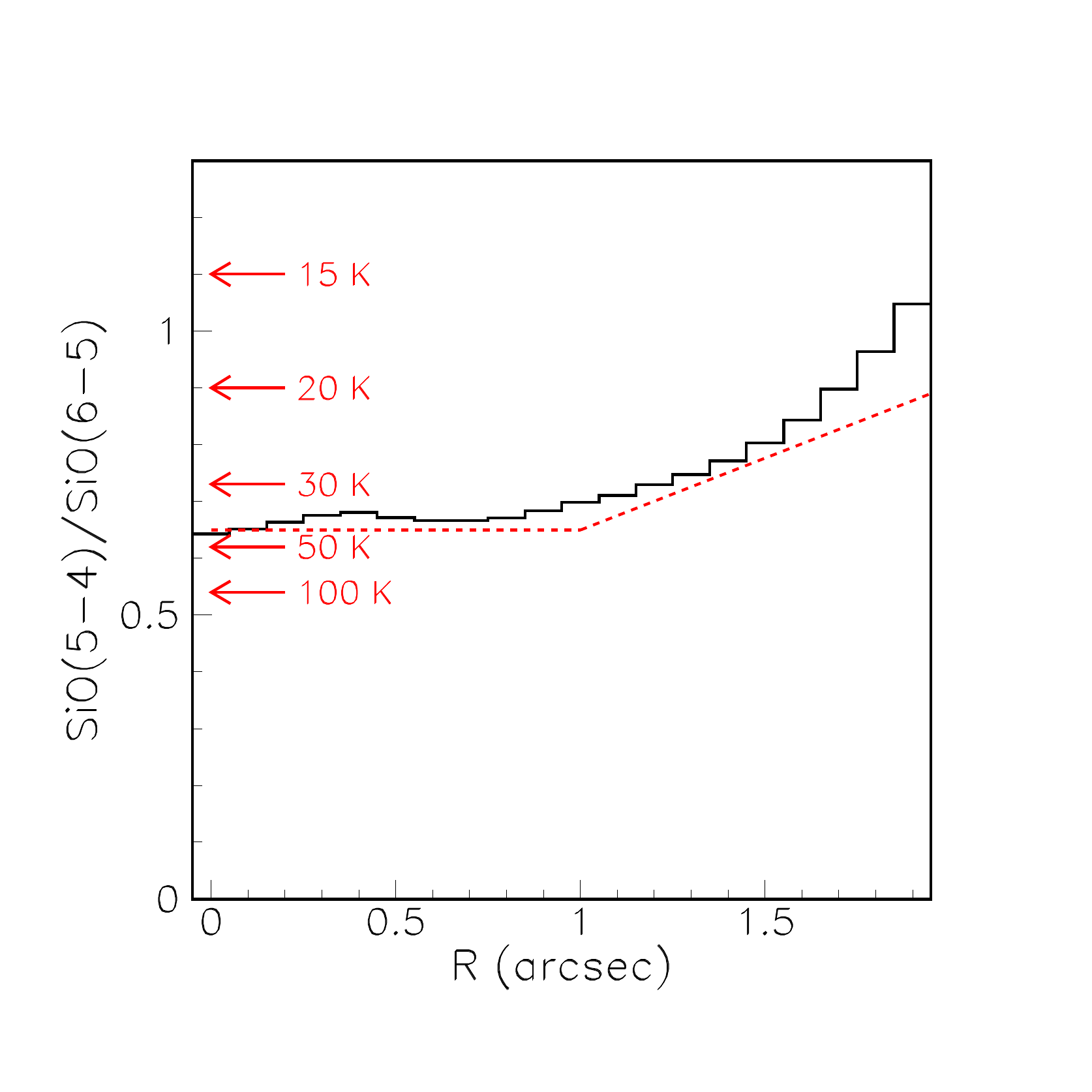}
 \includegraphics[height=5.cm,trim=.5cm .5cm 2.cm .5cm,clip]{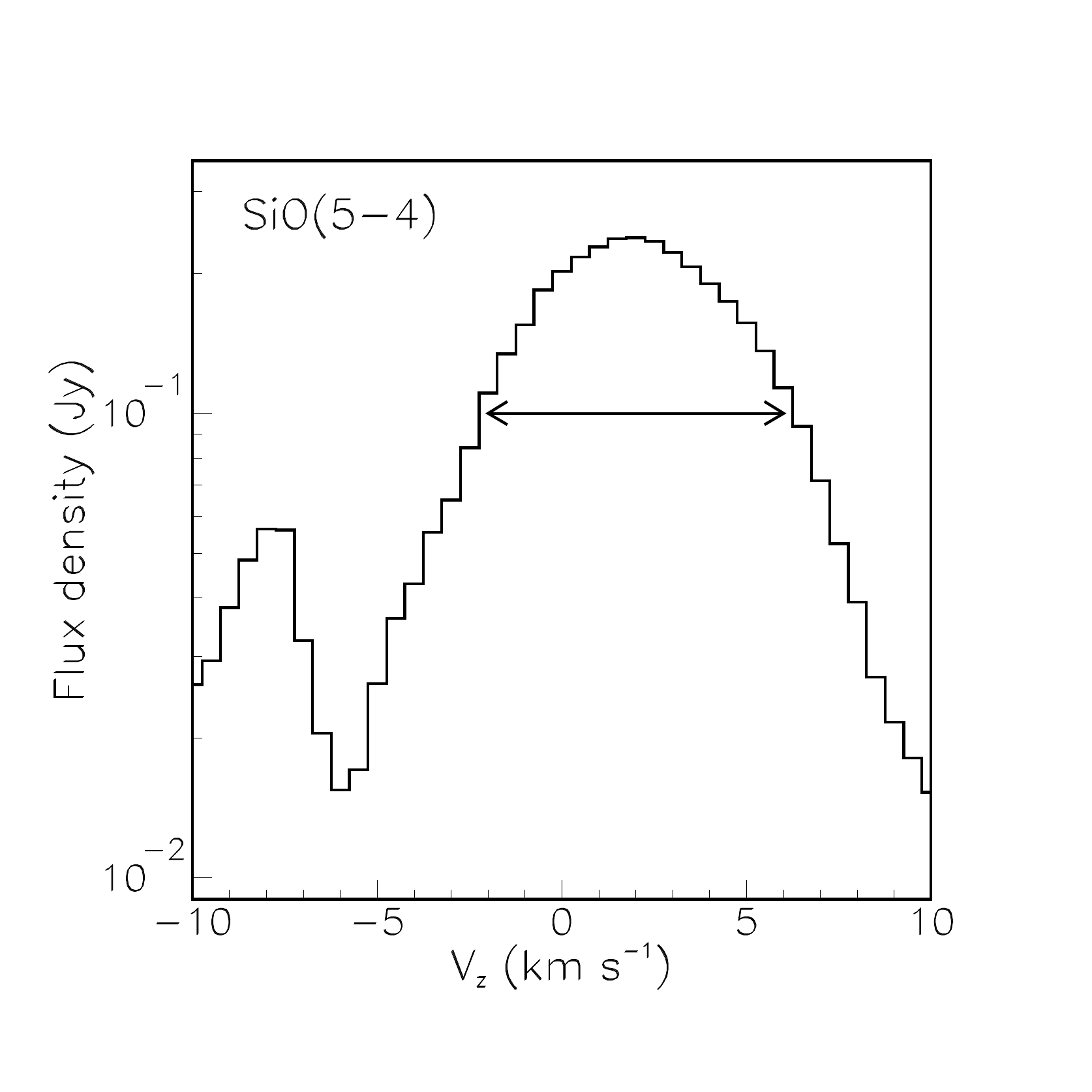}
 \includegraphics[height=5.cm,trim=.5cm .5cm 2.cm .5cm,clip]{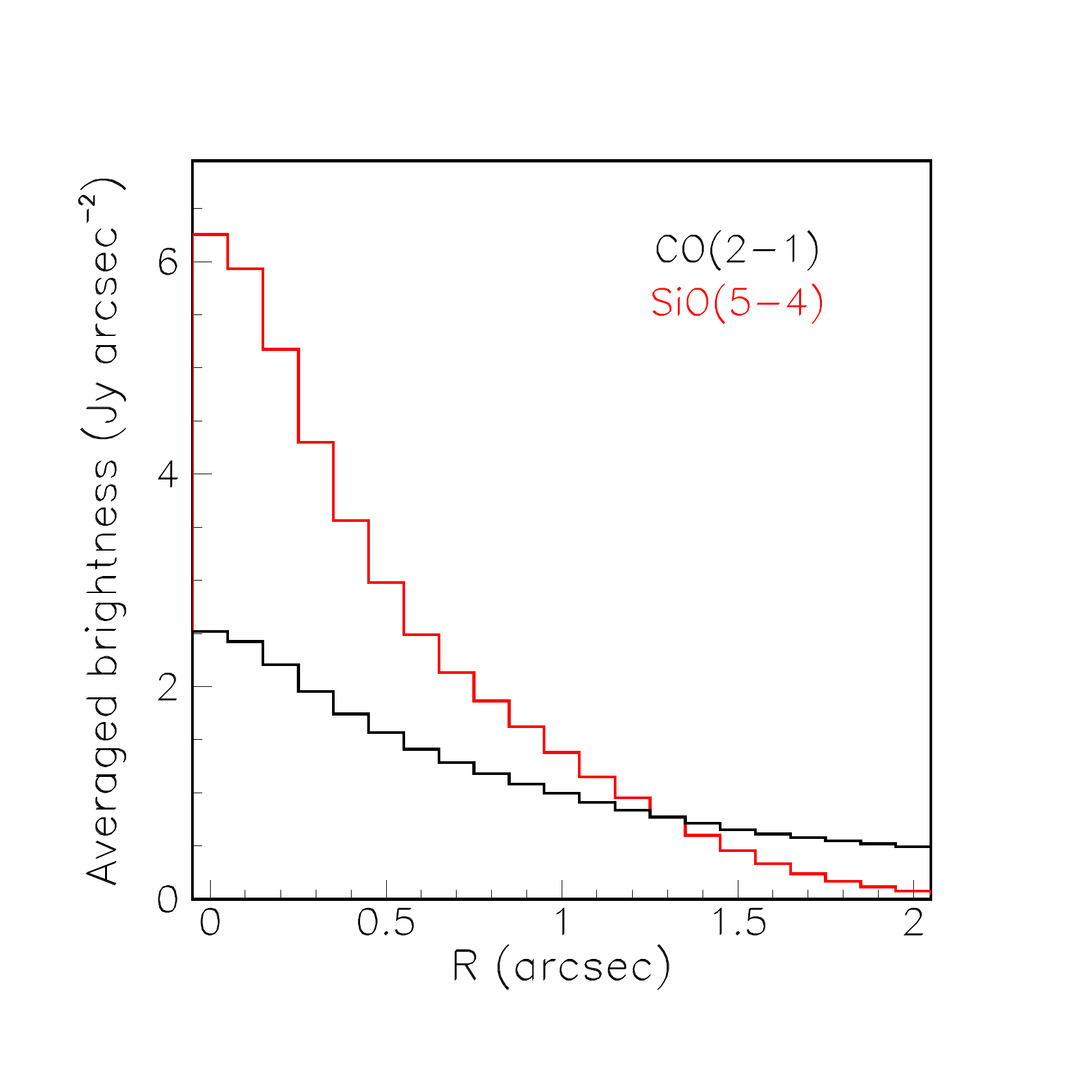}
 \includegraphics[height=5.cm,trim=.5cm .5cm 2.cm .5cm,clip]{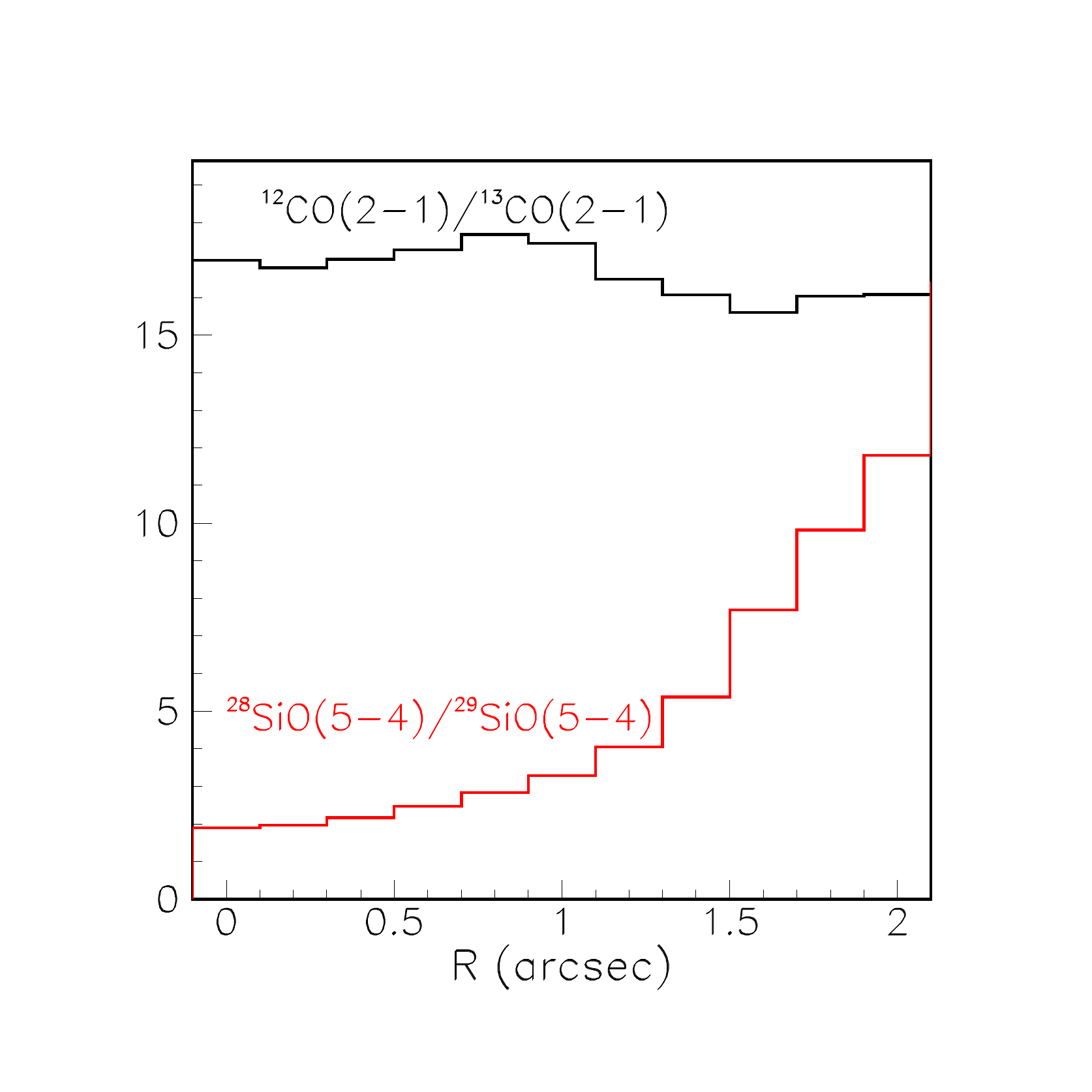}  
 \caption{Left: ratio $R_T$ of SiO(5-4) over SiO(6-5) line emitted
   fluxes as a function of $R$ for $|{\rm v_z}|<8$\kms. Corresponding
   temperature values using the relation $T{\rm [K]} =
   12.5/(\ln(R_T)+0.74$ are shown in red on the scale of the ordinate.
   The red dashed line indicates constant $R_T$ of 0.64 ($T \sim 42$\,K)
   for $R \leq 1\arcsec$
   and a linear increase of $R_T$ to 0.89  ($T \sim 20$\,K) from $1\arcsec$
   to $2\arcsec$.
   Center-left: Doppler velocity spectrum of the SiO(5-4) data for
   $R<0.1\arcsec$. Center-right: $R$ distribution of the SiO(5-4)
   (red) and $^{12}$CO(2-1) (black) emission averaged over position
   angle $\omega$, measured in Jy\,arcsec$^{-2}$ and averaged over
   $-2<{\rm v_z}<6$\kms as shown by the arrow in the center-left
   panel. Right: $R$ dependence of the
   $^{12}$CO(2-1)/$^{13}$CO(2-1) ratio (black) and of the
   $^{28}$SiO(5-4)/$^{29}$SiO(5-4) ratio (red).}
 \label{SiOCOisofig}
\end{figure*}

Figure~\ref{SiOCOisofig} (left) displays the observed dependence of
$R_T$  on $R$. If the SiO layer is optically thin, the value of $R_T$
provides a direct measure of the temperature, but if it is optically
thick, the emission probes only the outer part of the SiO layer and
measures a temperature that is representative of larger values of
$r$. When $R$ increases, the observed value of $R_T$ remains nearly
constant at $\sim0.65$ up to $R\sim1\arcsec$ and then increases to a
value of $\sim0.8$ at $1.5\arcsec$, corresponding to temperatures of
$\sim40$\,K and $\sim24$\,K, respectively. The value of $40$\,K,
associated with $R<1\arcsec$, is close to the lower value of the
temperature obtained from the CO model for $r=1\arcsec$ ($T_{\rm
CO}[{\rm K}]=70\exp(-r/2\arcsec)=42$\,K for $r=1\arcsec$) in
Sect.~\ref{cokinesec}. This suggests that for low values of the
projected distance from the star, $R$, the distance in space from the
star, $r$, effectively probed by the SiO emission stays approximately
constant and of the order of $1\arcsec$. Indeed, the strong
self-absorption causes the region probed by the SiO emission to be
confined beyond $r\sim1\arcsec$ (corresponding to $\tau_{\rm SiO} \sim
1$), but not to probe the volume beneath this surface. For larger
values of the projected distance $R>1\arcsec$, we expect the distance
in 3-dimensional space, $r$, of the effective SiO emitting surface to
increase progressively as a result of the elongation of the emission
volume along the line of poles, a pure geometrical effect
(Fig.~\ref{COomegdeproj}).

Qualitatively, we describe this trend in figure~\ref{SiOCOisofig}
(left) by assuming that $r$ stays at the $1\arcsec$ level for $R <
1\arcsec$ and increases approximately from $1\arcsec$ to $2.5\arcsec$
when the projected $R$ increases from $1\arcsec$ to $2\arcsec$. Such a
trend, when translated in terms of temperatures using the CO model
relation $T{\rm_{CO} [K]} = 70\exp(-r/2\arcsec)$, gives a fair
description of the observed dependence of $R_T$ on $R$ given the
important approximations that have been made (Fig.~\ref{SiOCOisofig}
left).

The ratio $\varepsilon/\tau$ of the emissivity to optical depth is a
measure of the lower limit of the emission of a self absorbing
layer. In the LTE approximation and to first order in $\Delta E/T$,
where $\Delta E$ is the energy of the transition,
$(\varepsilon/\tau)/(T\nu^2)$ is a constant, where $\nu$ is the
frequency of the emission. Therefore, at the same temperature,
$(\varepsilon/\tau)$[SiO(5-4)]
$=0.88(\varepsilon/\tau)$[$^{12}$CO(2-1)]. The value of
$(\varepsilon/\tau)$ is $T/26.3$ Jy\,arcsec$^{-2}$ for CO and $T/29.5$
Jy\,arcsec$^{-2}$ for SiO. Taking as reference $T=100$\,K for SiO and
50\,K for CO, the corresponding values of $(\varepsilon/\tau)$ are 1.9
Jy\,arcsec$^{-2}$ for CO and 3.4 Jy\,arcsec$^{-2}$ for SiO. The
observed values (Fig.~\ref{SiOCOisofig} center-right) are $\sim2.5$
and 6 Jy\,arcsec$^{-2}$, respectively at $R\sim 0$. This is less than
a factor 2 above the reference values, showing that the optical
thickness is close to that of the self-absorption regime. A similar
result has been observed in other AGB stars having mass-loss rates on
the order of $10^{-7}M_\odot{\rm yr}^{-1}$  (R Dor
\citep{nhungetal2021},  W Hya \citep{takigawaetal2017}).

\begin{figure*}
  \centering
  \includegraphics[height=6cm,trim=1.1cm .5cm .4cm .5cm,clip]{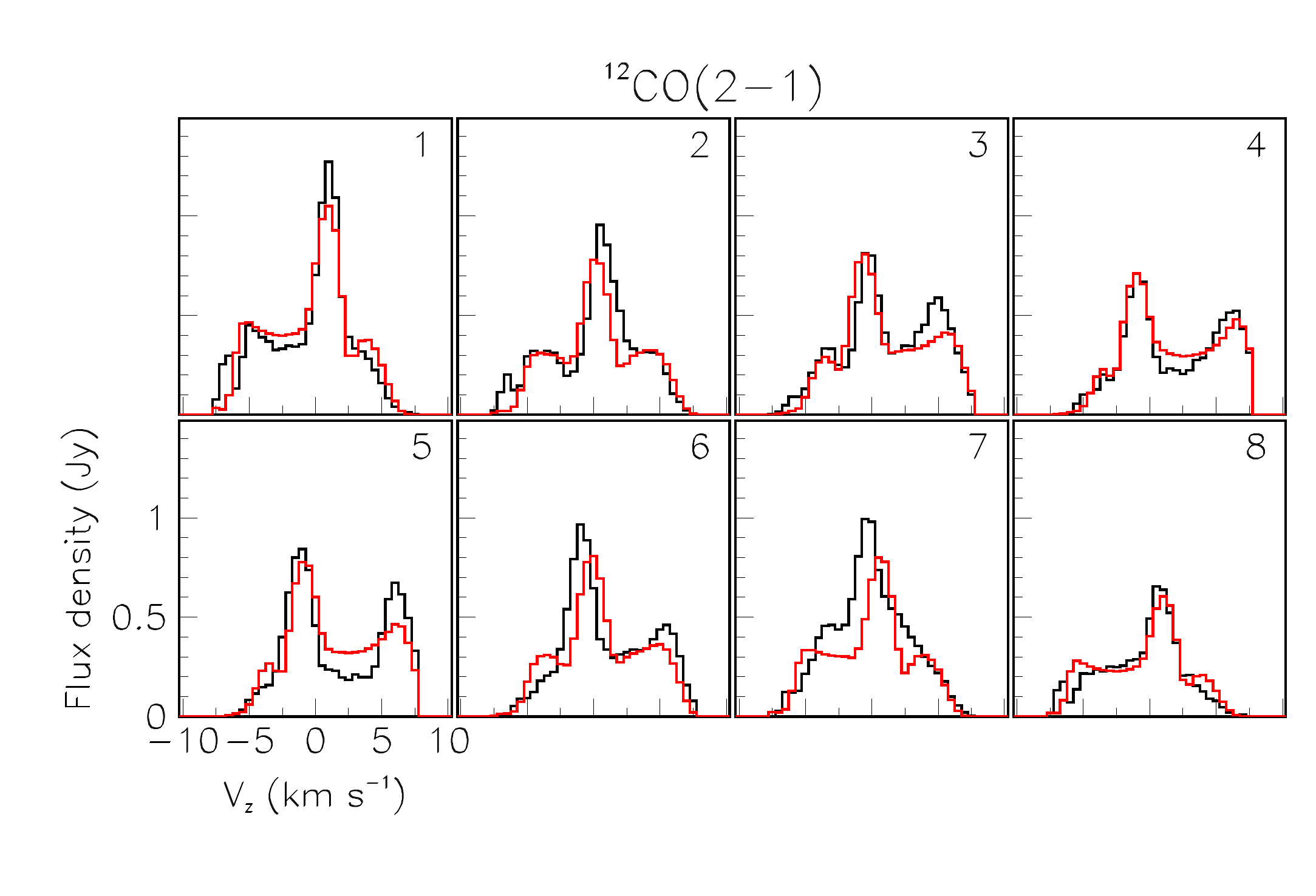}
  \includegraphics[height=6cm,trim=0.8cm .5cm .4cm .5cm,clip]{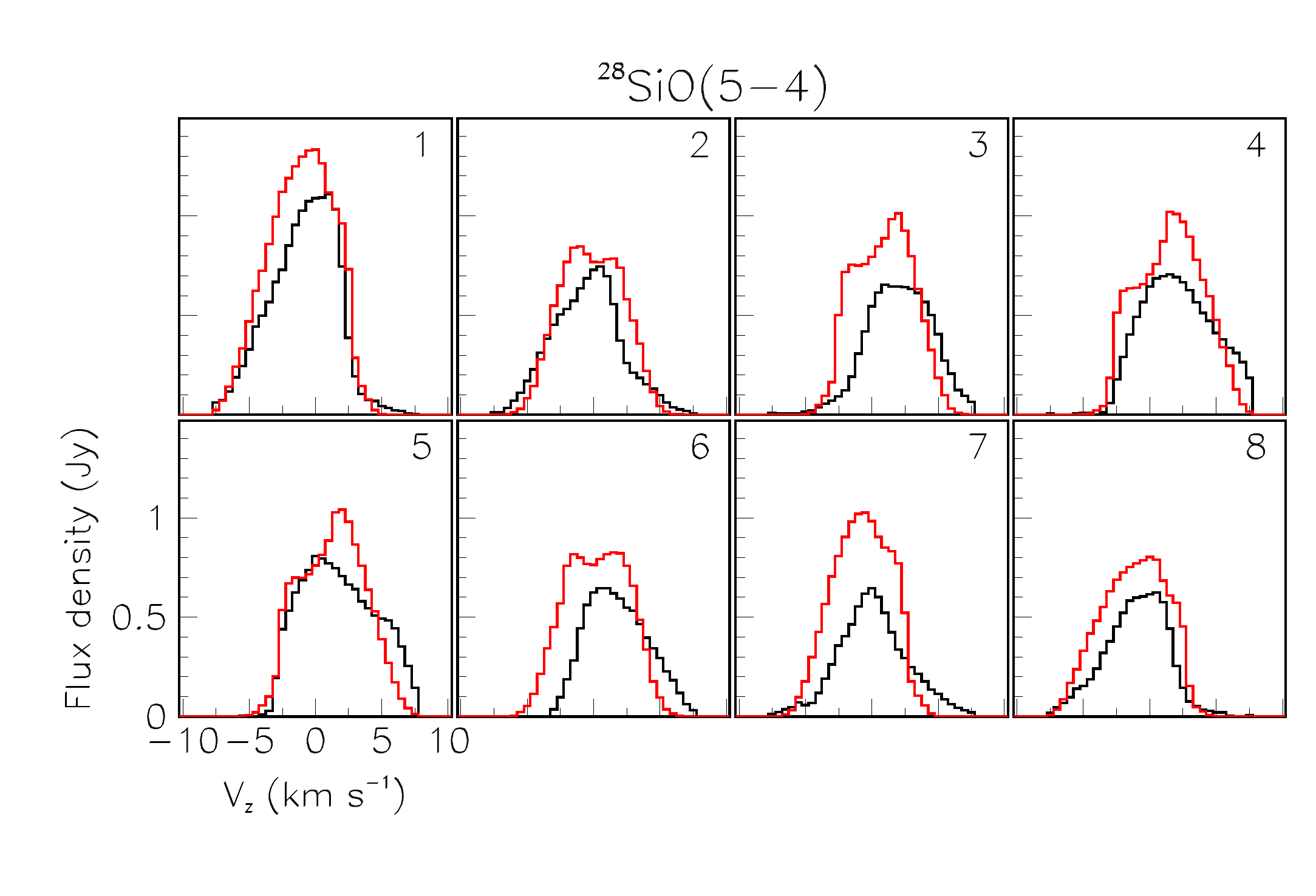}
  \caption{$^{12}$CO(2-1) and $^{28}$SiO(5-4) Doppler velocity spectra in 
    the ring
    $1\arcsec<R<1.5\arcsec$. The data are shown in black and the best fit
    model results are shown in red (see text). Octant intervals in
    position angle $\omega$ are numbered as $1=[7^\circ,52^\circ],
    2=[52^\circ,97^\circ], \dots 8=[322^\circ,7^\circ]$, these are indicated 
    on the left and central panels of Fig.~\ref{SiOwavefig}}.
  \label{COSiOobsmodfig}
\end{figure*}

To get deeper insight into this issue, having obtained qualitative
evidence for strong absorption, we need to account properly for
radiative transfer. To do so, we use the model developed in
Sect.~\ref{cokinesec} to describe the morpho-kinematics of the CO
component of the CSE: we compare in Fig.~\ref{COSiOobsmodfig} Doppler
velocity spectra of SiO and CO in octants of position angle (the first
one defined as $7^\circ<\omega<52^\circ$, then counter-clockwise for
increasing numbers, as indicated on Fig.~\ref{SiOwavefig}) in a ring
$1\arcsec<R<1.5\arcsec$.  This range of $R$ is far enough from the
star to be relatively independent of modeling the star neighborhood,
which contributes to the high-velocity wings of SiO spectra but our
current data do not offer much constraints on this region. We only
show SiO(5-4) but the results are essentially the same for
SiO(6-5). The CO data are seen to trace  the polar outflows from the
equatorial region, the SiO data do not. To adapt the CO model to a
description of the SiO data, we keep all parameters as listed in
Table~\ref{fitparatab} with the only exception of the radial profile:
we divide the CO density parameters $\rho_0$ by 3, and we further
multiply it by $\exp(-r/1.6)$, and set it to zero beyond
$r=1.6\arcsec$, in qualitative accordance with the distributions
displayed in the right panel of Fig.~\ref{SiOwavefig}. These
modifications  result in a ratio of SiO/CO $=0.18$ at $r=1\arcsec$, in
agreement with the estimate of \citet{vandesandeetal2018a}, who obtain
a SiO/CO ratio of 0.17-0.19 for this type of stars.
  
Apart from this major modification of the radial distribution of the
density, we use the same morpho-kinematics and same temperature
distribution as for the best fit to the CO data. Of course we also
change the parameters defining the transition, in particular with
Einstein coefficients three orders of magnitude larger than for the CO
transitions. The result, displayed in Fig.~\ref{COSiOobsmodfig},
describes the data surprisingly well given the crudeness of the
exercise. In addition to offering an understanding of the large
difference between CO and SiO emission, it provides a confirmation of
the validity of the description of the morpho-kinematics given by the
model. What happens is that the SiO emission in the modeled annular
ring between $1\arcsec$ and $1.5\arcsec$ probes only the outer layer
of the (confined) SiO volume because of the very large absorption,
while CO probes the lower density gas at much larger distances from
the star where the equatorial region and the polar outflows are well
separated. We refrain from attempting to adjust parameters to obtain a
better fit: modeling the inner region could not be done reliably with
the limited angular resolution of the present data.

Another illustration of the stronger absorption of the SiO line
compared to the CO line is displayed in the right panel of
Fig.~\ref{SiOCOisofig}. It compares the (projected) radial
distribution of the ratio $^{28}$SiO(5-4)/$^{29}$SiO(5-4) with that of
the ratio $^{12}$CO(2-1)/$^{13}$CO(2-1). The $^{29}$SiO data are
obtained with a similar beam size as the $^{28}$SiO and CO data, but
with a spectral resolution of only 3\kms, preventing us from a
detailed study  of their Doppler velocity spectra. We expect the
$^{28}$SiO/$^{29}$SiO isotopic ratio to be $\sim 12$ or lower
\citep{do2018,do2020}, but this line ratio is reached only at values
of $R$ large enough, where the optical depth has sufficiently
decreased for absorption to be less important. In the limit of
complete self-absorption (i.e., emission at the $\varepsilon/\tau$
level) the line ratio is unity. In contrast, the CO ratio is much less
affected by absorption, staying at $\sim 80\%$ of the value evaluated
in Sect.~\ref{cokinesec} ($\sim 20$) over the whole radial range.

\subsection{MHD wind models}\label{mhdmodelsec}

An important feature of the morpho-kinematics of RS Cnc is the
latitudinal dependence of velocities, being larger towards the pole
than along the equatorial plane. On the other hand, the modeling of
\citet{hmwng14} leads to an almost spheroidal distribution of matter
(their figure 6), which implies an isotropic radiation field within
the circumstellar shell (assuming position coupling between dust and
gas). In these conditions, one needs another cause than radiation to
accelerate matter preferentially along the polar axis.

Axi-symmetric models have been developed on the assumption that a
companion disturbs the gravitational field
\citep{tj93,mm99,dmrgh2020}.  These models have been successful in
explaining several observed features, such as the presence of spiral
structures, although the preferential acceleration along the polar
axis seems more difficult to reproduce.

In RS Cnc, we do not find evidence for the presence of a
companion. While this negative result does not necessarily mean that
there is no companion, we are considering here another mechanism that
could play a role: the presence of a magnetic field. In this
subsection, we argue that magnetic fields are one possible candidate
to explain a significant pole/equator velocity contrast $> 1$.

Indeed, magnetic fields are an appealing way to explain some of the
axi-symmetric shapes of planetary nebulae, such as polar jets or
equatorial winds.  Early simulations of winds with dipole fields and
rotation have shown that excess of magnetic pressure could be able to
repel the wind towards both the equator and/or the poles
\citep{1993MNRAS.262..936W}. Similar simulations also showed the
formation of equatorial disks with enhanced outflow velocities in the
equatorial region \citep{mbwg2000}. In a more toroidal configuration,
\citet{2004ApJ...615..921M} showed that both, jets and equatorial
disks,  could be produced depending on the magnitude of the
rotation. Finally \citet{2005ApJ...618..919G} showed that magnetically
driven winds yield strongly anisotropic outflows with highly
collimated polar jets.

Observational polarimetric studies \citep{2002A&A...392L...1G} have
revealed ordered magnetic fields in planetary nebulae, with various
degrees of toroidal configurations depending on the target.  Using a
handful of SiO masers \citep{2006A&A...450..667H} or CN Zeeman
measurements, \citet{dhwbetal2017} later attempted to characterize
further the magnitude and the radial dependence of the magnetic field
in the winds of AGB stars. They conclude that the field is on the
order of a few Gauss near the stellar surface, and consistent with a
$1/r$ dependence on the distance $r$ from the star (see their figure
6).

In this subsection, we report results of simplistic calculations which
integrate magnetized fluid parcel trajectories from the surface of an
AGB star up to 10 stellar radii. We integrate in time the acceleration
due to gravitational and Lorentz forces (pressure gradients become
quickly negligible after the sonic point is crossed):

\begin{equation}
\boldsymbol{\ddot{r}}=(\Gamma-1)\frac{GM_{*}}{r^{3}}\boldsymbol{r}+\frac{1}{\rho}\boldsymbol{J}\times\boldsymbol{B} \, ,
\end{equation}

where $r$ is the position of the fluid parcel, $\Gamma$ is the ratio
between the radiative and the gravitational force, $G$ is the
universal gravity constant, $M_{*}$ is the stellar mass, and
$\boldsymbol{J}=\frac{1}{4\pi}\boldsymbol{\nabla}\times\boldsymbol{B}$
is the current vector. The mass density $\rho$ and the magnetic field
$\boldsymbol{B}$ are prescribed, while we solve for the position and
velocity of the fluid parcels. We parametrize our equations with
$(\Gamma-1)GM_{*}={\rm v}_{\infty}R_{*}^{2}$ and
$\rho=\frac{\dot{M}}{4\pi r^{2}{\rm v}_{\infty}}$.  We assumed
$\dot{M}=1.24\times10^{-7}M_{\odot}{\rm yr}^{-1},$ and
$R_{*}=1.6\times10^{13}$\,cm as reasonable values for RS Cnc
\citep{hmwng14}. Our choice of parametrization allows us to easily
explore non-dimensional values of the parameters, independently of the
absolute observational constraints.  After we obtain a suitable
contrast between polar and equatorial velocities, we retrieve the
physical value for the velocity scale, here v$_{\infty}=5.6$\kms, in
order to obtain a given polar outflow velocity of 8\kms at
$r=10\,R_{*}$. The initial velocity vector is set with a small uniform
radial velocity and we probe starting trajectories at the surface with
a uniformly distributed initial latitude.

\begin{figure}
\includegraphics[width=0.9\columnwidth]{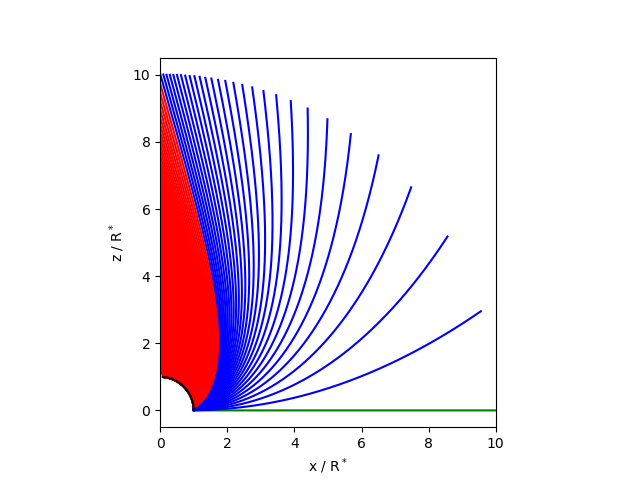}
\caption{Magnetized fluid parcel trajectories from a simplified model of RS Cnc
(see text). Red trajectories meet on the polar axis, where they will
likely generate a jet. \label{fig:Magnetised-fluid-parcels}}
\end{figure}
\begin{figure}
\includegraphics[width=0.9\columnwidth]{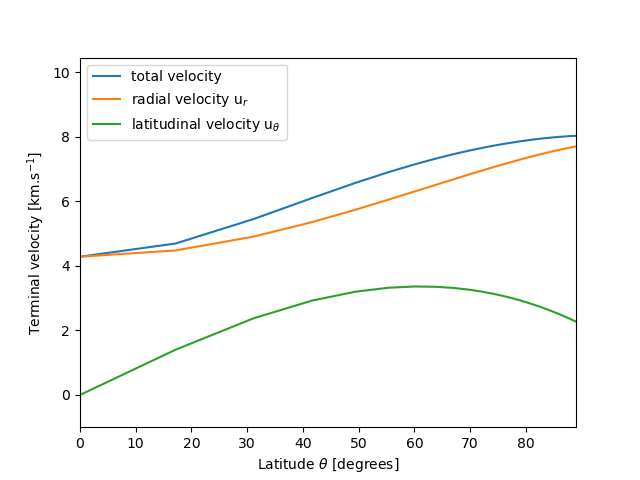}
\caption{Radial velocity at 10 stellar radii depending on the latitude for
the same simplified model of magnetized wind as shown in 
Fig.~\ref{fig:Magnetised-fluid-parcels}.
\label{fig:Radial-velocity-at}}
\end{figure}

This simple setup allows to quickly investigate various magnetic field
configurations. Figure~\ref{fig:Magnetised-fluid-parcels} displays the
trajectories in the meridional plane obtained for a toroidal magnetic
field with a $1/r^{1.1}$ decline from $B_{*}=0.5$\,G at the stellar
surface: $B_{\phi}=B_{*}\cos(\theta)/(r/R_{*})^{1.1}$ where $\theta$
is the latitude ($1/r^{1.1}$ gives a more pronounced velocity contrast
between pole and equator than $1/r$). The Lorentz force
$\boldsymbol{J}\times\boldsymbol{B}$ in this case is directed toward
the symmetry axis and acts as a focusing agent. All trajectories
eventually end up on the axis where they would presumably launch a
jet: this focuses mass-loss towards the poles.
Figure~\ref{fig:Radial-velocity-at} shows the resulting ``terminal''
velocity at 10 stellar radii, where we have separated the latitudinal
and the radial component. The flow velocities at this radius are
dominated by their radial component, but with a clearly slower wind at
the equator compared to the pole, as indicated by observations of,
for instance, RS Cnc and EP Aqr. Note that rotation tends to produce the
opposite effect (faster radial flow at equator compared to poles, due
to centrifugal acceleration). We found similar behavior for toroidal
magnetic field configurations closer to what
\citet{2004ApJ...615..921M} had found
($B_{\phi}=3B_{*}\cos(\theta)\sin(\theta)^{2}/(r/R_{*})^{2}$ with a
$1/r^{2}$ dependence and a concentration of $B$ at intermediate
latitudes). We investigated additional dipolar fields which are able
to generate some amount of rotation in the wind. Thanks to the
versatility of the present setting, we could quickly explore various
configurations, linear combinations between them, several  magnetic
field decay exponents, but have not investigated them systematically
yet.  The purpose of our investigation here is simply to show that a
magnetic field is a valid candidate to produce pole/equator velocity
ratios significantly greater than 1. Finally, note that our
termination radius of $r=10\,R_{*}$ is arbitrary, and a pertinent
match to the observations could be considered at various distances
depending on where the given tracer is expected to be concentrated.
These crude models are still far from matching quantitatively the
observational constraints from RS Cnc or EP Aqr, which require a
broader polar outflow and a thinner and denser equatorial disk. This
could be adjusted by providing a sharper toroidal magnetic barrier to
funnel the wind at the appropriate places. However, it would perhaps
overcome the limits of this crude exercise which still lacks
self-consistency as the density profile remains radial, unaffected by
the magnetic constraints (themselves blind to the wind), and shocks
generated by the crossings of trajectories at the polar axis are not
accounted for. We plan to investigate further with such simplified
models in future work though, as they might provide a useful means to
constrain the magnetic field configuration from the morpho-kinetics.

\subsection{HCN in M- to S-type stars}\label{nontechemistrysec}

\citet{che2006} recognized that the formation of CN/HCN
depends on the high activation barrier of the H+C$_{2}\rightarrow$
CH+C reaction, followed by rapid CN formation via N+CH$\rightarrow$CN+H.
Their abundance therefore depends on thermal excursions in shocks,
or inhomogeneities of temperature. In addition, both the total rate
of formation of the pair CN/HCN and the respective share between CN
and HCN depend on the H/H$_{2}$ ratio which is itself depending on 
out of equilibrium chemistry, due to the slow conversion between H
and H$_{2}$. \citet{che2006} was thus able to show that
the shocks produced by the pulsations close to the stellar photosphere
could produce highly increased yields of the HCN molecule in M or
S-type stars, despite their high O/C ratio. We note here that magnetic
fields produce shocks on the symmetry axis which might also help boosting
HCN production in the polar jet.

\begin{figure}
  \centering
  \includegraphics[width=90mm]{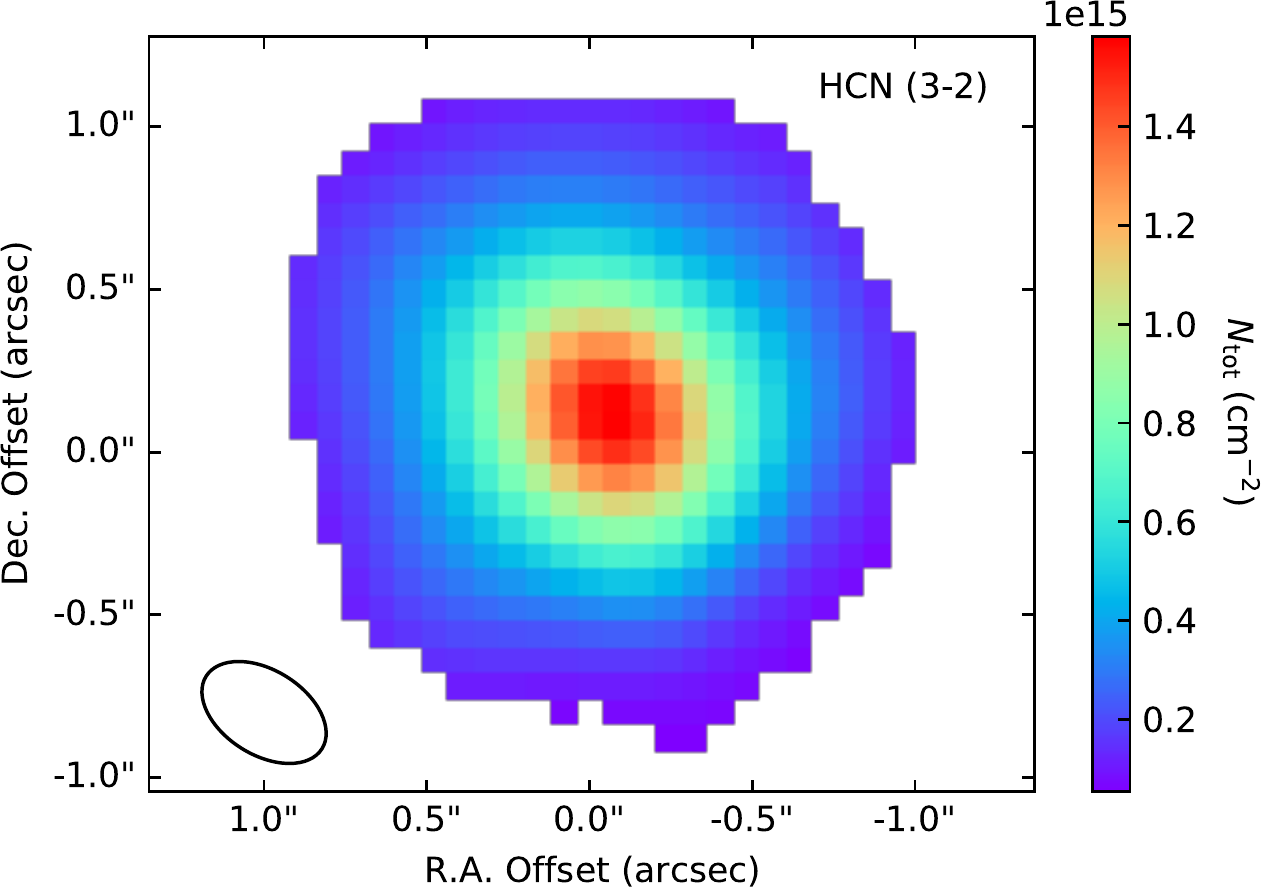}
  \caption{Map of the HCN column density as derived by an XCLASS modeling 
    (see Appendix~\ref{xclassmodels}).
    The black ellipse in the lower left corner indicates the synthesized beam}
    \label{hcnncolfig}
\end{figure}

HCN has long been detected and surveyed in M-type and S-type stars
\citep[e.g.,][]{dg1985,lnow88,bfo94,olnw98,sroetal2013}. RS Cnc,
however, has never been detected in ground-state or vibrationally
excited HCN despite various observational efforts with different
telescopes
\citep{1988A&A...194..230L,sojnz89,1989A&A...210..225N,1992A&A...263..183L,bfo94,bl94,olnw98}.
Adopting a mass-loss rate of $3\times10^{-7}\,M_\odot$yr$^{-1}$,
\citet{bfo94} estimated an upper limit to the HCN abundance in RS Cnc
of $4.5\times10^{-7}$.  This upper limit  becomes $1.35\times10^{-6}$
if we adopt $\dot{M}=1\times10^{-7}\,M_\odot$yr$^{-1}$.
\citet{sroetal2013} presented a comprehensive analysis of the HCN
abundance in a sample of 59 AGB stars, including 25 carbon-rich, 19
S-type, and 25 M-type stars, by means of a non-LTE radiative transfer
modeling. For M-type and S-type stars, they derive  a median HCN/H$_2$
abundance of order $1\times10^{-7}$ and $7\times10^{-7}$,
respectively, with a large spread between $\sim 5\times 10^{-8}$ and
$\sim 5\times 10^{-6}$.

By an XCLASS modeling of the HCN(3-2) line detected here (see
Appendix~\ref{xclassmodels}), and assuming a rotational temperature of
350\,K, we derive an HCN column density of $\sim1.6 \times
10^{15}$\,cm$^{-2}$, see Fig.~\ref{hcnncolfig}. With the same
assumptions made in Sect.~\ref{thermalsec}, this translates to HCN
abundances of $X$(HCN/<H>) = $3.3\times10^{-7}$, or, if hydrogen would
be completely bound in H$_2$, $X($HCN/H$_2) = 6.6\times10^{-7}$.  This
abundance fits perfectly in the range found by \citet{sroetal2013} for
M- to S-type stars and is also consistent with the upper limit derived
by \citet{bfo94} corrected for the mass-loss rate.

Clumpy and porous winds help UV photons to penetrate closer to the
star, thus photodissociating CO and N$_{2}$ to release more of the
C/NO and the N/CS pairs of reactants, which both produce CN: this
process was shown to considerably enhance the HCN abundances close to
the star \citep{vandesandeetal2018,vandesandeetal2020}. Rather than
uniformly distributed random clumpiness and porosity, one can also
imagine ordered density distributions in the wind, which could let UV
photons  penetrate through low density channels. These structures
could sometimes  be hard to witness due to line of sight confusion. In
fact, our simplified magnetized wind models (see
Fig.~\ref{fig:Magnetised-fluid-parcels} with scarcity of trajectories
around the equator) or more sophisticated magnetized wind models
\citep[e.g.][]{mbwg2000,1993MNRAS.262..936W} allow for lower
column-density channels at certain angles. These may result in
increased HCN abundance, but chemical post-processing in magnetized
models will be necessary to assess whether this is a viable
interpretation of the observations.

\section{Conclusions}\label{conclusec}

Using NOEMA equipped with PolyFiX, we have obtained high-spatial
resolution ($\sim$ 0.3\arcsec) images of RS Cnc in several lines of
different molecules.  We detect, and in most cases are able to map, 32
lines of 13 molecules and isotopologs (CO, $^{13}$CO, SiO,
$^{29}$SiO, SO, $^{34}$SO, SO$_2$, H$_2$O, HCN,  H$^{13}$CN, PN),
including several transitions from vibrationally excited states, and a
tentative identification of Si$^{17}$O and possibly $^{29}$Si$^{17}$O.
HCN, millimeter vibrationally excited H$_2$O, SO, SO$_2$, PN and their
isotopologs are first detections in RS Cnc.
 
From their first-moment maps, some of the lines, SiO(v=1,6-5), HCN,
SO, SO$_2$, show signs of rotation in the close vicinity of the star.

A population diagram analysis for the 11 observed SO$_2$ lines
provides a rotational temperature of about 320\,K in the region that
shows signs of rotation. Temperatures of this order are also  found
from an XCLASS modeling of the SO$_2$ lines.  For SO$_2$ and HCN, we
find column densities from the XCLASS modeling of $N_{\rm SO_2} \sim
3.5\times10^{15}$\,cm$^{-2}$ and $N_{\rm HCN} \sim 1.6 \times
10^{15}$\,cm$^{-2}$,  which translate to abundance ratios of
$X($SO$_2/$H$_2) = 1.5\times10^{-6}$ and $X($HCN/H$_2) =
6.6\times10^{-7}$, respectively, well within the range expected for an
MS-type star.

We find broad wings in the spectral line profiles of vibrational
ground state transitions of SiO and SO and in first vibrationally
excited transitions of SiO, that indicate radial velocities of about
twice the terminal outflow velocity as probed by CO.  As high
velocities very close to the star are also seen in similar objects,
like EP Aqr, $o$ Cet, and R Dor, the presence of these broad line wings
calls for a mechanism common to the class of pulsating AGB stars. We
interpret these high velocity line wings by the imprints of pulsation
shocks acting in the very inner region around these stars. 

The spatially resolved images allow us to trace the morpho-kinematics
of the wind around RS Cnc at different scales. In the inner part ($<$
0.5\arcsec, or 75\,AU), we find a rotating structure well traced by
the less abundant molecules (HCN, SO, SO$_2$), and by SiO in (v=1)
lines.  Outside 75\,AU, we find an expanding axi-symmetric outflow,
with velocities $\sim4$\kms in the equatorial plane, and $\sim9$\kms
along the polar axis. This polar axis is common with the axis of the
internal rotating structure. A model that fits the data cubes
obtained on the $^{12}$CO(2-1), $^{13}$CO(2-1) and SiO(v=0, 5-4 and
6-5) lines gives a mass-loss rate of $1 \times 10^{-7}M_\odot{\rm
yr}^{-1}$ for the equatorial region (latitude $< 30^\circ$) and of $2
\times 10^{-7}M_\odot{\rm yr}^{-1}$ for the polar outflows (latitude
$> 30^\circ$).

The $^{12}$CO/$^{13}$CO ratio is measured to be $\sim20$ on average,
$24\pm2$ in the polar outflows and $19\pm3$ in the equatorial region.

Although we cannot exclude the possibility that an unseen stellar or
sub-stellar companion shapes the circumstellar environment of RS Cnc,
we consider also the possibility of a magnetic field playing this
role. In particular, a toroidal magnetic field configuration would
provide a mechanism to produce the significant velocity contrast
between high polar outflow velocities and low expansion velocities in
the equatorial region that is observed in RS Cnc and other similar
stars.

\begin{acknowledgements}

We thank the staff at the NOEMA and Pico Veleta observatories for
their support of these observations. The authors are grateful to
the anonymous referee for a very detailed and valuable report that
helped improving the presentation of the material. This work is
based on observations carried out under project numbers W16BE, D17AE,
W19AX with the IRAM NOEMA interferometer and under project ID 136-19
with the IRAM 30m telescope. IRAM is supported by INSU/CNRS (France),
MPG (Germany) and IGN (Spain). The Ha Noi team acknowledges financial
support from the World Laboratory, the Odon Vallet Foundation and
VNSC.  This research is funded in part by the Vietnam National
Foundation for Science and Technology Development (NAFOSTED) under
grant number 103.99-2019.368. This work was supported by the Programme
National ``Physique et Chimie du Milieu Interstellaire'' (PCMI) of
CNRS/INSU with INC/INP co-funded by CEA and CNES.  This work has made
use of data from the European Space Agency (ESA) mission {\it Gaia}
(\url{https://www.cosmos.esa.int/gaia}), processed by the {\it Gaia}
Data Processing and Analysis Consortium (DPAC,
\url{https://www.cosmos.esa.int/web/gaia/dpac/consortium}). Funding
for the DPAC has been provided by national institutions, in particular
the institutions participating in the {\it Gaia} Multilateral
Agreement.

\end{acknowledgements}

%-------------------------------------------------------------------
% Please note that we have included the references to the file aa.dem in
% order to compile it, but we ask you to:
%
% - use BibTeX with the regular commands:
%   \bibliographystyle{aa} % style aa.bst
%   \bibliography{Yourfile} % your references Yourfile.bib
%
% - join the .bib files when you upload your source files
%-------------------------------------------------------------------
%
%
\bibliography{ref801}
\bibliographystyle{aa}

\begin{appendix}

\section{Resolved out flux}
\label{resolvedfluxsec}

\begin{figure}[h]
    \centering
    \vspace{-3.0cm}
    \includegraphics[width=110mm]{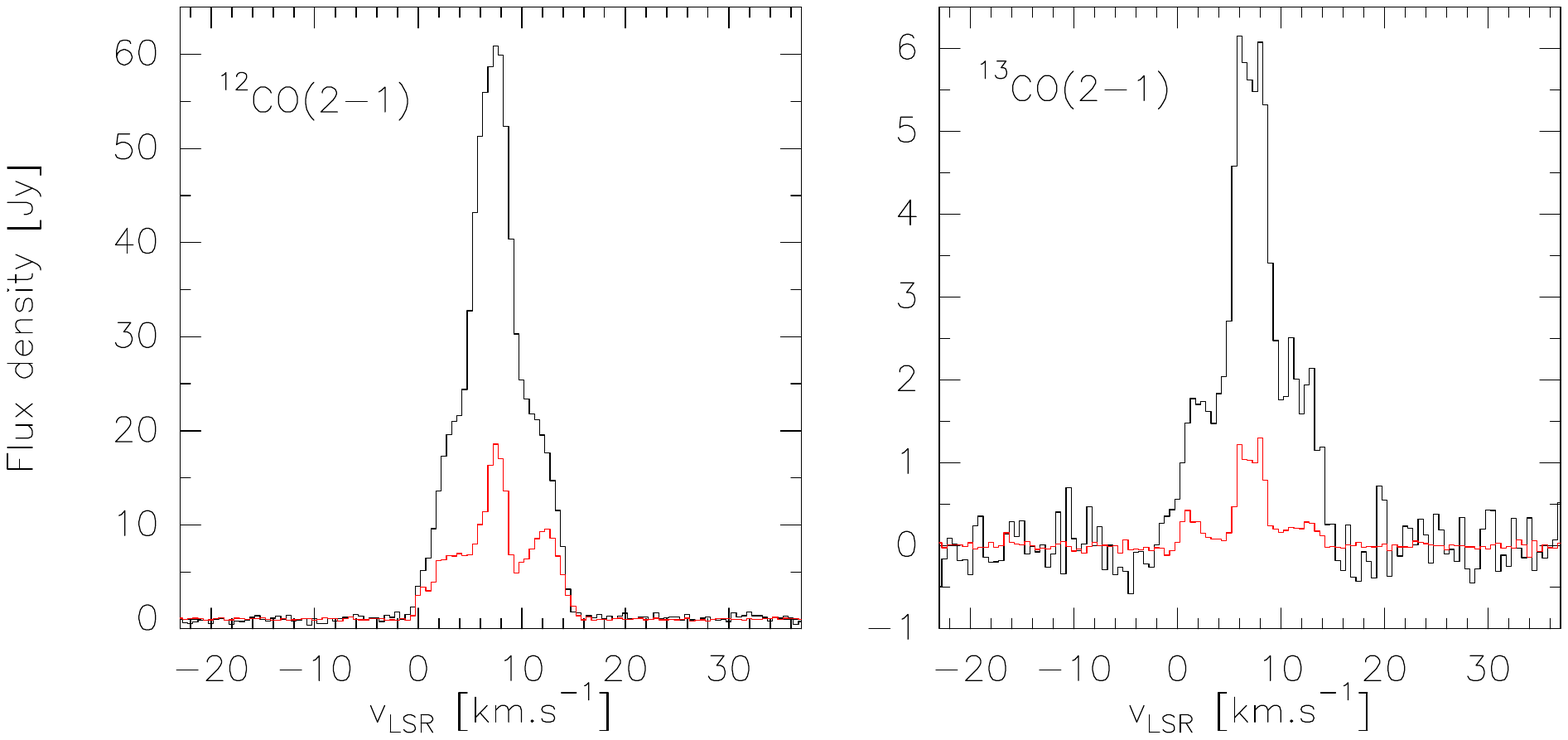}
    \caption{Left:  $^{12}$CO(2-1).
             Right: $^{13}$CO(2-1).
             A and D-configuration are merged, the
             spectral resolution is 0.5\kms and  
             the CO emission is integrated over the central 
             $22\arcsec \times 22\arcsec$, 
             i.e., over the full field of view of the NOEMA antennas at 
             230\,GHz.
             Black profiles: OTF data (i.e., the short-spacing information that 
                             is filtered out by the interferometer) are 
                             added, red profiles: A+D configuration 
                             interferometer data, only.} 
    \label{resolvedfluxfig}
% read-and-plot-COlines.greg
\end{figure}

The effect on the flux of filtering out large-scale structure with the
interferometer is shown in Fig.~\ref{resolvedfluxfig} for the CO and
$^{13}$CO lines. These two are the only lines discussed in this paper
that are affected by the short-spacing problem.

\section{Intensity maps}
\label{mapssec}

\begin{figure*}[h]
    \centering
    \includegraphics[width=60mm]{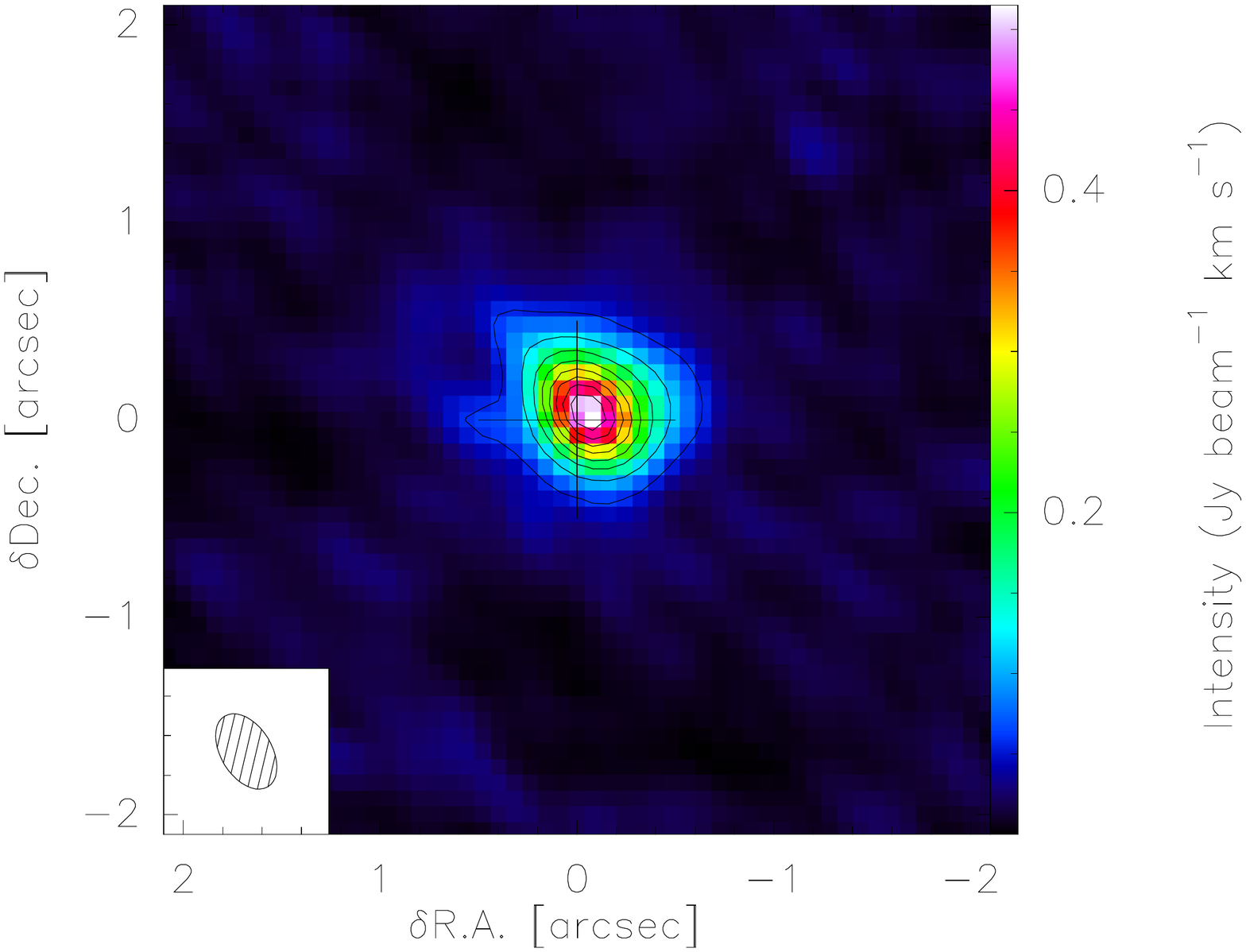}
    \includegraphics[width=60mm]{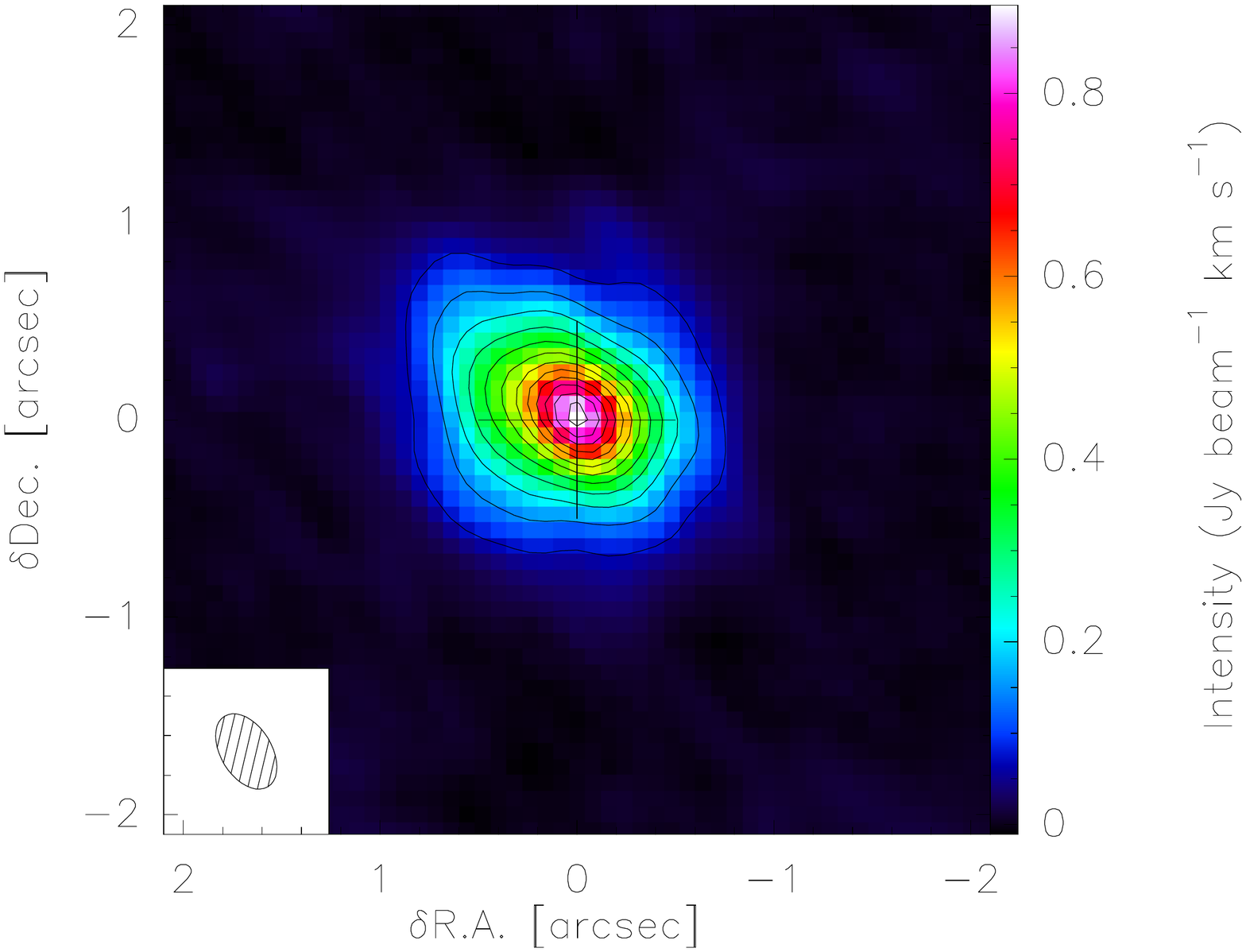}
    \includegraphics[width=60mm]{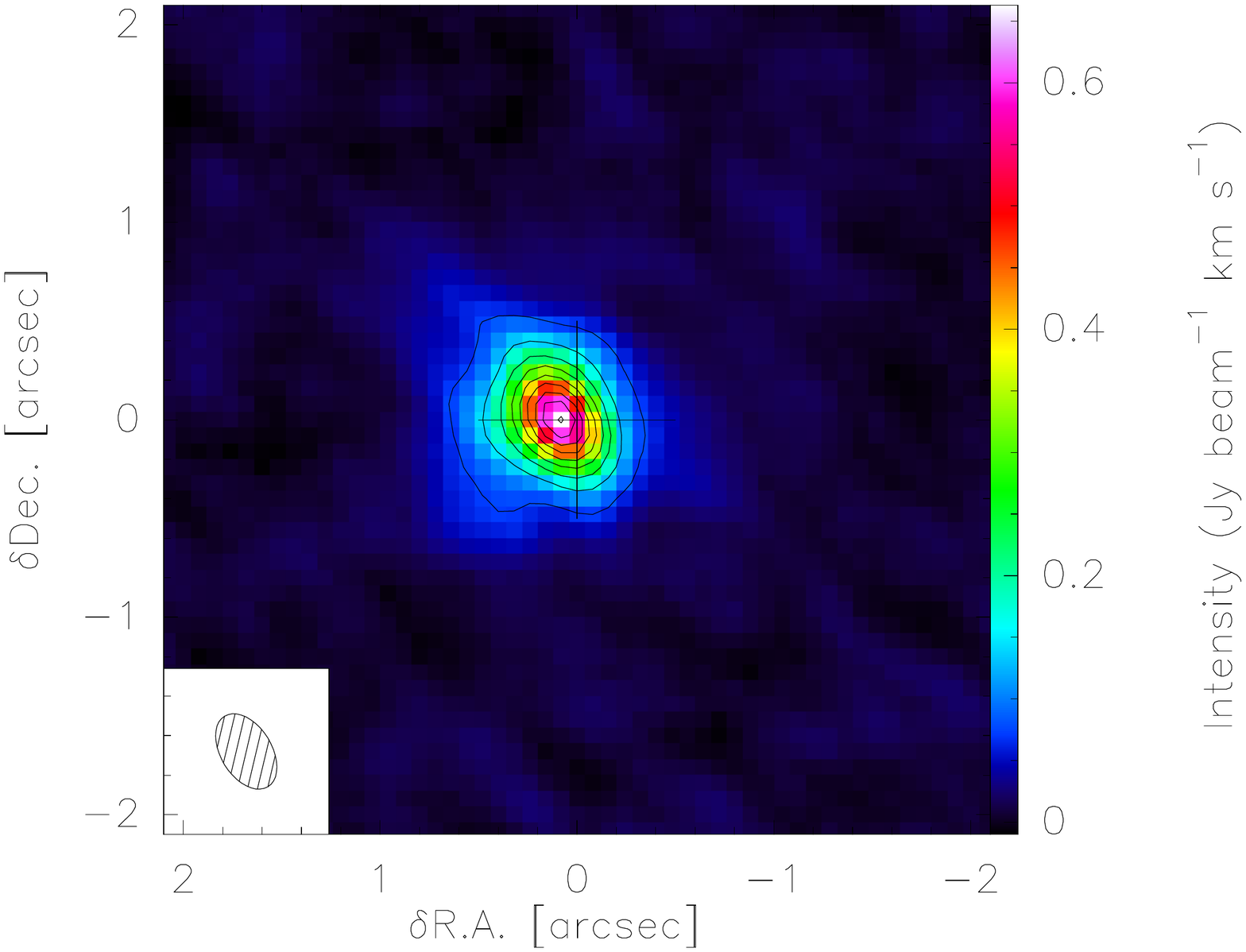}
    \includegraphics[width=60mm]{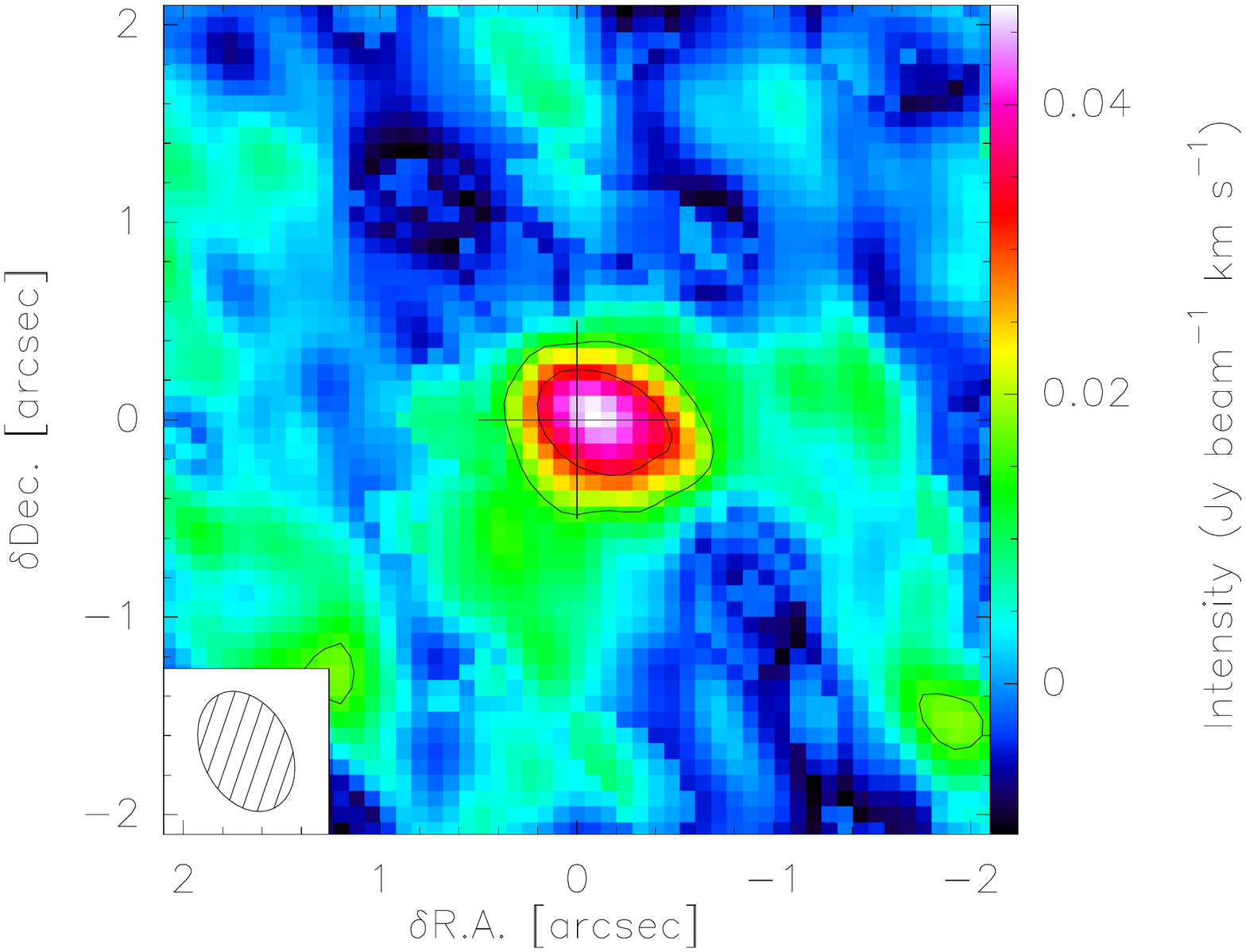}
    \includegraphics[width=60mm]{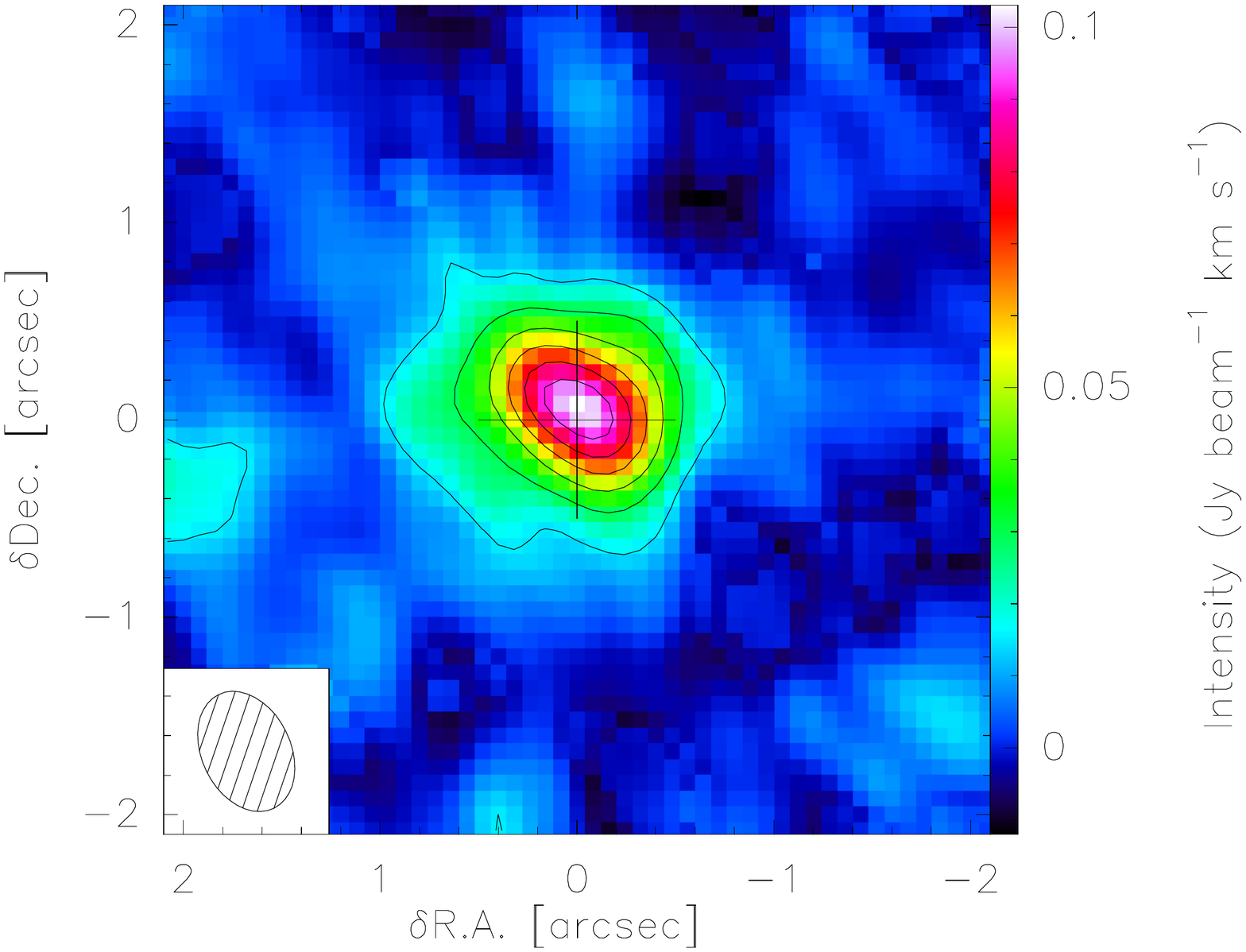}
    \includegraphics[width=60mm]{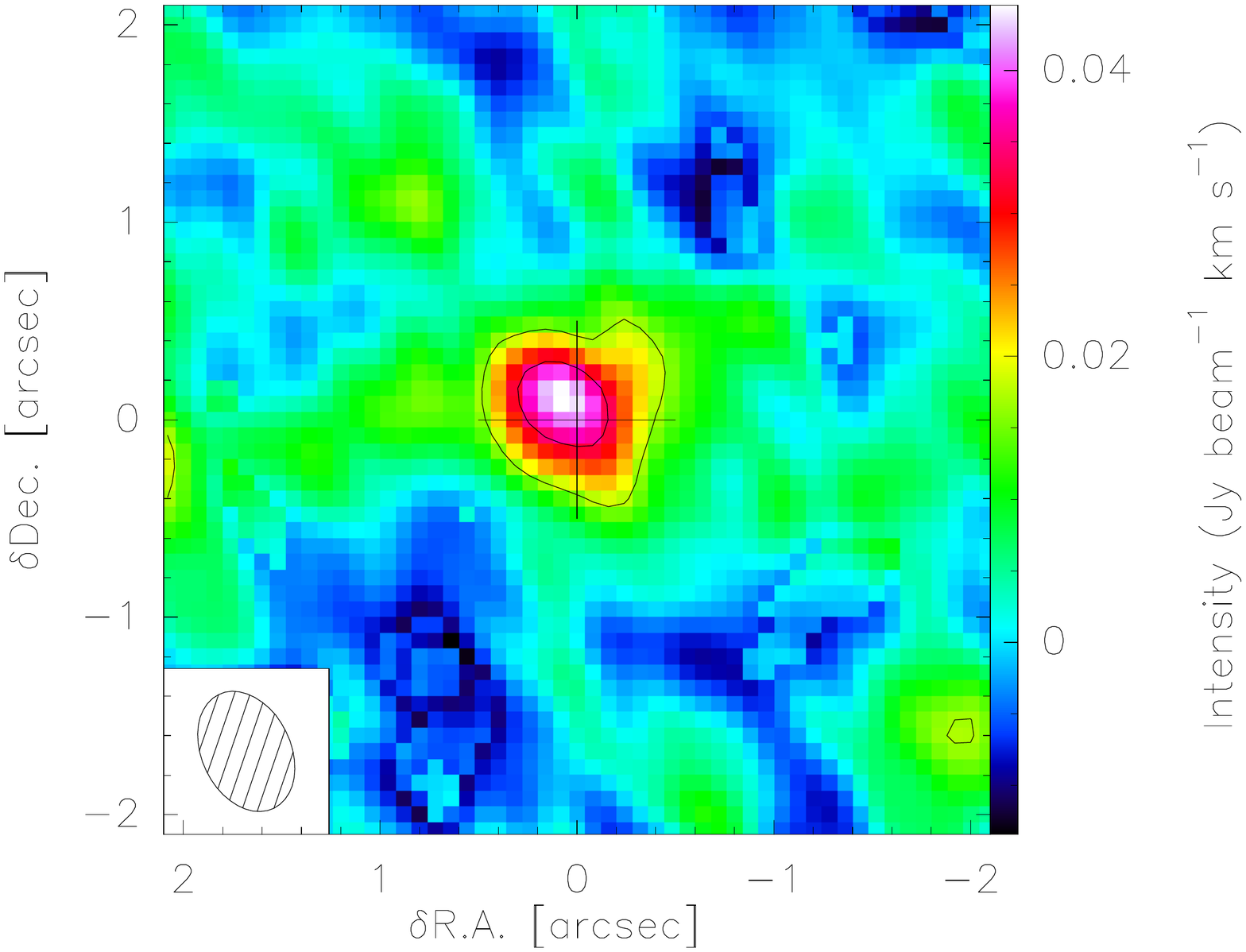}
    % make-blue-center-red_HCN.map
    \caption{Velocity-integrated intensity maps of the HCN(3-2)
      (upper row) and H$^{13}$CN(3-2) lines (lower row), covering three velocity
      intervals. Left: blue line wing [v$_{\rm lsr,*}-10$,v$_{\rm lsr,*}-2$] \kms,
      Middle: line center [v$_{\rm lsr,*}-2$,v$_{\rm lsr,*}+2$] \kms,
      Right: red line wing [v$_{\rm lsr,*}+2$,v$_{\rm lsr,*}+10$] \kms.
      North is up and east is to the left. Note the different color scales. 
      Contours are plotted every $10 \sigma$ for HCN and every $3\sigma$ 
      for H$^{13}$CN, where (from left to right) $1 \sigma = 5.7, 7.2,
      7.2$ mJy/beam$\cdot$\kms for HCN(3-2) and $1 \sigma = 5.2, 5.0,
      5.6$ mJy/beam$\cdot$\kms for H$^{13}$CN(3-2). 
      The black ellipse in the lower
      left corner indicates the synthesized beam. 
      Note that the HCN maps were produced using robust 
      weighting, whereas for the H$^{13}$CN maps, we applied natural weighting.}
    \label{HCNstructfig}
\end{figure*}

Here we present velocity-integrated intensity maps in three velocity
ranges for HCN and H$^{13}$CN (Fig.~\ref{HCNstructfig}), the four
detected SO lines  (Fig.~\ref{SOstructfig}), and three out of the 11
SO$_2$ lines (Fig.~\ref{SO2structfig}).  These are the SO$_2$ line
with the lowest upper level energy, the strongest SO$_2$ line detected
here, and the SO$_2$ line with the highest upper level energy,
respectively.  All these nine lines display kinematic structure in
east-west direction. 

\begin{figure*}[h]
    \centering
    \includegraphics[width=60mm]{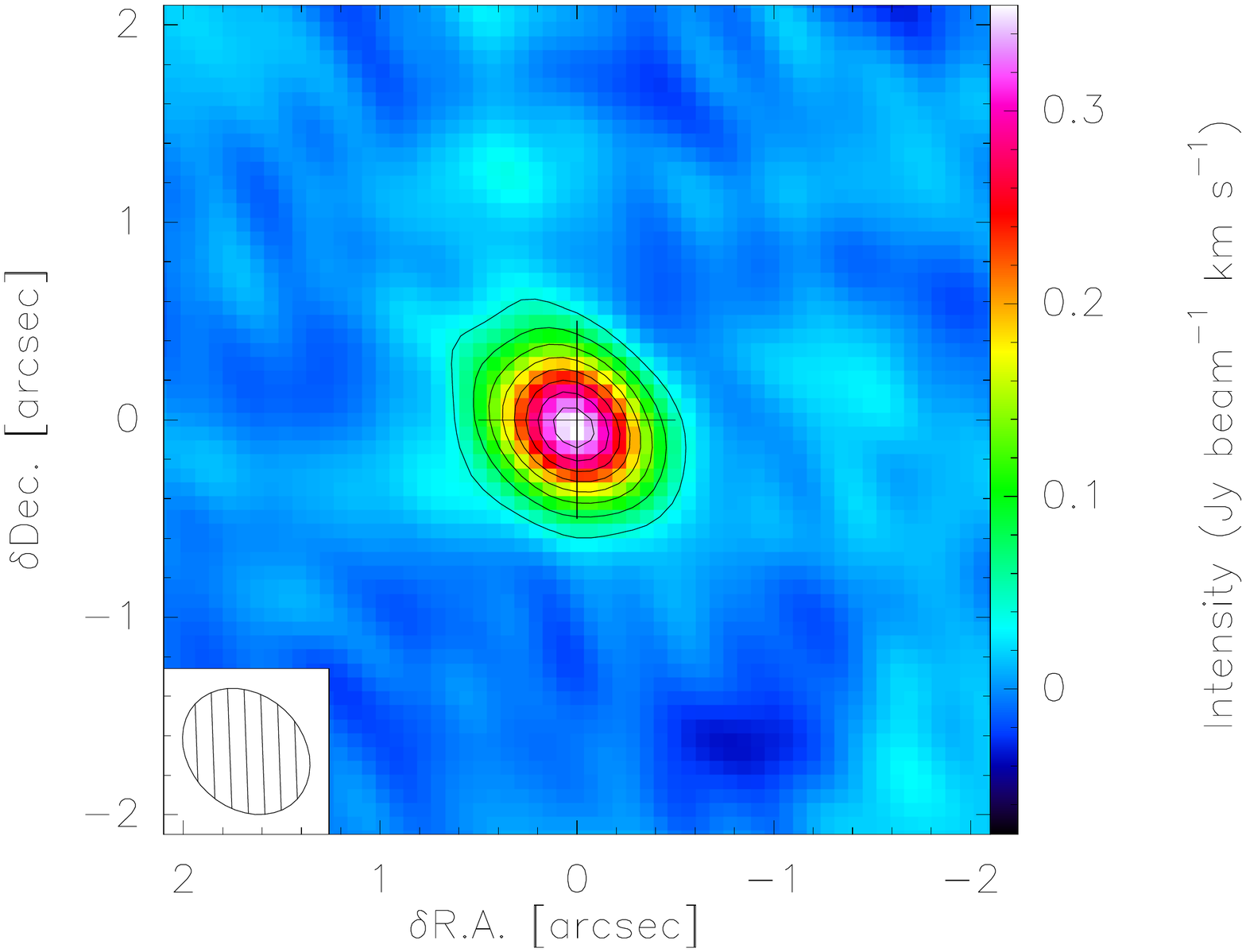}
    \includegraphics[width=60mm]{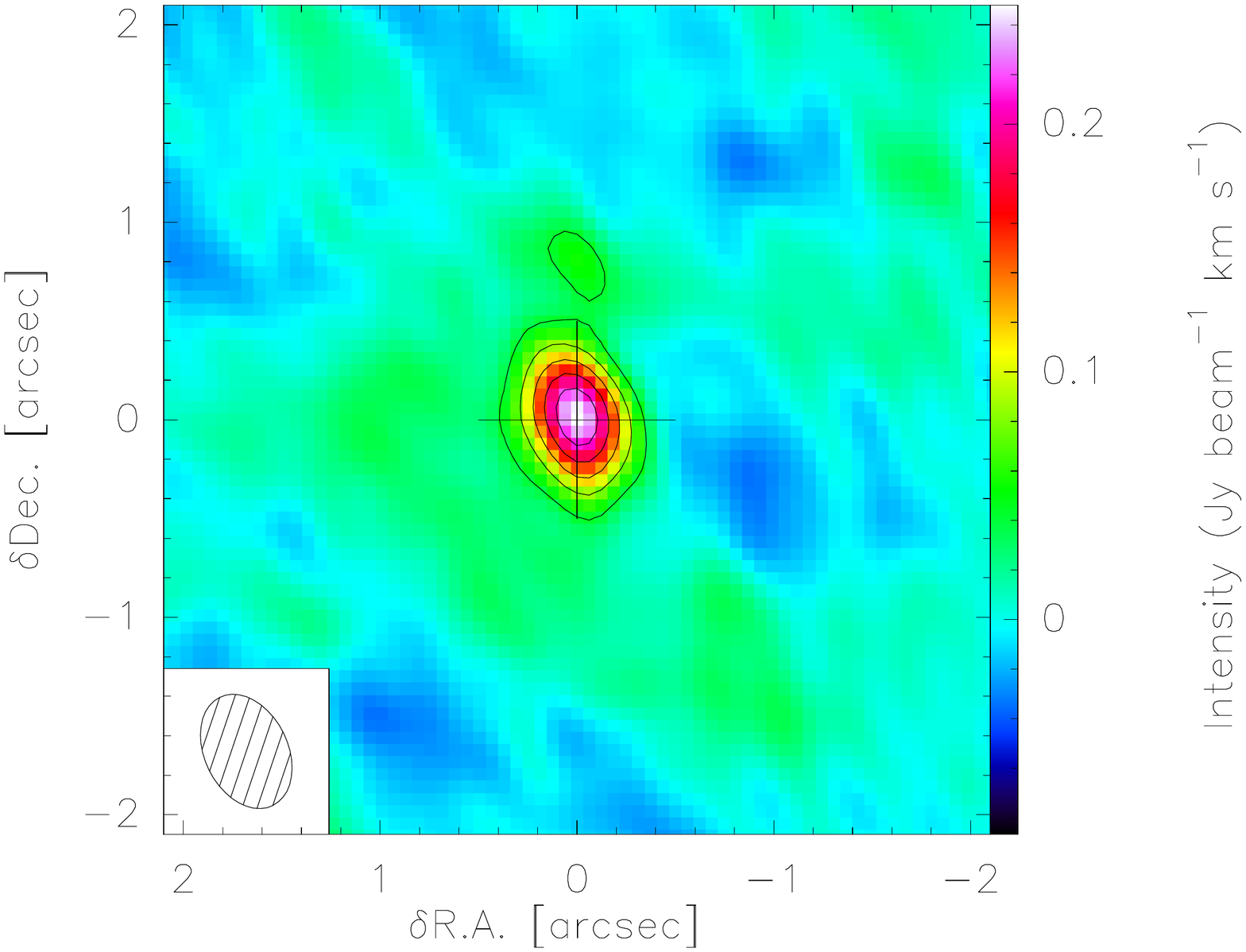}
    % make-intensity_H2O.map
    \caption{Zeroth moment maps of the H$_2$O 232\,GHz (left)  and
    H$_2$O 263\,GHz (right) lines. North is up and east is to the left.
    Note the different color scales.  Contours are plotted every $5
    \sigma$  where $1 \sigma = 13.8$ mJy/beam$\cdot$\kms for the 232\,GHz
    line and  $1 \sigma = 13.92$ mJy/beam$\cdot$\kms for the 263\,GHz
    line.  The black ellipse in the lower left corner indicates the
    synthesized beam.}
    \label{H2Ointensfig}
\end{figure*}

\begin{figure*}[h]
    \centering
    \includegraphics[width=60mm]{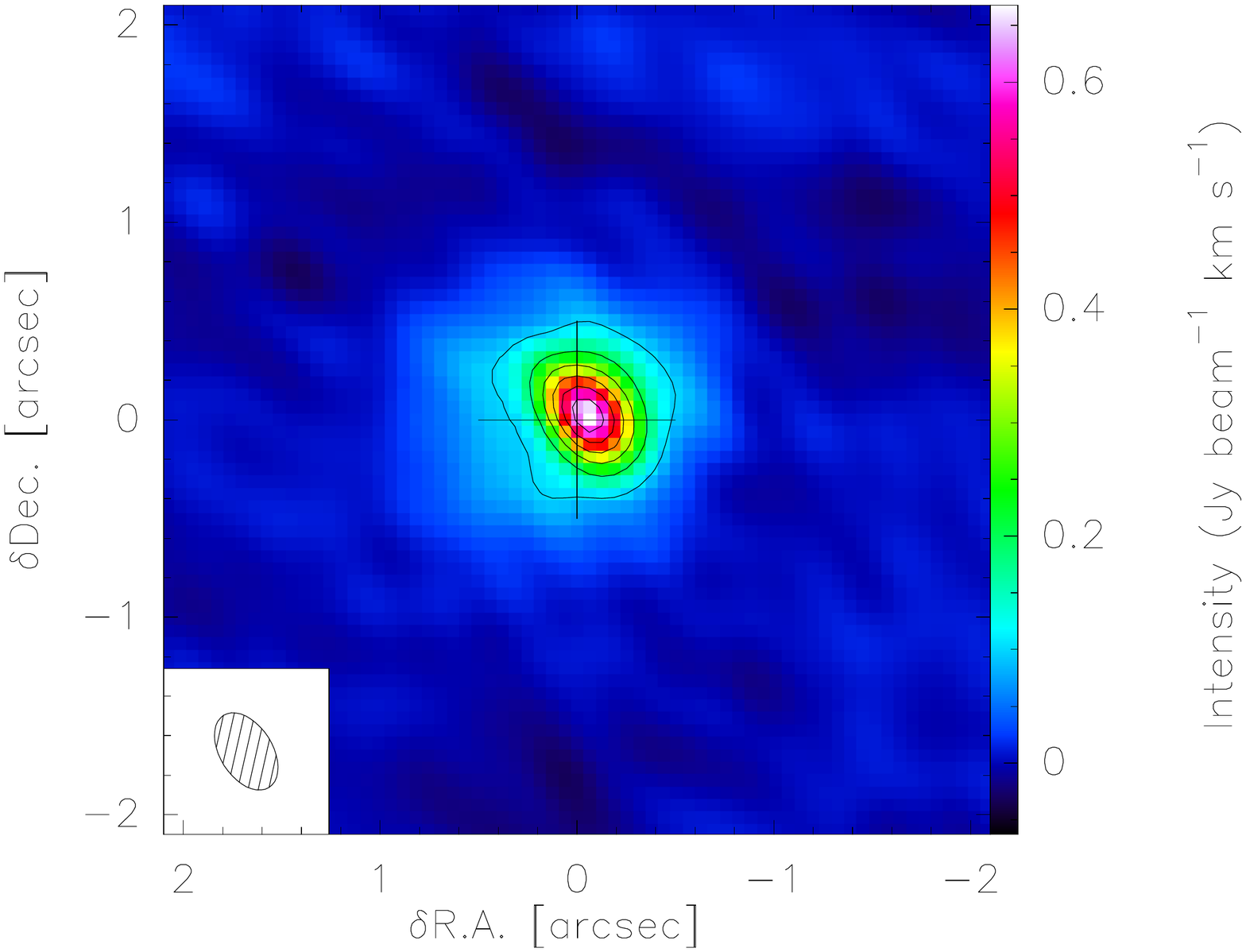}
    \includegraphics[width=60mm]{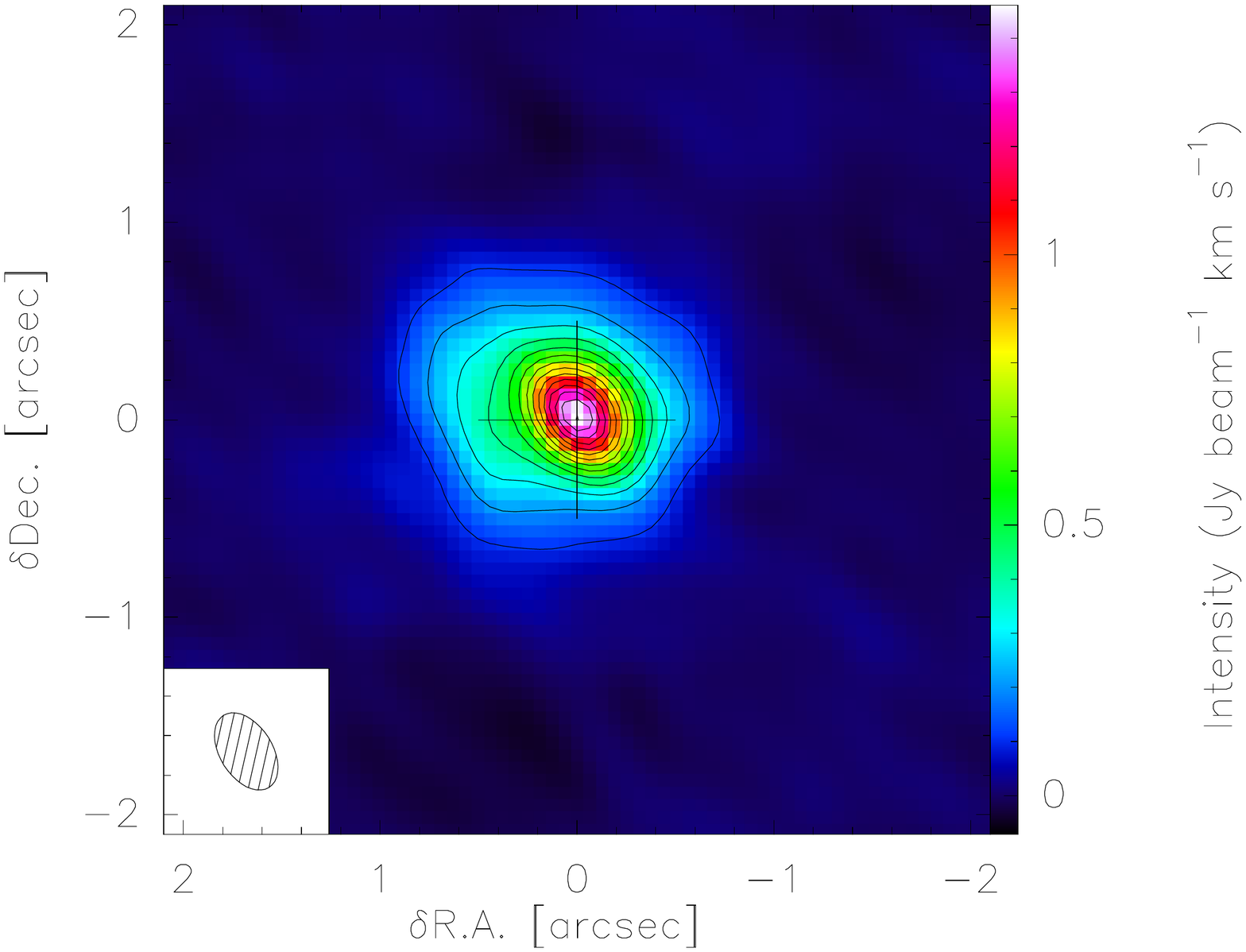}
    \includegraphics[width=60mm]{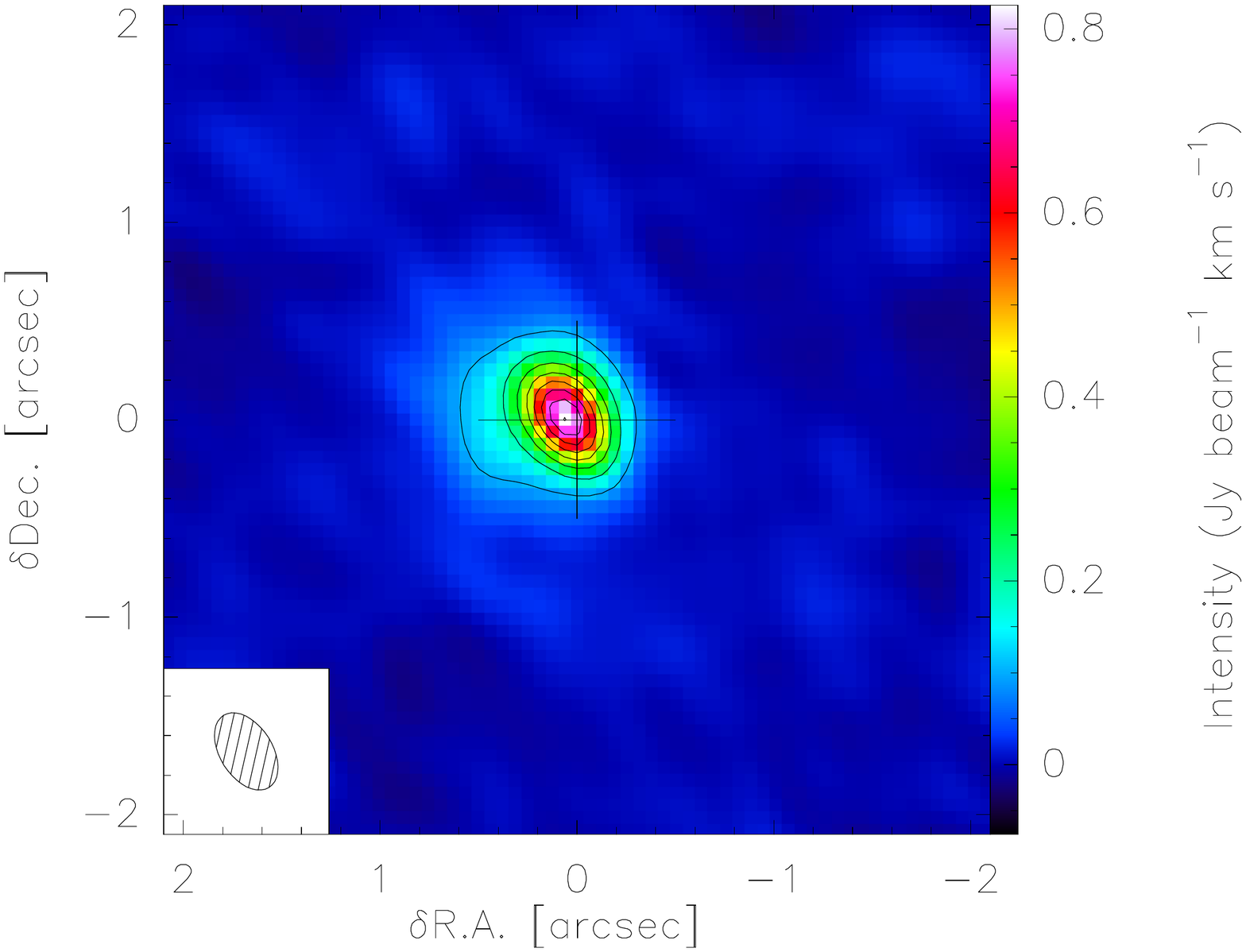}
    \includegraphics[width=60mm]{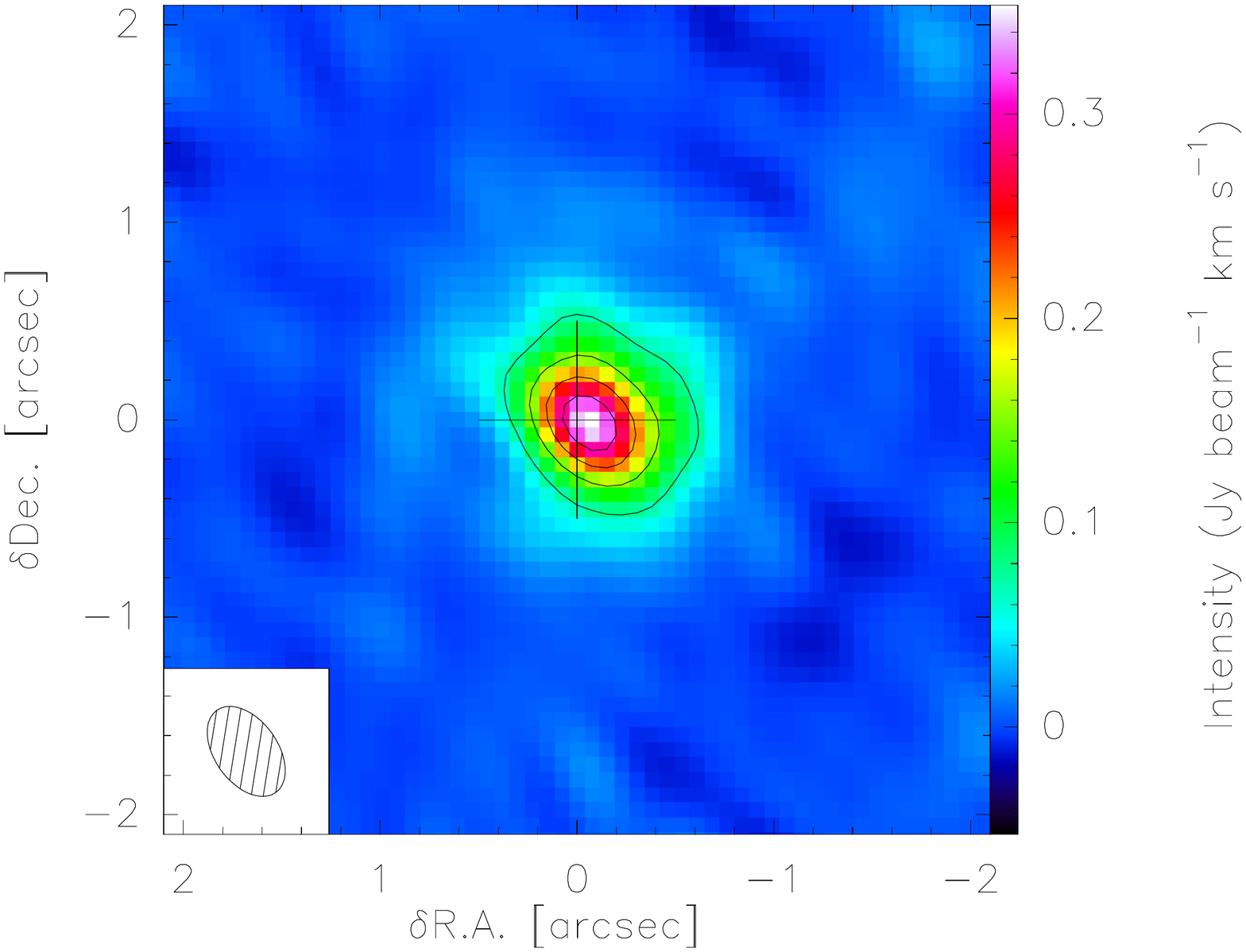}
    \includegraphics[width=60mm]{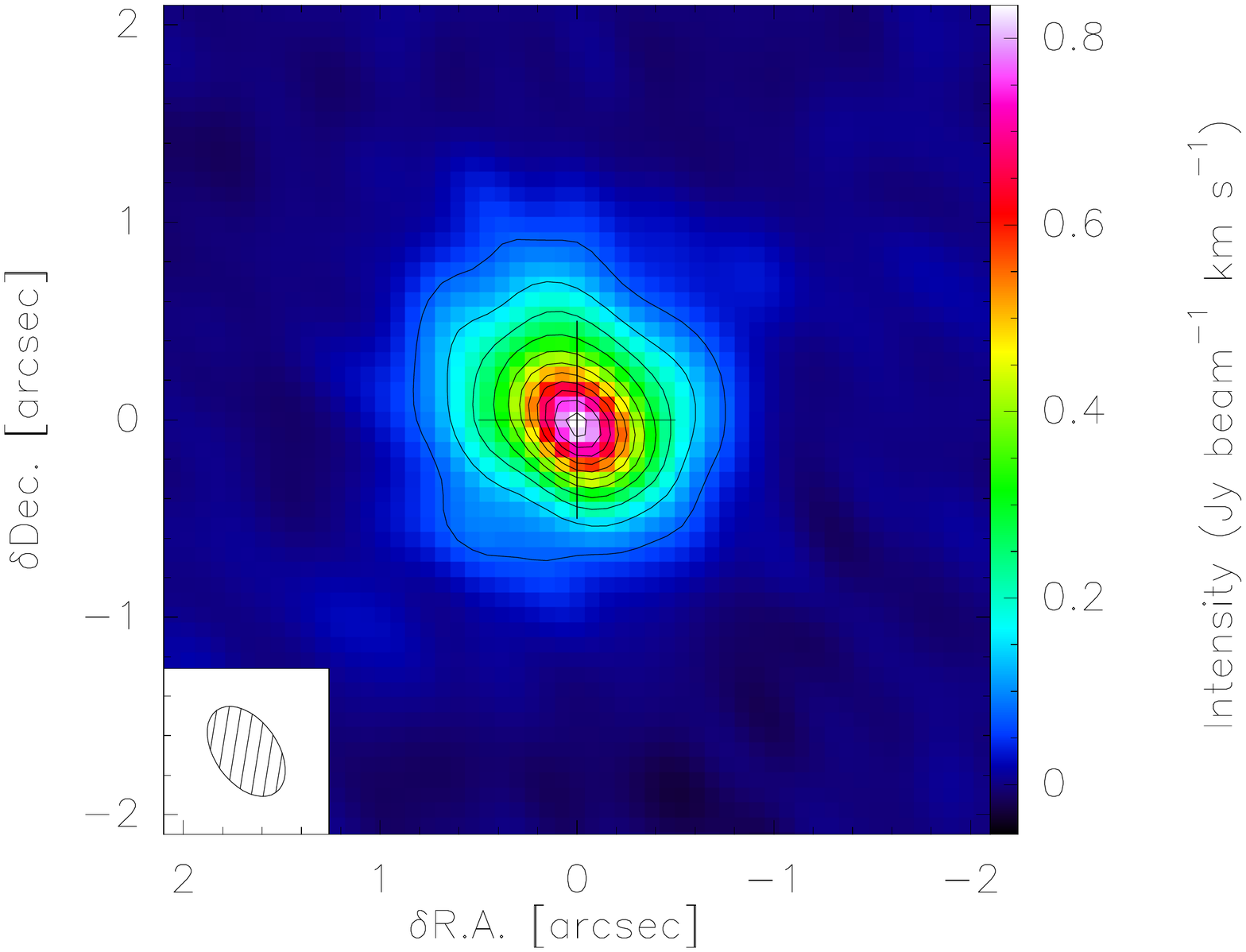}
    \includegraphics[width=60mm]{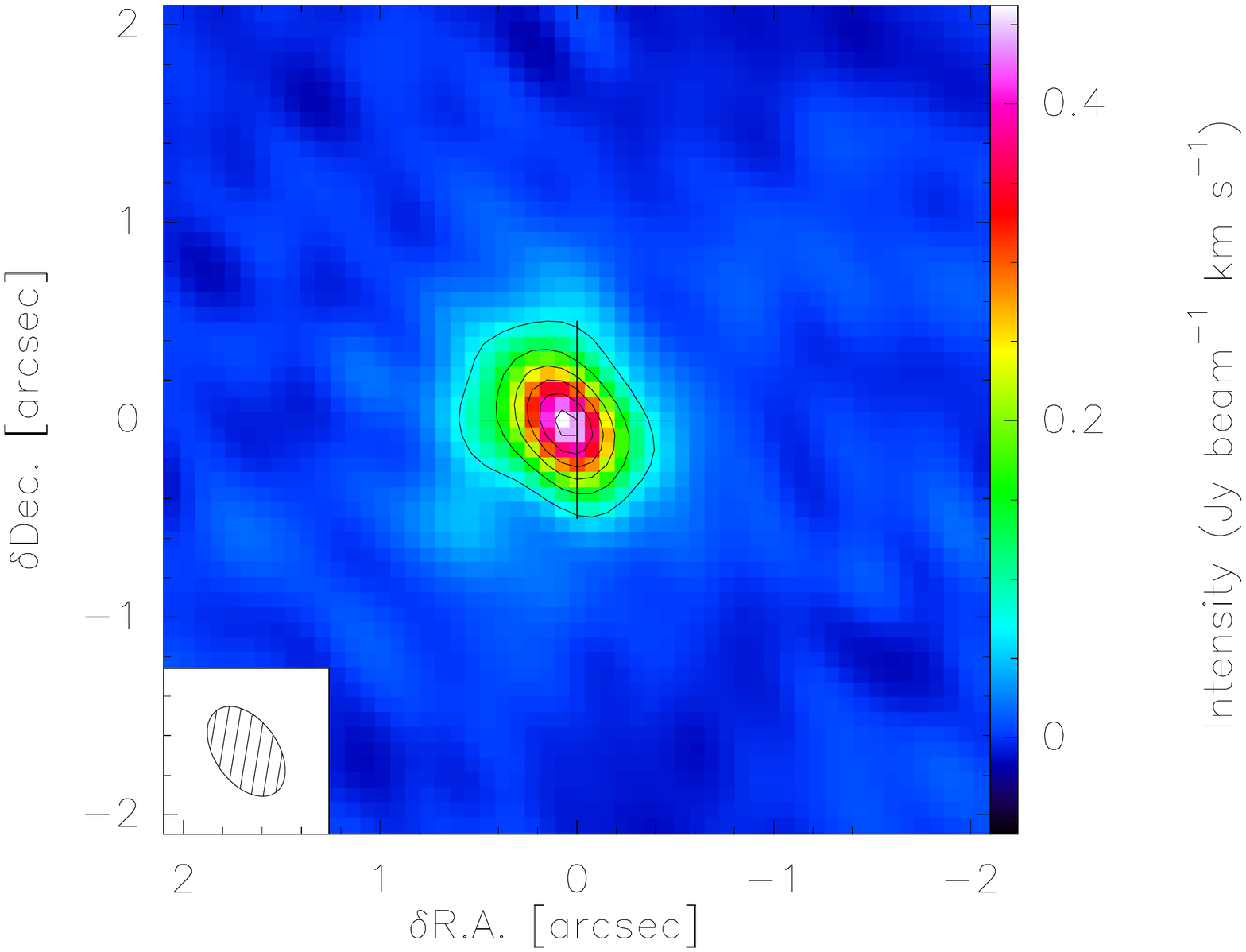}
    \includegraphics[width=60mm]{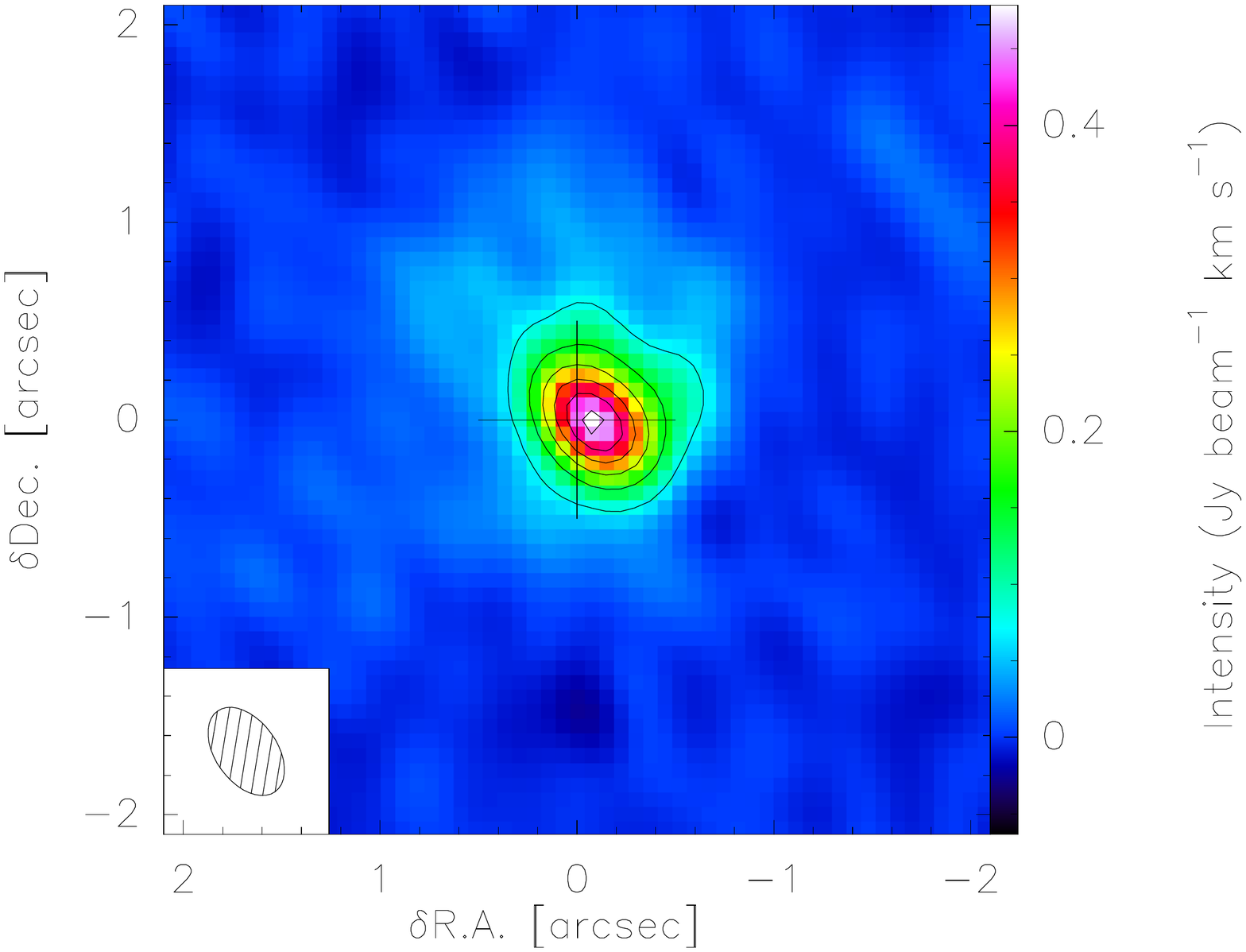}
    \includegraphics[width=60mm]{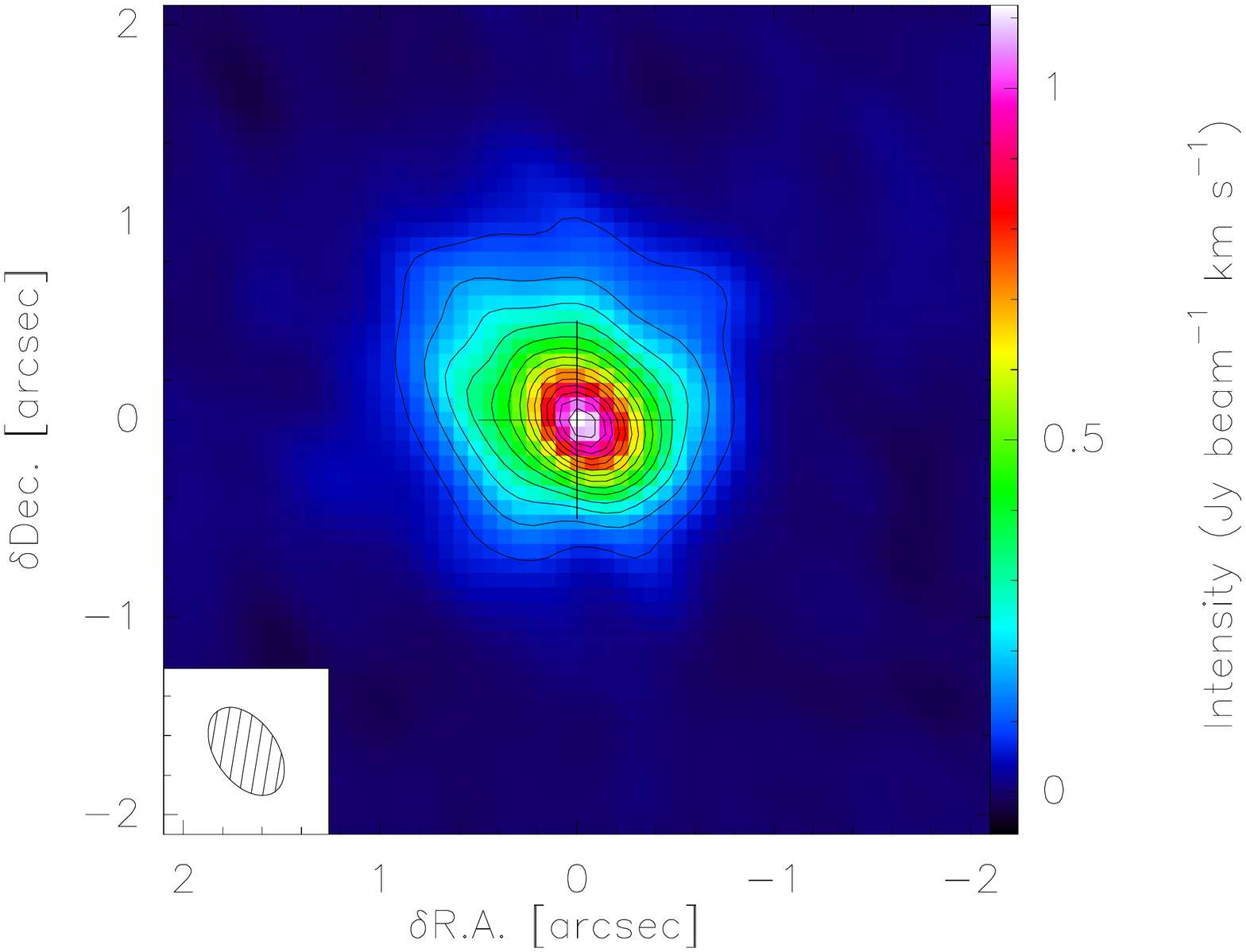}
    \includegraphics[width=60mm]{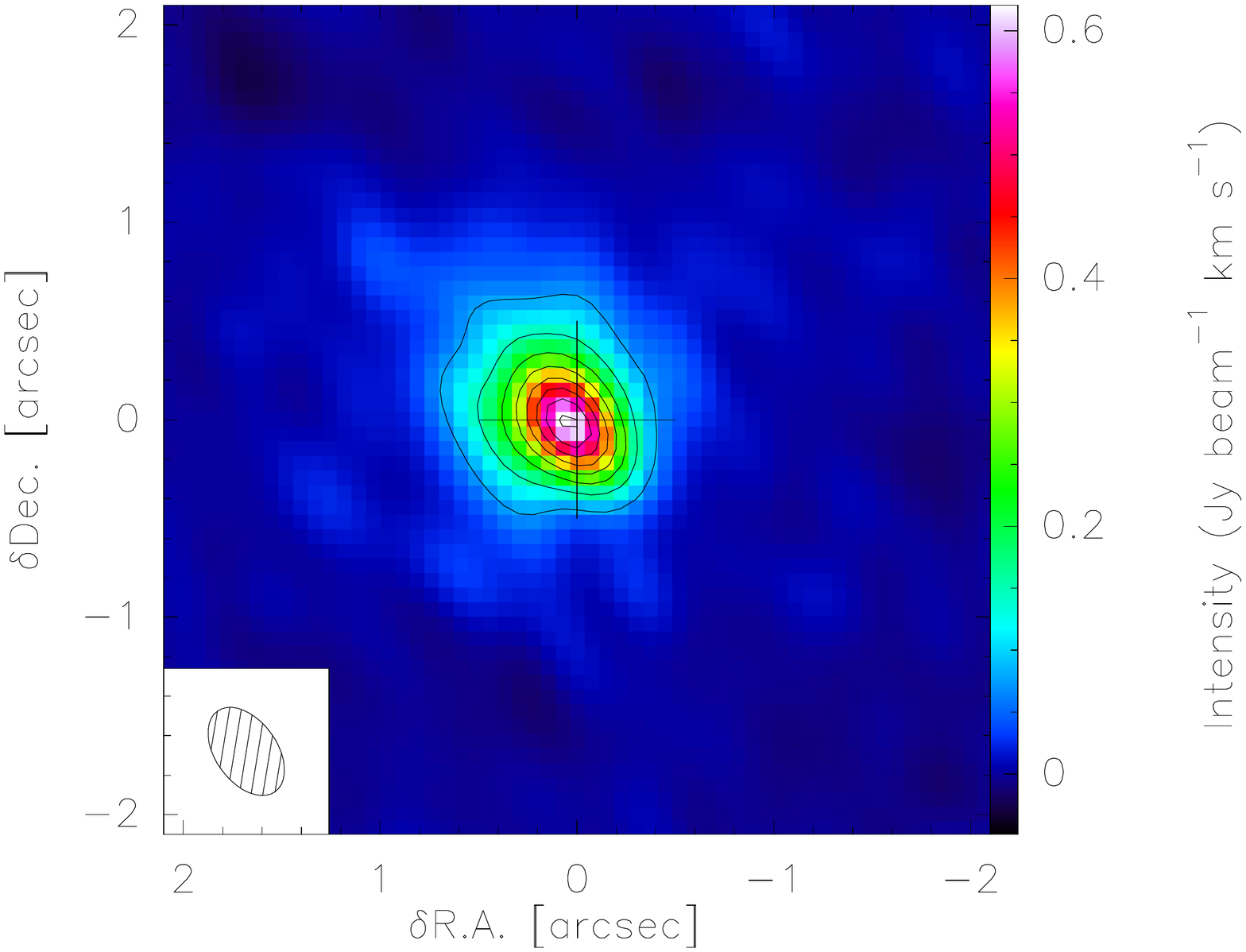}
    \includegraphics[width=60mm]{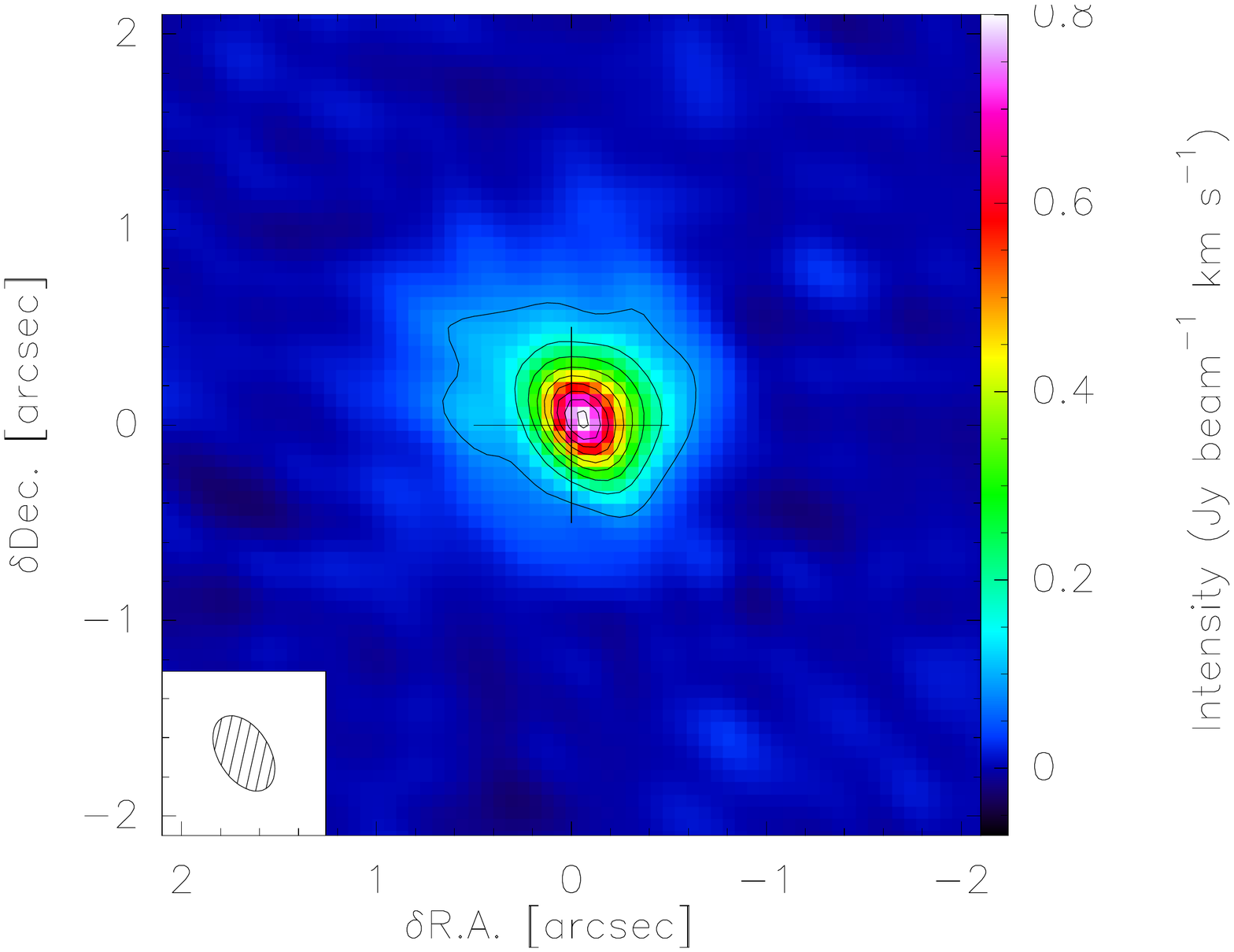}
    \includegraphics[width=60mm]{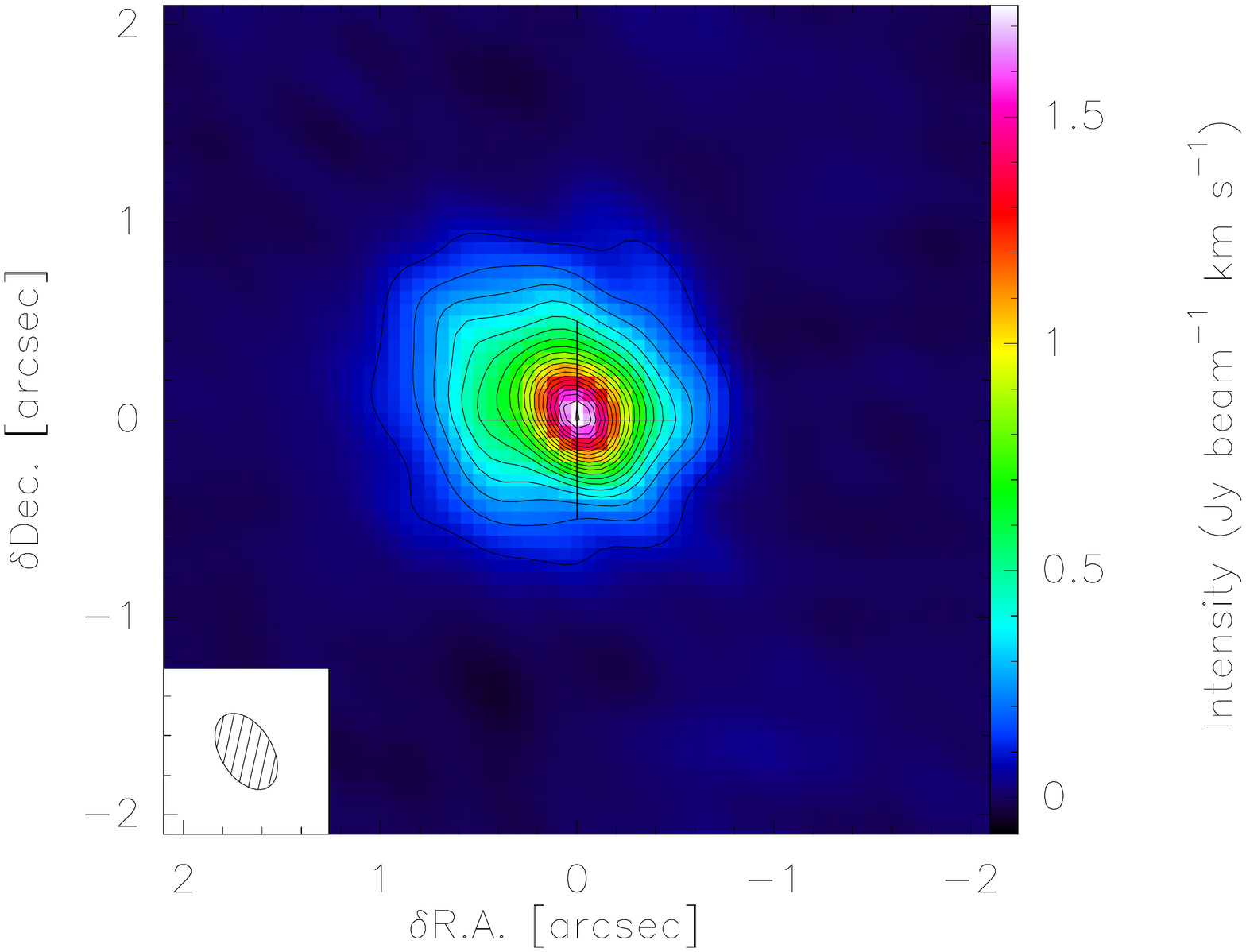}
    \includegraphics[width=60mm]{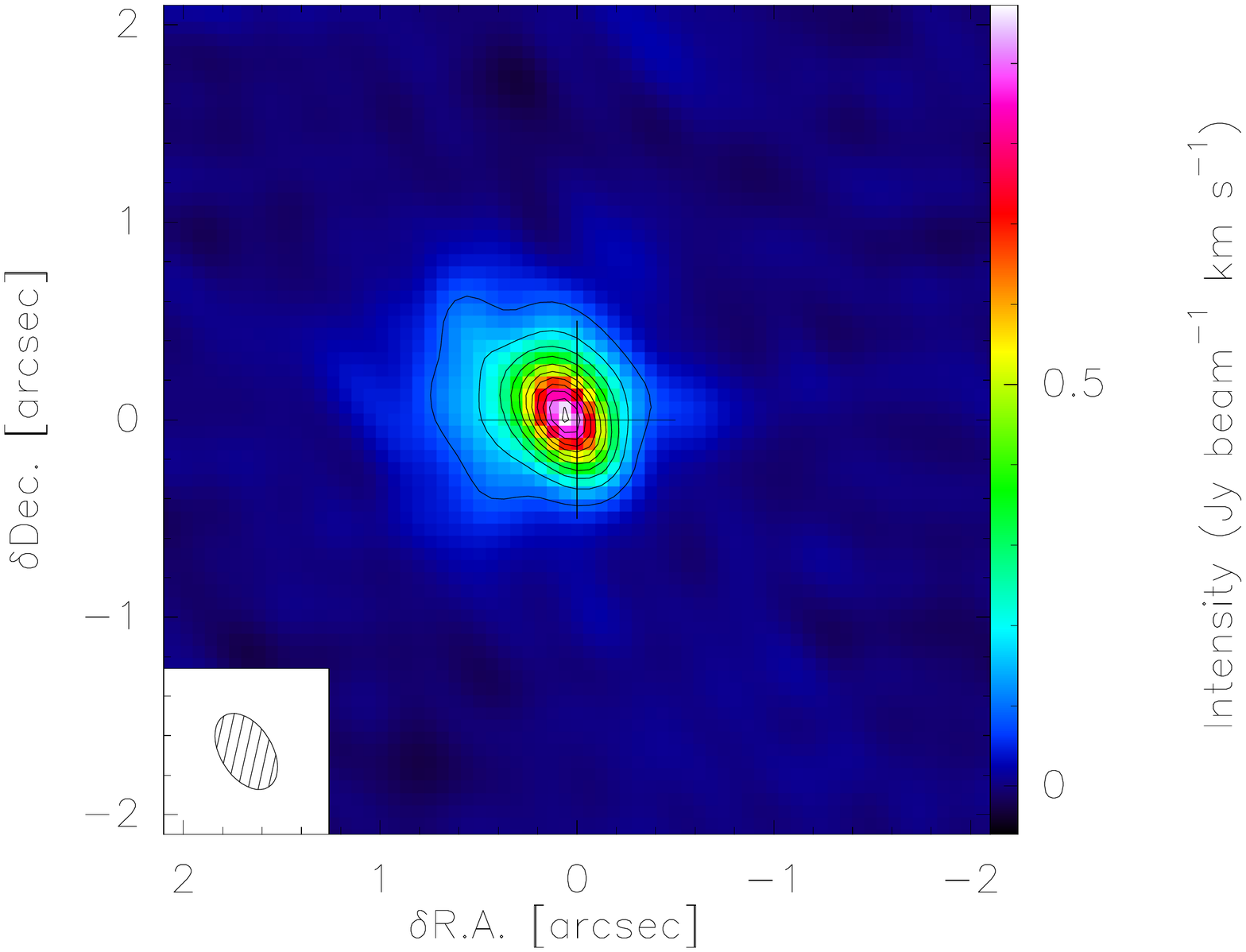}
    % make-blue-center-red_SO.map
    \caption{Velocity-integrated intensity maps of the SO(6(6)-5(5))
      (upper row), SO(5(5)-4(4)) (second row), SO(6(5)-5(4)) (third row)
      and SO(7(6)-6(5)) (lower row) lines covering three velocity
      intervals. Left: blue line wing [v$_{\rm lsr,*}-10$,v$_{\rm lsr,*}-2$] \kms,
      Middle: line center [v$_{\rm lsr,*}-2$,v$_{\rm lsr,*}+2$] \kms,
      Right: red line wing [v$_{\rm lsr,*}+2$,v$_{\rm lsr,*}+10$] \kms.
      North is up and east is to the left. Note the different color scales. 
      Contours are plotted every $10 \sigma$, where (from left to right)
      $1 \sigma = 10.0, 10.4, 10.3$ mJy/beam$\cdot$\kms for SO(6(6)-5(5)),
      $1 \sigma =  7.3,  7.6,  7.4$ mJy/beam$\cdot$\kms for SO(5(5)-4(4)),
      $1 \sigma =  7.6,  8.1,  7.4$ mJy/beam$\cdot$\kms for SO(6(5)-5(4)), and 
      $1 \sigma =  8.7,  9.6,  8.7$ mJy/beam$\cdot$\kms for SO(7(6)-6(5)).
      The black ellipse in the lower left corner indicates the synthesized beam.}
    \label{SOstructfig}
\end{figure*}

\begin{figure*}[h]
    \centering
    \includegraphics[width=60mm]{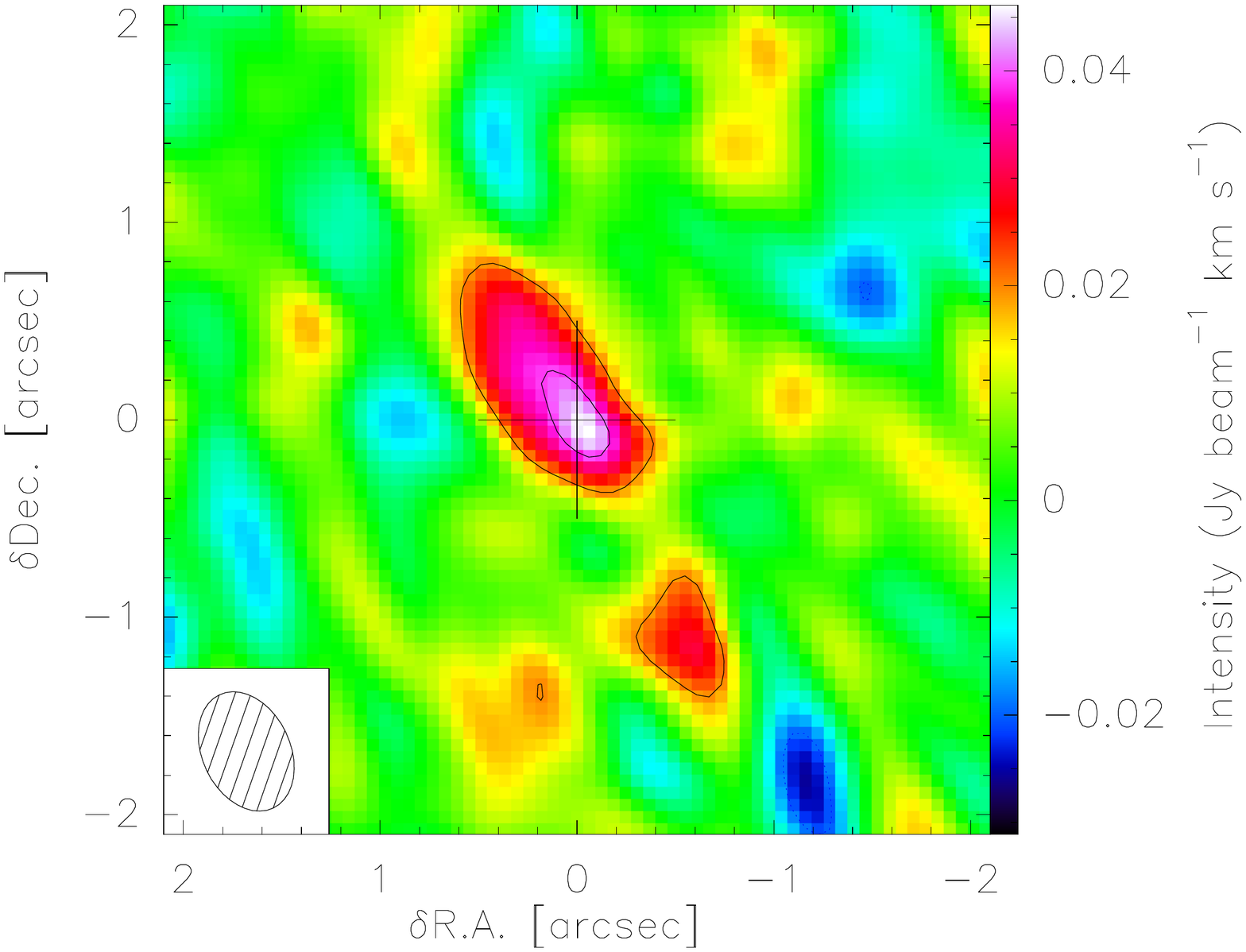}
    \includegraphics[width=60mm]{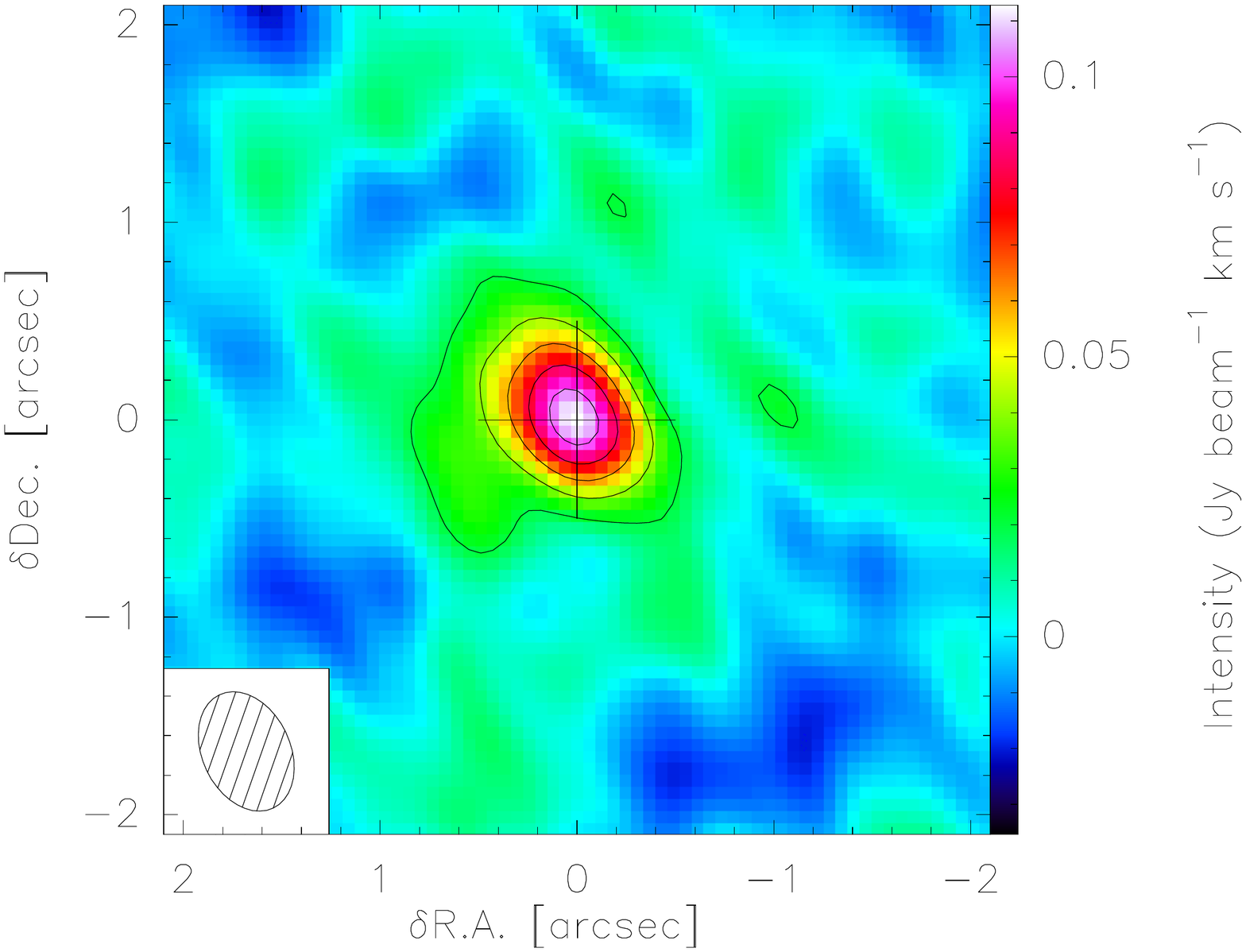}
    \includegraphics[width=60mm]{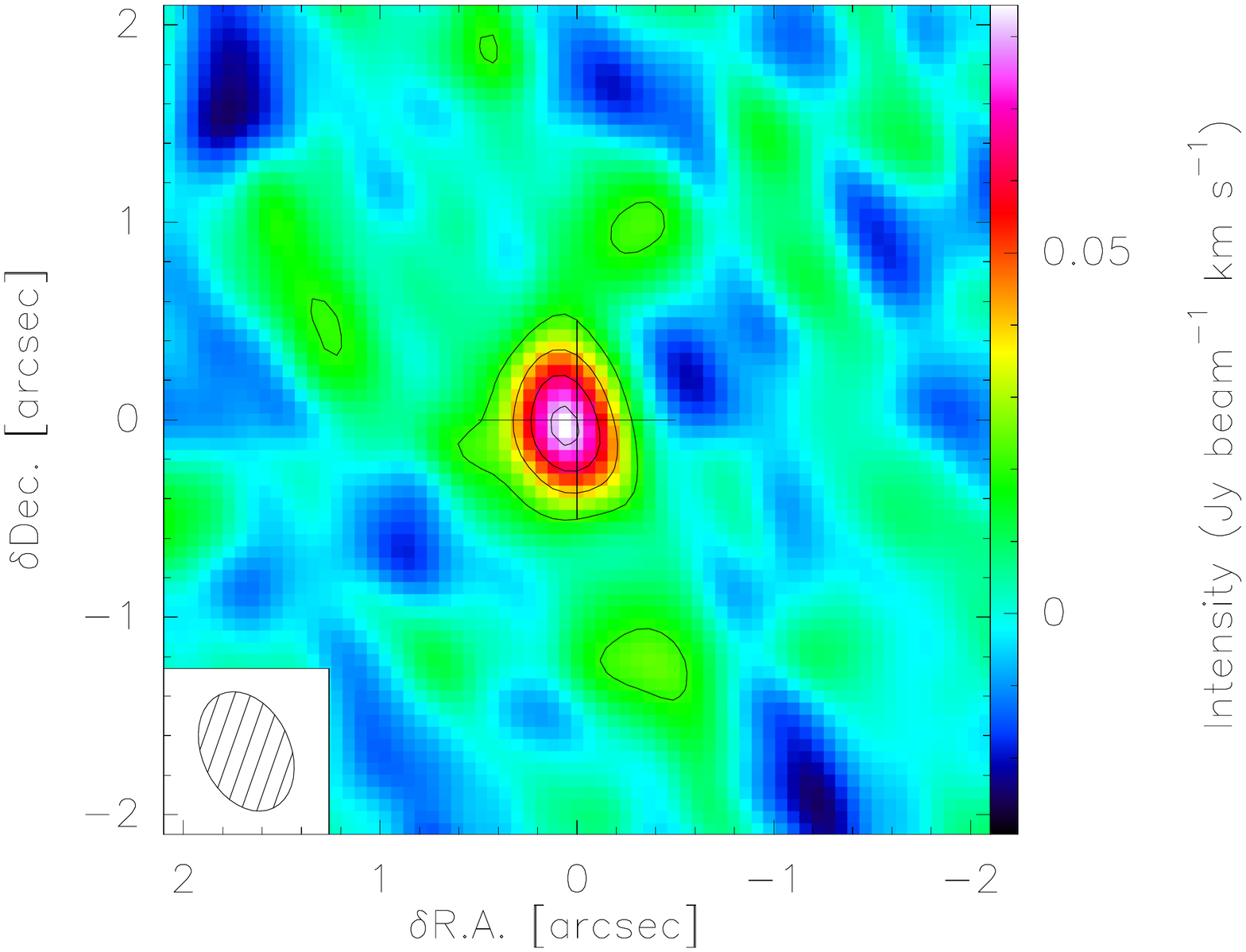}
    \includegraphics[width=60mm]{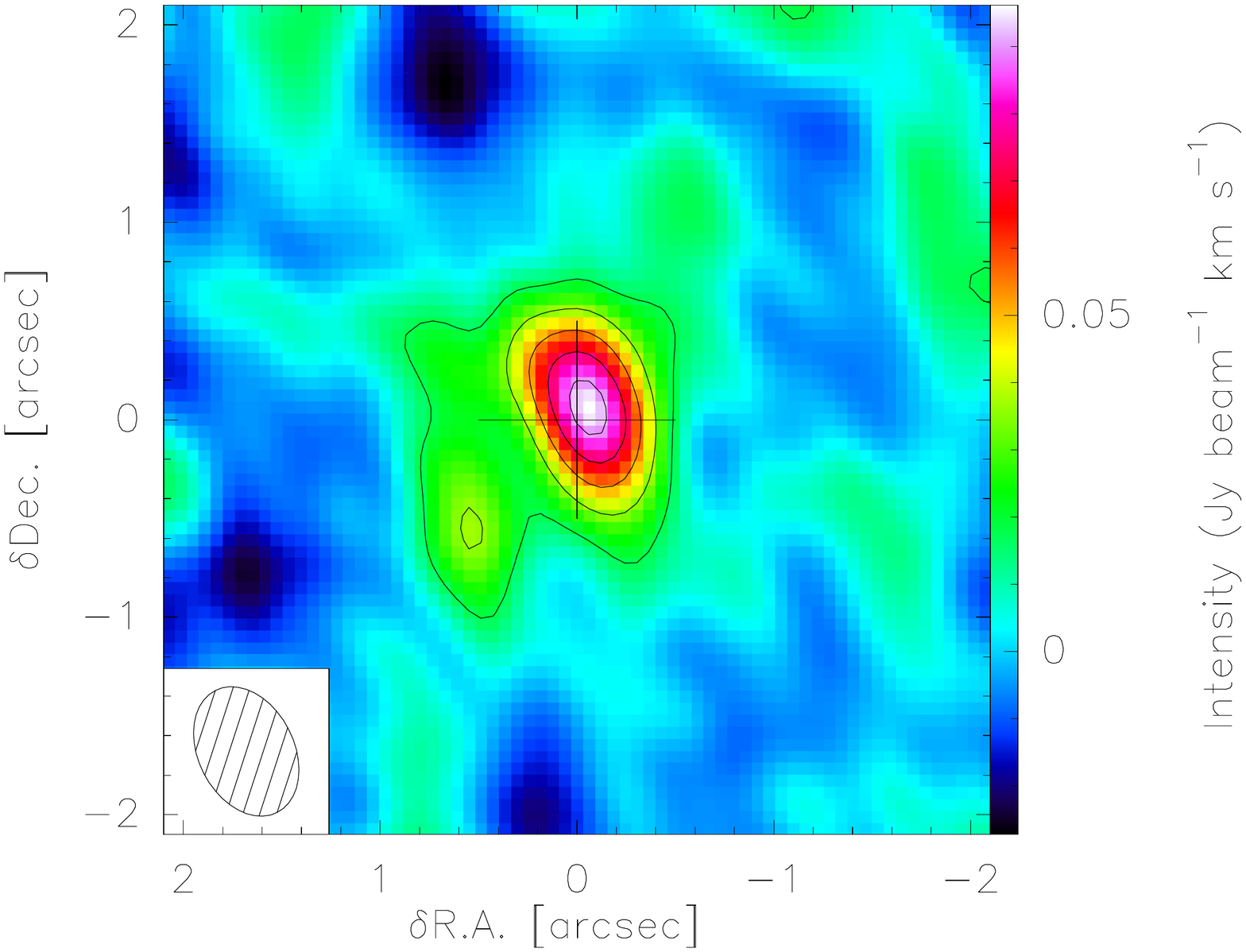}
    \includegraphics[width=60mm]{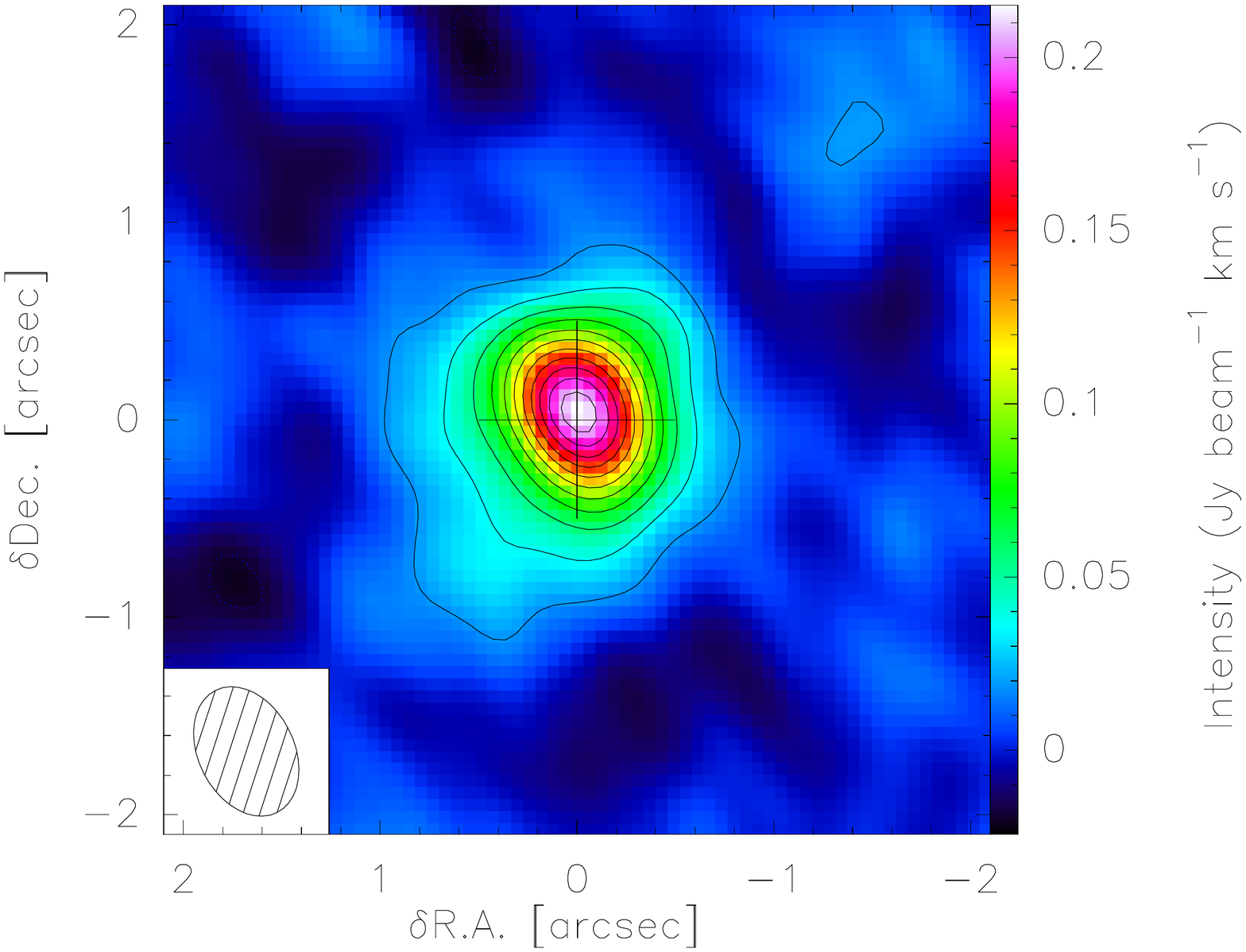}
    \includegraphics[width=60mm]{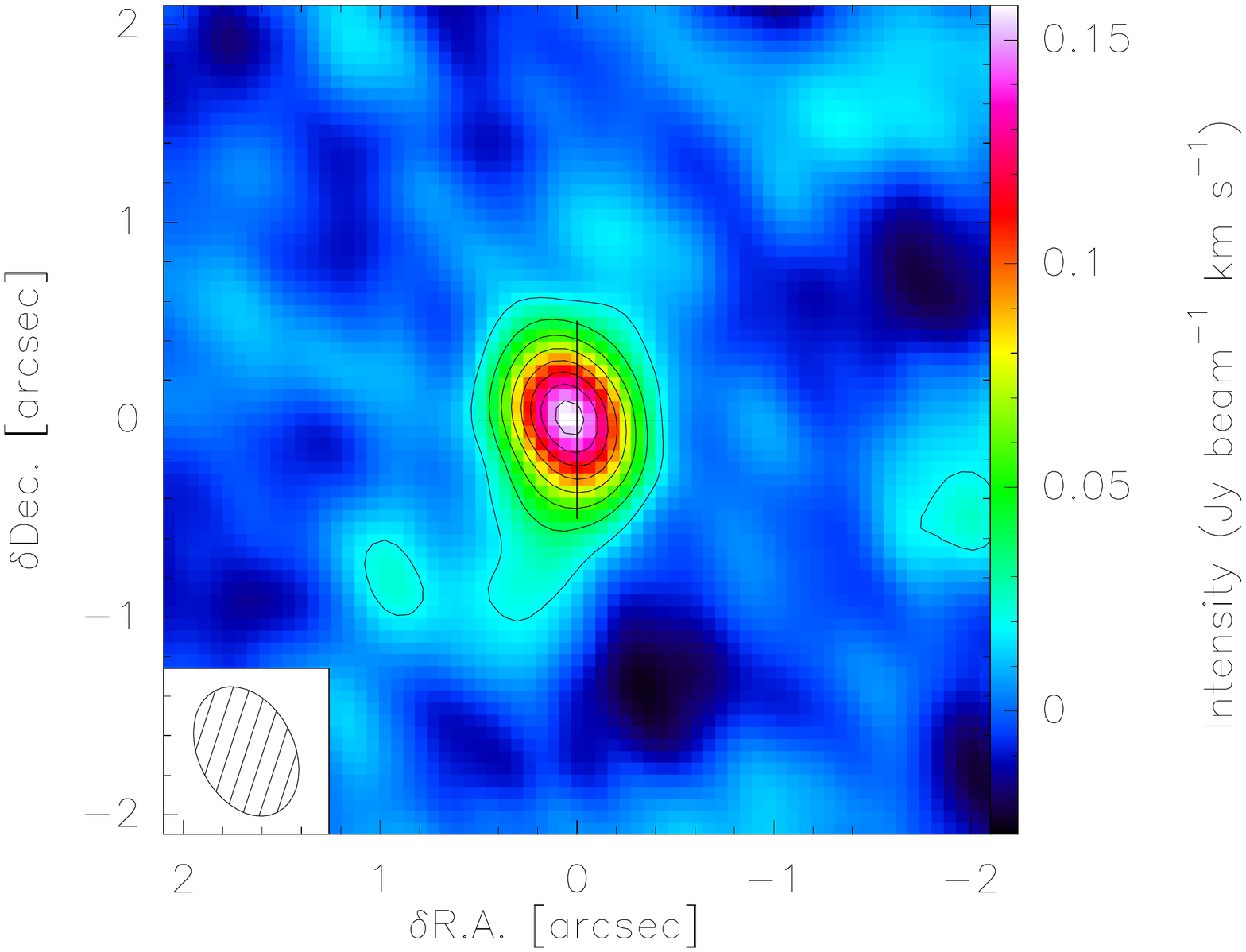}
    \includegraphics[width=60mm]{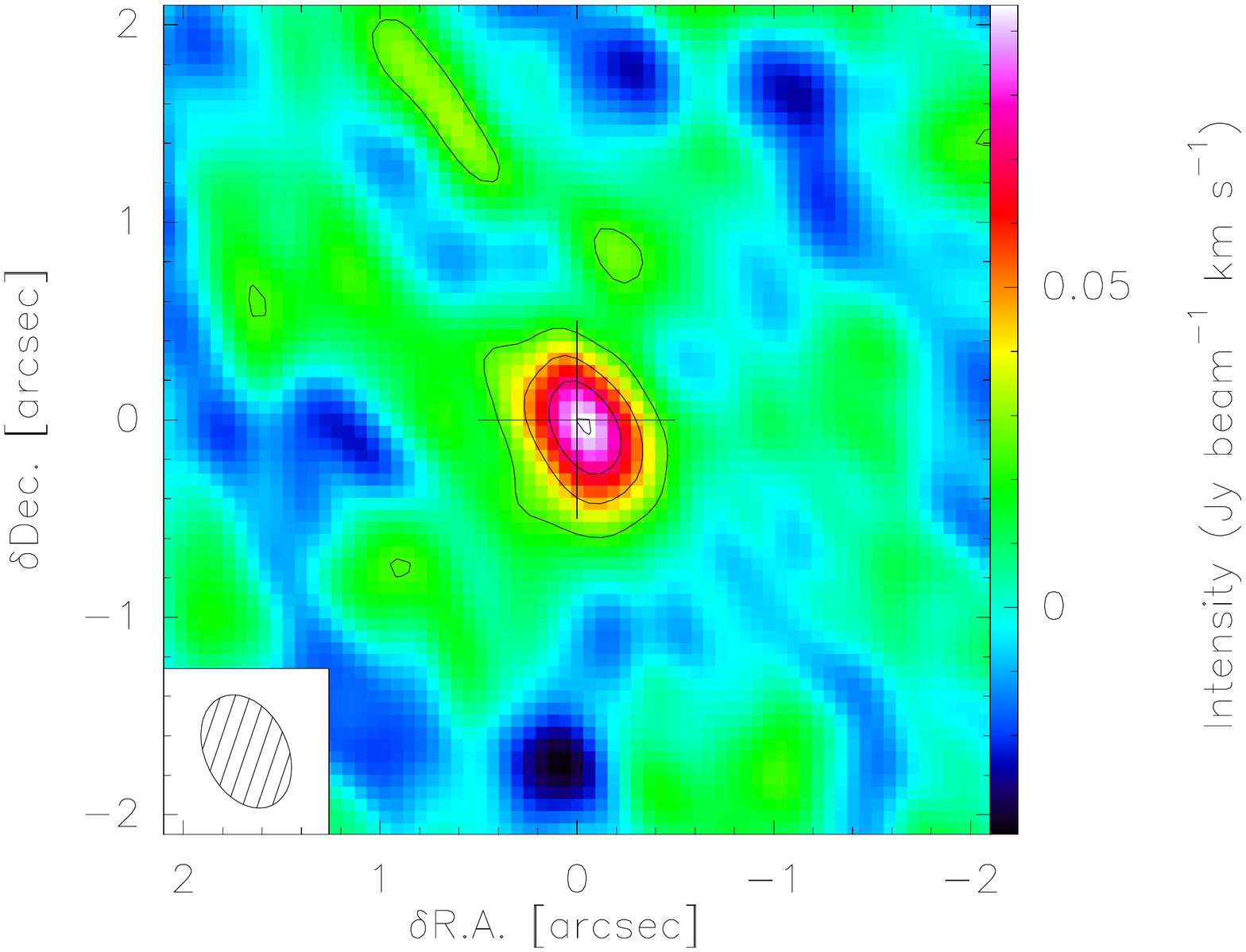}
    \includegraphics[width=60mm]{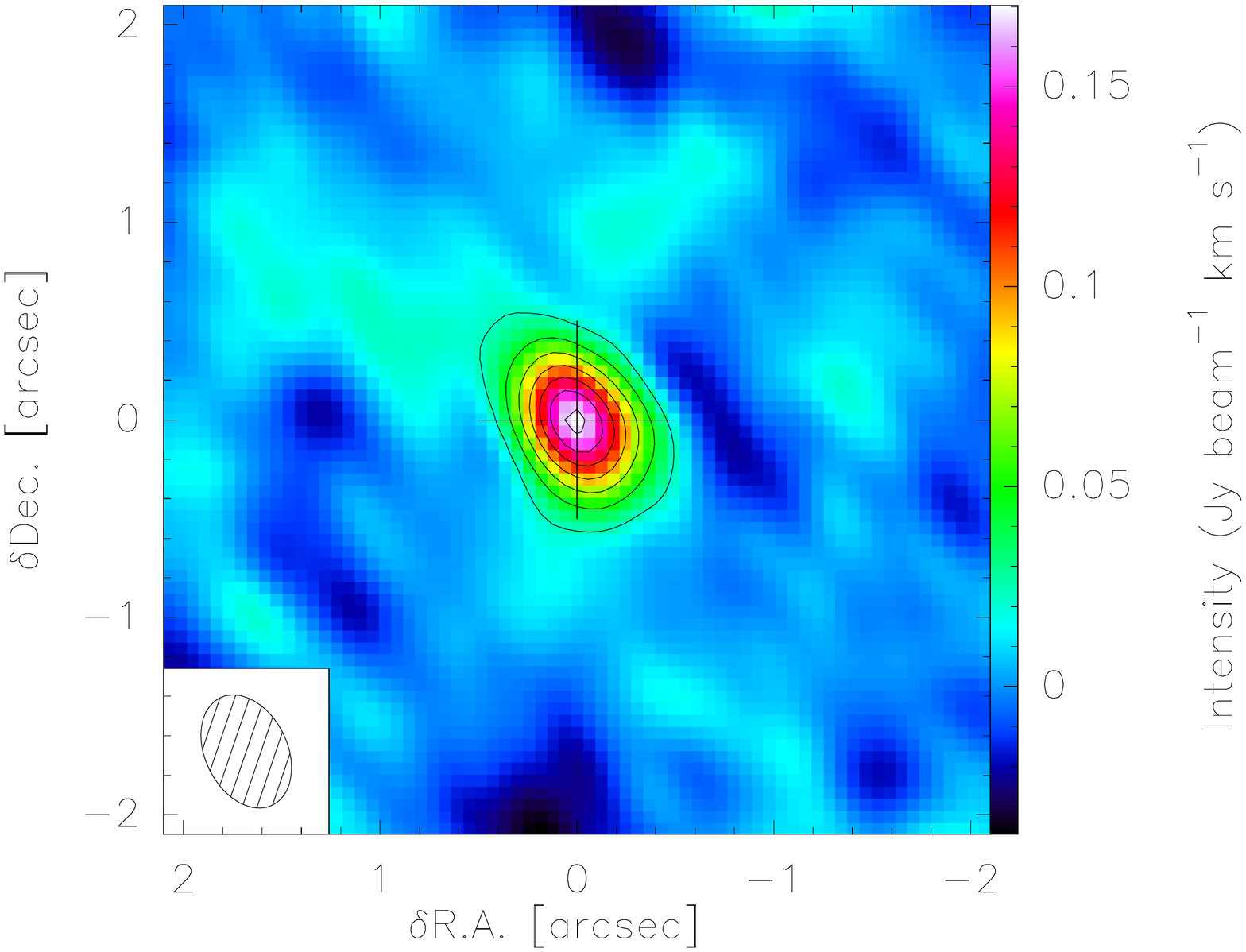}
    \includegraphics[width=60mm]{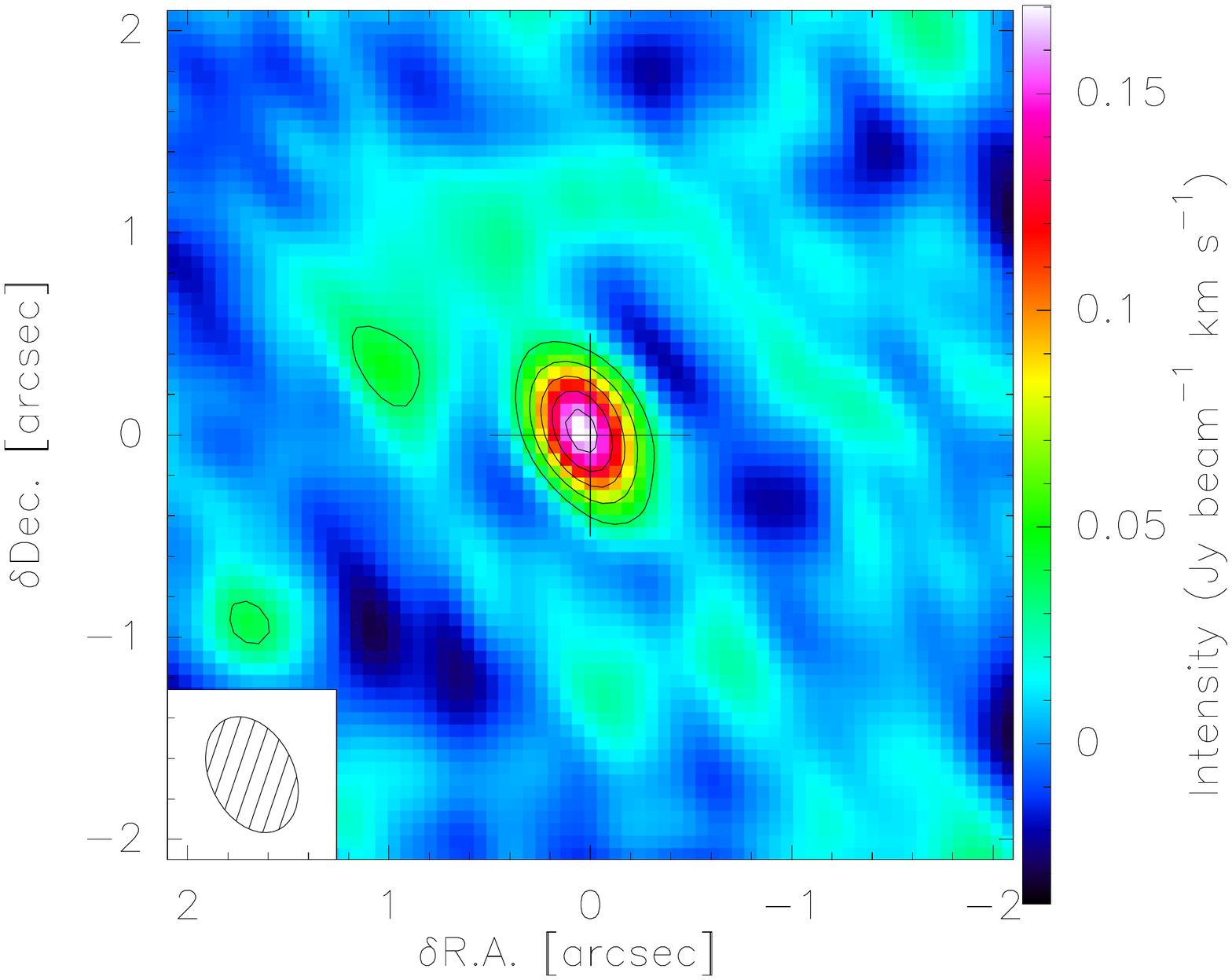}
    % make-blue-center-red_SO2.map
      \caption{Velocity-integrated intensity maps of SO$_2$ (9(3, 7)-9(2, 8))
        (lowest E$_u$, upper row),
        SO$_2$ (14(0,14)-13(1,13)) (strongest SO$_2$ line, second row), and 
        SO$_2$ (34(4,30)-34(3,31)) (highest E$_u$, lower row),  
      covering three velocity
      intervals. Left: blue line wing [v$_{\rm lsr,*}-10$,v$_{\rm lsr,*}-2$] \kms,
      Middle: line center [v$_{\rm lsr,*}-2$,v$_{\rm lsr,*}+2$] \kms,
      Right: red line wing [v$_{\rm lsr,*}+2$,v$_{\rm lsr,*}+10$] \kms.
      North is up and east is to the left. Note the different color scales. 
      Contours are plotted every $3 \sigma$, where (from left to right)
      $1 \sigma =  6.5,  6.7,  6.6$ mJy/beam$\cdot$\kms for SO$_2$( 9(3, 7)- 9(2, 8))
      $1 \sigma =  5.9,  6.2,  6.3$ mJy/beam$\cdot$\kms for SO$_2$(14(0,14)-13(1,13)),
      $1 \sigma =  7.8,  7.8,  7.7$ mJy/beam$\cdot$\kms for SO$_2$ (34(4,30)-34(3,31)) .
      The black ellipse in the lower left corner indicates the synthesized beam.}
    \label{SO2structfig}
\end{figure*}

\begin{figure*}[h]
    \centering
    \includegraphics[width=60mm]{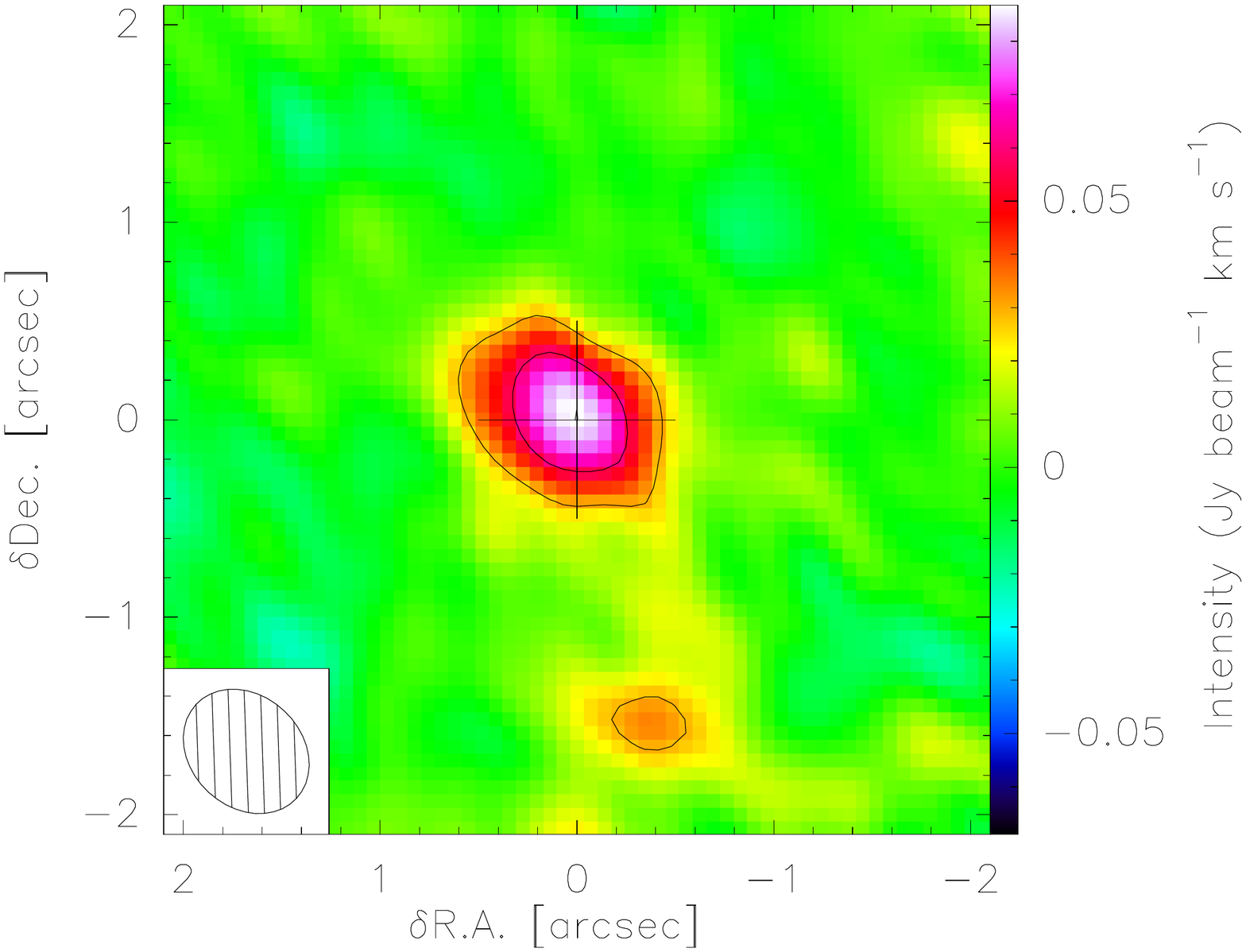}
    % make-intensity_PN.map
    \caption{Zeroth moment map of the PN(N=5-4,J=6-5) line.
      Contours are plotted every $3 \sigma$, where 
      $1 \sigma =  9.6$ mJy/beam$\cdot$\kms. North is up and 
      east is to the left. The black ellipse in the lower 
      left corner indicates the synthesized beam.}
    \label{PNstructfig}
\end{figure*}

Also shown are zeroth moment maps for the (unresolved) H$_2$O lines
(Fig.~\ref{H2Ointensfig}) and for the (weak) PN line
(Fig.~\ref{PNstructfig}).

\section{SO$_2$ line profiles}
\label{so2profilesec}
\begin{figure}[h]
    \includegraphics[width=110mm]{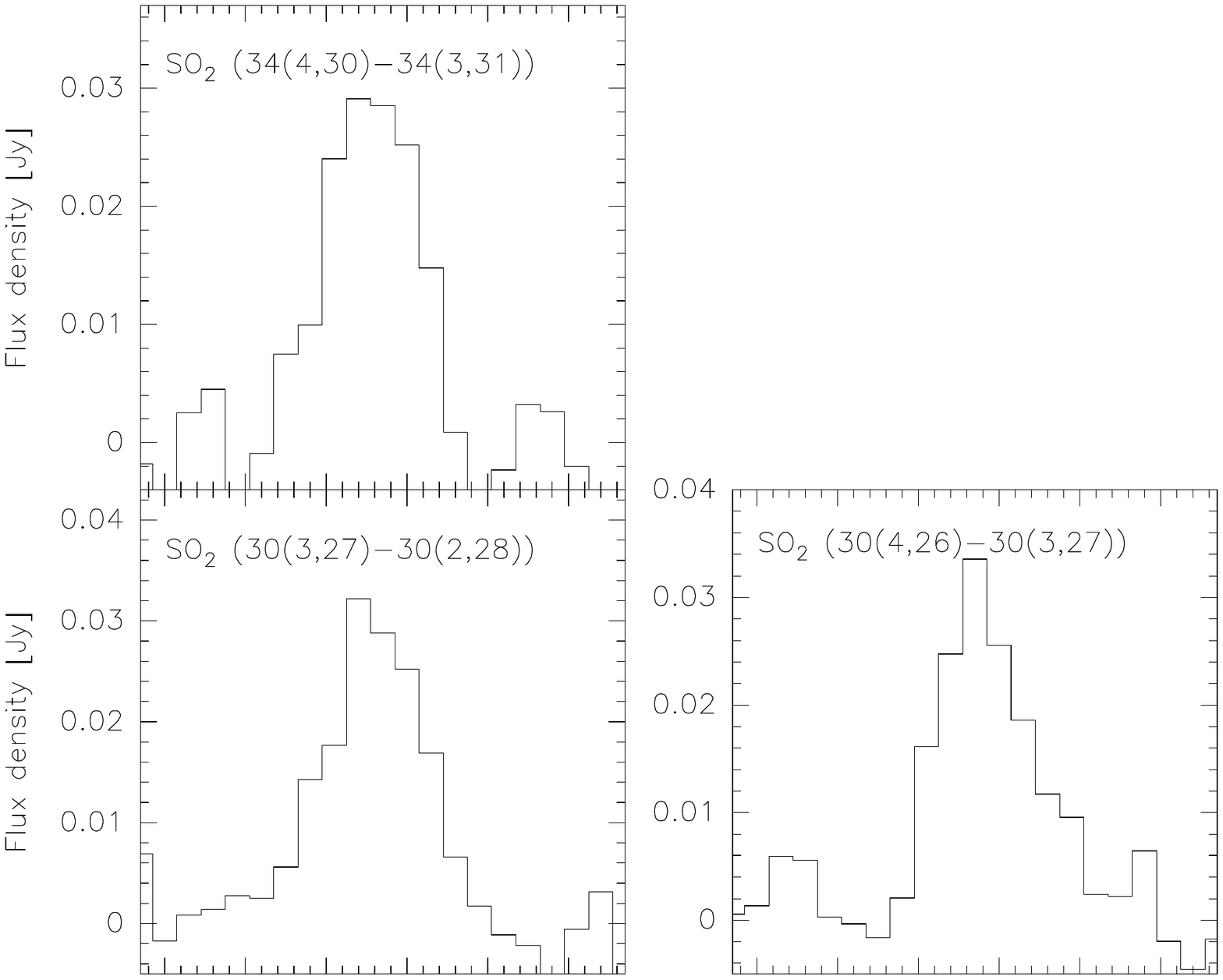}\vspace{-2.17cm}
    \includegraphics[width=110mm]{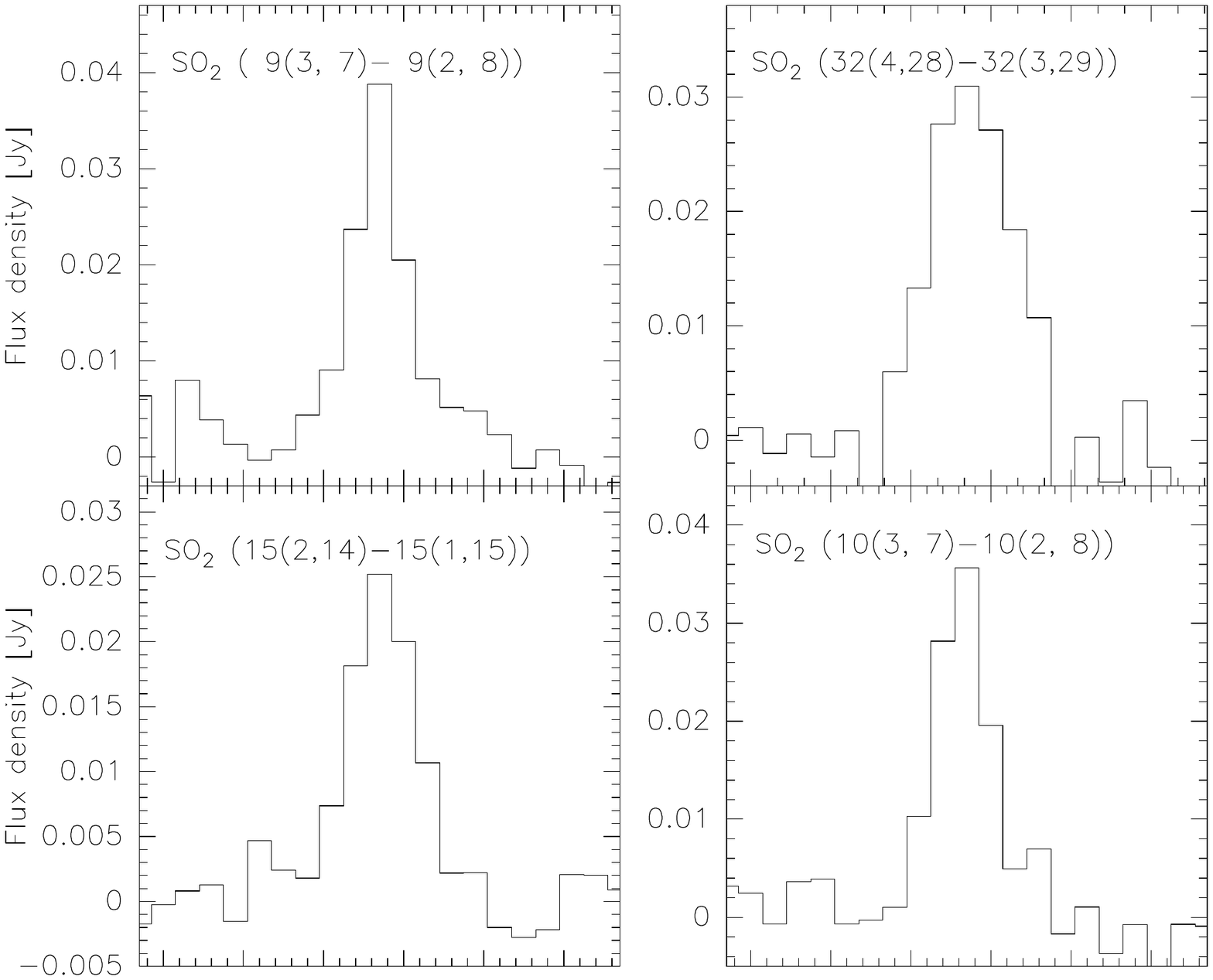}\vspace{-2.16cm}
    \includegraphics[width=110mm]{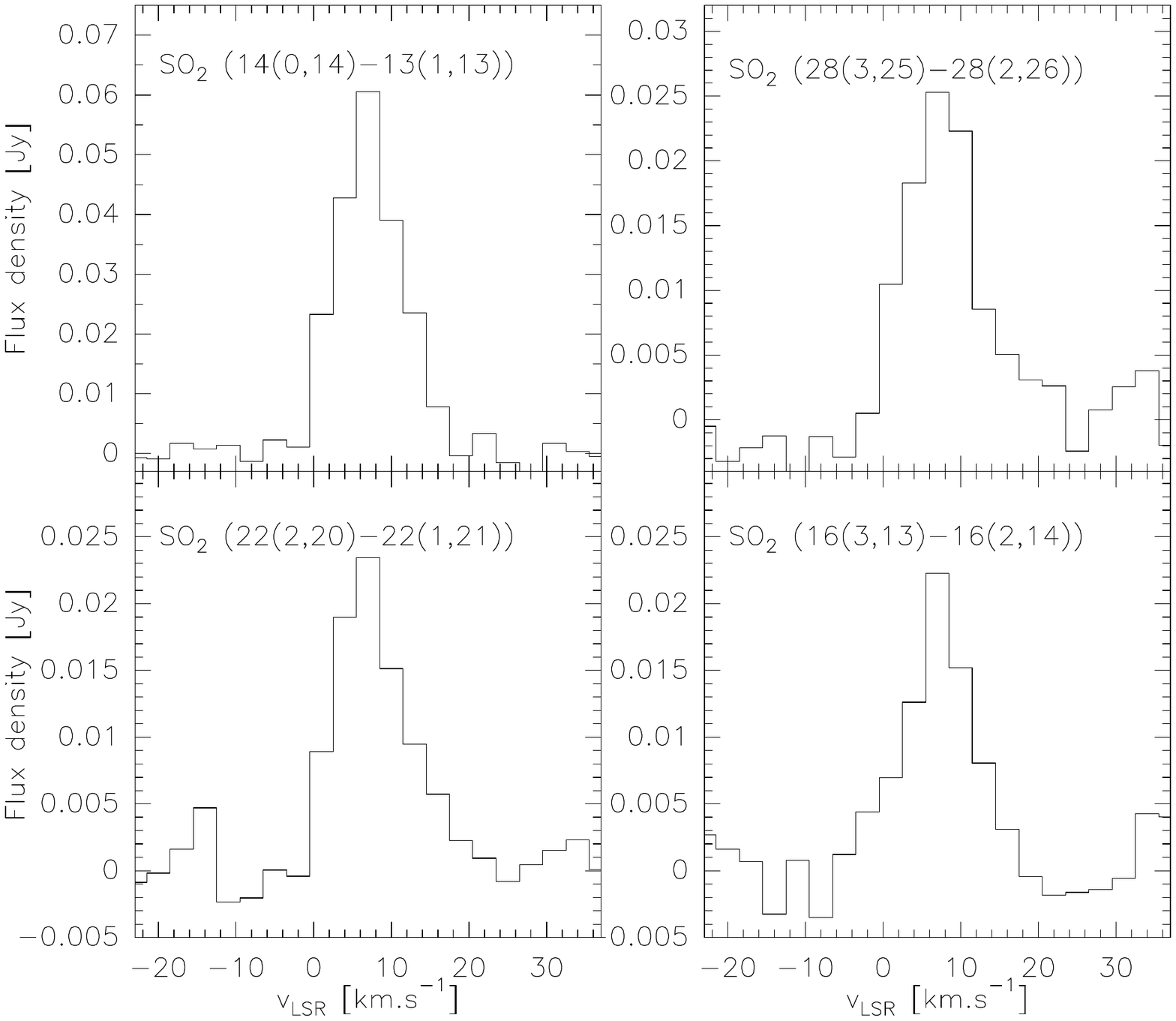}
    \caption{Profiles of all the 11 detected SO$_2$ lines. A and
      D-configuration data are merged, spectral resolution is 3\kms,
      and the emission is integrated over the central $2\arcsec
      \times 2\arcsec$ square aperture. The lines are ordered by 
      decreasing frequency from top  to bottom and left to right.}
    \label{so2profiles}
\end{figure}

In this section, we present the line profiles of all the 11 SO$_2$
lines detected with our setups (see Fig.~\ref{so2profiles}).

\section{XCLASS modeling of HCN and SO$_2$}
\label{xclassmodels}

The HCN(3-2) line and the 11 detected SO$_2$ emission lines were
modeled using the eXtended CASA Line Analysis Software Suite
(XCLASS\footnote{https://xclass.astro.uni-koeln.de},
\citealt{moeller2017}). XCLASS models and fits molecular lines by
solving the 1-D radiative transfer equation with the assumption of
local thermal equilibrium (LTE) and of an isothermal source. 1-D here
means that the radiative transfer equation is integrated along the
line of sight. Spectral lines are fitted with Gaussian profiles, and
optical depth effects and source size are considered in the
calculations. Molecular properties (e.g., Einstein coefficients,
partition functions, etc.) are taken from an embedded SQLite database
containing entries from the Cologne Database for Molecular
Spectroscopy (CDMS, \citealt{cdms2001,cdms2005}) and from the Jet
Propulsion Laboratory database (JPL, \cite{JPL}) using the Virtual
Atomic and Molecular Data Center (VAMDC, \citealt{endres2016}). The
fit parameter set for each line component consists of the source size
$\theta_{\rm source}$, the rotation temperature $T_{\rm rot}$, the
total column density $N_{\rm tot}$, the line width $\Delta \rm{v}$,
and the velocity offset v$_{\rm off}$ (given here in the LSR system).

The XCLASS package offers various algorithms to find the best-fit
parameters by minimizing the $\chi^2$ value, and here we utilized the
Levenberg-Marquardt (LM) method. To obtain maps of the physical
parameters, we use the \texttt{myXCLASSMapFIt} function to fit
HCN(3-2) and the 11 detected SO$_2$ emission lines (see
Fig.~\ref{so2profiles}) pixel by pixel.

\begin{figure}[h!]
\includegraphics[width=0.33\textwidth]{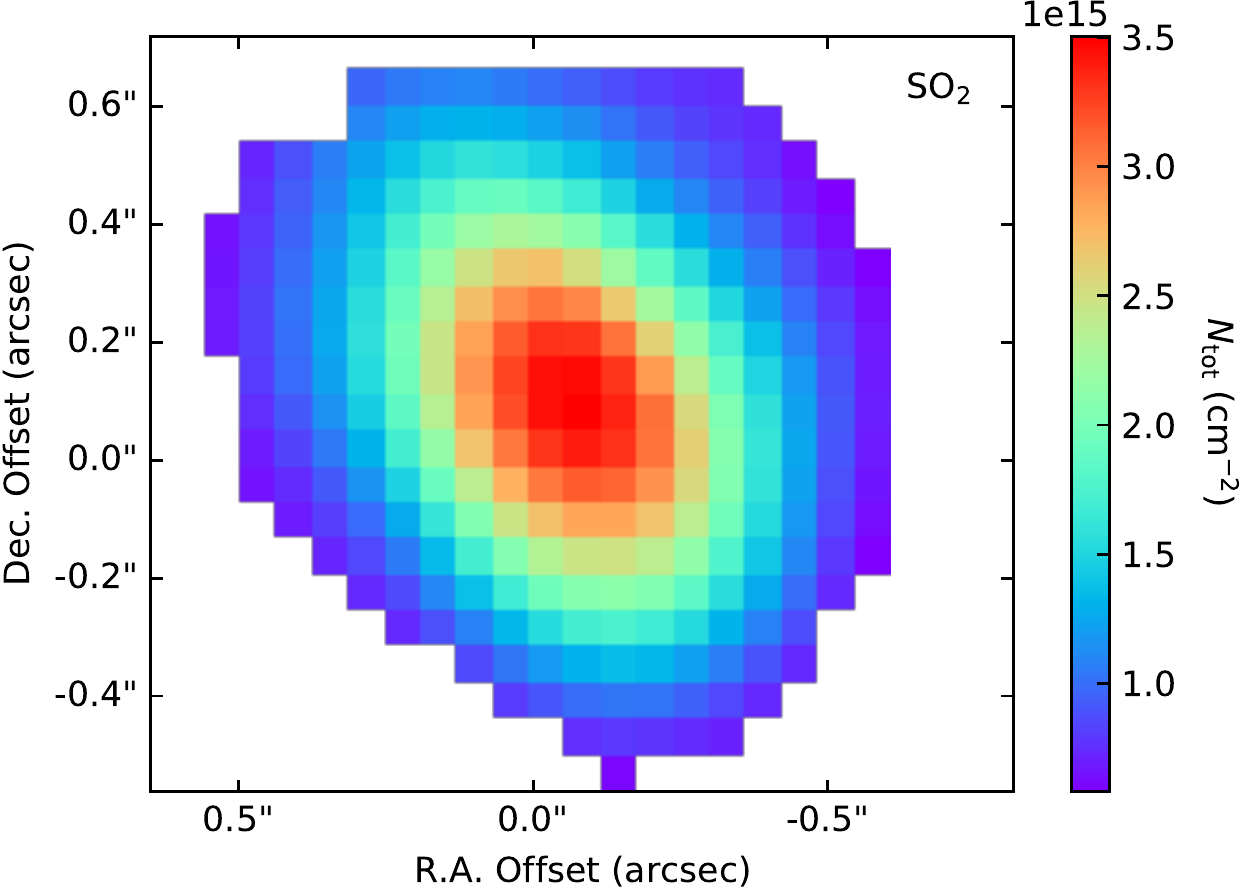}
\includegraphics[width=0.33\textwidth]{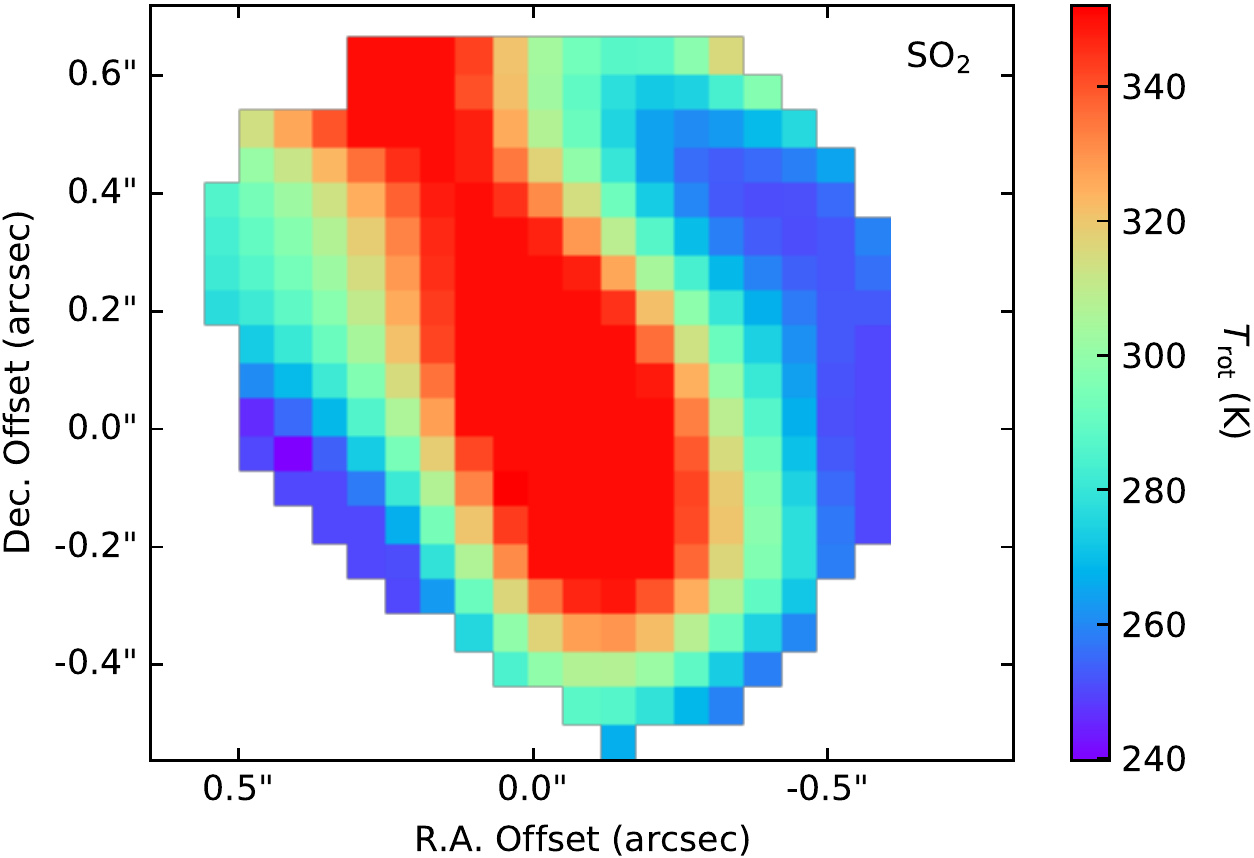}
\includegraphics[width=0.33\textwidth]{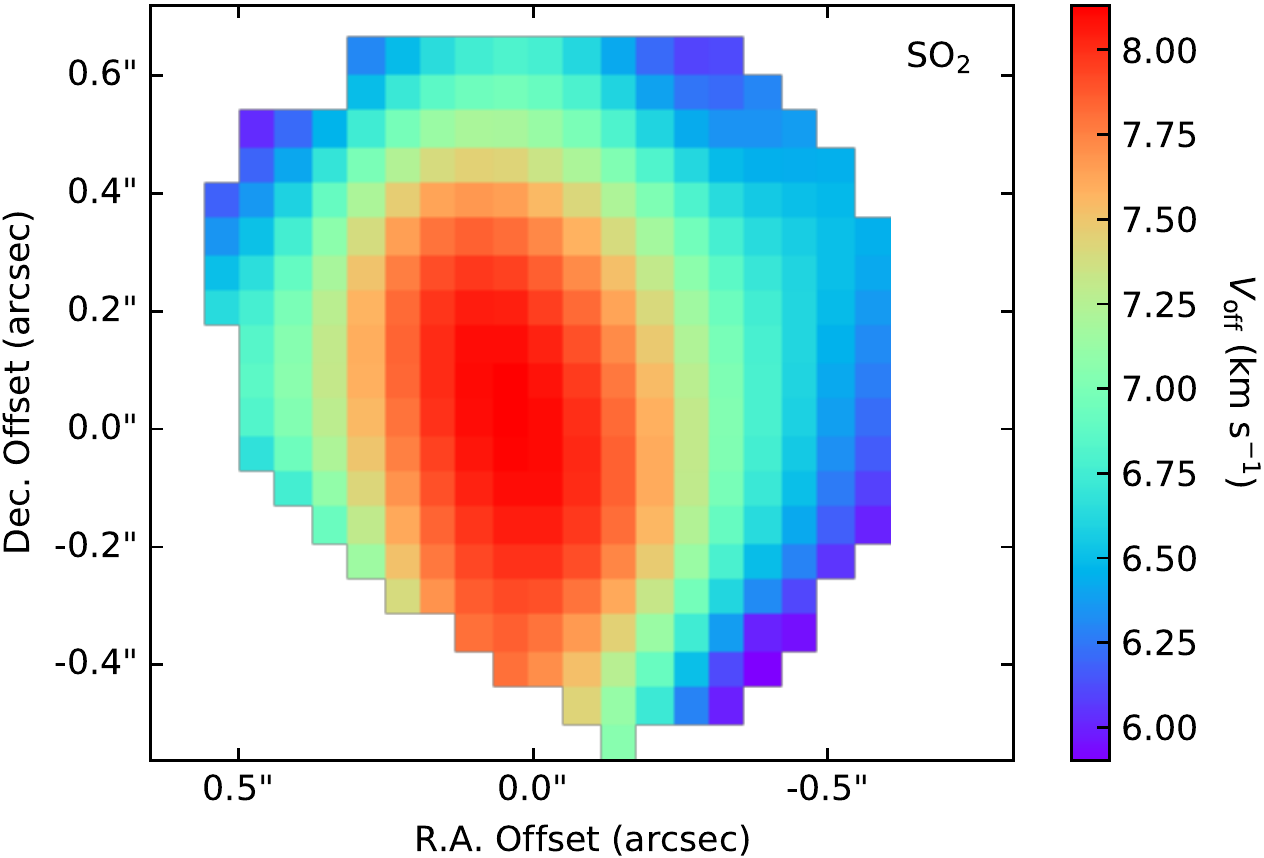}
\includegraphics[width=0.33\textwidth]{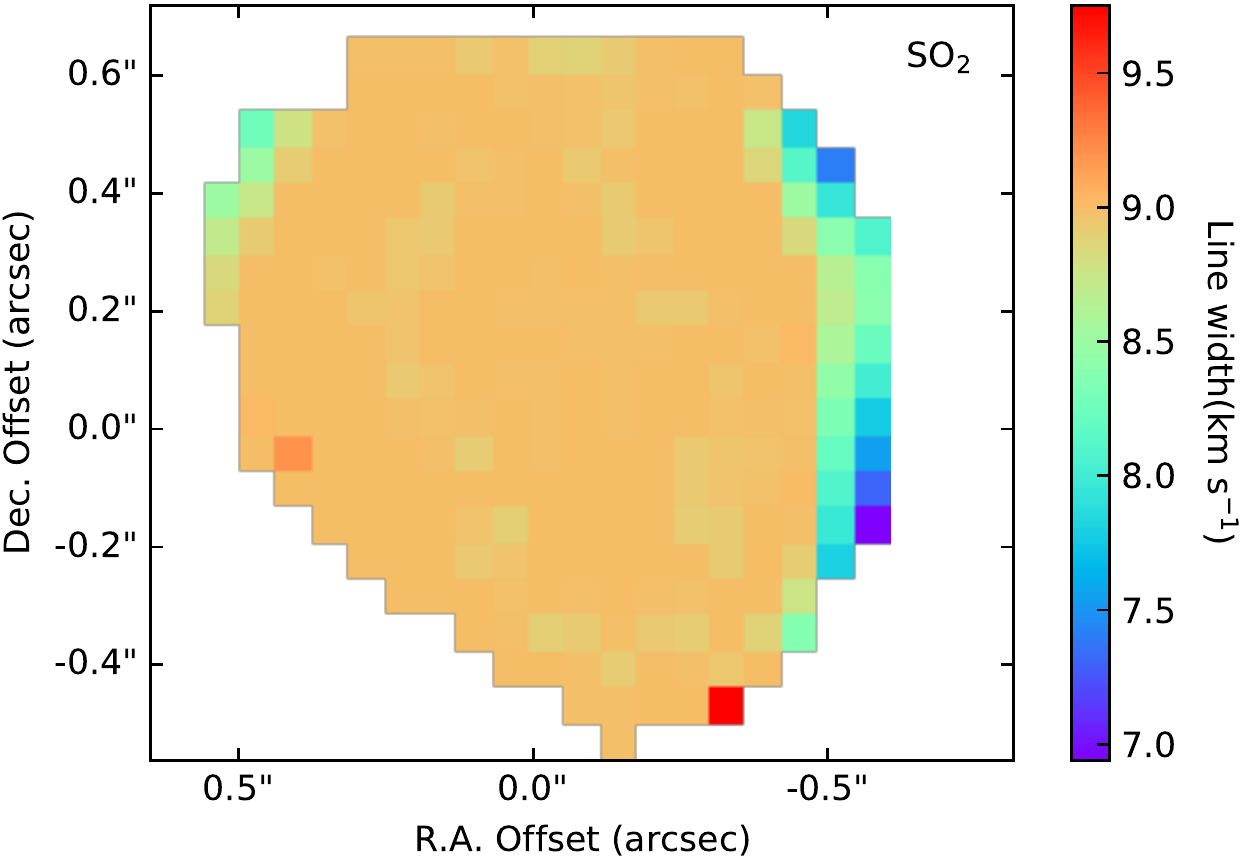}
\caption{Maps of total column density, rotation temperature, velocity
  offset (in the LSR system), and line width for SO$_2$ are
  derived with a threshold of 18$\sigma$ and fitting one Gaussian
  component to the line profiles.}
\label{fig:so2_xclass} 
\end{figure}

\begin{figure}[h!]
\includegraphics[width=0.33\textwidth]{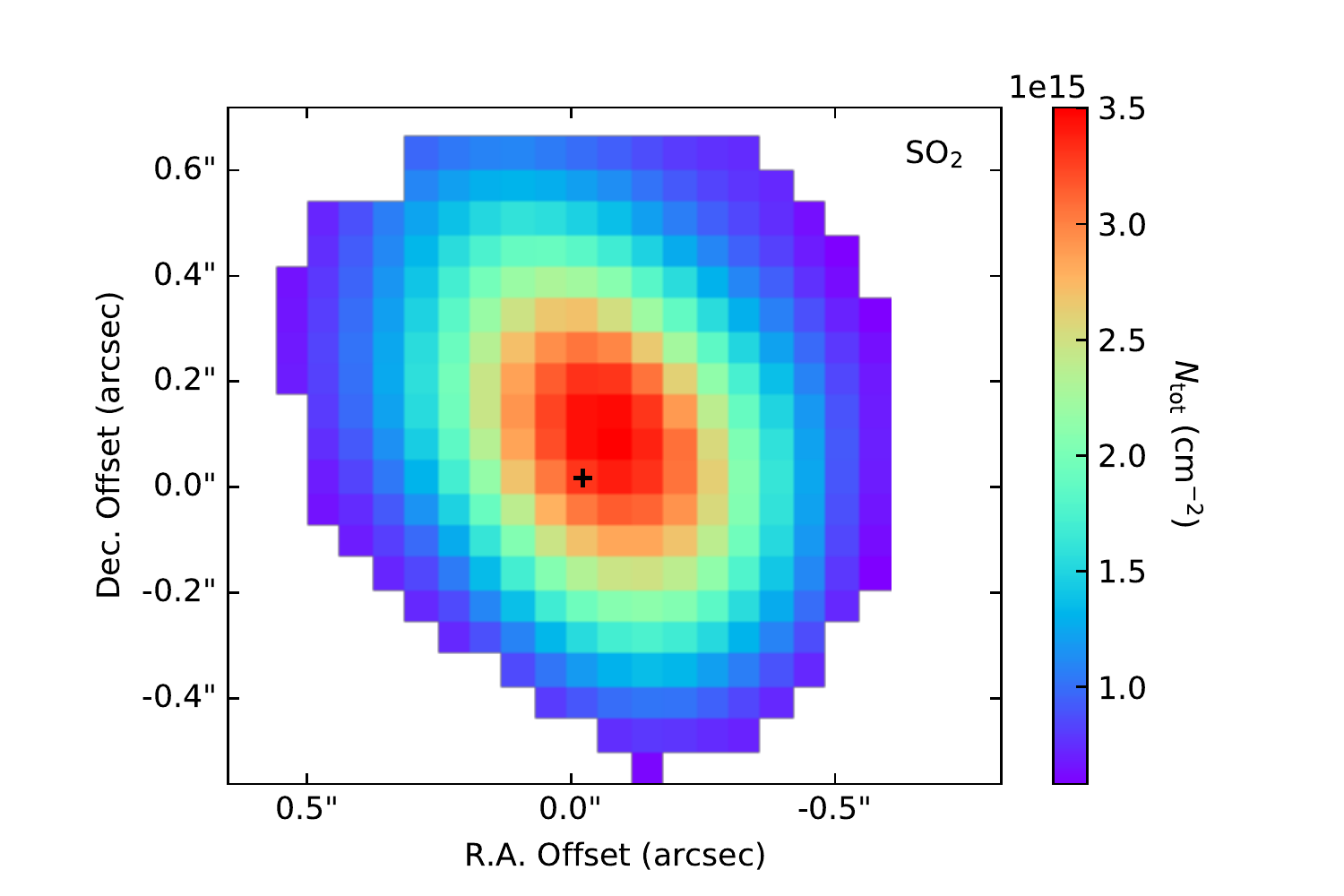}
\includegraphics[width=0.33\textwidth]{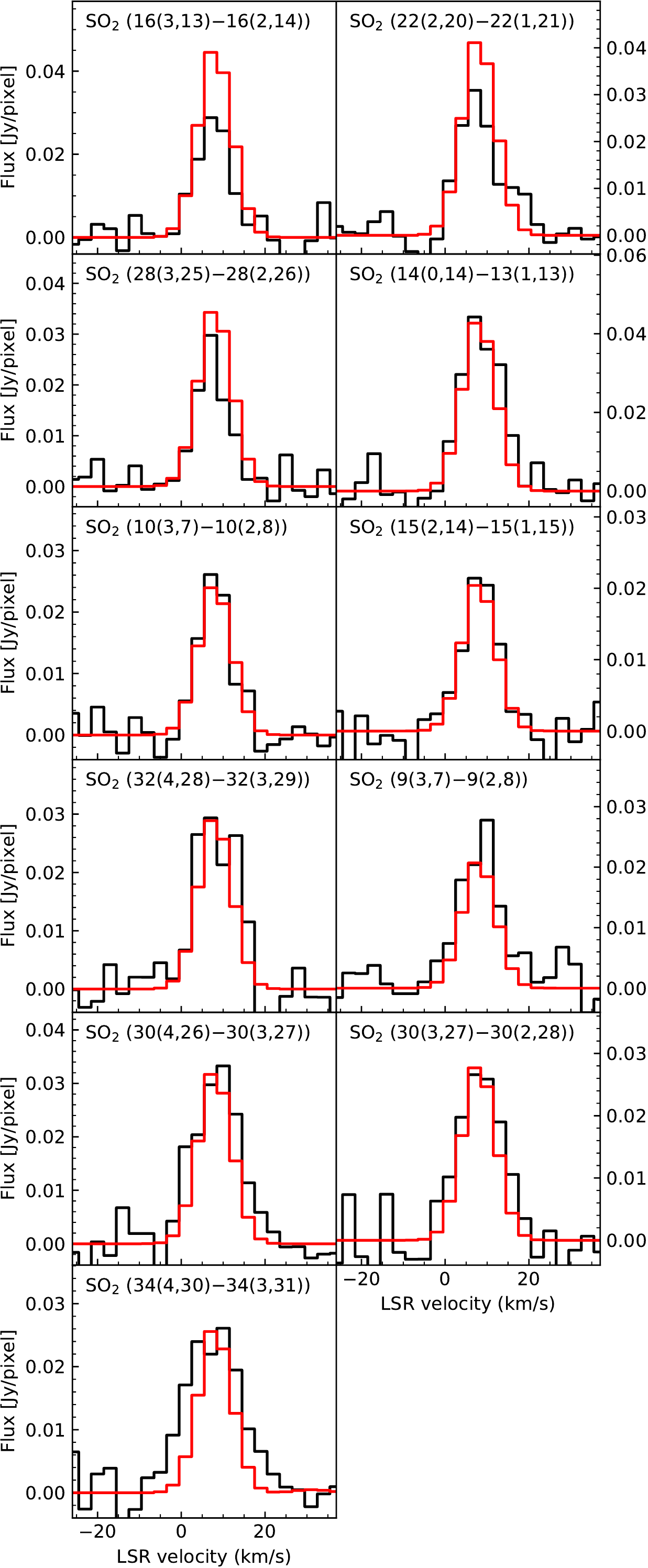}
\caption{Profiles of the 11 SO$_2$ emission lines (in black) extracted
  on the central pixel and XCLASS modeled lines (in red) at the position
  of the continuum at RA = 09:10:38.77 and Dec = $+$30:57:46.68 (see
  Sect.~\ref{contsec}) as marked on the upper diagram.}
\label{fig:so2lineft_xclass} 
\end{figure}

For SO$_2$, we modeled and fitted 11 lines simultaneously with a
threshold of 18$\sigma$ and a single Gaussian component. All fit
parameters are regarded as free parameters in the fitting process.
The XCLASS models for SO$_2$ result in a temperature of $T_{\rm rot}
\sim350$\,K (Fig.~\ref{fig:so2_xclass}), somewhat higher than the
result from our population diagram analysis, but within the error bars
(see Sect.~\ref{thermalsec}).  It also results in an average line
optical depth of 0.1, confirming that the assumption of the lines
being optically thin is justified when constructing the population
diagram.  Assuming LTE, the derived rotation temperature equals the
kinetic gas temperature $T_{\rm kin}$. For the SO$_2$ column density,
the XCLASS modeling results in a value of $N_{\rm SO_2}
\sim3.5\times10^{15}$\,cm$^{-2}$, almost identical to  the result from
the population diagram analysis presented in Sect.~\ref{thermalsec},
resulting in an abundance $X($SO$_2/$H$_2) = 1.5\times10^{-6}$.  The
respective results are shown in Figs.~\ref{fig:so2_xclass} and
\ref{fig:so2lineft_xclass}.

\begin{figure}[h!]
\includegraphics[width=0.33\textwidth]{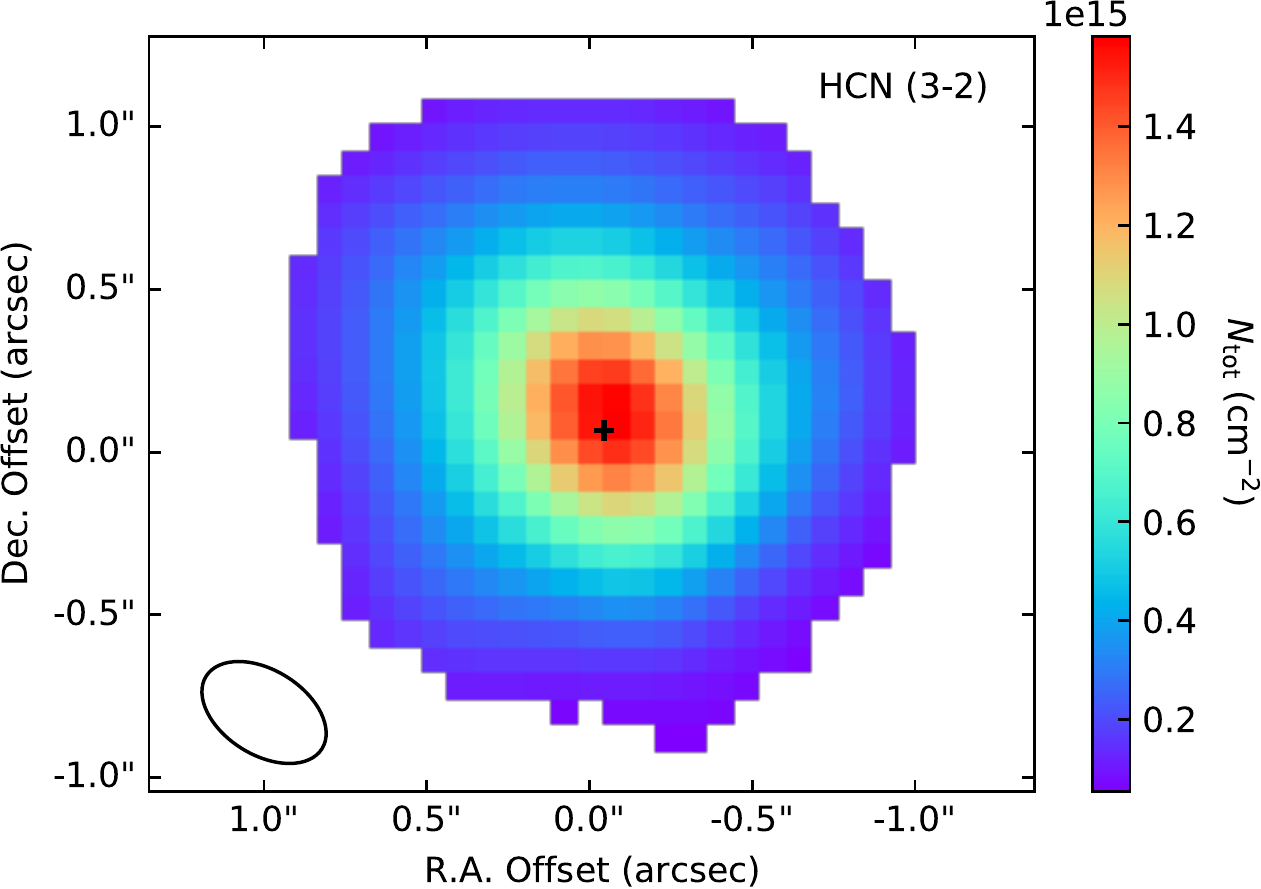}
\includegraphics[width=0.33\textwidth]{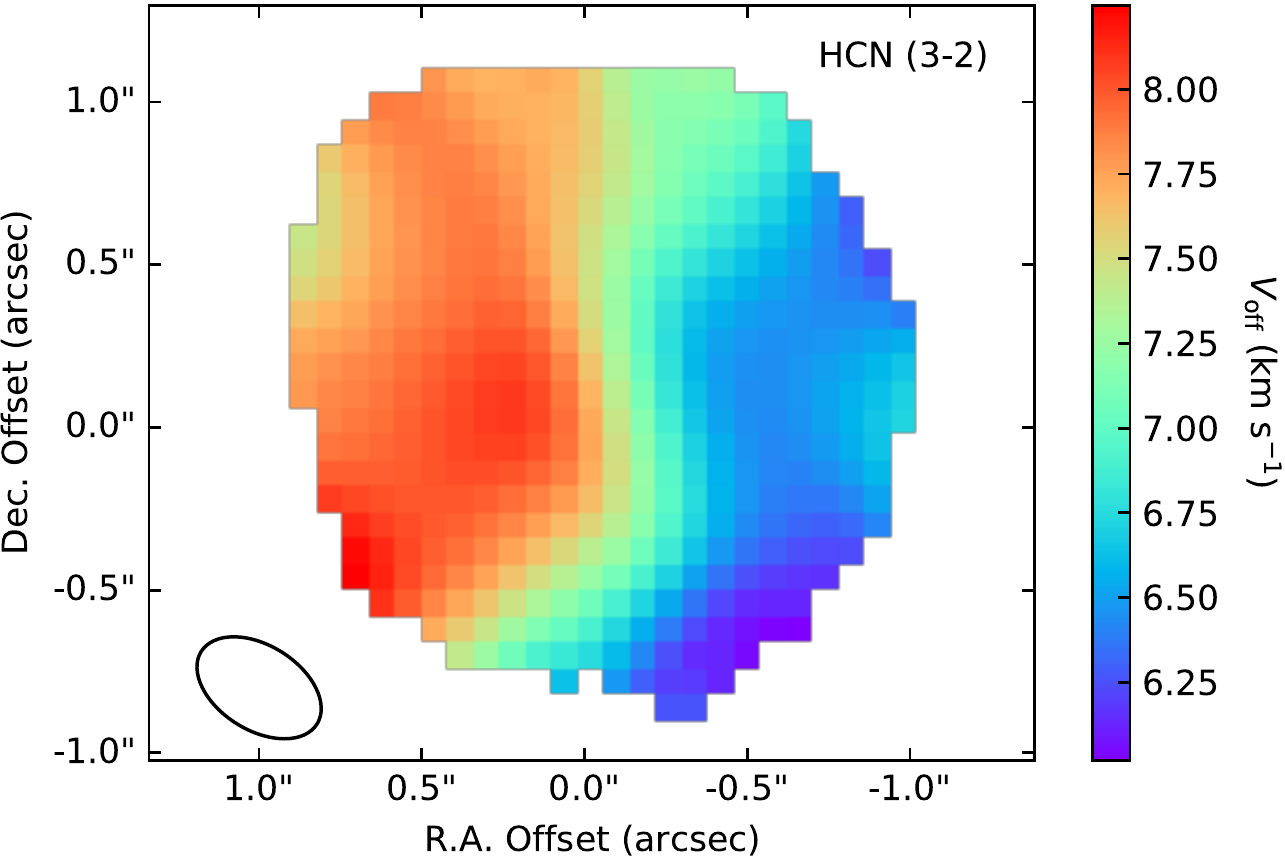}
\includegraphics[width=0.33\textwidth]{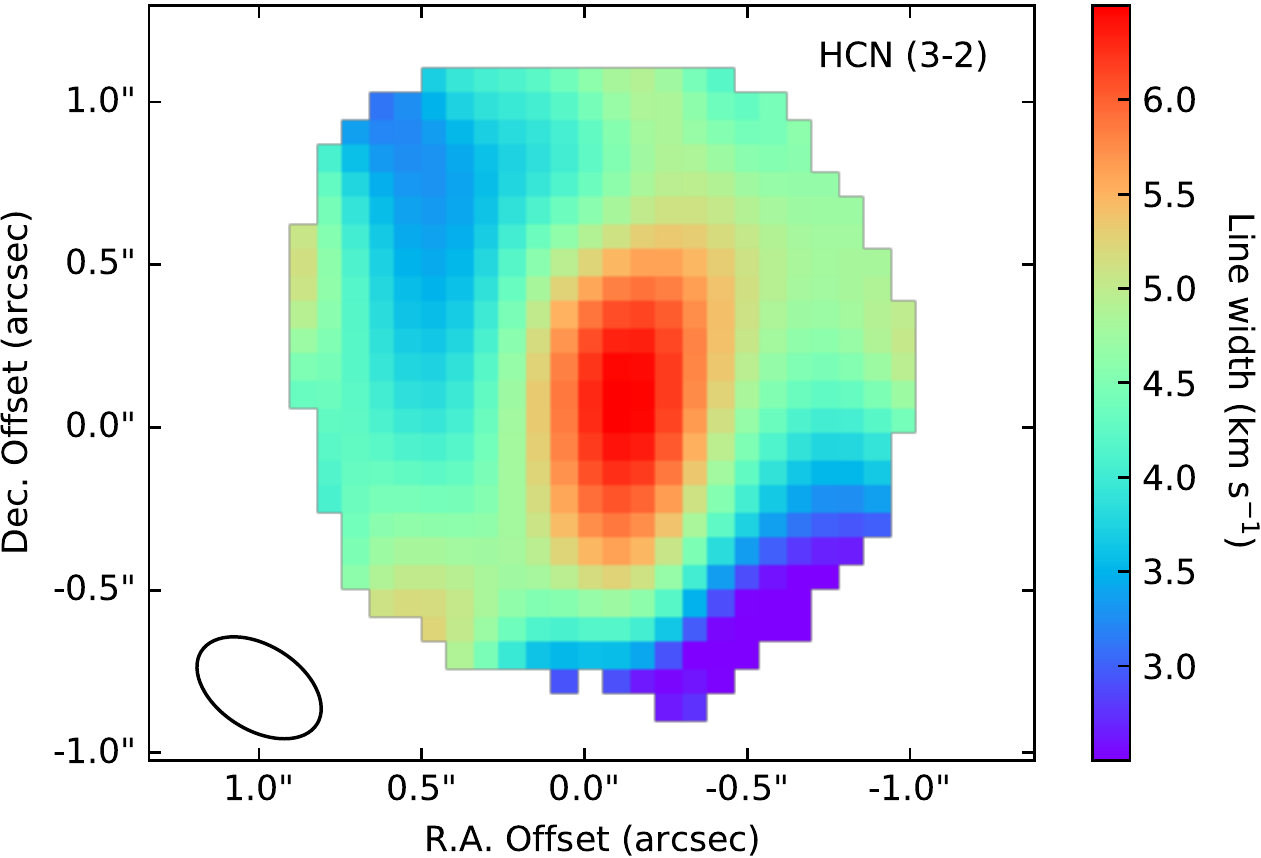}
 \caption{Maps of total column density, velocity offset (= v$_{\rm
          lsr}$), and line width of HCN(3-2) (see text for details). 
          The black ellipse in the lower left corner of each map 
          indicates the synthesized beam.}
\label{fig:hcn_xclass} 
\end{figure}

\begin{figure}[h!]
\includegraphics[width=0.33\textwidth]{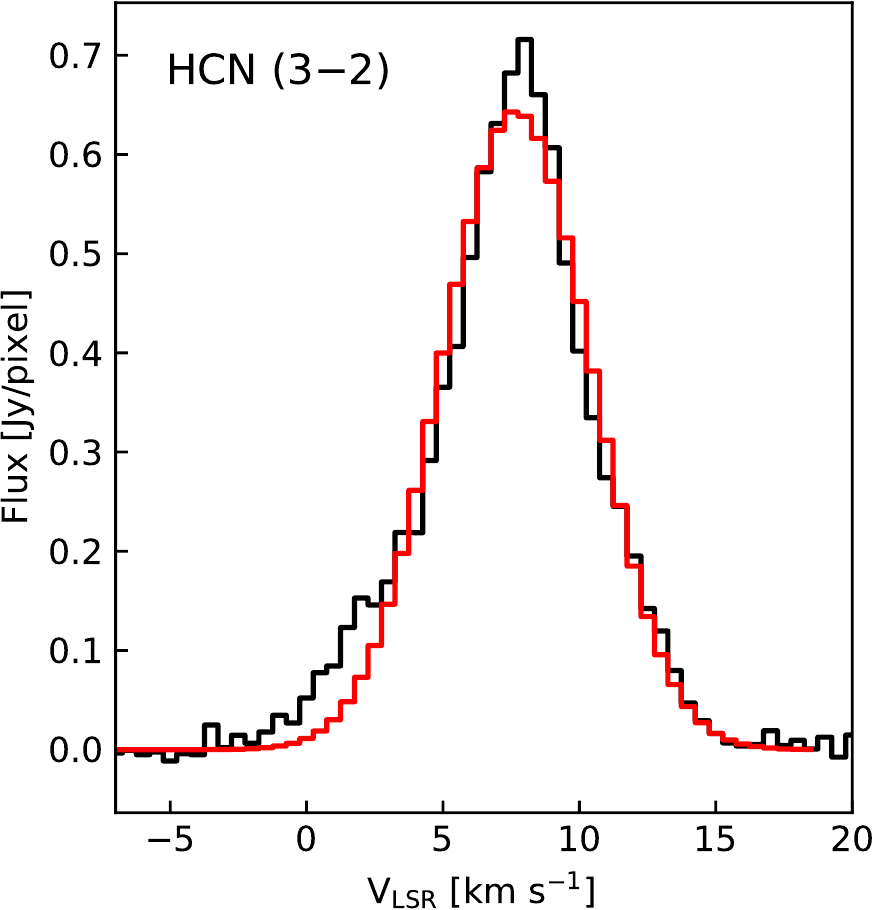}

\caption{Profile of the HCN(3-2) emission line (in black)  extracted
         on the central pixel and XCLASS modeled  line (in red) at 
         the position of the continuum at  RA = 09:10:38.77 and 
         Dec = $+$30:57:46.68 (see Sect.~\ref{contsec}) as marked on 
         the upper diagram of Fig.~\ref{fig:hcn_xclass}.}
\label{fig:hcnlineft_xclass} 
\end{figure}

For HCN(3-2), we applied the same threshold of 18$\sigma$ as for
SO$_2$ and fitted Gaussian components to the line profiles on each
pixel, taking into account the hyperfine structure of the line. The
velocity map and line widths displayed on Fig.~\ref{fig:hcn_xclass}
are due to the intrinsic (thermal and rotational) broadening only,
whereas the profile shown in Fig.~\ref{fig:hcnlineft_xclass}
represents the sum of the hyperfine components of HCN(3-2). As only
one HCN rotational line is available, the HCN rotational temperature
cannot be determined. We therefore fixed the rotation temperature at a
value of 350\,K, the same temperature as we find from the SO$_2$
modeling\, based on the similar emission region of SO$_2$ and HCN, see
Figs.~\ref{rotationfig},  \ref{HCNstructfig},
and~\ref{SO2structfig}. The HCN results are displayed in
Fig.~\ref{fig:hcn_xclass}. In particular, we derive an HCN column
density of $\sim 1.6 \times 10^{15}$\,cm$^{-2}$, which translates to
an HCN abundance of $X($HCN/H$_2) = 6.6\times10^{-7}$.

\end{appendix}

\end{document}